\documentclass[12pt,a4paper, twoside]{report}

\usepackage{import}
\usepackage{Preamble}

\begin{document}

\pagestyle{fancy}
\pagenumbering{Roman}

\begin{titlepage}
\begin{center}

\vspace*{1cm}

{\LARGE \bf Embedding formalism for \\[8pt] anti-de Sitter superspaces}

        \vspace{1.7cm}

       {\Large{\textbf{Nowar Eric Koning}}}\\
       \vspace{1mm}
         {\Large{BSc, MPhys}}\\
\large  
  \vspace{1cm}

  \vspace{3.5cm}

        \includegraphics[width=0.5\textwidth]{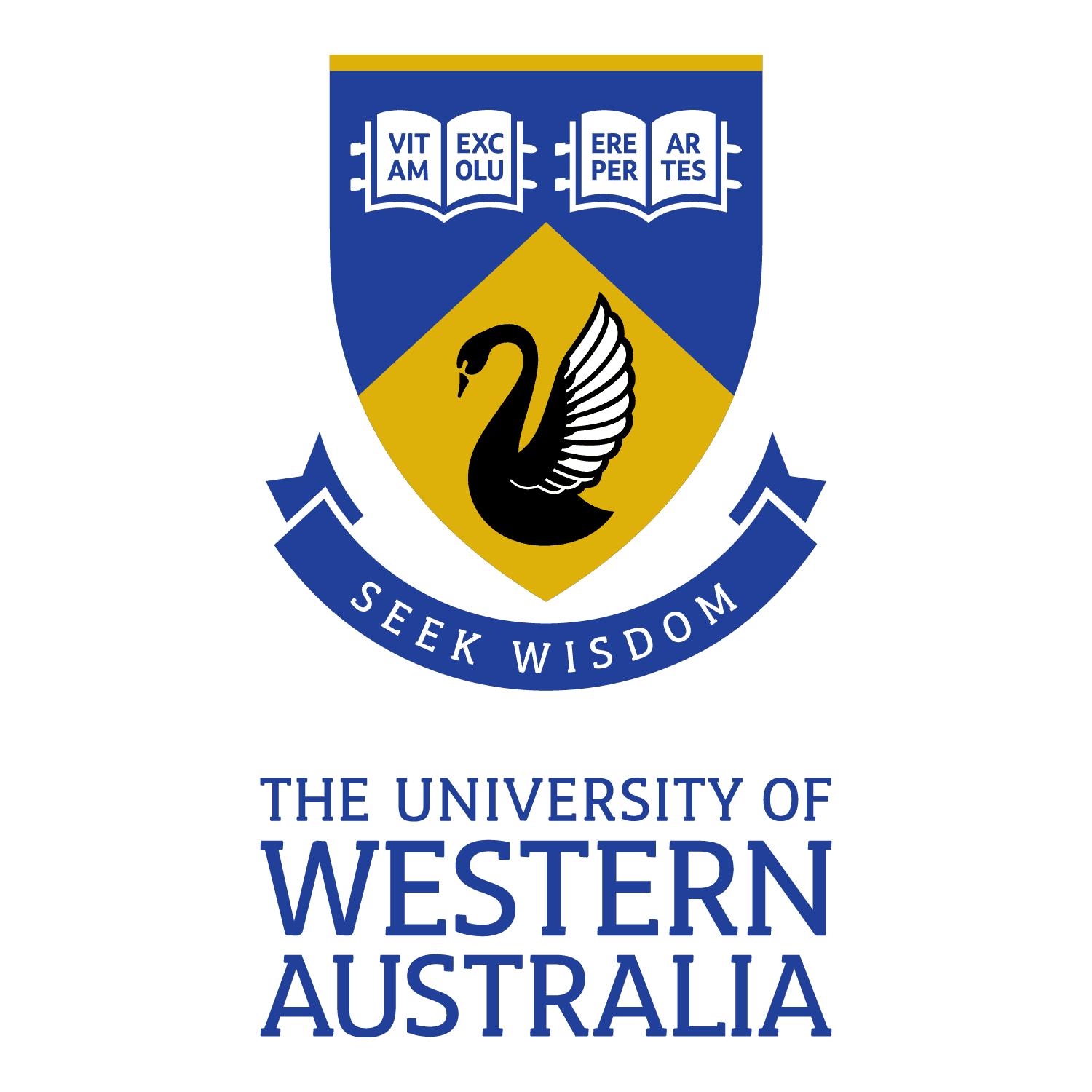}

            \vspace{1cm}

This thesis is presented for the degree of Doctor of Philosophy of\\
The University of Western Australia\\
School of Physics, Mathematics and Computing\\
Department of Physics\\ 
2026

\end{center}

\end{titlepage}
\chapter*{}
\thispagestyle{supervisors}
\addtocounter{page}{-1}
\chapter*{Abstract}%
This thesis is dedicated to the study of global realisations of the four and five-dimensional $\N$-extended anti-de Sitter superspaces, AdS$^{4|4\N}$ and AdS$^{5|8\N}$, respectively. 
Specifically, we develop embedding formalisms for both in terms of bi-supertwistors, which possess simple transformation properties under the AdS supergroup.
Our formalism has natural applications to models such as that for a superparticle, and we present several examples.

In the first part of this thesis, we review several key formalisms, including the existing bi-supertwistor realisations for the $\N$-extended compactified Minkowski and $\N$-extended AdS superspaces in four dimensions. 
Then we describe AdS$^{4|4\N}$ within the supergravity setting and derive two explicit conformally flat realisations for arbitrary $\N$, extending the previous results available within the literature. 
We then turn to develop the embedding formalism for AdS$^{4|4\N}$.
As an example application, we introduce new models for superparticles propagating in AdS$^{4|4\N}$. 
The specific feature of these new models is that they include two independent two-derivative terms.
We also show how such models are derived within the supergravity setting.

In the second part of this thesis, we propose an embedding formalism for the $\N$-extended AdS superspace in five dimensions. 
Specifically, we derive a bi-supertwistor realisation, extending the results available in the literature for the non-supersymmetric AdS$_5$ spacetime. 
Similarly to the four-dimensional case, our formalism offers a straightforward way to construct AdS super-invariants. 
As an example, we present new models for superparticles propagating in AdS$^{5|8\N}$, which contain two independent two-derivative terms and prove to be the five-dimensional analogues of those proposed for AdS$^{4|4\N}$. 
We show how such models originate within the supergravity setting in five dimensions, which is available for $\N=1\,.$

The final component of this thesis is devoted to extending the above considerations to AdS superpaces in other dimensions. 
Specifically, we generalise the deformation to the AdS$^{4|4\N}$ and AdS$^{5|8\N}$ supersymmetric intervals associated with the new superparticle models in each case to the three and two-dimensional AdS superspaces.
Finally, we show how these structures originate within the embedding formalism for AdS$_{(3|p\,,q)}$, which had been developed earlier in the literature.

\chapter*{Authorship Declaration}

This thesis is based on the published papers \cite{KKR, KKR2, KK, KRsp, KK3}. Their details are as follows:
\begin{enumerate}

\item N.~E.~Koning, S.~M.~Kuzenko and E.~S.~N.~Raptakis,  \\
{\it Embedding formalism for $\cN$-extended AdS superspace in four dimensions,}\\
JHEP {\bf 11}, (2023) 063
\href{https://arxiv.org/abs/2308.04135}{[arXiv:2308.04135 [hep-th]]}.
\begin{itemize}
\item {\bf Location in thesis:} Chapters \ref{bm} and \ref{ch3}.

\end{itemize}

\item N.~E.~Koning, S.~M.~Kuzenko and E.~S.~N.~Raptakis,  \\
{\it The anti-de Sitter supergeometry revisited,}\\
JHEP {\bf 02}, (2025) 175 
\href{https://arxiv.org/abs/2412.03172}{[arXiv:2412.03172 [hep-th]]}.
\begin{itemize}
\item {\bf Location in thesis:} Chapter \ref{ch3}.

\end{itemize}

\item N.~E.~Koning and S.~M.~Kuzenko,  \\
{\it Embedding formalism for AdS superspaces in five dimensions,}\\
JHEP {\bf 06}, (2025) 016 
\href{https://arxiv.org/abs/2406.10875}{[arXiv:2406.10875 [hep-th]]}.
\begin{itemize}
\item {\bf Location in thesis:} Chapter \ref{ch4}.

\end{itemize}

\item N.~E.~Koning and E.~S.~N.~Raptakis,  \\
{\it New superparticle models in AdS superspaces,}\\
JHEP {\bf 10}, (2025) 099
\href{https://arxiv.org/abs/2506.17897}{[arXiv:2506.17897 [hep-th]]}.
\begin{itemize}
\item {\bf Location in thesis:} Chapter \ref{ch5}.

\end{itemize}

\item N.~E.~Koning and S.~M.~Kuzenko,  \\
{\it Anti-de Sitter flag superspace,}\\
Phys. Rev. D {\bf 113}, (2026) 6, 065019 
\href{https://arxiv.org/abs/2512.14347}{[arXiv:2512.14347 [hep-th]]}.
\begin{itemize}
\item {\bf Location in thesis:} Chapter \ref{ch3}.

\end{itemize}

\end{enumerate}

Permission has been granted to use the above work.

\chapter*{Acknowledgements}

\nocite{KKR,KKR2,KK,KRsp,KK3}

This thesis has been completed on stolen land. 

I would not have made it to this point without the contributions and support of many of my family, friends, and peers. 
There are far too many to list in such a small setting. 

To my supervisor, Sergei. Thank you for the last six years. 
It seems like yesterday that I first walked into your office at twenty and struggled to answer your questions about unitary matrices and Jordan canonical form. 
Our many discussions on all ranges of topics have been inspiring and refreshing, and I consider myself extremely lucky to have been your student.

I would also like to thank my co-supervisor, A/Prof. Evgeny Buchbinder, who has provided me with excellent lectures and discussions over the course of my studies, as well as valuable feedback on parts of this thesis. 

This thesis has been completed over a period of time where the eyes of many people around the world have been opened. 
With that in mind, it would be disingenuous to not acknowledge the ridiculous privilege that has been afforded to me by the efforts of my parents. 
This is their achievement, much more so than it is mine. 

Aan mijn vader, Eric. 
Ik weet dat we het niet per se over alles eens zijn in de natuurkunde, maar ik heb dit geschreven met jouw aanpak in gedachten. 
Ik hoop dat het leuk is. 
The question is not `are you crazy?' It is `are you crazy enough?' 

To my mother, Hazar. That we are here now, and have achieved so much, is because of you. Seeing you flourish now, in your own way, is a great source of joy for me. 

To my older brother, Heidar. 
To have the two of us do PhDs simultaneously, and maintain such a close relationship throughout, is incredible. 
Here is to what may come in future. 

To my younger brother, Omar. 
You have been such a crazy reminder of what there is to life, outside of AdS, in so many ways that you may come to understand eventually. 
I hope that whatever it is that you do, you may experience wonder and joy in it. 

To the rest of my family. Hazem, Josee, Jos, Reyya, Sary, Nawar, Tete, and all the rest. 
We are all together! Tot straks!

To my friend and Great Collaborator, (Dr.) Emmanouil (Raptakis). 
I never would have reached the depths of AdS (and come back) that we did together without you. 
Thank you. 
Here I would like to also acknowledge the great influence that Kai Turner had on me at the beginning of my PhD. 
Your careful guidance helped me steer clear of many pitfalls in calculations and philosophy. 
Sometimes it took me years to realise what you meant, but your prophetic contributions always proved true. 

It is my pleasure to acknowledge the (present and former) members of the Field Theory and Quantum Gravity group at UWA, whom I have worked alongside over the course of this PhD. 
Dr. Darren Grasso, Dr. Michael Ponds, Dr. Benjamin Stone, Dr. Daniel Hutchings, Dr. Jessica Hutomo, Dr. Igor Samsonov, Jake, Josh, Arcadia, may we all proliferate the Kuzenko school of thought. 
In particular I would like to thank Michael and Ben, for vastly different influences over me. 
Michael, thank you for welcoming me into your tribe. 
Ben, the many hundreds (or thousands) of hours we have spent discussing the intricacies of various formalisms while on Hunt with Heidar will never be forgotten. See you in the Bayou. 

I have crossed paths (repeatedly) with many wonderful people in the last four years. 
Here I would like to highlight two of them who have had a great influence on me. 
Shon, I already miss the unique flair we shared in approach to life. Although we disagree on a great many things, arguing about all of these topics with you has been a lot of fun. 
I look forward to the next time we meet, wherever that may be. 
Kathi, thanks for sharing in our duck fascination. I am so grateful that we met in Anna's car that day. 

There are many more people who have contributed to this work in more ways than I can count. 
Kandy, Anna, Aston, Allana, Matt, Valeria, Chris, Sisyphus Sarah, Reuben, JP, Jade, Moody, Lavi, Celena, Ben Neil, the `Lantern's Reign' community. 
Thank you all.

I would like to thank my examiners, Professor Jeong-Hyuck Park and Professor Ulf Lindstr\"om, for invaluable comments and insight on my thesis.

It remains to acknowledge my two cats, Knuffel and Boef, although I'm not convinced that they know what AdS is. 

This research was supported by an Australian Government Research Training Program (RTP) Scholarship doi.org/10.82133/C42F-K220.


{\hypersetup{hidelinks}
\tableofcontents
}
\clearpage\pagenumbering{arabic}


\chapter{Introduction}

Our current understanding of the fundamental interactions is founded on symmetry principles.
These principles have guided the development of theoretical physics at both large scales in the universe, which are described by classical theories, in particular Einstein's general theory of relativity (GR); and small scales, which are encapsulated by the Standard Model of particle physics, a quantum field theory describing the electromagnetic, weak, and strong interactions. 
Despite their manifest differences in both phenomenology and mathematical structure, they share their roots in symmetry principles. 
Further, both can be formulated as gauge theories of their corresponding gauge groups.
The gauge group in the case of Einstein gravity is the Poincar\'e group,  see \cite{Blagojevic:2013xpa} for a review on the various gauge theory formulations of gravity. For the Standard Model\footnote{In what follows, we will refer to the Standard Model of particle physics simply as the Standard Model.} of particle physics it is $\sSU(3) \times \sSU(2) \times \sU(1)\,.$

General relativity has been remarkably successful at predicting large-scale effects, including the famous black hole solutions and gravitational wave propagation, see, e.g., \cite{W-grav,HE-grav}. 
The maximally symmetric solution to the vacuum field equations of GR without cosmological constant is the flat spacetime in $d$ dimensions, known as Minkowski space and denoted $\mathbb{M}^{d}$. 
Its symmetries are encoded in the Poincar\'e (also known as inhomogeneous Lorentz) algebra $\frak{iso}(d-1\,,1)$, which consists of translation generators $(P_{a})$ and Lorentz generators $(M_{ab})\,.$
Amongst themselves, they obey the relations 
\bsubeq \label{intro pc alg}
\begin{align}
	[M_{ab}\,, M_{cd}] &= 
	2\eta_{d[a} M_{b] c}
	- 2\eta_{c[a} M_{b] d}
	\,,
	\\
	[M_{ab},P_{c}] &= 2\eta_{c[b}P_{a]}\,,
	\\
	[P_{a},P_{b}] &= 0 \label{intro flat comm}
	\,,
\end{align}
\esubeq
with $a = 0\,,1\,,\ldots\,,d-1\,,$ and generate the transformations which preserve the flat spacetime interval
\begin{align} \label{intro mink met}
	\rd s^{2} = \eta_{ab}\rd x^{a} \rd x^{b}\,, \qquad \eta_{ab} = \text{diag}(-1\,, +1 \,, \ldots\,, +1)\,. 
\end{align}
The finite transformations preserving the interval \eqref{intro mink met} are elements of the Poincar\'e group, 
which proves to be the isometry group of Minkowski space. 

At the energy scales at which the Standard Model provides a valid description of particle physics, gravitational effects are neglibible.
This means that the theory is most conveniently formulated in Minkowski space.
In accordance with Einstein's relativity principle, the Poincar\'e group is a symmetry group of every closed relativistic system.
Indeed, the fields at play in the Standard Model furnish representations of the Poincar\'e algebra. 
Elementary particles are identified with the unitary irreducible representations, which were classified by Wigner \cite{Wigner:1939cj},
and are labelled by the eigenvalues of the Casimir operators $C_1$ and $C_2$,
\begin{align} \label{intro casimir}
	C_1 = P^{a}P_{a} = -m^{2}\id \,, \qquad C_2 = \mathbb{W}^{a}\mathbb{W}_{a} = s(s+1)m^{2}\id\,, 
\end{align}
where $\mathbb{W}^{a}$ is the Pauli-Lubanski vector, $m \geq 0$ is the mass\footnote{For $m=0$ the representations are labelled by helicity $\l$ which takes (half-)integer values.}, and $s$ is the spin taking (half-)integer values.
Despite the success of the Standard Model at describing the strong, weak, and electromagnetic interactions, it is not currently known how to extend it to describe gravitational effects in the regime where quantum gravity effects must be accounted for, see, e.g., \cite{W-1,W-2}. 
Further, the Standard Model faces conceptual issues such as the hierarchy problem, see \cite{Craig:2022eqo} for a review. 
In conjunction with the desire for a quantum theory of gravity, this leads to the widely-accepted conclusion that the Standard Model is incomplete.

\subsubsection{Conformal and anti-de Sitter symmetry}

It was realised in 1909 by Cunningham \cite{Cunningham:1910pxu} and Bateman \cite{Bateman:1909pyp,Bateman:1910mvi} that the maximal symmetry group of the free Maxwell equations was actually larger than the Poincar\'e group, and includes scale (and special conformal) transformations. 
The corresponding symmetry group is known as the conformal group, and it turns out that many massless theories enjoy conformal symmetry. 
Conformal symmetry has become a cornerstone of many areas of modern theoretical physics, including the AdS/CFT correspondence\footnote{Here, `AdS' refers to anti-de Sitter space, and `CFT' refers to conformal field theory.} \cite{Maldacena:1997re,Gubser:1998bc,Witten:1998qj} and string theory (see, e.g., \cite{Aharony:1999ti} and \cite{Blumenhagen:2013fgp,Green:2012oqa,Green:2012pqa,Polchinski:1998rq,Polchinski:1998rr} for reviews of each). 
Further, it was realised that certain quantum field theories are conformally invariant at sufficiently high energies, see, e.g., \cite{Schwartz:2014sze}. 

Conformal symmetry also plays a special role in general relativity.
Spacetimes in which the interval can be locally brought to the form 
\begin{align} \label{intro conf int}
	\rd s^{2} = \re^{2\s(x)}\eta_{ab}\rd x^{a} \rd x^{b}\,,
\end{align}
for some scale factor $\s(x)$ are known as conformally flat spacetimes.
Important examples include the $d$-dimensional (anti-)de Sitter space, denoted (A)dS$_d$.
AdS$_d$ is the maximally symmetric solution to the vacuum field equations of GR with negative cosmological constant, and dS$_d$ is the maximally symmetric solution with positive cosmological constant. 
Experimental observations suggest that our universe may correspond to one with positive cosmological constant, or that dark energy may actually exhibit dynamical behaviour, see \cite{DESI:2025wyn,Capozziello:2025qmh} for recent developments. 
Nevertheless, AdS plays an important role in modern approaches to quantum gravity such as the AdS/CFT correspondence, see, e.g., \cite{Aharony:1999ti}.
In addition, there are many other features of AdS that make it an interesting and rich field of study. 
For example, the AdS$_4$ algebra, $\frak{so}(3\,,2)$, has unitary positive energy representations which were classified in \cite{Fcp1,Fcp2,Fcp3,Fcp4}, see \cite{deWit:1999ui} for a review.\footnote{The AdS algebra in $d+1$ dimensions is $\frak{so}(d\,,2)\,,$ which is isomorphic to the conformal algebra in $d$ dimensions. This fact forms the geometric underpinning of the AdS/CFT correspondence.} 
This makes AdS$_4$ an ideal setting for developing quantum field theoretic techniques in a background of constant curvature, see, e.g., \cite{AdSqft1,AdSqft2,Birrell:1982ix,Bertan:2018afl}.

\subsubsection{Supersymmetry}

There are few mathematical developments which can rival both the beauty and impact of supersymmetry on theoretical and mathematical physics. 
It was introduced independently in the early 1970s in the East by Golfand and Likhtman \cite{Golfand:1971iw}, and Volkov and Akulov \cite{Volkov:1972jx,Volkov:1973ix}, and in the West by Wess and Zumino \cite{Wess:1974tw,Wess:1973kz}.\footnote{In 1971 Pierre Ramond introduced the first fermionic string model in \cite{ramondstring}. This work bridged the gap between models containing only bosons and those containing fermions, and laid the foundations for future work on worldsheet supersymmetry.} 
Supersymmetry has radically changed the landscape of quantum field theory.
In particular, eq. \eqref{intro casimir} shows us that elementary particles in the Standard Model are classified by mass and spin. 
Those particles with integer spin are known as bosons, and those with half-integer spin are known as fermions. 
Prior to the introduction of supersymmetry, bosons and fermions were distinct classes and did not mix under symmetry transformations. 
Supersymmetry offers a striking new option: symmetries between bosons and fermions.

Extensions to the Standard Model are severely restricted by the Coleman-Mandula theorem \cite{Coleman:1967ad}, see, e.g., \cite{W-3} for a review. 
This no-go theorem states that, in the presence of massive particles, the only Lie algebra of symmetries of the S-matrix that is consistent with quantum field theory is given by the direct sum of the Poincar\'e algebra and a compact Lie algebra, corresponding to internal symmetries.
For a massless theory, the Poincar\'e algebra is replaced with the conformal algebra. 
As supersymmetry is based on the use of superalgebras, it offers an elegant bypass to the Coleman-Mandula theorem. 
Further, it was shown by Haag, Lopuszanski, and Sohnius \cite{Haag:1974qh} (see \cite{WB} for a review), that the supersymmetric extension of the Poincar\'e algebra (or conformal algebra, in the massless case) is the only consistent extension of the symmetries of the S-matrix. 

In four dimensions, the supersymmetric extension of the Poincar\'e algebra is obtained by appending to the Lorentz and translation generators \eqref{intro pc alg} $\N$ mutually conjugate supersymmetry generators $Q_{\a}^{i}$ and $\bar{Q}_{\ad i}$, where $\a = 1\,,2$ and $\ad = \dot{1}\,, \dot{2}$ are Weyl spinor indices and $i = 1\,, \ldots \,, \N$. 
The supersymmetry generators obey the anticommutation relation 
\begin{align} \label{intro susy com}
	\{Q_{\a}^{i}\,, \bar{Q}_{\ad j}\} = 2\d^{i}_{j}P_{a}(\s^{a})_{\a\ad}\,,
\end{align}
where $\s^{a} = (\id_{2}\,, \vec{\s})$ and $\vec{\s}$ are the Pauli matrices. 
The result is known as the $\N$-extended super Poincar\'e algebra. 
It has a non-trivial group of outer automorphisms, which is known as the $R$-symmetry group.
The specific feature of the $R$-symmetry group is that its generators act on the supersymmetry generators $Q_{\a}^{i}$ and $\bar{Q}_{\ad i}$ and commute with the Lorentz and translation generators. 
Many other supersymmetry algebras also have non-trivial $R$-symmetry groups, but the structure of the $R$-symmetry group depends on both the spacetime dimension and geometry.\footnote{For a catalogue of $R$-symmetry groups for various superalgebras, we refer the reader to, e.g., \cite{deWit:2002vz}.}

The $\N$-extended superconformal algebra in four dimensions, introduced in \cite{Haag:1974qh},
is an extension of the conformal algbera and of the $\N$-extended super Poincar\'e algebra with vanishing central charges.
From either perspective, it can be introduced by including additional bosonic and fermionic generators. 
First, we note that the conformal algebra consists of the Lorentz and translation generators \eqref{intro pc alg} as well as generators of dilatations and special conformal transformations.
One can then introduce the supersymmetry generators $Q^{i}_{\a}$ and $\bar{Q}_{\ad i}$ as well as fermionic counterparts to the special conformal generators, known as the $S$-supersymmetry generators. 
It is important to point out that the anticommutator of $Q$- and $S$-supersymmetry generators involves terms containing the generators of the $R$-symmetry group.
These must then be included for the superconformal algebra to close. 

\subsubsection{Supergravity}

One of the many remarkable features of supersymmetry is that, in its gauged version, it implies gravity (see, e.g., the books \cite{BK,WB,Gates:1983nr,Fsg}, or \cite{D-history} for a historical account). 
The corresponding theory was first conceived for $\N=1$ in 1976 by Freedman, van Nieuwenhuizen, and Ferrara \cite{FvNF} and Deser and Zumino \cite{DZ}, and is known as supergravity. 
Supergravity was initially constructed as a theory featuring an on-shell\footnote{Here, `on-shell' means the supersymmetry transformations only form a closed algebra after imposing the equations of motion. In contrast, `off-shell' means that the algebra of such transformations closes without needing to impose the equations of motion.} supersymmetry between the gravitational sector, described by the vierbein $e_{m}{}^{a}$, and a Rarita-Schwinger gauge field $\psi_m^{\ah}$, which consists of the spinor $\psi_m^{\a}$ and its conjugate $\bar{\psi}_{m}^{\ad}$, and is associated with a spin-3/2 particle known as the gravitino. 
It is a very interesting topic for many reasons, a few of which we will illustrate below. 
In four dimensions, supergravity exists in various forms, e.g., the $\N=1$ supergravity theories; extended supergravities ($1<\N \leq 8$) for which the $\N=2$ theories were first constructed by Ferrara and van Nieuwenhuizen \cite{Ferrara:1976fu}; conformal supergravity \cite{Kaku:1978ea,CSG2}, see also \cite{Fradkin:1985am}; and supergravity theories with cosmological term \cite{Freedman:1976aw,Townsend:1977qa}. 
The $\N=2$ supergravity theory \cite{Ferrara:1976fu} achieved a long-standing goal of Einstein in the unification of electromagnetism and gravity.
Further, supergravity theories arise as low-energy limits of superstring theory, see \cite{Green:2012oqa,Green:2012pqa,Polchinski:1998rq,Polchinski:1998rr,Castellani:1991et,Castellani:1991eu,Castellani:1991ev}.

\subsubsection{Superspace}

With the growing interest in supersymmetric field theory (in particular, supergravity-matter coupled theories, see, e.g., \cite{VanProeyen:2025ylh} for a recent account), 
the necessity of an off-shell formulation for supersymmetry became clear. 
It was quickly realised that such an off-shell formulation requires the introduction of auxiliary fields, with the number of auxiliary fields growing with $\N$. 
Another approach was pioneered by Volkov and Akulov \cite{Akulov:1974xz}, and Salam and Strathdee \cite{Salam:1974yz}, which made use of a new concept known as superspace. 
The $\N$-extended Minkowski superspace in four dimensions, denoted $\mathbb{M}^{4|4\N}$, consists of the usual bosonic spacetime coordinates $x^{a}$ as well as $\N$ mutually conjugate fermionic coordinates $\q_{i}^{\a}$ and $\bar{\q}^{\ad i}$. 
Manifestly supersymmetric field theory can then be constructed in terms of superfields, which are functions of the superspace coordinates. 
Superspace techniques are utilised extensively in this thesis, and we refer the reader to the books \cite{BK,Gates:1983nr} for pedagogical introductions. For superspace approaches to $\N=1$ and $\N=2$ supergravity in four dimensions, we refer the reader to the review articles \cite{Kuzenko:2022skv,Kuzenko:2022ajd}.\footnote{The relationship between the component and superspace formulations of supergravity were first detailed in \cite{lindcomponents}. Further, fundamental contributions to the superspace formulation of supergravity were provided in \cite{Siegelsugra}.}

When the $R$-symmetry group is included, the most natural setting turns out to be a superspace further appended by the bosonic coordinates of the $R$-symmetry group. 
For eight supercharges, corresponding to $\N=2$ in four dimensions and $\N=1$ in five dimensions, one commonly has $\sSU(2)$ as the R-symmetry group, in which case the harmonic \cite{Galperin:1984av} and projective \cite{Karlhede:1984vr,Gates:1984nk,LR1,Lindstrom:1989ne} superspaces play an important role.\footnote{See also \cite{Rosly:1983ya}.} 
These superspaces are of the form 
\begin{align}
\mathbb{M}^{d|8} \times S^{2} \simeq \mathbb{M}^{d|8} \times \mathbb{C}P^{1}\,,
\end{align}
where $d$ is the spacetime dimension.
For a thorough introduction to harmonic superspace techniques, see the book \cite{Galperin:2001seg}, and for a review of projective superspace techniques, see \cite{Lindstrom:2009afn,K-lec}.

\subsubsection{Coset (super)spaces}
 
In this thesis we are primarily concerned with the study of a special class of superspaces, known as anti-de Sitter superspaces. 
These are supersymmetric extensions of anti-de Sitter space and arise in a number of important contexts, which we will elaborate on below. 
In order to develop the notion of an AdS superspace, however, it is necessary to begin from the group-theoretic origins of superspace in general.
For this purpose, we make use of a formalism known as the coset construction. 
 
The coset construction, or the method of non-linear realisations, was initially developed in the physics literature by Coleman, Wess, and Zumino \cite{Coleman:1969sm,Callan:1969sn} to describe the low-energy dynamics in systems with spontaneously broken symmetry.\footnote{Early developments of the method of non-linear realisations were also provided in \cite{Isham:1969ci,Salam:1969rq}. The extension to include spacetime symmetries was provided by Volkov \cite{Volkov:1973vd} (see also Ogievetsky \cite{ogiev-non}). The non-linear realisation of conformal symmetry was given in \cite{Salam:1969bwb,Isham:1970gz,Isham:1971dv}.} 
Its applications also relate to the construction of gauge theories, including gravity.
In addition, it can be a useful tool to develop a purely group-theoretic realisation of a given homogeneous space, see, e.g., \cite{Kleppe}. Indeed, let us consider $\mathbb{M}^{d}$, which is a homogeneous space of $\sISO_{0}(d-1\,,1)$, the connected component of the Poincar\'e group in $d$ dimensions. 
Then, according to the coset construction, $\mathbb{M}^{d}$ can be realised as the coset space\footnote{The $d$-dimensional Minkowski space can also be realised as $\mathbb{M}^{d} = \sIO(d-1\,,1)/ \sO(d-1\,,1)$, but here we prefer to use the connected component of the Poincar\'e and Lorentz groups.} 
\begin{align} \label{intro Mink coset}
\mathbb{M}^{d} = \frac{\sISO_{0}(d-1\,,1)}{\sSO_{0}(d-1\,,1)}\,,
\end{align} 
where $\sSO_{0}(d-1\,,1)$ denotes the connected component of the Lorentz group in $d$ dimensions. 
There is a one-to-one correspondence between points in Minkowski space and the cosets in \eqref{intro Mink coset}, which are determined by the translation component of the group elements $g \in \sISO_{0}(d-1\,,1)\,.$
Such translations are parametrised by $d$-vectors $x^{a}$ which are interpreted as coordinates on Minkowski space.

The method outlined above readily generalises to other homogeneous (super)spaces. 
For example, AdS$_{d+1}$ and dS$_{d+1}$ can be defined as 
\begin{align}
\text{AdS}_{d+1} = \frac{\sSO_{0}(d\,,2)}{\sSO_{0}(d\,,1)}\,, \qquad \qquad \text{dS}_{d+1} = \frac{\sSO_{0}(d+1\,,1)}{\sSO_{0}(d\,,1)}\,. \label{ads ds coset}
\end{align}
Superspaces can be defined in this way by extending the spacetime symmetry group in the coset realisation to a supersymmetry group. 
Let us denote the $\N=1$ super Poincar\'e group in four dimensions as $\cS\P(4|1)$.
Then, $\mathbb{M}^{4|4}$ can be defined as\footnote{Strictly speaking, the stabiliser of $\mathbb{M}^{4|4}$ should be taken as $\sSpin(3\,,1)\simeq\sSL(2\,,\mathbb{C})$.} 
\begin{align} \label{intro mss}
	\mathbb{M}^{4|4} = \frac{\cS\P(4|1)}{\sSO_{0}(3\,,1)}\,.
\end{align}
This group-theoretic approach is used extensively in this thesis. 

\subsubsection{Anti-de Sitter supersymmetry}

We are now in a position to discuss anti-de Sitter superspace. 
The simplest AdS superspace was introduced in the early years of supersymmetry by Keck \cite{Keck} and Zumino \cite{Zumino} as the coset superspace
\begin{align} \label{intro ads1}
	\text{AdS}^{4|4} = \frac{\sOSp(1|4;\mathbb{R})}{\sSpin(3\,,1)}\,,
\end{align}
where $\sOSp(1|4;\mathbb{R})$ is the $\N=1$ supersymmetric extension of the AdS$_4$ group, $\sSO_{0}(3\,,2)\,.$ 
A major development for AdS superfield techniques was provided by the work of Ivanov and Sorin \cite{IS} (see also \cite{Ivanov:1979ft}) in which superfield representations and supermultiplets were thoroughly studied. 
In the supergravity setting, it was realised that AdS$^{4|4}$ arises as a maximally supersymmetric solution to the off-shell formulations for $\N=1$ supergravity, see \cite{Siegel:1977hn,WZ,Stelle:1978ye,Ferrara:1978em,Townsend:1977qa,Kaku:1978ea,Butter:2011vg}. 
The description of the $\N=2$ AdS superspace, 
\begin{align} \label{intro ads2}
	\text{AdS}^{4|8} = \frac{\sOSp(2|4;\mathbb{R})}{\sSpin(3\,,1)\times\sO(2)}\,,
\end{align}
where $\sO(2)$ is the $R$-symmetry group, as a maximally supersymmetric solution to the off-shell formulations for $\N=2$ supergravity came later, see \cite{Kuzenko:2008ep,Kuzenko:2008qw,Butter:2011ym,Butter:2012jj}. 
Recent attention has been directed towards the AdS$^{4|8}$ harmonic superspace, see \cite{Ivanov:2025jdp,Gargett:2025xcg}. 

It is not just the four-dimensional case which is of interest; supergravity theories in diverse dimensions have AdS solutions, see, e.g., \cite{Freedman:1976uk,deWit:1981yv,Gates:1982ct,Pernici:1984xx,Gunaydin:1984qu,Romans:1985tw,BILS,KLT-M12,KR}. 
AdS superspaces also play an exceptional role in superstring theories and the AdS/CFT correspondence, most notably in the duality between type IIB superstring theory in AdS$_{5}\times S^{5}$ and $\N=4$ super Yang-Mills theory in four dimensions, see, e.g., \cite{Aharony:1999ti}. 
The study of AdS superspaces is therefore a very interesting topic.
In four and five dimensions, the $\N$-extended AdS superspaces\footnote{According to Nahm's classification \cite{Nahm:1977tg}, superconformal algebras (and hence AdS$_{d+1}$ superalgebras) exist in spacetime dimensions $d \leq 6\,.$ The group-theoretic definitions \eqref{intro ads coset} can be generalised to AdS superspaces in other dimensions provided the appropriate AdS supergroup exists.} can be defined on group-theoretic grounds as follows 
\begin{align} \label{intro ads coset}
\text{AdS}^{4|4\N} = \frac{\sOSp(\N|4;\mathbb{R})}{\sSpin(3\,,1)\times\sO(\N)}\,, ~~~ \text{AdS}^{5|8\N} = \frac{\sSU(2,2|\N)}{\sSpin(4\,,1)\times\sU(\N)}\,,
\end{align}
see, e.g., \cite{BILS}.
Here, $\sOSp(\N|4;\mathbb{R})$ and $\sSU(2,2|\N)$ are identified with the $\N$-extended AdS supergroups\footnote{The supergroups $\sOSp(\N|4;\mathbb{R})$ and $\sSU(2,2|\N)$ are also identified with the $\N$-extended superconformal groups in three and four dimensions, respectively.} in four and five dimensions, respectively,
and the $\sO(\N)$ and $\sU(\N)$ factors correspond to the $R$-symmetry groups in AdS$^{4|4\N}$ and AdS$^{5|8\N}$. 
Despite their importance to a multitude of topics within the literature, there is still not much known about the $\N$-extended AdS superspaces in diverse dimensions. 
A significant step forward was provided by ref. \cite{BILS}, in which the conformal flatness of AdS superspaces with bodies of the form AdS$_{m} \times S^{n}$ was studied. 
The primary focus of this thesis is to bridge this gap by studying the AdS superspaces \eqref{intro ads coset}.

\subsubsection{Embedding formalisms}

When considering symmetries such as (super)conformal symmetry or AdS (super)symmetry, it is often powerful to work with an embedding formalism. 
An embedding formalism is a framework in which a geometry is realised as a surface in a higher-dimensional `embedding space'.
Famous examples of embedding formalisms include those for compactified\footnote{It is well known that the action of the conformal group is not globally defined on Minkowski space, see, e.g., \cite{BK}. Instead, the appropriate setting when considering conformal symmetry is compactified Minkowski space.}  Minkowski space $\overline{\mathbb{M}}{}^{d}$ \cite{Dirac:1936fq} (also known as Dirac's conformal space, see \cite{K-compactified12} for a review) and (A)dS. 
In particular, both the AdS$_{d+1}$ algebra and the conformal algebra in $d$ dimensions are isomorphic to $\frak{so}(d\,,2)$. 
Then, in this framework, both AdS$_{d+1}$ and $\overline{\mathbb{M}}{}^{d}$ can be realised as surfaces in $\mathbb{R}^{d\,,2}$, defined by coordinates $X^{\ua}$ satisfying the constraints
\bsubeq \label{intro embed equations}
\begin{align}
	\text{AdS}_{d+1}: ~~~ X^{\ua}X^{\ub}\eta_{\ua\ub} &= -\ell^{2}  \label{intro ads embed}
	\\
	\overline{\mathbb{M}}{}^{d}: ~~~ X^{\ua}X^{\ub}\eta_{\ua\ub} &= 0\,, 
\end{align}
where 
\begin{align}
	\eta_{\ua\ub} &= \text{diag}(-1\,, \underbrace{+1\,, \ldots \,, +1}_{d} \,, -1)\,, ~~~~ \ua = 0\,,1\,,\ldots\,,d+1\,,
\end{align}
and $\ell > 0$ is a constant parameter taken as the AdS radius.
De Sitter space can be realised in a similar fashion, via embedding in $\mathbb{R}^{d+1\,,1}$ as 
\begin{align}
	\text{dS}_{d+1}: ~~~ X^{\ua}X^{\ub}\tilde{\eta}_{\ua\ub} = \ell^{2}\,, ~~~ \tilde{\eta}_{\ua\ub} = \text{diag}(-1\,, \underbrace{+1\,, \ldots \,, +1}_{d+1})\,. \label{intro ds embed}
\end{align}
\esubeq
The surfaces \eqref{intro ads embed} and \eqref{intro ds embed} constitute global realisations of (A)dS, and are closely related to their definitions as homogeneous spaces, eq. \eqref{ads ds coset}.

Embedding formalisms have proven to have a broad range of applications. 
For example, 
they provide a straightforward approach to constructing two-point functions
which are useful for determining propagators in maximally symmetric spacetimes, see, e.g., \cite{Allen:1985wd,Allen:1986qj}.
Embedding methods can also be applied to the dynamics of extended supersymmetric objects, such as superparticles and superstrings, see, e.g., \cite{DS,BDS} for pedagogical reviews. 
Such an approach is also known as superembedding, and it has found great success in deriving the equations of motion for the M5-brane for the first time \cite{Howe:1996yn} as well as uncovering the nature of key features of these models, including the so-called $\k$-symmetry, see \cite{Szg} for a review.
While the embedding approach which leads to the relations \eqref{intro embed equations} and the superembedding approach are related, they differ in several key aspects. 
For example, in the superembedding approach, the dynamics of a super-$p$-brane are determined by its embedding into a target (embedding) superspace, the geometry of which corresponds to an appropriate theory of supergravity. 
In the embedding approach, it is the supergravity background itself, e.g., AdS superspace, which is embedded into an embedding superspace.
Regarding the embedding of AdS superspace, as of yet, such a formalism has only been fully developed in the three-dimensional case \cite{KTM,KT}. 
Although ref. \cite{KTM} proposed an embedding formalism for AdS$^{4|4\N}$, it did not develop its applications beyond the general structure of two-point functions, or study its geometry. 
The present thesis aims to extend the development and applications of the embedding formalism for AdS$^{4|4\N}$, as well as derive an embedding formalism for the $\N$-extended AdS superspace in five dimensions, AdS$^{5|8\N}$, and its harmonic and projective extensions.

\subsubsection{Supertwistor techniques: the general approach}

Our goal raises the natural question: what tools can one make use of to develop an embedding formalism for AdS superspace? 
Due to the specific nature of supersymmetry, and the vastly different properties of spinors across different spacetime dimensions, one cannot expect that an embedding formalism for AdS superspaces will have a universal form analogous to eq. \eqref{intro ads embed}. 
Instead, such formalisms should be developed on a case-by-case basis making use of the properties of the AdS supergroup. 

The approach to this problem is best described in the language of twistors, which were
first introduced by Penrose in the 1960s \cite{Penrose:1967wn,Penrose:1972ia} and have become a very rich field of study, see, e.g., the books \cite{PR,Huggett:1986fs,Ward:1990vs}, the review \cite{A-twist}, and the historical account \cite{Atiyah:2017erd} for more on twistor theory. 
The marriage of twistors and supersymmetry was introduced by Ferber \cite{Ferber:1977qx}, and supertwistors have since found numerous applications in theoretical and mathematical physics. 
These include the supertwistor formulations for conformal supergravity theories in diverse dimensions \cite{Howe:2020xrg,Howe:2020hxi}, as well as the (super)twistor descriptions of (super)particles in both flat and AdS (super)spaces, see, e.g., \cite{BENGTSSON198881, CGKRZ, CRZ, CKR, BLPS, Z, Cm, Cm2, ABGT, ABGT2, U, U2, Adamo:2016rtr}. 
Most notably for our purposes, however, supertwistors have been used to describe the $\N$-extended compactified Minkowski superspaces in three and four dimensions \cite{ManinNi,Kotrla:1984ky,Howe:1994ms,Kuzenko:2010rp}, as well as their harmonic and projective extensions \cite{Howe:1994ms,Kuzenko:2010rp,Rosly:1985nyf,Lukierski:1988vw,Howe:1995md,Kuzenko:2006mv,K-compactified12,Buchbinder:2015qsa}, and the $\N$-extended AdS superspaces in three and four dimensions \cite{KTM,KT}, see also \cite{Kuzenko:2014yia}. 

Supertwistors prove ideal for a global realisation of AdS superspace due to their transformation properties. 
For our purposes, the space of even (odd) supertwistors can be identified with $\mathbb{C}^{4|\N}$ $(\mathbb{C}^{\N|4})$ and is endowed with the linear action of the supergroups $\sOSp(\N|4;\mathbb{R})$ and $\sSU(2,2|\N).$
From this perspective, the problem of developing a global realisation of AdS superspace becomes that of identifying an appropriate constrained surface in the supertwistor space. 
In the case of AdS$^{4|4\N}$, the appropriate supertwistor constraints were identified in ref. \cite{KTM}. 
For AdS$^{5|8\N}$, they are derived in the present work. 

Closely related to the supertwistor realisations of the $\N$-extended compactified Minkowski and AdS superspaces in four dimensions, are the so-called bi-supertwistor realisations for the same superspaces. 
The bi-supertwistor realisation for the $\N$-extended compactified Minkowski superspace in four dimensions was first introduced by Siegel in \cite{SGLBST1,SGLBST2}, and its extension to AdS$^{4|4\N}$ was derived in \cite{KTM}. 
In this realisation, AdS$^{4|4\N}$ is described by graded antisymmetric supermatrices obeying certain constraints. These supermatrices, known as bi-supertwistors, prove to be the supersymmetric analogues of the embedding coordinates $X^{\ua}$ in \eqref{intro ads embed}, which will be elaborated on in the main body. 
The interplay between the supertwistor and bi-supertwistor approaches to AdS superspace is central to this thesis.

\subsubsection{Thesis structure}

This thesis is organised as follows. 
In chapter \ref{bm} we will review necessary background material, establishing our notation and conventions for the following chapters. 
This includes relevant formalism regarding homogeneous spaces in section \ref{bm homo}. The $\N=1$ and $\N=2$ AdS superspaces in four dimensions are reviewed in section \ref{bm super ads}, and supertwistor and bi-supertwistor realisations for both four-dimensional compactified Minkowski superspace and AdS superspace are reviewed in section \ref{st realisations}.

Chapters \ref{ch3}, \ref{ch4}, and \ref{ch5} contain the bulk of the original material in this thesis, and are based on the papers \cite{KKR,KKR2,KK,KRsp,KK3}.
A description of the included material and their location in this thesis is provided in the authorship declaration, as well as outlined at the beginning of each chapter. 
The general form and flow of each chapter varies on pedagogical grounds, and is also described at the beginning of each. 
The original results of \cite{KKR,KKR2,KK,KRsp,KK3} 
which are contained in this thesis are largely presented as if they appeared here for the first time. 
However, where appropriate, relevant citations are included. 

The focus of chapter \ref{ch3} is on the four dimensional AdS superspace, AdS$^{4|4\N}$.  Two distinct approaches to the AdS supergeometry are described. Specifically, AdS$^{4|4\N}$ is realised within the $\sSU(\N)$ superspace setting, as well as within the (bi-)supertwistor picture. The precise relationship between these two approaches is also detailed. 
In chapter \ref{ch4}, these considerations are extended to the $\N$-extended five-dimensional AdS superspace, AdS$^{5|8\N}$. A supertwistor realisation is derived for arbitrary $\N \geq 0$, and its relationship to the supergravity realisation, which is available for $\N=1$, is described. 
The methods which have been made use of in chapter \ref{ch4} are inspired by those developed in chapter \ref{ch3} for the four-dimensional case. 
Finally, chapter \ref{ch5} is concerned with superparticle models in AdS superspaces in three and two dimensions. Building on the results of chapters \ref{ch3} and \ref{ch4}, we derive new models for superparticles propagating in these AdS superspaces.


\chapter{Background material}\label{bm} 

\thispagestyle{knuff1}

This chapter is dedicated to reviewing essential formalism for the later stages of this thesis. 
It is accompanied by several technical appendices.
Appendix \ref{spinor conventions} details our spinor conventions in diverse dimensions.
Appendices \ref{5d appendix} and \ref{Supertwistors} detail our conventions for the supergroups $\sSU(2,2|\N)$ and $\sOSp(\N|4;\mathbb{R})$ and their supertwistors.

\section{Homogeneous spaces} \label{bm homo}

In this section we will provide a brief overview of useful formalism regarding homogeneous spaces, and then go through various examples. 
We present only those elements of the formalism which are useful for the analysis of later chapters in this thesis, and refer the reader to sources such as \cite{WB,s-diff,f-diff,n-diff} for comprehensive discussion.
 
We begin by considering a homogeneous space $\cM$ of Lie group $G$. 
By definition, $\cM$ is a manifold on which the group $G$ acts transitively. 
This homogeneous space can always be realised as a left coset space 
\begin{align}
\cM = G/H\,,	
\end{align}
for a suitable closed subgroup $H$ of $G$. 
For our purposes, it is useful to introduce a marked point (or origin) in $\cM$, then the subgroup $H$ is chosen as that which leaves the origin invariant. 
It is known by many names, including the Little group, the stability subgroup (or stabiliser), the isotropy group, amongst others. We will use these names interchangeably. 

We will now introduce the coset representative.  
A global coset representative for a homogeneous space $\cM = G / H$ is an injective map $\cS: \cM \rightarrow G$ such that $\p \circ \cS = \text{id}_{\cM}$, where $\p$ denotes the canonical projection $\p: G \rightarrow G/H$.  
For many homogeneous spaces, a global coset representative does not exist. 
In such cases, local coset representatives $\cS_{A}: U_{A} \rightarrow G$ with the property $\p \circ \cS_{A} = \text{id}_{U_{A}}$ can be defined on open charts $\{U_{A}\}$ that provide an atlas for $\cM$. 
In the intersection of two charts, $U_{A}$ and $U_{B}$, $U_{A} \cap U_{B} \neq \emptyset$, the corresponding coset representatives $\cS_{A}$ and $\cS_{B}$ are related by a point-dependent little group transformation, $\cS_{B}(x) = \cS_{A}(x)h_{AB}(x)\,,$ with $h_{AB}(x) \in H$.
Here, $x$ denotes a point in $\cM$.

For the moment, let us assume the existence of a global coset representative. 
Then, each element in $G$ can be uniquely decomposed as 
\begin{align}
	g = \cS(x)h\,, \qquad g \in G\,, \qquad h \in H\,.
\end{align}
Now the action of the group $G$ on the homogeneous space $\cM$ can be expressed as 
\begin{align} \label{crep trf rule}
	g\cS(x) = \cS(g\cdot x)h(g\,,x) \equiv \cS(x')h(g\,,x)\,, \qquad h(g\,,x) \in H.
\end{align}
Here, $h(g\,,x)$ obeys the important property 
\begin{align}
	h(g_1 g_2 \,, x) = h(g_1 \,, g_2 x) h(g_2 \,, x)\,. 
\end{align}
The right-multiplication by the element $h(g\,,x)$ is used as a compensating transformation to preserve the parametrisation of the coset representative.

Let us denote the Lie algebra of $G$ as $\frak{g}$ and that of $H$ as $\frak{h}$. 
Then, we introduce the complement of $\frak{h}$ in $\frak{g}$ as $\frak{k}$. 
That is, 
\begin{align}
	\frak{g} = \frak{k} \oplus \frak{h}\,.
\end{align}
Now, let $\{\cT_{a}\}$ be a basis in $\frak{k}$ and $\{\cH_{i}\}$ be a basis in $\frak{h}\,.$ 
These generators are assumed to obey an algebra of the form 
\bsubeq \label{homo example alg}
\begin{align}
	[\cT_{a}\,,\cT_{b}] &= f_{ab}{}^{c}\cT_{c} + f_{ab}{}^{i}\cH_{i}\,,
	\\
	[\cH_{i}\,,\cT_{a}] &= f_{i a}{}^{b}\cT_{b}\,,
	\\
	[\cH_{i}\,,\cH_{j}] &= f_{ij}{}^{k}\cH_{k}\,. 
\end{align}
\esubeq
In the above, the structure constants obey natural reality and (anti)symmetry conditions. 
Indeed, all of the homogeneous spaces of interest considered in this work have an algebra of the form \eqref{homo example alg}.\footnote{The pair $\frak{g}$ and $\frak{h}$ form what is known as a reductive Klein pair, see \cite{s-diff}.}
This can be compactly represented as 
\begin{align}
	[\frak{h}\,,\frak{h}] \in \frak{h}\,, \qquad [\frak{k}\,,\frak{h}] \in \frak{k}\,.
\end{align}
It is often useful to use an exponential parametrisation of the coset representative, in which we have the following 
\begin{align}
\cS(\f) = \exp(\f^{a}\cT_{a}) \implies g = \exp(\f^{a}\cT_{a})\exp(u^{i}\cH_{i})\,.	
\end{align}
The fields $\f^{a}$ are taken to depend on the coordinates $x\,.$

We now introduce the Maurer-Cartan one-form, as 
\begin{align}
	\o = \cS^{-1}\rd\cS\,. 
\end{align}
It takes its values in the Lie algebra $\frak{g}$, and it is useful to decompose it into $\o|_{\frak{k}}$ and $\o|_{\frak{h}}$. 
Indeed, let us denote the following 
\begin{align}
	\o|_{\frak{k}} &= {\bf{E}} = {\bf{E}}^{a}\cT_{a}\,, \qquad \o|_{\frak{h}} = {\bm{\O}} = {\bm{\O}}^{i}\cH_{i}\,.
\end{align}
The one-form $\bf{E}$ is known as the vielbein, and $\bm{\O}$ is known as the connection. 
It is useful to further decompose them into the basis $\rd x^{m}$, where $x^{m}$ are the local coordinates on $\cM$, as 
\bsubeq
\begin{align}
	{\bf{E}}^{a}\cT_{a} &= \rd x^{m} E_{m}{}^{a}\cT_{a}\,,
	\\
	{\bm{\O}}^{i}\cH_{i} &= \rd x^{m} \O_{m}{}^{i}\cH_{i}\,.
\end{align}
\esubeq
The vielbein matrix $E_{m}{}^{a}$ is assumed to be non-singular, and its inverse $E_{a}{}^{m}$, defined by 
\begin{align}
	E_{m}{}^{a}E_{a}{}^{n} = \d_{m}{}^{n} \Longleftrightarrow E_{a}{}^{m}E_{m}{}^{b} = \d_{a}{}^{b}\,, 
\end{align}
can be used to define the fields $E_{a}$ and $\O_{a}$ as 
\bsubeq
\begin{align}
	E_{a} &= E_{a}{}^{m}\partial_{m}\,, \qquad \partial_{m} = \frac{\partial}{\partial x^{m}}\,, 
	\\
	\O_{a} &= E_{a}{}^{m}\O_{m}{}^{i}\cH_{i}\,.
\end{align}
\esubeq
The fields $E_{a}$ and $\O_{a}$ are used to define coviariant derivatives, as 
\begin{align}
	\cD_{a} = E_{a} + \O_{a}\,. 
\end{align}

There are two tensorial objects we can construct from the vielbein and the connection. 
They are given by the following 
\bsubeq
\begin{align}
	{\bf{R}} &= \rd \bm{\O} - \bm{\O}\wedge\bm{\O}\,, \label{homo curve} \\
	{\bf{T}} &= \rd \bf{E} - \bf{E}\wedge\bm{\O} - \bm{\O}\wedge\bf{E} \label{homo tors} \,.
\end{align}
\esubeq
Expression \eqref{homo curve} is known as the curvature tensor, and \eqref{homo tors} is known as the torsion. 
From the definition of the Maurer-Cartan form, it is clear that $\o$ satisfies the Maurer-Cartan structure equation
\begin{align}
	\rd \o - \o \wedge\o &= 0 \implies \rd \bf{E} + \rd \bm{\O} = (\bf{E} + \bm{\O})\wedge(\bf{E} + \bm{\O})\,.
\end{align}
We then find that the tensors $\bf{R}$ and $\bf{T}$ are given by the following equivalent expressions 
\bsubeq
\begin{align}
	(\bf{E} \wedge \bf{E})|_{\frak{h}} &= \bf{R}\,, 
	\\
	(\bf{E}\wedge\bf{E})|_{\frak{k}} &= \bf{T}\,.
\end{align}
\esubeq
Since the curvature and torsion are two-forms, they can be decomposed into the basis $\{\bf{E}^{a}\}$ as follows 
\bsubeq
\begin{align}
	{\bf{R}} &= \frac{1}{2}{\bf{E}}^{b}\wedge{\bf{E}}^{a}\left\{R_{ab}{}^{i}\cH_{i}\right\}\,, \\
	{\bf{T}} &= \frac{1}{2}{\bf{E}}^{b}\wedge{\bf{E}}^{a}\left\{T_{ab}{}^{c}\cT_{c}\right\}\,,
\end{align}
\esubeq
to obtain the components $R_{ab}{}^{i}$ and $T_{ab}{}^{c}\,.$
Finally, as shown in, e.g., \cite{WB} and \cite{Kleppe}, the covariant derivatives $\cD_{a}$ obey the commutation relations 
\begin{align}
	[\cD_{a}\,,\cD_{b}] = -T_{ab}{}^{c}\cD_{c} + R_{ab}{}^{i}\cH_{i}\,.
\end{align}

\subsection{Minkowski space}

We will first illustrate this formalism with a simple example, the $d$-dimensional Minkowski spacetime. 
For this purpose we consider a cartesian coordinate system, in which the metric takes the standard form 
\begin{align}
	\eta_{ab} = \text{diag}(- ~ + ~ + ~\ldots~ +)\,.
\end{align}
As is well-known, Minkowski space is a homogeneous space of the connected component of the Poincar\'e group in $d$ dimensions, $\sISO_0(d-1\,,1)$.
A Poincar\'e transformation acts on the coordinates $x^{a}$ as 
\begin{align}
	g: x^{a} \rightarrow x'^{a} = \L^{a}{}_{b}x^{b} + a^{a}\,, \qquad g = (\L\,,a) \in \sISO_0(d-1\,,1)\,.
\end{align}
In the above, $\L = (\L^{a}{}_{b})$ is an element of the connected component of the Lorentz group, $\sSO_0(d-1\,,1)\,.$ This means it obeys the master equation 
\begin{align}
	\eta_{cd}\L^{c}{}_{a}\L^{d}{}_{b} = \eta_{ab}\,, ~~~ \det \L = 1\,, ~~~ \L^{0}{}_{0} \geq 1\,.
\end{align}

We will now introduce a preferred point and consider its stabiliser. 
The simplest choice for this is the origin in our coordinate system, 
\begin{align}
	x_{(0)}^{a} = (0\,, \ldots \,, 0)\,.  
\end{align}
This is invariant under Lorentz transformations, and so we identify the stabiliser with the Lorentz group. 
Then, following the above considerations, the $d$-dimensional Minkowski spacetime is realised as the coset space 
\begin{align} \label{bm mink coset}
	\mathbb{M}^{d} = \frac{\sISO_0(d-1\,,1)}{\sSO_0(d\,,1)}\,.
\end{align}
Indeed, given two elements $g_1\,, g_2 \in \sISO_0(d-1\,,1)$, their product is given by 
\begin{align}
	(\L_1\,,a_1)(\L_2\,,a_2) = (\L_1\L_2 \,, a_1 + \L_1 a_2)\,.
\end{align}
We can therefore globally factorise the element $(\L\,,a)$ as 
\begin{align}
	g = (\L\,,a) = \underbrace{(\id_{d}\,,a)}_{\cS(a)}\underbrace{(\L\,,0)}_{h(\L)}\,.
\end{align}
The element $\cS(a)$ can be identified with a global coset representative for $\mathbb{M}^{d}$, and the parameters $a^{a}$ of the translations can be interpreted as coordinates on $\mathbb{M}^{d}$.

\subsection{$\N$-extended Minkowski superspace in four dimensions}

The starting point for our foray into Minkowski superspace will be a description of the super Poincar\'e group.
Let us denote the $\N$-extended super Poincar\'e group in four dimensions as $\cS\P(4|\N)$.  
It is convenient to realise $\cS\P(4|\N)$ as a subgroup of the superconformal group in four dimensions, $\sSU(2,2|\N)$, then its elements can be represented by supermatrices of the form
\begin{align} \label{bm spg elements}
	g(x\,,\q\,,n\,,u) = \underbrace{\left(\begin{array}{c|c||c}
		\id_{2} & ~0~ & ~0~ \\
		\hline
		-\ri \tilde{x}_{+} & \id_{2} & 2\bar{\q} \\
		\hline\hline
		2\q & 0 & \id_{\N}
	\end{array}\right)}_{\cS(x\,,\q\,,\bar{\q})}
	\underbrace{\left(\begin{array}{c|c||c}
		~n~ & ~0~ & ~0~\\
		\hline
		0 & (n^{\dag})^{-1} & 0 \\
		\hline \hline 
		0 & 0 & u
	\end{array}\right)}_{h(n\,,u)}\,, 
\end{align}
with 
\bsubeq
\begin{align}
	\tilde{x}_{\pm} &= x_{\pm}^{m}(\tilde{\s}_{m})\,, ~~~  x_{\pm}^{m} = x^{m} \pm \ri \q_{i}^{\a}(\s^{m})_{\a\bd}\bar{\q}^{\bd i}\,, ~~~ \q = (\q_{i}^{\a})\,, 
	\\ n &\in \sSL(2\,,\mathbb{C})\,, ~~~ u \in \sSU(\N)\,,
\end{align}
\esubeq
see, e.g., \cite{K-lec,Park:1999pd}. 
Here we have the global coset representative 
\begin{align} \label{smink global crep}
	\cS(x\,,\q\,,\bar{\q}) = \left(\begin{array}{c|c||c}
		\id_{2} & ~0~ & ~0~ \\
		\hline
		-\ri \tilde{x}_{+} & \id_{2} & 2\bar{\q} \\
		\hline\hline
		2\q & 0 & \id_{\N}
	\end{array}\right) = \exp\left(\begin{array}{c|c||c}
		~0~ & ~0~ & ~0~ \\
		\hline 
		-\ri \tilde{x} & 0 & 2\bar{\q} 
		\\
		\hline \hline 
		2\q & 0 & 0
	\end{array}\right)\,.
\end{align}
It should be pointed out that the coset representative is determined completely in terms of its first two columns, 
\begin{align} \label{bm reduced coset rep}
\underline{\cS}(x_{+}\,,\q) = \left(\begin{array}{c}
\id_{2}\\
\hline 
-\ri\tilde{x}_{+} \\
\hline \hline 
2\q
\end{array}\right)\,. 
\end{align}
As we will see in section \ref{st realisations}, the columns \eqref{bm reduced coset rep} coincide with the two-plane in the Minkowski chart of compactified Minkowski superspace, see eq. \eqref{mink 2plane}.

It follows from the general considerations above that Minkowski superspace is realised as the coset\footnote{Sometimes the stabiliser is taken to be just $\sSL(2,\mathbb{C})$, or $\sSL(2\,,\mathbb{C})\times\sU(\N)$, see \cite{K-lec}. This depends on the definition of the super Poincar\'e group \eqref{bm spg elements}.} superspace 
\begin{align}
	\mathbb{M}^{4|4\N} = \frac{\cS\P(4|\N)}{\sSL(2\,,\mathbb{C})\times\sSU(\N)}\,. 
\end{align}
We can see that Minkowski superspace is parametrised by local coordinates $z^{M} = (x^{m}\,, \q_{i}^{\a}\,, \bar{\q}^{i}_{\ad})$. 
The coordinates $\q_{i}^{\a}$ are Grassmann-odd, $\q_{i}^{\a}\q_{j}^{\b} = -\q_{j}^{\b}\q_{i}^{\a}\,,$ and are related to $\bar{\q}^{i}_{\ad}$ as $(\q_{i}^{\a})^{*} = \bar{\q}^{\ad i}$. 
Inspecting the transformation rule \eqref{crep trf rule} yields the most general finite super Poincar\'e transformation on the coordinates $z^{M}$
\bsubeq
\begin{align}
	x^{\ad\a} &\rightarrow (n^{\dag})^{\ad}{}_{\bd}x^{\bd\b}n_{\b}{}^{\a} + a^{\ad\a} + 4\ri \bar{\e}^{\ad i}\e_{i}^{\a}\,,
	\\
	\q_{i}^{\a} &\rightarrow  u_{i}{}^{j} \q_{j}^{\b} n_{\b}{}^{\a} + \e_{i}^{\a} \,,
	\\
	\bar{\q}^{\ad i} &\rightarrow (n^{\dag})^{\ad}{}_{\bd}\bar{\q}^{\bd j}(u^{-1})_{j}{}^{i} + \bar{\e}^{\ad i}\,.
\end{align}
\esubeq
Although we have explicitly included the indices in the above expressions, it is often useful and convenient to suppress them. 
In section \ref{4d compact str} we will consider infinitesimal superconformal transformations on (compactified) Minkowski superspace and express them with their indices suppressed. 

The super Poincar\'e algebra is spanned by supermatrices of the form 
\begin{align}
	\U = \left(\begin{array}{c|c||c}
		\o_{\a}{}^{\b} & 0 & 0 \\
		\hline
		-\ri a^{\ad\b} & - \bar{\o}^{\ad}{}_{\bd} & 2\bar{\e}^{\ad j} \\
		\hline \hline 
		2\e_{i}{}^{\b} & 0 &  \r_{i}{}^{j}
	\end{array}\right)\,, 
\end{align}
with 
\begin{align}
	\tr\o = 0\,, ~~~ \tr\r = 0\,,~~~ \r = -\r^{\dag}\,, ~~~ \tilde{a} = (a^{\ad\b}) = \tilde{a}^{\dag}\,.
\end{align}
We can therefore identify the elements of $\frak{t}$ as 
\begin{align}
		\frak{t}^{A}P_{A} = \left(\begin{array}{c|c||c}
		0 & ~0~ & 0 \\
		\hline
		-\ri a^{\ad\b} & 0 & 2\bar{\e}^{\ad j} \\
		\hline \hline 
		2\e_{i}{}^{\b} & 0 & 0
	\end{array}\right)\,, \qquad 
	P_{A} = (P_{a}\,,Q^{i}_{\a}\,,\bar{Q}_{i}^{\ad})\,, 
\end{align}
where we have introduced the generators of super-translations $P_{A}$. 
The generators $Q_{\a}^{i}$ and $\bar{Q}_{i}^{\ad}$ obey the important relation 
\begin{align}
	\{Q_{\a}^{i}\,,\bar{Q}_{j \ad}\} = 2 \d^{i}_{j}P_{\a\ad}\,. 
\end{align}
Further, the elements of $\frak{h} = \frak{sl}(2\,,\mathbb{C}) \oplus \frak{su}(\N)$ take the form 
\begin{align}
	h = \left(\begin{array}{c|c||c}
		\o_{\a}{}^{\b} & 0 & 0 \\
		\hline
		0 & - \bar{\o}^{\ad}{}_{\bd} & 0 \\
		\hline \hline 
		0 & 0 &  \r_{i}{}^{j}
	\end{array}\right)\,. 
\end{align}

For the supervielbein and connection, $\cS^{-1}\rd\cS$ yields the following 
\begin{align}
	{\bf{E}} = \left(\begin{array}{c|c||c}
	~0~ & ~0~ & ~0~ \\
	\hline
	-\ri\tilde{\P} & 0 & 2\rd\bar{\q}\\
	\hline \hline 
	2\rd\q & 0 & 0
	\end{array}\right)\,, \qquad {\bm{\O}} = 0\,,
\end{align}
where $\P^{a} = -\frac{1}{2}\tr(\tilde{\P}\s^{a}) = \rd x^{a} + \ri(\q_{i}\s^{a}\rd\bar{\q}^{i} - \rd\q_{i}\s^{a}\bar{\q}^{i})$ is the $\N$-extended Volkov-Akulov one-form \cite{Volkov:1972jx, Volkov:1973ix, Akulov:1974xz}. 
In particular, the supervielbein one-forms are 
\begin{align}
	{\bf{E}}^{A} = (\P^{a}\,, \rd\q_{i}^{\a}\,,\rd\bar{\q}^{i}_{\ad}) = \rd z^{M} E_{M}{}^{A}\,.
\end{align}
This implies the flat superspace covariant derivatives are given by 
\bsubeq \label{flat mink cd}
\begin{align}
	D_{A} = E_{A}{}^{M}\partial_{M} = (\partial_{a}\,,D_{\a}^{i}\,,\bar{D}^{\ad}_{i})\,,
\end{align}
with
\begin{align}
	D_{\a}^{i} = \partial_{\a}^{i} + \ri(\s^b)_{\a\bd}\bar{\q}^{\bd i}\partial_{b}\,, ~~~
	 \bar{D}_{\ad i} = - \bar{\partial}_{\ad i} - \ri\q_{i}^{\b}(\s^{b})_{\b\ad}\partial_{b}\,.
\end{align}
\esubeq
They obey the algebra 
\begin{align}
	\{D_{\a}^{i}\,,\bar{D}^{\ad}_{j}\} = -2\ri\d^{i}_{j}\partial_{\a}{}^{\ad}\,.
\end{align}

The key property of the covariant derivatives \eqref{flat mink cd} is that they map tensor superfields into tensor superfields. 
Let us introduce the representation for the supersymmetry generators 
\begin{align}
Q_{\a}^{i} = \ri \partial_{\a}^{i} + (\s^{b})_{\a\bd}\bar{\q}^{\bd i}\partial_{b}\,, ~~~ \bar{Q}_{\ad i} = -\ri\bar{\partial}_{\ad i} - \q_{i}^{\b}(\s^{b})_{\b\ad}\partial_{b}\,. 	
\end{align}
Then, an infinitesimal supersymmetry transformation on a superfield $\vf(z)$ (with its indices suppressed) yields 
\begin{align}
	\d \vf(z) = \ri(\e_{i}^{\a}Q_{\a}^{i} + \bar{\e}^{i}_{\ad}\bar{Q}_{i}^{\ad})\vf(z)\,,
\end{align}
for the parameters $(\e_{i}^{\a}\,, \bar{\e}^{i}_{\ad})$. 
That the covariant derivatives map tensor superfields into tensor superfields means that $D_{A}\vf(z)$ is a tensor superfield. 
This implies that the covariant derivatives commute with supersymmetry transformations
\begin{align}
	[D_{A}\,, \e^{\b}_{j}Q_{\b}^{j}] = [D_{A}\,,\bar{\e}_{\bd}^{j}\bar{Q}^{\bd}_{j}] = 0\,.
\end{align}
An important further implication of this is the possibility to impose covariant differential constraints on superfields. 
A superfield $\f(z)$ satisfying the constraint 
\begin{align}
	\bar{D}_{\ad i}\f(z) = 0 
\end{align}
is said to be chiral. 
Finally, if we consider the first two columns of our coset representative $\cS(x\,,\q\,,\bar{\q})$, see eq. \eqref{smink global crep} and \eqref{bm reduced coset rep}, we find that they are parametrised by the chiral coordinates $\tilde{x}_{+}$ and $\q$. 
We will return to this in section \ref{4d compact str}.

\subsection{Anti-de Sitter space} \label{bm ads sec}

The $d$-dimensional anti-de Sitter space can be realised as a hypersurface in the pseudo-Euclidean space $\mathbb{R}^{d-1\,,2}$ defined by 
\begin{align} \label{bm ads def}
	\eta_{\ua\ub}X^{\ua}X^{\ub} = -\ell^{2}\,, \qquad \ua = 0\,, 1\,, \ldots \,, d-1\,, d\,,
\end{align}
for some positive constant $\ell$. 
The metric $\eta_{\ua\ub}$ is given by $\eta_{\ua\ub} = \text{diag}(-1\,, +1 \,, \ldots \,, +1 \,, -1)\,.$
It follows that AdS$_{d}$ is a homogeneous space of the group $\sSO_0(d-1\,,2)\,.$ 
Indeed, choosing a preferred point as $X_{(0)}^{\ua} = (\ell\,, 0 \,, \ldots\,, 0)$, we find that the stabiliser is given by $\sSO_0(d-1\,,1)$. 
Then, following the general considerations described above, we can identify AdS$_{d}$ as the coset space 
\begin{align} \label{standard ads coset}
	\text{AdS}_{d} = \frac{\sSO_0(d-1\,,2)}{\sSO_0(d-1\,,1)}\,. 
\end{align}
One can easily verify the dimensions of the above coset space as $\frac{1}{2}(d+1)(d)-\frac{1}{2}(d)(d-1) = d\,.$
Realisation \eqref{standard ads coset} constitutes the standard realisation of AdS$_d$ as a coset space. 
For the purposes of this thesis, in which embedding formalisms are developed for AdS superspaces, it is necessary to consider realisations in terms of \textit{spin groups}. 
That is, we can consider the homogeneous space 
\begin{align} \label{bm ads spin coset}
	\text{AdS}_{d} = \frac{\sSpin(d-1\,,2)}{\sSpin(d-1\,,1)}\,. 
\end{align}
For the remainder of this chapter, however, we will consider the standard realisation \eqref{standard ads coset}.
Useful isomorphisms and spinor conventions in diverse dimensions are described in appendix \ref{spinor conventions}. 

We will now sketch out various details of the coset construction illustrated above.
We can pick a basis for the AdS algebra as follows. 
We denote the generators of $\frak{so}(d-1\,,1)$ by $M_{ab} = -M_{ba}$, with $a = 0\,,\ldots\,,d-1$. 
These generators are associated with Lorentz transformations.
The basis for the complement of $\frak{so}(d-1\,,1)$ in $\frak{so}(d-1\,,2)$ are denoted $P_{a}$. 
These are the generators associated with translations.  
Amongst themselves, they obey the commutation relations 
\bsubeq
\begin{align}
	[M_{ab}\,,M_{cd}] &= 2\eta_{d[a}M_{b]c} - 2\eta_{c[a}M_{b]d}
	\,, \\
	[M_{ab}\,,P_{c}] &= 2\eta_{c[b}P_{a]}
	\,,
	\\
	[P_{a}\,,P_{b}] &= \frac{1}{\ell^{2}}M_{ab}\,.
\end{align}
\esubeq
The parameter $\ell$ is taken to be the length parameter of AdS. In the coset construction it is convenient to set $\ell = 1$. 

The Maurer-Cartan one-form is given by 
\begin{align}
	\cS^{-1}\rd\cS = \rd x^{m}E_{m}{}^{a}P_{a} + \frac{1}{2} \rd x^{m}E_{m}{}^{a}\O_{a}{}^{cd}M_{cd}\,,
\end{align}
for local coordinates $x^{m}$ parametrising AdS$_{d}$.
The covariant derivatives, $\cD_{a}$ are then given by 
\begin{align}
	\cD_{a} = E_{a}{}^{m}\partial_{m} + \frac{1}{2}\O_{a}{}^{cd}M_{cd}\,.
\end{align}
Their algebra is given by the familiar algebra of covariant derivatives in AdS$_d$, 
\begin{align}
	[\cD_{a}\,,\cD_{b}] = -\frac{1}{\ell^{2}}M_{ab}\,. 
\end{align}

Let us now briefly turn away from the coset construction, and introduce a conformally flat atlas for AdS$_{d}$. 
For this purpose we introduce two coordinate charts:
(i) the north chart, in which 
$X^d> -\ell$; and 
(ii) the south chart, in which $X^d<\ell$.

For a given point $X^{\ua}$ in the north chart, it is useful to introduce the straight line $\G^\ua_{\mathfrak N} (t)$ connecting $X^{\ua}$ with the `north pole' $X^{\ua}_{{\rm north}} = (0\,, \ldots\,, -\ell)$. 
This straight line will intersect the plane $X^{d} = 0$ at the point $x^{\ua} = (x^{a}\,, 0)$. 
The coordinates $x^{a}$ are then taken to be the local coordinates of $X^{\ua}$ in the north chart. 
Analogous definitions hold for the local coordinates $y^{a}$ in the south chart. 
This method of parametrising AdS$_d$ is known as stereographic projection. 

The straight line $\G^A_{\mathfrak N} (t)$ can be parametrised in terms of an evolution parameter t as
\bea
\G^{\ua}_{\mathfrak N} (t) = (1-t) X^{\ua}_{\rm north} + t x^\ua~.
\eea
Then $X^\ua \in {\rm AdS}_d$ corresponds to some value $t'$ of the evolution parameter, $\G^{\ua}_{\mathfrak N} (t') = X^\ua$.
We then find the following relations
\bsubeq \label{bm nc relations}
\begin{align}
x^a = \frac{\ell X^a}{\ell +X^d} ~ &\implies ~
x^2 := \eta_{ab}x^a x^b = - \ell^2 \frac{\ell - X^d}{\ell + X^d}
\\
&\implies ~ \ell^2 - x^2 = \frac{2\ell^3}{\ell + X^d} >0~.
\end{align}
\esubeq
We can express the embedding coordinates $X^{\ua}$ in terms of $x^{a}$ as
\bea
X^a = \frac{2\ell^2}{\ell^2 - x^2} x^a~, \qquad X^d = \ell \frac{\ell^2 +x^2} {\ell^2 - x^2}~.
\eea
Finally, the we can determine the induced metric in the north chart. 
From 
\begin{align}
	\rd s^{2} = \eta_{\ua\ub} \rd X^{\ua} \rd X^{\ub} = \eta_{ab} \rd X^{a} \rd X^{b} - (\rd X^{d})^{2}\,, \qquad \rd X^{\ua} = \frac{\partial X^{\ua}}{\partial x^{a}} \rd x^{a}\,, 
\end{align} 
we find 
\bea \label{north chart induced metric appendix}
\rd s^2_{\rm north} = \frac{4\ell^4}{(\ell^2 - x^2)^2} \eta_{ab} \rd x^a \rd x^b~.
\eea

Repeating the above analysis for the south chart, we parametrise the straight line $\G^{\ua}_{\mathfrak S} (t)$ as
\bea
\G^\ua_{\mathfrak S} (t) = (1-t) X^{\ua}_{\rm south} + t y^\ua~, \qquad y^\ua = (y^a, 0)~,
\eea
We obtain analogous relations to \eqref{bm nc relations}
\begin{align}
y^a = \frac{\ell X^a}{\ell -X^d} ~ &\implies ~
y^2 := \eta_{ab}y^a y^b = - \ell^2 \frac{\ell + X^d}{\ell - X^d}
\\
&\implies ~ \ell^2 - y^2 = \frac{2\ell^3}{\ell - X^d} >0~.
\end{align}
Expressing the embedding coordinates $X^{\ua}$ in terms of the local coordinates $y^{a}$ yields
\bea
X^a = \frac{2\ell^2}{\ell^2 - y^2} y^a~, \qquad X^d = -\ell \frac{\ell^2 +y^2} {\ell^2 - y^2}~.
\eea
Finally, the induced metric in the south chart is given by 
\bea
\rd s^2_{\rm south} = \frac{4\ell^4}{(\ell^2 - y^2)^2} \eta_{ab} \rd y^a \rd y^b~.
\eea

In the intersection of the two charts, with $-\ell <X^d < \ell$, the transition functions are given by 
\bea
y^a = - \frac{\ell^2}{x^2} x^a~~\implies ~ ~y^2 x^2 = \ell^4~. \label{B.9}
\eea
Further, we can see that in this intersection $x^2 <0 \Longleftrightarrow y^2 <0$.


Another local coordinate system for AdS$_d$ which has found many applications in the literature, in particular for use with the AdS/CFT correspondence, is known as \textit{Poincar\'e coordinates}.
In this coordinate system, the spacetime metric is invariant under Poincar\'e transformations of one dimension less, that is transformations of $\sISO(d-2\,,1)$. 
The local coordinates in the Poincar\'e patch, $(z\,,x^{\ha})$ with $\ha = 0\,, \ldots\,, d-2$, are related to the embedding coordinates as follows 
\bsubeq
\begin{align}
	X^{\ua} = \frac{1}{z}\left(x^{\ha}\,, - \frac{1}{2}(1-x^{2}-(\ell z)^{2})\,, - \frac{1}{2}(1+x^{2}+(\ell z)^{2})\right)\,.
\end{align}
\esubeq
Then, the AdS metric takes the form 
\begin{align}
	\rd s^{2} = \frac{1}{(\ell z)^{2}}\left( \eta_{\ha\hb}\rd x^{\ha}\rd x^{\hb} +\rd z^{2} \right)
\end{align}


Regarding the coset realisation \eqref{standard ads coset}, a global coset representative exists, see, e.g., \cite{Kuzenko:1995aq}. 
The story is different for the realisation based on the use of spin groups, eq. \eqref{bm ads spin coset}. 
In the four-dimensional case, such a realisation is given by 
\begin{align}
	\text{AdS}_{4} = \frac{\sSp(4;\mathbb{R})}{\sSL(2\,,\mathbb{C})}\,,
\end{align}
and local coset representatives based on the use of stereographic coordinates (as described above) are derived in chapter \ref{ch3}. 
This realisation is crucial to the definition of the $\N$-extended AdS superspace in four dimensions 
\begin{align}
	\text{AdS}^{4|4\N} = \frac{\sOSp(\N|4;\mathbb{R})}{\sSL(2\,,\mathbb{C})\times\sO(\N)}\,.
\end{align}
For AdS$_{5}$, eq. \eqref{bm ads spin coset} yields 
\begin{align} \label{bm ads 5 spin coset}
	\text{AdS}_5 = \frac{\sSU(2,2)}{\sUSp(2,2)}\,. 
\end{align}
A local coset representative based on the use of Poincar\'e coordinates was derived for the realisation \eqref{bm ads 5 spin coset} in \cite{CRZ}. 
Similarly to the four-dimensional case, realisation \eqref{bm ads 5 spin coset} is the starting point for the $\N$-extended AdS superspace in five dimensions, 
\begin{align}
	\text{AdS}^{5|8\N} = \frac{\sSU(2,2|\N)}{\sUSp(2,2)\times\sU(\N)}\,, 
\end{align}
which is discussed in more detail in chapter \ref{ch4}.

\section{Anti-de Sitter superspace in four dimensions} \label{bm super ads}

In this section we will turn away from the discussion regarding homogeneous spaces of the previous section, and focus our attention to the $\N=1$ and $\N=2$ anti-de Sitter superspaces in four dimensions. 
These superspaces are denoted AdS$^{4|4}$ and AdS$^{4|8}$, respectively, and we will elaborate on specific features of their supergeometry within the supergravity setting.  
The description of both as a homogeneous space, as well as further development of the geometry within the supergravity setting, is reserved for chapter \ref{ch3}.

\subsection{AdS$^{4|4}$} \label{bm n=1 ads sec}

The $\N=1$ AdS superspace was introduced in the 1970s by Keck \cite{Keck} and Zumino \cite{Zumino} as the coset superspace 
\begin{align}
	\text{AdS}^{4|4} = \frac{\sOSp(1|4;\mathbb{R})}{\sSpin(3\,,1)}\,,
\end{align}
and the comprehensive study of general supermultiplets on AdS$^{4|4}$ was carried out by Ivanov and Sorin \cite{IS}, arguably one of the most important works on AdS supersymmetry. 
Further, it was realised that AdS$^{4|4}$ arises as a maximally supersymmetric solution in off-shell formulations for $\N=1$ supergravity, see, e.g., \cite{WZ,Stelle:1978ye,Ferrara:1978em,Townsend:1977qa,Kaku:1978ea,Gates:1983nr,BK,Butter:2011vg}.

Let us introduce local coordinates for AdS$^{4|4}$ as $z^{M} = (x^{m}\,,\q^{\m}\,,\bar{\q}_{\dmu})$, where the fermionic coordinates $\q^{\m}$ and $\bar{\q}_{\dmu}$ are related as $(\q^{\m})^{*} = \bar{\q}^{\dmu}$. 
The geometry of AdS$^{4|4}$ is described by the covariant derivatives $\cD_{A}$, which take the form 
\begin{align}
	\cD_{A} = (\cD_{a}\,, \cD_{\a}\,, \cDB^{\ad}) = E_{A}{}^{M}\partial_{M} + \frac{1}{2}\O_{A}{}^{bc}M_{bc}\,.
\end{align}
In the above, $M_{ab} \Longleftrightarrow (M_{\ab}\,, \bar{M}_{\ad\bd})$ are the Lorentz generators, and their action on the covariant derivatives is given by 
\begin{align}
	[M_{ab}\,, \cD_{c}] = 2\eta_{c[a}\cD_{b]}\,, \qquad [M_{\ab}\,,\cD_{\g}] = \ve_{\g(\a}\cD_{\b)}\,, \qquad [\bar{M}_{\ad\bd}\,, \cDB_{\gd}] = \ve_{\gd(\ad}\cDB_{\bd)}\,. 
\end{align} 
The covariant derivatives $\cD_{A}$ obey the algebra 
\bsubeq
\begin{align}
	\{\cD_{\a}\,,\cD_{\b}\} &= -4\bar{\m}M_{\ab}\,, \qquad \{\cDB_{\ad}\,,\cDB_{\bd}\} = 4\m\bar{M}_{\ad\bd}
	\\
	\{\cD_{\a}\,,\cDB_{\bd}\} &= -2\ri\cD_{\a\bd}\,, 
	\\
	[\cD_{a}\,,\cD_{\b}] &= -\frac{\ri}{2}\bar{\m}(\s_{a})_{\b\gd}\cDB^{\gd}\,, \qquad [\cD_{a}\,,\cDB_{\bd}] = \frac{\ri}{2}\m(\s_{a})_{\g\bd}\cD^{\g}\,,
	\\
	[\cD_{a}\,,\cD_{b}] &= -|\m|^{2}M_{ab}\,,
\end{align}
\esubeq
where $\m\neq 0$ is a constant complex parameter, related to the AdS scalar curvature $\cR$ as $\cR = -12\m\bar{\m}$. 

The infinitesimal isometries of AdS$^{4|4}$ are generated by Killing supervectors\footnote{In chapter \ref{ch3} we will determine the Killing supervectors of AdS$^{4|4\N}$ making use of the conformal Killing supervectors of $\mathbb{M}^{4|4\N}$. This is why we use the bold notation $\bm{\x}^{A}$ to denote the components of the AdS Killing supervector with respect to the basis $E_{A}$.} defined as 
\begin{align} \label{bm n=1 killing}
	\x = \bm{\x}^{A}E_{A}\,, \qquad [\bm{\x}^{B}\cD_{B} + \frac{1}{2}\L^{cd}M_{cd}\,, \cD_{A}] = 0\,,  
\end{align}
for some Lorentz transformation determined by the parameters $\L^{cd}(z)$. 
The conditions \eqref{bm n=1 killing} turn out to completely determine $\L^{cd}(z)$ in terms of $\x$, and were solved in \cite{BK}.
They are equivalent to 
\bsubeq
\begin{align}
	\cD_{(\a}\bm{\x}_{\b)\bd} &= 0\,, \qquad ~\, \cDB^{\bd}\bm{\x}_{\a\bd} + 8\ri\bm{\x}_{\a} = 0\,, 
	\\
	\cD_{\a}\bm{\x}^{\a} &= 0\,, \qquad \cDB_{\ad}\bm{\x}_{\a} + \frac{\ri}{2}\m\bm{\x}_{\a\ad} = 0\,, 
	\\
	\cD_{\a}\bm{\x}_{\b} &= \L_{\ab}\,. 
\end{align}
\esubeq
The set of all Killing supervectors of AdS$^{4|4}$ can be shown to span the superalgebra $\frak{osp}(1|4;\mathbb{R})$. 
More details on the supergeometry of AdS$^{4|4}$ are provided in chapter \ref{ch3}.

\subsection{AdS$^{4|8}$} \label{bm n=2 ads sec}

The $\N=2$ AdS superspace can be defined\footnote{Whether the $R$-symmetry group is taken as $\sSO(2)$ or $\sO(2)$ is a matter of considering only the connected component $\sOSp_0(\N|4;\mathbb{R})$ of the $\N=2$ AdS supergroup.} as the coset superspace
\begin{align} \label{bm n=2 ads definition}
	\text{AdS}^{4|8} = \frac{\sOSp_0(2|4;\mathbb{R})}{\sSL(2\,,\mathbb{C})\times\sSO(2)}\,.
\end{align}
Here, $\sOSp_0(2|4;\mathbb{R})$ is the connected component of the $\N=2$ AdS supergroup in four dimensions, $\sOSp(2|4;\mathbb{R})$.
Its description as a maximally supersymmetric solution in the minimal off-shell formulation for $\N=2$ supergravity with a cosmological term was given in \cite{Kuzenko:2008ep, Kuzenko:2008qw, Butter:2011ym}.
In this section we will give a brief overview of certain aspects of its supergeometry.

Let us introduce the local coordinates $z^{M} = (x^{m}\,, \q_{\imath}^{\m}\,, \bar{\q}^{\imath}_{\dmu})\,,$ with $\imath = \underline{1}\,,\underline{2}$. 
The Grassmann-odd coordinates are related by the rule $(\q_{\imath}^{\m})^{*} = \bar{\q}^{\imath \dmu}$. 
The AdS$^{4|8}$ covariant derivatives then take the form
\begin{align}
	\cD_{A} = (\cD_{a}\,, \cD_{\a}^{i}\,, \cDB^{\ad}_{i}) = E_{A}{}^{M}\partial_{M} + \frac{1}{2}\O_{A}{}^{cd}M_{cd} + \F_{A}{}^{jk}\mathbb{J}_{jk}\,,
\end{align}
where $\mathbb{J}_{ij} = \mathbb{J}_{ji}$ are the $\sSU(2)$ generators. 
$\sSU(2)$ indices are raised and lowered by the $\sSU(2)$-invariant tensors $\ve^{ij}$ and $\ve_{ij}$, with $\ve_{ij} = -\ve_{ji}$ and $\ve^{12} = -\ve_{12} = +1$, by the rule 
\begin{align}
	\psi^{i} = \ve^{ij}\psi_{j}\,, \qquad \psi_{i} = \ve_{ij}\psi^{j} ~~~ \Longleftrightarrow ~~~ \ve_{ik}\ve^{kj} = \d_{i}^{j}\,.
\end{align}
Finally, the action of the $\sSU(2)$ generators on the covariant derivatives is given by 
\begin{align}
	[\mathbb{J}_{kl}\,,\cD_{\a}^{i}] = -\d^{i}_{(k}\cD_{\a l)}\,, \qquad [\mathbb{J}_{kl}\,, \cDB^{\ad}_{i}] = -\ve_{i(k}\cDB_{l)}^{\ad}\,. 
\end{align}

In the supergravity setting\footnote{In such a setting, the local structure group is taken to be $\sSL(2\,,\mathbb{C})\times\sSU(2)$. In chapter \ref{ch3} we consider the $\N$-extended version of such a setting, which we refer to as $\sSU(\N)$ superspace. Its relation to Howe's $\sU(\N)$ superspace (see  \cite{Howe:1980sy,Howe:1981gz}) is given in refs \cite{KKR, KKR2}. In the $\N=2$ case, the relation to the $\sU(2)$ superspace formulation for supergravity was given earlier in \cite{Kuzenko:2008ep,Kuzenko:2009zu}.} of \cite{Kuzenko:2008ep} (see also \cite{Grimm1980}), the AdS$^{4|8}$ algebra of covariant derivatives is given by
\bsubeq \label{bm n=2 ads alg}
\begin{align}
	\{\cD_{\a}^{i}\,,\cD_{\b}^{j}\} &= 4S^{ij}M_{\ab} +2\ve_{\ab}\ve^{ij}S^{kl}\mathbb{J}_{kl}\,, \qquad 
	\{\cD_{\a}^{i}\,, \cDB^{\bd}_{j}\} = -2\ri\d_{i}^{j}\cD_{\a}{}^{\bd}\,,
	\\
	[\cD_{a}\,,\cD_{\b}^{j}] &= \frac{\ri}{2}(\s_{a})_{\b\gd}S^{jk}\cDB^{\gd}_{k}\,, \qquad [\cD_{a}\,,\cDB_{j}^{\bd}] = \frac{\ri}{2}(\tilde{\s}_{a})^{\bd\g}S_{jk}\cD_{\g}^{k}\,, \\
	[\cD_{a}\,,\cD_{b}] &= -S^{2}M_{ab}\,,
\end{align}
\esubeq
where $S^{ij}$ obeys the following 
\bsubeq
\begin{align}
	S^{ij} = S^{ji}\,, \qquad \cD_{A}S^{jk} = 0\,, \qquad \overline{S^{ij}} = S_{ij} = \ve_{ik}\ve_{jl}S^{kl}
\end{align}
and $S^{2}$ is defined as 
\begin{align}
	S^{2} = \frac{1}{2}S^{ij}S_{ij} = \text{const.}
\end{align}
\esubeq
As pointed out in \cite{Kuzenko:2008qw}, despite the fact that the covariant derivatives have an $\sSU(2)$ connection, the $\sSU(2)$ generators appear in the algebra \eqref{bm n=2 ads alg} only in the combination $S^{kl}\mathbb{J}_{kl}$. 
It follows that the corresponding curvature is generated by a $\sU(1)$ subgroup of $\sSU(2)$, and there exists a gauge in which the $\sSU(2)$ connection takes the form 
\begin{align}
	\F_{A}{}^{kl}\mathbb{J}_{kl} = \F_{A}S^{kl}\mathbb{J}_{kl}\,. 
\end{align}
This is consistent with the definition of the coset superspace \eqref{bm n=2 ads definition}. Further implications of this, for arbitrary $\N > 1$, are discussed in chapter \ref{ch3}.

Finally, we point out that the infinitesimal isometries of AdS$^{4|8}$ are generated by Killing supervectors defined by 
\begin{align} \label{bm n=2 killing}
	\x = \bm{\x}^{A}E_{A}\,, \qquad [\bm{\x}^{B}\cD_{B} +\frac{1}{2}\L^{cd}M_{cd} + \L^{kl}\mathbb{J}_{kl}  \,, \cD_{A}] = 0\,.  
\end{align}
for some Lorentz transformation determined by the parameters $\L^{cd}(z)$ and $\sSU(2)$ transformation determined by $\L^{kl}(z)$, satisfying $\overline{\L^{kl}} = \L_{kl}$. 
As in the $\N=1$ case, the conditions \eqref{bm n=2 killing} prove to completely determine $\L^{cd}(z)$ and $\L^{kl}(z)$ in terms of $\x$. 
Their solution is given in ref. \cite{Kuzenko:2008qw}. 
In chapter \ref{ch3} we will derive the Killing supervectors of AdS$^{4|4\N}$, for arbitrary $\N$, via a different method.

\section{Supertwistor realisations} \label{st realisations}

In this section we will review the supertwistor and bi-supertwistor realisations of compactified Minkowski superspace in four dimensions, see, e.g. \cite{K-compactified12, Kuzenko:2006mv} and references therein, and of $\N$-extended AdS superspace in four dimensions, see \cite{KTM}.
We omit a comprehensive and pedagogical introduction to twistors and twistor theory, and instead focus only on those elements which are relevant to the analysis presented later in this work. 
For a pedagogical introduction, see, e.g., \cite{A-twist,PR,Huggett:1986fs,Ward:1990vs}. 

\subsection{The (bi-)supertwistor realisation of compactified Minkowski superspace in four dimensions} \label{4d compact str}

The formalism presented in this section makes use of supertwistors of $\sSU(2,2|\N)$, for which our conventions are contained in appendix \ref{5d appendix}.

Our starting point is to consider the space of even two-planes associated with supertwistor space $\mathbb{C}^{4|\N}$.
Such a two-plane is generated by two supertwistors $\bm{T}^{\m} = (\bm{T}_{A}{}^{\m})$, with $\m = 1\,,2\,,$ such that the bodies\footnote{The terminology ``body'' and ``soul'' follows \cite{DeWitt}. We remind the reader that a supernumber may be written as $z = z_{B} + \sum\limits_{k=1}^{\infty} \frac{1}{k!}C_{i_1 \ldots i_k} \x_{i_1}\ldots \x_{i_k}$, with $i_{1},i_{2} = 1\,,2\,,\ldots$. Here, $z_{B} \in \mathbb{C}$ and is known as the body of the supernumber $z$, $C_{i_{1}\ldots i_{k}} \in \mathbb{C}$, and $\x_{i}$ are the generators of the (infinite-dimensional) Grassmann algebra, $\x_{i}\x_{j} = - \x_{j}\x_{i}$. The soul of $z$ is the `bodiless' component, given by $z_{S} = z - z_{B} = \sum\limits_{k=1}^{\infty} \frac{1}{k!}C_{i_1 \ldots i_k} \x_{i_1}\ldots \x_{i_k}$. It is the component of $z$ which is expanded purely in terms of the generators of the Grassmann algebra.} of $\bm{T}_{\ah}^{1}$ and $\bm{T}_{\ah}^{2}$ are linearly independent. 
Compactified Minkowski superspace can be identified with the space of null-two planes through the origin in supertwistor space, $\mathbb{C}^{4|\N}$, see, e.g., \cite{ManinNi}. 
In particular, the null condition implies the following constraints on the supertwistors $\bm{T}^{\m}$
\begin{align} \label{compact null}
	\braket{\bm{T}^{\m}}{\bm{T}^{\n}} \equiv \bar{\bm{T}}^{\m}\bm{T}^{\n} = 0\,.
\end{align}
The two supertwistors $\bm{T}^{\m}$ form a basis for the null two-plane. 
As such, they are defined modulo the equivalence relation 
\begin{align} \label{compact equiv}
	\bm{T}^{\m} \sim \widetilde{\bm{T}}{}^{\m} = \bm{T}^{\n}M_{\n}{}^{\m}\,, \qquad M \in \sGL(2\,,\mathbb{C})\,,
\end{align}
as the basis $\{\widetilde{\bm{T}}{}^{\m}\}$ spans the same null two-plane. 
Given such a two-plane, it is convenient to express it in the following form 
\begin{align}
	(\bm{T}_{A}{}^{\m}) = \left(\begin{array}{c}
		F_{\a}{}^{\m} \\
		G^{\ad\m}
		\\
		\hline\hline
		\Q_{i}{}^{\m}
	\end{array}\right)\,, \qquad i = 1\,, \ldots\,, \N\,.
\end{align}
Then, the null condition eq. \eqref{compact null} implies the following constraints on the matrices $F = (F_{\a}{}^{\m})\,, ~ G = (G^{\ad\m})\,,$ and $\Q = (\Q_{i}{}^{\m})$
\begin{align} \label{compact null blocks}
	F^{\dag}G + G^{\dag}F = \Q^{\dag}\Q\,.
\end{align}

Let us now consider the open domain where the $2\times2$ block $F_{\a}{}^{\m}$ is non-singular. 
As we will see shortly, Minkowski superspace can be identified with this open domain. 
Making use of the equivalence relation \eqref{compact equiv}, one can choose $F_{\a}{}^{\m} = \d_{\a}{}^{\m}$. 
The conditions \eqref{compact null blocks} are then solved by two-planes of the form 
\begin{align} \label{mink 2plane}
	(\bm{T}_{A}{}^{\m}) = \left(\begin{array}{c}
	\d_{\a}{}^{\m} 
	\\
	-\ri x_{+}^{\ad\m}
	\\
	\hline\hline
	2\q_{i}{}^{\m}
\end{array}\right)\,, \qquad x_{+}^{\ad\m} = x_{+}^{m}(\tilde{\s}_{m})^{\ad\m}\,, \qquad x_{+}^{m} = x^{m} + \ri\q_{i}\s^{m}\bar{\q}^{i}\,.
\end{align}
We will refer to this region as the Minkowski chart. 
We can see from eq. \eqref{mink 2plane} that, in the Minkowski chart, the two-planes are parametrised by the chiral coordinates $x_{+}^{m}$ and $\q_{i}^{\m}$. 
Further, we find that the two-planes in the Minkowski chart coincide with the first two columns of the coset representative for the $\N$-extended Minkowski superspace, see eq. \eqref{smink global crep} and eq. \eqref{bm reduced coset rep}.

An infinitesimal $\sSU(2,2|\N)$ transformation, $g = \id_{4+\N} + \mathfrak{g}$, acts on $\bm{T}_{A}{}^{\m}$ as 
\begin{align}
	\bm{T}_{A}{}^{\m} \rightarrow \bm{T}'_{A}{}^{\m} = \big(\d_{A}{}^{B} + \mathfrak{g}_{A}{}^{B}\big)\bm{T}_{B}{}^{\g} N_{\g}{}^{\m}\,, \qquad N \in \sSL(2,\mathbb{C})\,, 
\end{align}
where $\mathfrak{g}$ is defined in \eqref{su22n element}.
One can show that the coordinates $x_{+}^{m}$ and $\q_{i}^{\m}$ transform as 
\bsubeq \label{scf trf bm}
\begin{align}
	\d \tilde{x}_{+} &= \tilde{a} + \frac{1}{2}(\D + \bar{\D})\tilde{x}_{+} - \bar{\o}\tilde{x}_{+} - \tilde{x}_{+}\o + \tilde{x}_{+}b\tilde{x}_{+} + 4\ri\bar{\e}\q - 4\tilde{x}_{+}\eta\q\,, \label{scf trf 1}
	\\
	\d \q &= \e + \frac{1}{2\N}\big((\N-2)\D + 2\bar{\D}\big)\q - \q\o + \L\q + \q b\tilde{x}_{+} - \ri\bar{\eta}\tilde{x}_{+} - 4\q\eta\q\,, \label{scf trf 2}
\end{align}
\esubeq
where the parameters $\D$, $a^{a}$, and $b^{a}$ are related to those in \eqref{su22n element} by the rule
\bsubeq
\begin{align}
	\D &= (\o^{45} - \frac{2\ri\N}{\N-4}\t)\,,
	\\
	a^{a} &= - \frac{1}{2}(\o^{a4} - \o^{a5})\,,
	\\
	b^{a} &= \frac{1}{2}(\o^{a4} + \o^{a5})\,. 
\end{align}
\esubeq
They correspond to 4D Lorentz transformations $(\o\,, \bar{\o})$, translations $(a)$, special conformal transformations $(b)$, Q-supersymmetry $(\e\,, \bar{\e})$ and S-supersymmetry $(\psi\,, \bar{\psi})$ transformations, combined scale and chiral  transformations $(\D)$, and $\sSU(\N)$ transformations $(\L)$. 
The expressions \eqref{scf trf bm} coincide with the standard superconformal transformations on Minkowski superspace, see, e.g., \cite{Park:1999pd,Kuzenko:2006mv}. 
To help illustrate this situation, we will switch off the Grassmann variables and consider infinitesimal Poincar\'e transformations. 
We first point out that, under finite Poincar\'e transformations, the coordinates $\tilde{x} = x^{m}\tilde{\s}_{m}$ of Minkowski space transform as follows
%
\begin{align}
	\tilde{x} \rightarrow N^{\dag}\tilde{x} N + \tilde{a}\,, \qquad N \in \sSL(2\,,\mathbb{C})\,, \qquad \tilde{a} = \tilde{a}^{\dag}\,.
\end{align}
Taking $N_{\a}{}^{\m}$ and $a^{\ad\m}$ to be infinitesimal, $N_{\a}{}^{\m} = \d_{\a}{}^{\m} - \o_{\a}{}^{\m}$, leads to the following 
\begin{align} \label{infini pp trf}
	x^{\ad\m} \rightarrow - \bar{\o}^{\ad}{}_{\bd}\tilde{x}^{\bd\m} - x^{\ad\a}\o_{\a}{}^{\m} + a^{\ad\m}\,,
\end{align}
which is recreated in the relation \eqref{scf trf 1} by setting all parameters other than $\o$ and $a$ to zero. 

Closely related to the supertwistor realisation described above is the bi-supertwistor realisation for compactified Minkowski superspace. 
In the space of graded antisymmetric supermatrices $\bm{X} = (\bm{X}_{AB})$, with
$\bm{X}_{AB} = -(-1)^{\e_{A}\e_{B}}\bm{X}_{BA}$, let us consider a surface $\frak{L}$ defined by the constraints 
\bsubeq
\begin{align}
	\bar{\bm{X}}{}^{AB}\bm{X}_{BC} &= 0\,,
	\\
	\bm{X}_{[AB} \bm{X}_{CD\}} &= 0\,, \label{g antisym mink bst}
\end{align}
\esubeq
where $[\ldots\}$ denotes the graded antisymmetrisation of indices.
In the above, the dual supermatrix $\bar{\bm{X}} = (\bar{\bm{X}}{}^{AB})$ is defined as $\bar{\bm{X}} = \bm{\O}\bm{X}^{\dag}\bm{\O}$.
An important implication of relation \eqref{g antisym mink bst} is that the body of the bosonic block $\bm{X}_{ij}$, defined in 
\begin{align}
	\bm{X} = \left(\begin{array}{c||c}
		\bm{X}_{\ah\bh} & \bm{X}_{\ah j} \\
		\hline \hline 
		\bm{X}_{i \bh} & \bm{X}_{ij}
	\end{array}\right)\,,
\end{align}
vanishes.
Finally, we require that $\bm{X}_{\ah\bh}$ is a non-zero $4\times 4$ antisymmetric matrix. 
On the surface $\frak{L}$, we introduce the equivalence relation 
\begin{align}
	\bm{X}_{AB} \sim \l \bm{X}_{AB}\,, \qquad \l \in \mathbb{C} - \{0\}\,.
\end{align}
It turns out that the space $\frak{L} / \sim$ is equivalent to the supertwistor realisation of compactified Minkowski superspace introduced above, see \cite{K-compactified12} for the proof. 
The bi-supertwistors $\bm{X}_{AB}$ and the supertwistors $\bm{T}_{A}{}^{\m}$ are related by the rule 
\begin{align}
	\bm{X}_{AB} = \ve_{\m\n}\bm{T}_{A}{}^{\m}\bm{T}_{B}{}^{\n}\,.
\end{align} 
In particular, in the Minkowski chart \eqref{mink 2plane}, the bi-supertwistor $\bm{X} = (\bm{X}_{AB})$ takes the form 
\begin{align} \label{bm mink bst}
		(\bm{X}_{AB}) = 
	\left(
	\begin{array}{c|c||c}
		\ve_{\ab} 
		& 
		-\ri  x_{+}{}_{\a}{}^{\bd} 
		& 
		2 \q_{\a j}
		\\
		\hline 
		\ri  x_{+}{}^{\ad}{}_{\b}
		&
		- x_{+}^{2} \ve^{\ad\bd} 
		&
		-2 \ri x_{+}{}^{\ad \g}\q_{\g j} 
		\\
		\hline \hline 
		-2 \q_{i \b}
		&
		2 \ri \q_{i \g} x_{+}{}^{\g \bd} 
		&
		4  \q_{i}{}^{\a}\q_{\a j}
	\end{array}
	\right)\,.
\end{align}

\subsection{The (bi-)supertwistor realisation of AdS$^{4|4\N}$} \label{ads4 bst}

This section is dedicated to a review of the (bi-)supertwistor realisation for AdS$^{4|4\N}$ developed in \cite{KTM}. 
The situation here differs slightly to that of the previous subsection, as the AdS$^{4|4\N}$ supertwistors\footnote{The situation is again different for AdS$^{5|8\N}$ as described in chapter \ref{ch4}. These differences are based on the well-known fact that the conformal group in $d$ dimensions and the AdS group in $d+1$ dimensions are locally isomorphic. It is then appropriate that column vectors of the conformal group in $d$ dimensions are used when describing a global realisation of AdS$_{d+1}$.} are defined to transform in the fundamental representation of $\sOSp(\N|4;\mathbb{R})$, rather than $\sSU(2,2|\N)$. 
Technical details for the supertwistors utilised in this section and the supergroup $\sOSp(\N|4;\mathbb{R})$ can be found in appendix \ref{Supertwistors}.

In analogy with the story for compactified Minkowski superspace described above, let us consider the space of even two-planes $\cP = (\bm{T}_{A}{}^{\m})$ in the space of even complex supertwistors $\mathbb{C}^{4|4\N}$.
Our point of departure from the above analysis is in the constraints which define our surface of interest. 
In contrast to the null two-planes considered above, we consider those two-planes satisfying the conditions
\begin{subequations}\label{planecon}
\bea
\det \big( \cP^{\rm sT} {\mathbb J} \cP \big) &\neq &0~, \label{planecon.a}\\
\cP^\dagger {\mathbb J} \cP \equiv (* \cP)^{\rm sT} {\mathbb J} \cP &=& 0~.\label{planecon.b}
\eea
\end{subequations}
In the above, $*\cP = (*\bm{T}_{A}{}^{1}\,, *\bm{T}_{A}{}^{2})$ denotes the conjugate of the even supertwistors $\bm{T}_{A}{}^{\m}$, see eq. \eqref{A.14}.  
We emphasise that these conditions are preserved under $\sOSp(\N|4;\mathbb{R})$ transformations, rather than those of $\sSU(2,2|\N)$, and under the equivalence transformations 
\begin{align} \label{equiv}
	\bm{T}^\mu \sim \widetilde{\bm{T}}{}^\m=\bm{T}^\nu M_\nu{}^\mu\,, \qquad M=(M_\m{}^\n)  \in \sGL(2,\mathbb{C})\,.
\end{align}
The basis $\{\widetilde{\bm{T}}{}^{\m}\}$ defines the same two-plane $\cP$. 
The conditions \eqref{planecon} imply that the bodies of the four even supertwistors $\bm{T}^{\hat \m} =(\bm{T}^\m,*\bm{T}^{\dot\m} )$ form a basis for ${\mathbb C}^4$.
In what follows, the supertwistor $*\bm{T}$ will be denoted $\bar{\bm{T}}$.

Any pair of complex even supertwistors $\cP$ satisfying the constraints \eqref{planecon} will be referred to as a frame. 
The space of such frames is denoted $\frak{F}_{\N}$, and $\sOSp(\N|4;\mathbb{R})$ acts on $\frak{F}_{\N}$ by the rule 
\bea
g (\bm{T}^\m , \bar{\bm{T}}^{\dot \m} ) = (g \bm{T}^\m , g  \bar{\bm{T}}^{\dot \m} )~, 
\qquad g \in \sOSp(\cN|4; {\mathbb R})~.
\label{2.4}
\eea
This action extends naturally to the quotient space $\frak{F}_{\N}/\sim$. 
As was shown in \cite{KTM}, it turns out that the quotient space $\frak{F}_{\N}/\sim$ can be identified with the $\N$-extended anti-de Sitter superspace in four dimensions, 
\begin{align}
	\text{AdS}^{4|4\N} = \frak{F}_{\N}/\sim \,. \label{3.11}
\end{align}
This will be proven in chapter \ref{ch3}, based on an alternative realisation of the AdS supergroup.

A powerful application of a global realisation of AdS$^{4|4\N}$ such as that described above lies in the construction of $\sOSp(\N|4;\mathbb{R})$-invariant functions. 
Indeed, reference \cite{KTM} developed such invariant functions of two points making use of this formalism. 
For any two frames $\cP\,, \tilde{\cP} \in \frak{F}_{\N}$, the following functions are invariant under $\sOSp(\N|4;\mathbb{R})$ transformations
\begin{subequations}\label{two-point}
\bea
\frac{1}{\ell^2}
\o( \cP, \widetilde{\cP}) &:=& -2\frac{ \langle \bar{\bm{T}}^{\dot \m} | \widetilde{\bm{T}}{}^\n \rangle 
 \langle  \bar{\bm{T}}_{\dot \m} | \widetilde{\bm{T}}_\n \rangle   } 
{ 
\langle  \bar{\bm{T}}{}^{\dot \s} | \bar{\bm{T}}_{\dot \s} \rangle 
\langle  \widetilde{\bm{T}}{}^\r | \widetilde{\bm{T}}_\r \rangle
}~, \label{two-point.a}
\\
\frac{1}{\ell^2}
\o_{(+)} (\cP, \widetilde{\cP}) &:=& 2\frac{ \langle  \bm{T}^{ \m} | \widetilde{\bm{T}}{}^\n \rangle  
 \langle  \bm{T}_{ \m} | \widetilde{\bm{T}}_\n \rangle   } 
{ 
\langle  \bm{T}^{ \s} |  \bm{T}_{ \s} \rangle
\langle  \widetilde{\bm{T}}{}^\r | \widetilde{\bm{T}}_\r \rangle } -1~,  \label{two-point.b}\\
\frac{1}{\ell^2}
\o_{(-)}( \cP, \widetilde{\cP}) &:=& 2\frac{ \langle  \bar{\bm{T}}{}^{\dot \m} | \widetilde{\bar{\bm{T}}}{}^{\dot \n} \rangle
 \langle  \bar{\bm{T}}_{\dot \m} | \widetilde{\bar{\bm{T}}}_{\dot \n} \rangle   } 
{ 
\langle  \bar{\bm{T}}{}^{\dot \s} | \bar{\bm{T}}_{\dot \s} \rangle
\langle  \widetilde{\bar{\bm{T}}}{}^{\dot \r} | \widetilde{\bar{\bm{T}}}_{\dot \r} \rangle 
}-1 ~,  \label{two-point.c}
\eea
\end{subequations}
for a fixed positive parameter $\ell$.
They are also invariant under arbitrary equivalence transformations of the form \eqref{equiv}, and so they are well defined on the quotient space \eqref{3.11}.
In the non-supersymmetric case, $\N=0$, the three functions coincide. 
This means that for $\N\geq 1$ they differ only in the fermionic sector. We will elaborate on this in chapter \ref{ch3}. 
The two-point functions \eqref{two-point} can be viewed as the AdS-analogues of those introduced in the context of superconformal symmetry in \cite{Park:1997bq,Park:1998nra,Park:1999pd,Park:1999cw}.

As we saw in subsection \ref{4d compact str}, there exists an alternative realisation of four-dimensional compactified Minkowski superspace in terms of bi-supertwistors. 
This was also developed for AdS$^{4|4\N}$ in \cite{KTM}. 
In the space of graded antisymmetric supermatrices $\bm{X}_{AB} = -(-1)^{\e_{A}\e_{B}}\bm{X}_{BA}$ let us consider the surface defined by the following constraints
\begin{subequations} \label{bi-super2}
\bea
\bm{X}_{[AB} \bm{X}_{CD \}} &=&0~,  \label{bi-super2.a}\\
(-1)^{\e_B} \bm{X}_{AB} {\mathbb J}^{BC} \bm{X}_{CD} &=& \ell \bm{X}_{AD} ~,\\
 {\mathbb J}^{BA}  \bm{X}_{AB}&=& 2\ell~, \\
(-1)^{\e_B} \bm{X}_{AB} {\mathbb J}^{BC} \bar{\bm{X}}_{CD} &=& 0~.
\eea
\end{subequations}
It turns out that this surface is equivalent to the quotient space \eqref{3.11}.
Indeed, the bi-supertwistors \eqref{bi-super2} and supertwistors $\bm{T}_{A}{}^{\m}$ are related by the rule
\begin{subequations}\label{bi-super}
\bea
\bm{X}_{AB} &:=& -2 \ell \frac{ \bm{T}_A{}^{ \m}  {\bm{T}}_{B}{}^\n \ve_{\m\n} } 
{ \langle   \bm{T}^{ \g} |  \bm{T}^{\d} \rangle \ve_{\g \d} }
=- (-1)^{\e_A \e_B}  \bm{X}_{BA}~,
\\
\bar{\bm{X}}_{AB} &:=& -2 \ell\frac{ \bar{\bm{T}}_A{}^{ \dmu}  \bar{\bm{T}}_B{}^{\dnu}  \ve_{\dmu \dnu}  } 
{ \langle   \bar{\bm{T}}^{ \dot \g} |  \bar{\bm{T}}^{\dot \d} \rangle  \ve_{\dot \g \dot \d}}
=- (-1)^{\e_A \e_B}  \bar{\bm{X}}_{BA}~.
\eea
\end{subequations} 
We point out that these supermatrices are invariant under arbitrary equivalence transformations of the form \eqref{equiv}. 

The two-point functions \eqref{two-point} can be conveniently expressed in terms of bi-supertwistors.
In terms of the supermatrices $\bm{X} = (\bm{X}_A{}^B) $ and $\bar{\bm{X}}= (\bar{\bm{X}}_A{}^B)$ defined by\footnote{Here we emphasise that the index structure of $\bm{X}$ differs compared to the case for compactified Minkowski superspace, as, when considering the supergroup $\sOSp(\N|4;\mathbb{R})$, one can raise and lower indices according to the rules \eqref{bm lowered bst}.} 
\bea \label{bm lowered bst}
\bm{X}_A{}^B = (-1)^{\e_C} \bm{X}_{AC} {\mathbb J}^{CB}~,\qquad 
\bar{\bm{X}}_A{}^B = (-1)^{\e_C} \bar{\bm{X}}_{AC} {\mathbb J}^{CB}~,
\eea 
these expressions have the form:
\begin{subequations} \label{str two-points}
\bea
\o( \cP, \widetilde{\cP}) &=& -\frac 12 {\rm Str} \big( \bar{\bm{X}} \widetilde{\bm{X}} \big)~, \\
\o_{(+)}( \cP, \widetilde{\cP}) &=& \frac 12 {\rm Str} \big(\bm{X} \widetilde{\bm{X}} \big)-\ell^2~,\\
\o_{(-)}( \cP, \widetilde{\cP}) &=&\frac 12 {\rm Str} \big( \bar{\bm{X}} \widetilde{\bar{\bm{X}}} \big)-\ell^2~.
\eea
\end{subequations}
We point out that the $ \sOSp(\cN|4; {\mathbb R})$ transformation \eqref{2.4} acts on $\bm{X}$ and $\bar{\bm{X}}$ as follows 
\bea
\bm{X}~ \to ~ g \bm{X}g^{-1} ~, \qquad \bar{\bm{X}}~ \to ~ g \bar{\bm{X}}g^{-1} ~.
\eea

Below we will elaborate on certain details of the non-supersymmetric case, $\N=0$, of the formalism described above. 
This bi-twistor formalism proves to be equivalent to the bi-spinor formalism for AdS$_{4}$ introduced in \cite{Binder:2020raz}. 
In this case, the bi-twistors $X_{\ah\bh}$ can be shown to satisfy the reality condition
\bea
X_{\langle \hal \hbe \rangle} 
+ {\bar X}_{\langle \hal \hbe \rangle } =0 ~,\qquad 
X_{\langle \hal \hbe \rangle  } := X_{ \hal \hbe } - \frac{\ell}{2} J_{\hal \hbe}~, \quad 
\bar X_{\langle \hal \hbe \rangle  } := \bar X_{ \hal \hbe } - \frac{\ell}{2} J_{\hal \hbe}~,
\eea
where $X_{\langle \hal \hbe \rangle} $ and $\bar X_{\langle \hal \hbe \rangle} $ denote 
the $J$-traceless parts of $X_{ \hal \hbe } $ and $\bar X_{ \hal \hbe   } $, respectively,
\bea
J^{ \hal \hbe } X_{\langle \hal \hbe \rangle} =0~, \qquad 
J^{ \hal \hbe } \bar X_{\langle \hal \hbe \rangle} =0~.
\eea
Making use of $X_{\ah\bh}$, we can construct a real 5-vector
\bea
X_{\hat a} := \frac{1}{2} (J\G_{\hat a} )^{\hal \hbe} X_{\langle \hal  \hbe \rangle } 
=\frac{1}{2} (J\G_{\hat a} )^{\hal \hbe} X_{ \hal \hbe } 
~, \qquad \hat a = 0,1,2,3,4 ~.
\eea
Here $\G_{\hat a}= \big( (\G_{\hat a})_\hal{}^\hbe\big)$ are {\it real} $4\times 4$ matrices which obey the anti-commutation relations 
\bea
\{ \G_{\hat a}  , \G_{\hat b}  \} = 2\eta_{\hat a \hat b} {\mathbbm 1}_4~, \qquad 
\eta_{\hat a \hat b}= {\rm diag} \, (-1,+1,+1,+1,-1)~, 
\eea
and are characterised by the property
\bea
\G_{\hat a}^{\rm T} = J  \G_{\hat a}  J^{-1} \quad \Longleftrightarrow \quad
(J  \G_{\hat a})^{\rm T} = -J  \G_{\hat a} ~. 
\eea
They constitute a Majorana representation of the gamma-matrices for $\mathbb{R}^{3\,,2}$, see appendix \ref{spinor conventions}. 
The explicit realisation of $\G_{\hat a}$ is given by eq. \eqref{majorana gamma 3 2}.
Making use of the completeness relation
\bea
(J\, \G^{\hat a})^{\hat \a \hat \b} (J\, \G_{\hat a} )^{\hat \g \hat \d} 
= - J^{\hat \a \hat \b} J^{\hat \g \hat \d} 
+ 2 (J^{\hat \a \hat \g} J^{\hat \b \hat \d} - J^{\hat \a \hat \d} J^{\hat \b \hat \g} )~,
\eea 
we obtain 
\bea
X^{\hat a} X_{\hat a} = - \ell^2~.
\eea

Let us discuss the technical and conceptual differences between the twistor/bi-twistor, and bi-spinor formalisms for AdS$_4$. As pointed out above, in the non-supersymmetric case, $\N=0,$ the technical details of the bi-twistor formalism and the bi-spinor formalism introduced in \cite{Binder:2020raz} coincide. 
Specifically, the basic variables are constrained objects $T_{\ah}{}^{\m}$ which have an $\sSp(4;\mathbb{R})$ index $\ah$ and an $\sSL(2\,,\mathbb{C})$ index $\m$. 
The two approaches differ in their origins and interpretation. 
On the one hand, the bi-spinor formalism is introduced by constructing objects which transform covariantly in the spinor representation of both the AdS$_4$ group and the Lorentz group. This leads naturally to $T_{\ah}{}^{\m}$ as described above. 
As we will see in chapter \ref{ch4}, this is similar to how the supertwistor realisation of AdS$^{5|8\N}$ is introduced. 
On the other hand, the above situation is a gauge-fixed version of the twistor/bi-twistor formalism. 
In the twistor/bi-twistor picture, AdS$_4$ is identified with a certain surface in the twistor space. Specifically, points in AdS$_4$ are associated with null two-planes in the twistor space, satisfying the conditions \eqref{planecon}.
An advantage of this approach is that the relationship between the (super)twistor realisation of AdS (super)space and the compactified Minkowski (super)space in four dimensions is made explicit, as derived in appendix \ref{ads and mink}.

Finally, we will discuss the introduction of a local coordinate system for AdS$^{4|4\N}$. 
For this purpose it is convenient to express the two-plane in the form 
\begin{align}
	\cP = (\bm{T}_{A}{}^{\m}) = \left(\begin{array}{c}
		F_{\a}{}^{\m}
		\\
		G^{\a\m}
		\\
		\hline \hline 
		\Q_{I}{}^{\m}
	\end{array}\right)\,. 
\end{align}
Then, the conditions \eqref{planecon} imply that the blocks $F\,, G$ and $\Q$ satisfy the following conditions
\bsubeq
\begin{align}
	F^{\T}G-G^{\T}F - \ri\Q^{\T}\Q &= \ri \ell \ve\,,
	\\
	F^{\dag}G-G^{\dag}F + \ri\Q^{\dag}\Q &= 0\,,
\end{align}
\esubeq
for some parameter $\ell\,.$
Making use of the equivalence relations \eqref{equiv}, one can choose $\ell >0\,.$ 
Let us now consider an open domain where the $2\times 2$ block $F_{\a}{}^{\m}$ is non-singular.
Again, making use of the equivalence relation \eqref{equiv}, one can choose $F_{\a}{}^{\m} \sim \d_{\a}{}^{\m}$. 
We then find the general solution for two-planes
\begin{align}
	\cP = \frac{1}{\sqrt{z_{-}}} \left(\begin{array}{c}
		\d_{\a}{}^{\m}\\
		-x_{-}^{\b\m} + \frac{\ri}{2}(\ell z_{-} + \q^{2})\ve^{\b\m}
		\\ \hline \hline
		\ri\sqrt{2}\q_{I}{}^{\m}
	\end{array}\right)\,,
\end{align}
with 
\begin{align}
	z_{\pm} = z\pm \frac{1}{2\ell}(\q-\bar{\q})^{2}\,, \qquad \q^{2} = \q_{I}{}^{\a}\q_{I\a}\,, \qquad \bar{\q}^{2} = \bar{\q}_{I}{}^{\a}\bar{\q}_{I\a}\,. 
\end{align}
This coordinate system was derived in \cite{KTM}, and is the four-dimensional analogue of the Poincar\'e like coordinates detailed in chapter \ref{ch4} for AdS$^{5|8\N}$ and those for AdS$_{(3|p\,,q)}$ derived in \cite{KT}. 
In chapter \ref{ch3} we derive another coordinate system, based on the use of stereographic coordinates, for which we make use of the alternative realisation of the AdS supergroup detailed in appendix \ref{adsgroup}.

 
\begin{subappendices}

\section{Spinor conventions} \label{spinor conventions}

This appendix is dedicated to outlining our spinor conventions in various dimensions. 

\subsection{$\sSpin(3,1)$} \label{3,1 conv}
Our two-component spinor conventions follow \cite{BK} and are similar to those used in \cite{WB}.

Consider the space $\mathbb{R}^{3,1}$ equipped with metric
\begin{align}
	\eta_{ab} = \text{diag}(-1\,,+1\,,+1\,,+1)\,, \qquad \qquad a = 0,1,2,3\,. 
\end{align}
The gamma matrices in $3+1$ dimensions satisfy 
\begin{align}
	\{\g_{a}\,,\g_{b}\} = -2\eta_{ab}\id_{4}\,,
\end{align}
and can be chosen as
\begin{align} \label{4d gm}
	\g_{a} = \left(  (\g_{a})_{\ah}{}^{\bh} \right) = 
	\left(\begin{array}{cc}
		0 & \s_{a} \\
		\tilde{\s}_{a} & 0
	\end{array}\right)\,, \qquad \ah = 1\,,2\,,3\,,4\,,
\end{align}
with 
\begin{align}
	\s_{a} &= (\id_{2}\,, \vec{\s}) \equiv \big((\s_{a})_{\a\ad}\big)\,, \quad \tilde{\s}_{a} = (\id_{2}\,, -\vec{\s}) \equiv \big((\tilde{\s}_{a})^{\ad\a}\big)\,, \quad \a = 1\,,2\,, \quad \ad= \dot{1} \,, \dot{2}\,.
\end{align}
The matrix $\g_{5} = -\ri\g_0\g_1\g_2\g_3$ then takes the form 
\begin{align}
	\g_{5} = \left(\begin{array}{cc}
		\id_{2} & 0 \\
		0 & -\id_{2}
	\end{array}\right)\,.
\end{align}
The Lorentz generators in the spinor representation are 
\begin{align}
	\S_{ab} = -\frac{1}{4}[\g_{a}\,,\g_{b}] = \left((\S_{ab})_{\ah}{}^{\bh}\right) \,, 
\end{align}
They are given by 
\bsubeq 
\begin{align}
	\S_{ab} &= \left(\begin{array}{cc}
		(\s_{ab})_{\a}{}^{\b} & 0
		\\
		0 & (\tilde{\s}_{ab})^{\ad}{}_{\bd}
	\end{array}\right)\,, 
	\\
	(\s_{ab})_{\a}{}^{\b} &= -\frac{1}{4}(\s_{a}\tilde{\s}_{b}-\s_{b}\tilde{\s}_{a})_{\a}{}^{\b} \,,
	\\
	(\tilde{\s}_{ab})^{\ad}{}_{\bd} &= -\frac{1}{4}(\tilde{\s}_{a}\s_{b}-\tilde{\s}_{b}\s_{a})^{\ad}{}_{\bd}\,.
\end{align}
\esubeq
Two-component spinor indices are raised and lowered using the spinor metrics
\begin{subequations} \label{epsilon def}
\bea
\ve^{\ab} &=& - \ve^{\b \a} ~, \qquad \ve_{\ab} = - \ve_{\b \a} ~, \qquad \ve^{12}= \ve_{21} =1~; \\
\ve^{\ad \bd} &=& - \ve^{\bd \ad} ~, \qquad \ve_{\ad \bd} = - \ve_{\bd \ad} ~, \qquad \ve^{\dot 1 \dot 2}= \ve_{\dot 2 \dot 1} =1~,
\eea
\end{subequations}
by the rules: 
\begin{align} \label{raise and lower}
	\psi^{\a} := \ve^{\ab}\psi_{\b}\,, \qquad \psi_{\a} = \ve_{\ab}\psi^{\b}\,; \qquad 
	\bar \phi^{\ad} := \ve^{\ad \bd}\bar \phi_{\bd}\,, \qquad \bar \phi_{\ad} = \ve_{\ad \bd} \bar \phi^{\bd}\,.
\end{align}

The charge conjugation matrix, $\mathcal{C} $, is defined by 
\bsubeq \label{4d charge conjugation}
\begin{align} 
	\mathcal{C}^{-1}\g_{a}\mathcal{C} = -\g_{a}^{\T}
	\quad \implies \quad 
	\mathcal{C}^{-1}\g_{5}\mathcal{C} = \g_{5}^{\T}\,.
\end{align}
It has the properties
\begin{align}
	\mathcal{C}^{\dag} = \mathcal{C}^{\T} = -\mathcal{C} = \mathcal{C}^{-1}\,,
\end{align}
and can be chosen as 
\begin{align}
	\mathcal{C} = \left(
	\begin{array}{cc}
		\ve_{\ab} & 0 \\
		0 & \ve^{\ad\bd}
	\end{array}
	\right)\,. 
\end{align}
\esubeq

A number of useful properties are satisfied by the matrices $\s_{a}$ and $\tilde{\s}_{a}$. 
They include: 
\bsubeq 
\begin{align}
	(\s_a \tilde{\s}_{b} + \s_{b}\tilde{\s}_{a})_{\a}{}^{\b} &= -2\eta_{ab}\d_{\a}{}^{\b}\,,
	\\
	(\tilde{\s}_{a}\s_{b} + \tilde{\s}_{b}\s_{a})^{\ad}{}_{\bd} &= -2\eta_{ab}\d^{\ad}{}_{\bd}\,,
	\\
	\text{Tr}(\s_{a}\tilde{\s}_{b}) &= -2\eta_{ab}\,,
	\\
	(\s^{a})_{\a\ad}(\tilde{\s}_{a})^{\bd\b} &= -2\d_{\a}{}^{\b}\d_{\ad}{}^{\bd}\,. 
\end{align}
\esubeq

Associated with a four-vector $x^{a}$ are the Hermitian matrices 
\begin{align}
	x_{\a\ad} = x^{a}(\s_{a})_{\a\ad}\,, \quad x^{\ad\a} = x^{a}(\tilde{\s}_{a})^{\ad\a}\,. 
\end{align}
It is sometimes convenient to use the condensed notation 
\begin{align}
	x := (x_{\a\ad})\,, \qquad \tilde{x}:= (x^{\ad\a})\,. 
\end{align}
\subsection{$\sSpin(3,2)$} \label{3,2 conv}
Consider the space $\mathbb{R}^{3,2}$ equipped with metric 
\begin{align}
	\eta_{\ha\hb} = \text{diag}(-1\,,+1\,,+1\,,+1\,,-1)\,, \qquad \ha = 0\,,1\,,2\,,3\,,4\,.
\end{align}
The gamma matrices satisfy 
\begin{align}
	\{\g_{\ha}\,,\g_{\hb}\} = -2\eta_{\ha\hb}\id_{4}\,, \qquad (\g_{\ha})^{\dag} = J\g_{\ha}J\,, 
\end{align}
with 
\begin{align}
\g_{\ha} = \left((\g_{\ha})_{\ah}{}^{\bh}\right)\,,
\end{align}
and
\begin{align}
	J := \g_{0}\g_{4} = -J^{-1}\,. 
\end{align}
They can be chosen as 
\begin{align}
	\g_{a} = \left(\begin{array}{cc}
		0 & (\s_{a})_{\a\bd} \\
		(\tilde{\s}_{a})^{\ad\b} & 0
	\end{array}\right)\,, \qquad 
	\g_{4} = \left(\begin{array}{cc}
		\d_{\a}{}^{\b} & ~0~ \\
		~0~ & -\d^{\ad}{}_{\bd}
	\end{array}\right) \,.
\end{align}
The charge conjugation matrix satisfies 
\begin{align}
	C (\g_{\ha}) C^{-1} = (\g_{\ha})^{\T}\,,
\end{align}
and can be chosen as 
\begin{align} \label{3,2 cc matrix}
	C = (\ve^{\ah\bh}) = \left(\begin{array}{cc}
		\ve^{\a\b} & 0 \\
		0 & -\ve_{\ad\bd}
	\end{array}\right)\,. 
\end{align}
In chapter \ref{ch3}, the charge conjugation matrix will be denoted $\hat{\ve}$.

We can introduce the $\mathfrak{so}(3,2)$ generators in the spinor representation 
\begin{align}
	\S_{\ha\hb} = -\frac{1}{4}[\g_{\ha}\,,\g_{\hb}] = \left(  (\S_{\ha\hb})_{\ah}{}^{\bh} \right)\,, 
\end{align}
and they satisfy the following properties 
\bsubeq
\begin{align}
	(\S_{\ha\hb})^{\dag} &= -J(\S_{\ha\hb})J^{-1}\,,
	\\
	(\S_{\ha\hb})^{\T} &= -C(\S_{\ha\hb})C^{-1}\,. 
\end{align}
\esubeq
An element of the Lie algebra $\mathfrak{so}(3,2)$, given by the following 
\begin{align}
	\o = \frac{1}{2}\o^{\ha\hb}\S_{\ha\hb}\,, \qquad (\o^{\ha\hb})^{*} = \o^{\ha\hb}\,,
\end{align}
then satisfies the master equations 
\bsubeq
\begin{align}
	J\o + \o^{\dag}J &=0\,, \\
	C\o + \o^{\T}C &= 0\,.
\end{align}
\esubeq
These imply the group equations 
\bsubeq \label{bm appendix extra group rel}
\begin{align}
	g^{\dag}Jg &= J\,,
	\\
	g^{\T}Cg &= C\,. 
\end{align}
\esubeq
As we will see in the main body, this is an isomorphic realisation of the group $\sSp(4;\mathbb{R})$, which we will denote $\sSp(4;\mathbb{R})_{C}$. In the relations \eqref{bm appendix extra group rel}, $g \in \sSp(4;\mathbb{R})_{C}\,.$
Detailed in appendix \ref{Supertwistors} and \ref{adsgroup} is another realisation defined by the conditions 
\bsubeq
\begin{align}
	g^{\dag}Jg &= J\,,
	\\
	g^{\T}Jg &= J~.
\end{align}
\esubeq
These conditions imply the reality condition $g^{\T} = g^{\dag}$. 
Making use of the real 3D gamma-matrices $((\g_{\imath})_{\ab}) = (\id_{2}\,, \s_{1}\,, \s_{3})$, with $\imath = 0\,,1\,,2$, we can introduce a Majorana representation for the gamma-matrices for $\mathbb{R}^{3\,,2}$ as 
\begin{align} \label{majorana gamma 3 2}
	\G_{\imath} = \left(\begin{array}{cc}
		(\g_{\imath})_{\a}{}^{\b} & ~0~ 
		\\
		~0~ & (\g_{\imath})^{\g}{}_{\d}
	\end{array}\right)\,, \qquad 
	\G_{3} = \left(\begin{array}{cc}
	~0~ & \ve_{\a\b}
	\\
	\ve^{\g\d} & ~0~
\end{array}\right)\,,
\qquad 
\G_{4} = \left(\begin{array}{cc}
	~0~ & -\ve_{\ab}
	\\
	\ve^{\g\d} & ~0~
\end{array}
\right)\,.
\end{align}
We emphasise that the specific notation used here differs from that in section \ref{cf2}. 
Each context has specific applications for which particular conventions prove more convenient.

\subsection{$\sSpin(4,1)$} \label{4,1 conv}
Consider the space $\mathbb{R}^{4,1}$ equipped with metric
\begin{align}
	\eta_{\ha\hb} = \text{diag}(-1\,,+1\,,+1\,,+1\,,+1)\,, \qquad \qquad \ha = 0,1,2,3,4\,. 
\end{align}
The corresponding gamma matrices satisfy
\begin{align} \label{5d gm}
	\{\g_{\ha}\,, \g_{\hb}\} = -2\eta_{\ha\hb} \id_{4}\,, \qquad (\g_{\ha})^{\dag} = \g_{0}\g_{\ha}\g_{0}\,, 
\end{align}
and can be chosen as 
\begin{align} \label{5d gamma}
	\g_{a} = \left(\begin{array}{cc}
		0 & (\s_{a})_{\a\bd} \\
		(\tilde{\s}_{a})^{\ad\b} & 0
	\end{array}\right) \,, 
	\qquad 
	\g_{4} = \left(\begin{array}{cc}
		-\ri \d_{\a}{}^{\b} & 0
		\\
		0 & \ri \d^{\ad}{}_{\bd} 
	\end{array} \right) \,.
\end{align}
They are related to the Lorentz spinor generators in ${\mathbb R}^{4,2}$, \eqref{6d Lorentz gen}, as 
\begin{align}
	(\S_{\ha5})_{\ah}{}^{\bh} &= -\frac{\ri}{2}(\g_{\ha})_{\ah}{}^{\bh}\,.
\end{align}
The charge conjugation matrix is defined as 
\begin{align} \label{5d cc}
	C\g_{\ha}C^{-1} = (\g_{\ha})^{\T}\,, \qquad C = (\ve^{\ah\bh}) = \left(\begin{array}{cc}
		\ve^{\a\b} & 0 \\
		0 & -\ve_{\ad\bd}
	\end{array}\right)\,. 
\end{align}
The antisymmetric matrices $\ve^{\ah\bh}$ and its inverse, $\ve_{\ah\bh}$, are used to raise and lower the four-component spinors indices in ${\mathbb R}^{4,1}$.

Now we can define the Lorentz generators for the spinor representation,
\begin{align}
	\S_{\ha\hb} := -\frac{1}{4}[\g_{\ha}\,, \g_{\hb}] = \left((\S_{\ha\hb})_{\ah}{}^{\bh}\right) \,,
\end{align}
and choose them in the form
\begin{align}
	\S_{ab}= \left(\begin{array}{cc}
		(\s_{ab})_{\a}{}^{\b} & 0 \\
		0 & (\tilde{\s}_{ab})^{\ad}{}_{\bd}
	\end{array}\right)\,, 
	\qquad 
	\S_{a4} = \left(\begin{array}{cc}
		0 & -\frac{\ri}{2}(\s_{a})_{\a\bd} \\
		\frac{\ri}{2}(\tilde{\s}_{a})^{\ad\b} & 0
	\end{array}\right)\,. 
\end{align}
We will see later that these generators coincide with the $4+1$ subset of those given in $4+2$ dimensions in \eqref{underlined 5d generators}.
We therefore find an explicit form for a general element of the algebra by `switching off' the $\o^{\ha5}$ components of \eqref{su(2,2) generic algebra element}:
\begin{align} \label{so(4,1) generic algebra element}
	\frac{1}{2}\o^{\ha\hb}\S_{\ha\hb} = 
	\left(\begin{array}{c|c}
		\frac{1}{2}\o^{ab}(\s_{ab})_{\a}{}^{\b} & - \frac{\ri}{2}(\o^{a4})(\s_{a})_{\a\bd}\\
		\hline 
		\frac{\ri}{2}(\o^{a4})(\tilde{\s}_{a})^{\ad\b} & \frac{1}{2}\o^{ab}(\tilde{\s}_{ab})^{\ad}{}_{\bd}
	\end{array}\right)\,.
\end{align}

Given \eqref{5d gm} and \eqref{5d cc}, we have the following properties
\begin{align}
	(\S_{\ha\hb})^{\dag} = -\g_{0}(\S_{\ha\hb})\g_{0}\,, \qquad (\S_{\ha\hb})^{\T} = - C(\S_{\ha\hb})C^{-1}\,.
\end{align}
Writing an element of the Lie algebra $\mathfrak{so}(4,1)$ as \eqref{so(4,1) generic algebra element}:
\begin{align}
	\o := \frac{1}{2}\o^{\ha\hb}\S_{\ha\hb}\,, \qquad \o^{\ha\hb} = (\o^{\ha\hb})^{*}\,,
\end{align}
we find the following master equations
\bsubeq
\begin{align} \label{so(4,1) master algebra}
	C\o + \o^{\T}C &= 0\,, 
	\\
	\g_{0}\o + \o^{\dag}\g_{0} &= 0\,.
\end{align}
\esubeq
These correspond to the group equations 
\bsubeq \label{so(4,1) master group}
\begin{align}
	g^{\T}Cg &= C \,,
	\\
	g^{\dag}\g_{0}g &= \g_{0}\,.
\end{align}
\esubeq
This group is known as $\sUSp(2,2)$. In the main body, $\g_{0}$ is often denoted by $\O\,.$

A Dirac spinor in ${\mathbb R}^{4,1}$, $\Psi = (\Psi_{\ah})\,,$ and its Dirac conjugate, $\bar{\Psi} = (\bar{\Psi}^{\ah}) = \Psi^{\dag}\g_{0}\,,$ look like
\begin{align}
	\Psi = \left(\begin{array}{c}
		\psi_{\a} \\
		\bar{\f}^{\ad}
	\end{array}\right) \,, \qquad \bar{\Psi} = \left(\begin{array}{cc}
		\f^{\a}\,, & \bar{\psi}_{\ad}
	\end{array}\right)\,. 
\end{align}
We can combine $(\bar{\Psi}^{\ah}) = \left(\f^{\a}\,, \bar{\psi}_{\ad}\right)$ and $(\Psi^{\ah}) = (\ve^{\ah\bh}\Psi_{\bh}) = \left(\psi^{\a}\,, - \bar{\f}_{\ad}\right)$ into an $\sSU(2)$ doublet:
\begin{align}
	(\Psi_{\ui}^{\ah}) = \left(\Psi^{\a}_{\ui}\,, - \bar{\Psi}_{\ad \ui}\right) \,,
	\qquad \left(\Psi_{\ui}^{\a}\right)^{*} = \bar{\Psi}^{\ad \ui}\,, \qquad \ui = \underline{1}\,,\underline{2}\,,
\end{align}
such that $\Psi_{\underline{1}}^{\a} = \f^{\a}$ and $\Psi_{\underline{2}}^{\a} = \psi^{\a}\,.$ The $\sSU(2)$ indices are raised and lowered in the usual way: $\Psi^{\ah \ui} = \ve^{\ui\uj}\Psi^{\ah}_{\uj}\,.$ The spinor $\Psi^{\ui} = (\Psi_{\ah}^{\ui})$ satisfies the pseudo-Majorana condition 
\begin{align} \label{pseudo majorana}
	\bar{\Psi}_{\ui}^{\T} = C\Psi_{\ui}\,. 
\end{align}
\subsection{$\sSpin(4,2)$} \label{4,2 conv}
In this appendix we will detail the spinor conventions for $\sSO_0(4,2)\,.$
Consider the space $\mathbb{R}^{4,2}$ parametrised by coordinates $x^{\ua}\,, ~ \ua = 0,1,2,3,4,5\,,$ and with metric 
\begin{align}
	\eta_{\ua\ub} = \text{diag}(-1,+1,+1,+1,+1,-1)\,. 
\end{align}
The corresponding gamma matrices, $\G_{\ua}$, obey the anti-commutation relations
\begin{align} \label{6D clifford}
	\{\G_{\ua},\G_{\ub}\} = -2\eta_{\ua\ub}\id_{8}\,.
\end{align}
We can choose $\G_{\ua}$ in the so-called Weyl representation
\begin{align}
	\G_{\ua} = \left(\begin{array}{cc}
		0 & \underline{\S}_{\ua} \\
		\underline{\tilde{\S}}_{\ua} & 0
	\end{array}\right)\,,
\end{align}
with 
\bsubeq
\begin{align}
	\un{\S}_{\ua} &= \left(\ri\g_{a}\,, \g_{5}, \id_{4}\right) \equiv \left((\underline{\S}_{\ua})_{\ah\underline{\bh}} \right)\,, ~~~ \ah = 1\,,2\,,3\,,4\,, ~~~ \underline{\ah} = \underline{1}\,, \underline{2}\,, \underline{3}\,, \underline{4}\,,
	 \\
	\un{\tilde{\S}}_{\ua} &= \left(-\ri\g_{a}\,, -\g_{5}\,,\id_{4}\right) \equiv \left((\underline{\tilde{\S}}_{\ua})^{\underline{\ah}\bh}\right)\,.
\end{align}
\esubeq
Here $\g_{a}$ and $\g_{5}$ are the gamma matrices in $3+1$ dimensions, see appendix \ref{3,1 conv}.

In accordance with \eqref{6D clifford}, the $\underline{\S}$ and $\underline{\tilde{\S}}$ matrices obey the following relations 
\begin{align} \label{reduced 6d clifford}
	\underline{\S}_{\ua}\tilde{\underline{\S}}_{\ub} + \underline{\S}_{\ub}\tilde{\underline{\S}}_{\ua} = -2\eta_{\ua\ub}\id_{4}\,,
	\qquad 
	\tilde{\underline{\S}}_{\ua}\underline{\S}_{\ub} + \tilde{\underline{\S}}_{\ub}\underline{\S}_{\ua} = -2\eta_{\ua\ub}\id_{4}\,. 
\end{align}
The Hermitian conjugation properties of $\G_{\ua}$ are 
\begin{align} \label{big gam conj}
	\G_{\ua}^{\dag} = \G_0\G_5\G_{\ua}\G_0\G_5\,,
\end{align}
which implies the following Hermitian conjugation properties for the $\underline{\S}$ and $\underline{\tilde{\S}}$ matrices
\begin{align} \label{u sigma conj}
	(\underline{\S}_{\ua})^{\dag} = \g_{0}\underline{\tilde{\S}}_{\ua}\g_{0}\,, \qquad (\underline{\tilde{\S}}_{\ua})^{\dag} = \g_{0}\underline{\S}_{\ua}\g_{0}\,. 
\end{align}

The Dirac spinor representation is generated by 
\begin{align} \label{Lorentz gens}
	\mathfrak{J}_{\ua\ub} = -\frac{1}{4}[\G_{\ua},\G_{\ub}] = \left(\begin{array}{cc}
		\underline{\S}_{\ua\ub} & 0 \\
		0 & \tilde{\underline{\S}}_{\ua\ub}
	\end{array}\right)\,.
\end{align}
Here we have defined 
\bsubeq \label{underlined lorentz gen}
\begin{align}
	\underline{\S}_{\ua\ub} &:= -\frac{1}{4}\left(\underline{\S}_{\ua}\tilde{\underline{\S}}_{\ub} - \underline{\S}_{\ub}\tilde{\underline{\S}}_{\ua}\right) \equiv \left((\underline{\S}_{\ua\ub})_{\ah}{}^{\bh}\right) \,, \label{underlined twistor gen}
	\\
	\underline{\tilde{\S}}_{\ua\ub} & := 
	-\frac{1}{4}\left(\tilde{\underline{\S}}_{\ua}\underline{\S}_{\ub} - \tilde{\underline{\S}}_{\ub}\underline{\S}_{\ua}\right) \equiv \left((\tilde{\underline{\S}}_{\ua\ub})^{\underline{\ah}}{}_{\underline{\bh}}\right)\,. \label{underlined dual gen}
\end{align}
\esubeq
Given \eqref{big gam conj}, the Hermitian conjugation properties of $\mathfrak{J}_{\ua\ub}$ are
\begin{align}
	(\mathfrak{J}_{\ua\ub})^{\dag} = \G_0\G_5(\mathfrak{J}_{\ua\ub})\G_0\G_5\,.
\end{align}
For the matrices $\underline{\S}_{\ua\ub}$ and $\tilde{\underline{\S}}_{\ua\ub}$ we have the following
\begin{align}
	(\underline{\S}_{\ua\ub})^{\dag} = -\g_0(\underline{\S}_{\ua\ub})\g_{0}\,, \qquad 
	(\tilde{\underline{\S}}_{\ua\ub})^{\dag} = -\g_0(\tilde{\underline{\S}}_{\ua\ub})\g_{0}\,. 
\end{align}

A Dirac spinor looks like 
\begin{align} \label{diracspinor}
	\Psi = \left(\begin{array}{c}
		\psi \\
		\f
	\end{array} \right)\,, \qquad \psi = (\psi_{\ah})\,, \quad \f = (\f^{\underline{\ah}})\,. 
\end{align}
Its conjugate is defined as 
\begin{align}
	\bar{\Psi} := -\ri\Psi^{\dag}\G_{0}\G_{5} = 
	\left(\begin{array}{cc}
		\psi^{\dag}\g_{0} \,, & - \f^\dag\g_0 
	\end{array}\right) 
	\,, 
	\quad \psi^{\dag}\g_0 \equiv (\bar{\psi}^{\ah})\,, 
	\quad \phi^\dag\g_0 \equiv (\bar{\phi}_{\underline{\ah}})\,.
\end{align}
An infinitesimal $\sSO_0(4,2)$ transformation acts on $\Psi$ and $\bar{\Psi}$ as 
\bsubeq \label{infinitesimal trf}
\begin{align}
	\d \Psi &= \frac{1}{2}\o^{\ua\ub}\mathfrak{J}_{\ua\ub}\Psi\,, \label{infini 1}
	\\
	\d \bar{\Psi} &= -\frac{1}{2}\bar{\Psi}\o^{\ua\ub}\mathfrak{J}_{\ua\ub}\,. \label{infini 2}
\end{align}
\esubeq
This representation is the sum of two irreducible representations, as is clear from \eqref{Lorentz gens}.
The twistor (left Weyl spinor) representation is associated with spinors of the form
\begin{align}
	\Psi_{\text{L}} = \left(\begin{array}{c}
		\psi \\
		0
	\end{array}\right)\,, \qquad \psi = (\psi_{\ah})\,,
\end{align}
such that
\begin{align}
	\G_{7}\Psi_{\text{L}} = \Psi_{\text{L}}\,,
\end{align}
where $\G_{7}$ is defined as 
\begin{align}
	\G_{7} := -\ri\G_0\G_1\G_2\G_3\G_4\G_5 \equiv \left(\begin{array}{cc}
		\id_{4} & 0\\
		0 & -\id_{4}
	\end{array}\right)\,. 
\end{align}
The Dirac conjugate of $\Psi_{\text{L}}$ is given by
\begin{align}
	\bar{\Psi}_{\text{L}} = \left(\begin{array}{cc}
		\bar{\psi} \,, & 0
	\end{array}\right)\,, \qquad \bar{\psi} := \psi^\dag\g_{0} = (\bar{\psi}^{\ah})\,,
\end{align}
and transforms according to the dual twistor representation. Following \eqref{infinitesimal trf}, the infinitesimal $\sSO(4,2)$ transformation laws of $\Psi_{\text{L}}$ and $\bar{\Psi}_{\text{L}}$ are given by
\bsubeq
\begin{align}
	\d \psi_{\ah} &= \frac{1}{2}\o^{\ua\ub}(\underline{\S}_{\ua\ub})_{\ah}{}^{\bh}\psi_{\bh}\,,
	\\
	\d \bar{\psi}^{\ah} &= -\frac{1}{2}\bar{\psi}^{\bh}\o^{\ua\ub}(\underline{\S}_{\ua\ub})_{\bh}{}^{\ah}\,.
\end{align}
\esubeq
In the main body of this thesis, we use the matrices $\S_{\ua}$ and $\tilde{\S}_{\ua}$. These are related to the matrices $\underline{\S}_{\ua}$ and $\underline{\tilde{\S}}_{\ua}$ as described below. We introduce the $8\times8$ charge conjugation matrix, $\mathscr{C}$, satisfying
\begin{align}
	\mathscr{C}^{-1}\G_{\ua}\mathscr{C} &= -\G_{\ua}^{\T}\,.
\end{align}
It can be chosen as
\begin{align}
	\mathscr{C} = \left(\begin{array}{cc}
		0 & \g_{5}\mathcal{C} \\
		-\g_{5}\mathcal{C} & 0
	\end{array}\right) &\equiv \left(\begin{array}{cc}
		0 & \mathscr{C}_{\ah}{}^{\underline{\bh}} \\
		\mathscr{C}^{\underline{\ah}}{}_{\bh} & 0
	\end{array}\right)\,,
\end{align}
with $\mathcal{C}$ as the $3+1$ dimensional charge conjugation matrix, \eqref{4d charge conjugation}. 
The inverse of $\mathscr{C}$ is 
\begin{align}
	\mathscr{C}^{-1} = 
	\left(
	\begin{array}{cc}
		0 & -\mathcal{C}^{-1}\g_{5} \\
		\mathcal{C}^{-1}\g_{5} & 0 
	\end{array}\right)
	\equiv 
	\left(\begin{array}{cc}
		0 & (\mathscr{C}^{-1})^{\ah}{}_{\underline{\bh}} \\
		(\mathscr{C}^{-1})_{\underline{\ah}}{}^{\bh} & 0
	\end{array}\right) \,.
\end{align}
We also have
\begin{align} \label{conjugate lorentz}
	\mathscr{C}^{-1}\mathfrak{J}_{\ua\ub}\mathscr{C} = -\mathfrak{J}_{\ua\ub}^{\T}\,. 
\end{align}
Using $\mathscr{C}$, we can define the charge conjugate spinor 
\begin{align}
	\Psi_{\mathscr{C}} := \mathscr{C}\bar{\Psi}^{\T}\,. 
\end{align}
Its infinitesimal transformation rule is
\begin{align} \label{conj trf}
	\d \Psi_{\mathscr{C}} = \frac{1}{2}\o^{\ua\ub}\mathfrak{J}_{\ua\ub}\Psi_{\mathscr{C}}\,,
\end{align}
which coincides with \eqref{infini 1}. 
Consider the contents of the charge conjugate spinor
\begin{align}
	\Psi_{\mathscr{C}} &= \mathscr{C}\bar{\Psi}^{\T} 
	\\
	\notag 
	&= \left(\begin{array}{cc}
		0 & \mathscr{C}_{\ah}{}^{\underline{\bh}} \\
		\mathscr{C}^{\underline{\ah}}{}_{\bh} & 0
	\end{array}\right)
	\left(\begin{array}{c}
		\bar{\psi}^{\bh} \\
		-\bar{\f}_{\underline{\bh}}
	\end{array}\right) 
	\\
	\notag 
	&= \left(\begin{array}{c}
		-\mathscr{C}_{\ah}{}^{\underline{\bh}}\bar{\f}_{\underline{\bh}} \\
		\mathscr{C}^{\underline{\ah}}{}_{\bh}\bar{\psi}^{\bh}
	\end{array}\right)\,.
\end{align}
Since $\Psi_{\mathscr{C}}$ transforms as \eqref{conj trf}, $\bar{\f}_{\ah} := \mathscr{C}_{\ah}{}^{\underline{\bh}}\bar{\f}_{\underline{\bh}} = \bar{\f}_{\underline{\bh}}\mathscr{C}^{\underline{\bh}}{}_{\ah}$ transforms like a twistor. 
This means we can convert underlined indices into not-underlined indices. 

We have so far considered the twistor representation. 
The other (right Weyl spinor) representation is equivalent to the dual twistor representation, as 
\begin{align}
	\f^{\ah} := (\mathscr{C}^{-1})^{\ah}{}_{\underline{\bh}}\f^{\underline{\bh}} = \f^{\underline{\bh}}(\mathscr{C}^{-1})_{\underline{\bh}}{}^{\ah}
\end{align} 
transforms like a dual twistor. This also follows from \eqref{conjugate lorentz}, which yields 
\begin{align}
	-(\underline{\S}^{\T}_{\ua\ub})^{\ah}{}_{\bh} 
	=
	(\mathscr{C}^{-1})^{\ah}{}_{\underline{\gh}}(\tilde{\underline{\S}}_{\ua\ub})^{\underline{\gh}}{}_{\underline{\dhat}}\mathscr{C}^{\underline{\dhat}}{}_{\bh}\,. 
\end{align}

Now, the matrices $\S_{\ua}$ and $\tilde{\S}_{\ua}$ (with twistor indices) are defined as 
\begin{align} \label{new sigma def}
	(\S_{\ua})_{\ah\bh} := (\underline{\S}_{\ua})_{\ah\underline{\gh}}\mathscr{C}^{\underline{\gh}}{}_{\bh}\,, 
	\qquad (\tilde{\S}_{\ua})^{\ah\bh} := (\mathscr{C}^{-1})^{\ah}{}_{\underline{\gh}}(\underline{\tilde{\S}}_{\ua})^{\underline{\gh}\bh}\,.  
\end{align}
They are antisymmetric
\begin{align}
	(\S_{\ua})_{\ah\bh} = -(\S_{\ua})_{\bh\ah}\,, \qquad (\tilde{\S}_{\ua})^{\ah\bh} = -(\tilde{\S}_{\ua})^{\bh\ah}\,,
\end{align}
and satisfy the relations 
\begin{align} \label{s stilde clifford}
	\S_{\ua}\tilde{\S}_{\ub} + \S_{\ub}\tilde{\S}_{\ua} = -2\eta_{\ua\ub}\id_{4}\,,
	\qquad 
	\tilde{\S}_{\ua}\S_{\ub} + \tilde{\S}_{\ub}\S_{\ua} = -2\eta_{\ua\ub}\id_{4}\,. 
\end{align}
It is useful here to give the explicit form of these matrices:
\begin{align} \label{bt basis def}
	\begin{split}
		\S_{a} &= \left(
		\begin{array}{cc}
			0 & \ri\s_{a}\ve \\
			-\ri \tilde{\s}_{a}\ve^{-1} & ~0~
		\end{array}
		\right)\,,
		\\
		\S_{4} &= \left(
		\begin{array}{cc}
			-\ve^{-1} & ~0~ \\
			0 & -\ve
		\end{array}
		\right)\,,
		\\
		\S_{5} &= \left(
		\begin{array}{cc}
			-\ve^{-1} & ~0~ 
			\\
			0 & \ve
		\end{array}
		\right)\,,
	\end{split}
	\quad 
	\begin{split}
		\tilde{\S}_{a} &= \left(
		\begin{array}{cc}
			0 & \ri\ve\s_{a} \\
			-\ri\ve^{-1}\tilde{\s}_{a} & ~0~
		\end{array}\right)\,,
		\\
		\tilde{\S}_{4} &= \left(
		\begin{array}{cc}
			\ve & ~0~ \\
			~0~ & \ve^{-1}
		\end{array}
		\right)\,,
		\\
		\tilde{\S}_{5} &= \left(
		\begin{array}{cc}
			-\ve & ~0~ \\
			~0~ & \ve^{-1}
		\end{array}
		\right)\,. 
	\end{split}
\end{align}
Since the Lorentz generators \eqref{underlined twistor gen} are invariant under the replacement $\underline{\S}_{\ua} \rightarrow \S_{\ua}\,, \tilde{\underline{\S}}_{\ua}\rightarrow \tilde{\S}_{\ua}\,,$ we can unambiguously write them without an underline.
Explicitly, they take the form 
\bsubeq\label{6d Lorentz gen}
\begin{align}
	\S_{ab}
	&= \left(\begin{array}{cc}
		(\s_{ab})_{\a}{}^{\b} & 0
		\\
		0 & (\tilde{\s}_{ab})^{\ad}{}_{\bd}
	\end{array}\right)\,, & \S_{a4} = \left(\begin{array}{cc}
		0 & -\frac{\ri}{2}(\s_{a})_{\a\bd} \\
		\frac{\ri}{2}(\tilde{\s}_{a})^{\ad\b} & 0
	\end{array}\right)\,, \label{underlined 5d generators}
	\\
	\S_{a5} &= \left(\begin{array}{cc}
		0 & -\frac{\ri}{2}(\s_{a})_{\a\bd} \\
		-\frac{\ri}{2}(\tilde{\s}_{a})^{\ad\b} & 0
	\end{array}\right)\,, 
	& \S_{45} = 
	\left(\begin{array}{cc}
		-\frac{1}{2}\d_{\a}{}^{\b} & 0 \\
		0 & \frac{1}{2}\d^{\ad}{}_{\bd} 
	\end{array}\right) \,.
\end{align}
\esubeq
Here we note that we can express a generic Lie algebra element as 
\bsubeq
\begin{align}
	\o = \frac{1}{2}\o^{\ua\ub}\S_{\ua\ub}\,, \qquad \o^{\ua\ub} &= (\o^{\ua\ub})^{*}\,,
	\\
	\g_{0}\o^{\dag} + \o\g_{0} &= 0\,.
\end{align}
\esubeq
We then have the following
\begin{align} \label{su(2,2) generic algebra element}
	\frac{1}{2}\o^{\ua\ub}\S_{\ua\ub}= 
	\left(\begin{array}{c|c}
		\frac{1}{2}\o^{ab}(\s_{ab})_{\a}{}^{\b} - \frac{1}{2}\o^{45}\d_{\a}{}^{\b} & - \frac{\ri}{2}(\o^{a4} + \o^{a5})(\s_{a})_{\a\bd}\\
		\hline 
		\frac{\ri}{2}(\o^{a4} - \o^{a5})(\tilde{\s}_{a})^{\ad\b} & \frac{1}{2}\o^{ab}(\tilde{\s}_{ab})^{\ad}{}_{\bd} + \frac{1}{2}\o^{45}\d^{\ad}{}_{\bd}
	\end{array}\right)\,.
\end{align}

The matrices \eqref{new sigma def} satisfy a number of additional useful properties.
Their Hermitian conjugation properties can be read off using those of $\underline{\S}_{\ua}$ and $\underline{\tilde{\S}}_{\ua}$ as well as the properties of $\mathcal{C}$ and $\g_{5}$:
\begin{align} \label{new hermit}
	(\S_{\ua})^{\dag} 
	&= \g_{0} \tilde{\S}_{\ua} \g_{0}\,. 
\end{align}
We have used $\mathcal{C}^{\dag} = \mathcal{C}^{-1}$, $\g_{0}^{\T} = \g_{0}$ and \eqref{4d charge conjugation}. 
They also satisfy the following completeness relations
\bsubeq \label{completeness relations}
\begin{align}
	\frac{1}{2}\ve_{\ah\bh\gh\dhat}(\tilde{\S}^{\ua})^{\gh\dhat} &= 
	(\S^{\ua})_{\ah\bh}\,,
	\\
	\frac{1}{2}\ve^{\ah\bh\gh\dhat}(\S^{\ua})_{\gh\dhat} &= 
	(\tilde{\S}^{\ua})^{\ah\bh}  \,,
	\\
	(\S^{\ua})_{\ah\bh}(\S_{\ua})_{\gh\dhat} &= 2\ve_{\ah\bh\gh\dhat}\,.
\end{align}
\esubeq
We can now establish the one-to-one correspondence between complex vectors $V^{\ua}$ in 
${\mathbb R}^{4,2}$
and antisymmetric bi-twistors $V_{\ah\bh} = -V_{\bh\ah}\,,$
\bsubeq \label{o-to-o}
\begin{align}
	V_{\ah\bh} &= V^{\ua}(\S_{\ua})_{\ah\bh}\,, \qquad V^{\ua} = \frac{1}{4}(\tilde{\S}^{\ua})^{\ah\bh}V_{\ah\bh}\,,
	\\
	V^{\ah\bh} &= V^{\ua}(\tilde{\S}_{\ua})^{\ah\bh}\,, \qquad V^{\ua} = \frac{1}{4}(\S^{\ua})_{\ah\bh}V^{\ah\bh}\,. 
\end{align}
\esubeq
In the above, we have 
\begin{align}
	V^{\ah\bh} = \frac{1}{2}\ve^{\ah\bh\gh\dhat}V_{\gh\dhat}\,,
\end{align}
consistent with \eqref{completeness relations}. 
We will now introduce the condensed notation 
\begin{align}
	V := (V_{\ah\bh})\,, \qquad \tilde{V} := (V^{\ah\bh})\,. 
\end{align}
We wish to investigate how the bi-twistor associated with $V^{\ua}$ is related to that associated with $\bar{V}^{\ua}$, the complex conjugate of the six-vector $V^{\ua}\,.$ 
Taking the Hermitian conjugate of $V$, using \eqref{new hermit}, we find
\begin{align} \label{bt real step 1}
	V = V^{\ua}(\S_{\ua}) \rightarrow V^{\dag}
	&= \bar{V}^{\ua}\g_{0}(\tilde{\S}_{\ua})\g_{0}
	\\
	\notag 
	&= \g_{0}\tilde{\bar{V}}\g_{0}\,.
\end{align}
Inserting $V^{\ua} = \bar{V}^{\ua}$ into the above, we find
\bsubeq \label{reality cond for six vector}
\begin{align} 
	V^{\dag} = \g_{0}\tilde{V}\g_{0} \implies \tilde{V} = \g_{0}V^{\dag}\g_{0}\,. 
\end{align}
Now, identifying $\g_{0}$ with $\O$ in \eqref{su(2,2) master} and expressing the reality condition in index notation, we have
\begin{align} \label{index reality}
	\O^{\ah\gh}(V^{\dag})_{\gh\dhat}\O^{\dhat\bh} = \frac{1}{2}\ve^{\ah\bh\gh\dhat}V_{\gh\dhat}\,. 
\end{align}
\esubeq
\subsection{$\sSpin(5)$} \label{5,0 conv}
Consider the space $\mathbb{R}^{5}$, equipped with metric 
\begin{align}
	\d_{\hI\hJ}\,, \qquad \hI = 0\,,1\,,2\,,3\,,4\,.
\end{align}
The gamma matrices satisfy 
\begin{align}
	\{\g_{\hI}\,, \g_{\hJ}\} = -2\d_{\hI\hJ}\id_{4}\,,
\end{align}
and can be chosen as 
\bsubeq
\begin{align}
	\g_{0} &= \ri\left(\begin{array}{cc}
		0 & \id_{2} \\
		\id_{2} & 0
	\end{array}\right)
	\,,
	\\
	\g_{i} &= \left(\begin{array}{cc}
		0 & \s_{i} \\
		-\s_{i} & 0	
	\end{array}\right) 
	\,, \quad i = 1\,,2\,,3\,,
	\\
	\g_{4} &= -\ri\left(\begin{array}{cc}
		\id_{2} & 0 \\
		0 & -\id_{2}	
	\end{array}\right)
	\,.
\end{align}
\esubeq
Their Hermitian conjugation properties are 
\begin{align}
	\g_{\hI}^{\dag} = -\g_{\hI}\,,
\end{align}
and the charge conjugation $C$ matrix satisfies 
\begin{align}
	C\g_{\hI}C^{-1} = (\g_{\hI})^{\T}\,. 
\end{align}
It can be chosen as 
\begin{align}
	C = \left(\begin{array}{cc}
		\ve & ~0~ \\
		~0~ & \ve	
	\end{array}\right)
	\,.
\end{align}

Now we can introduce the `Lorentz' generators in the spinor representation as 
\begin{align}
	\S_{\hI\hJ} := -\frac{1}{4}[\g_{\hI}\,,\g_{\hJ}]\,,
\end{align}
and an element of the Lie algebra $\mathfrak{so}(5)$ is then given by 
\begin{align} \label{so5 alg element}
	\o = \frac{1}{2}\o_{\hI\hJ}\S_{\hI\hJ}\,, \qquad (\o_{\hI\hJ})^{*} = \o_{\hI\hJ}\,.
\end{align}
The $\mathfrak{so}(5)$ generators satisfy the following properties 
\bsubeq
\begin{align}
	(\S_{\hI\hJ})^{\dag} &= -\S_{\hI\hJ}\,, 
	\\
	(\S_{\hI\hJ})^{\T} &= - C(\S_{\hI\hJ})C^{-1}\,. 
\end{align}
\esubeq
It follows that the element \eqref{so5 alg element} satisfies 
\bsubeq
\begin{align}
	\o + \o^{\dag} &= 0\,,
	\\
	C\o + \o^{\T}C &= 0\,. 
\end{align}
\esubeq
These are the infinitesimal forms of the group equations 
\bsubeq
\begin{align}
	g^{\dag}g &= \id_{4}\,, \\
	\qquad g^{\T}Cg &= C
\end{align}
\esubeq
which define the group $\sUSp(4)$. It follows that the Lie algebras are isomorphic, 
\begin{align}
	\mathfrak{so}(5) \simeq \mathfrak{usp}(4)\,. 
\end{align}
\subsection{$\sSpin(6)$} \label{6,0 conv}
Consider the space $\mathbb{R}^{6}$ equipped with metric 
\begin{align}
	\d_{\uI\uJ} \,, \qquad \uI = 0\,,1\,,2\,,3\,,4\,,5\,.
\end{align}
The gamma-matrices satisfy 
\begin{align}
	\{\G_{\uI}\,, \G_{\uJ}\} = -2\d_{\uI\uJ}\id_{8}\,,
\end{align}
and can be chosen as 
\begin{align}
	\G_{\uI} = \left(\begin{array}{cc}
		~0~ & \S_{\uI} \\
		\tilde{\S}_{\uI} & ~0~
	\end{array}\right)\,,
\end{align}
with 
\bsubeq
\begin{align}
	\S_{\uI} &= \left(-\g_{0}\,, \ri\g_{i}\,,\g_{5}\,, \ri\id_{4}\right)\,, 
	\\
	\tilde{\S}_{\uI} &= \left(\g_{0}\,, -\ri\g_{i}\,, -\g_{5}\,, \ri\id_{4}\right)\,,
\end{align}
\esubeq
where $\g_{a} = (\g_{0}\,, \g_{i})$ are the gamma-matrices in $3+1$ dimensions, eq. \eqref{4d gm}. 
The matrices $\S_{\uI}$ and $\tilde{\S}_{\uI}$ satisfy the relations 
\bsubeq
\begin{align}
	\S_{\uI}\tilde{\S}_{\uJ} + \S_{\uJ}\tilde{\S}_{\uI} &= -2\d_{\uI\uJ}\,, \\
	\tilde{\S}_{\uI}\S_{\uJ} + \tilde{\S}_{\uJ}\S_{\uI} &= -2\d_{\uI\uJ}\,. 
\end{align}
\esubeq

The Dirac spinor representation is generated by 
\begin{align} \label{so6 gen}
	\mathcal{J}_{\uI\uJ} &= -\frac{1}{4}[\G_{\uI}\,, \G_{\uJ}] \\
	\notag 
	&= \left(
	\begin{array}{cc}
		\S_{\uI\uJ} & 0 \\
		0 & \tilde{\S}_{\uI\uJ}
	\end{array}
	\right)\,. 
\end{align}
In the above, $\S_{\uI\uJ}$ and $\tilde{\S}_{\uI\uJ}$ are given by 
\bsubeq
\begin{align}
	\S_{\uI\uJ} &= - \frac{1}{4}\left(\S_{\uI}\tilde{\S}_{\uJ} - \S_{\uJ}\tilde{\S}_{\uI}\right) \\
	\tilde{\S}_{\uI\uJ} &= - \frac{1}{4}\left(\tilde{\S}_{\uI}\S_{\uJ} - \tilde{\S}_{\uJ}\S_{\uI}\right)\,.
\end{align}
\esubeq
Clearly, this representation is reducible. 
The matrices $\S_{\uI}$ and $\tilde{\S}_{\uI}$ are related by the following rule
\begin{align}
	\left(\S_{\uI}\right)^{\dag} = -\tilde{\S}_{\uI}\,. 
\end{align}
This means that the generators $\S_{\uI\uJ}$ satisfy 
\begin{align} \label{so6 gen conj}
	(\S_{\uI\uJ})^{\dag} = -\S_{\uI\uJ}\,. 
\end{align}

Now, we can introduce an element of the Lie algebra $\frak{so}(6)$ 
\begin{align}
	\o = \frac{1}{2}\o_{\uI\uJ} \S_{\uI\uJ}\,, \qquad (\o_{\uI\uJ})^{*} = \o_{\uI\uJ}\,.  
\end{align}
Due to the relation \eqref{so6 gen conj}, it satisfies 
\begin{align}
	\o^{\dag} + \o = 0\,,
\end{align}
which proves the following well-known isomorphism $\frak{so}(6) \simeq \frak{su}(4)\,.$
	
\section{The supergroup $\sSU(2,2|\N)$ and supertwistors} \label{5d appendix}

This appendix is dedicated to reviewing the technical details of the (super)twistor formalism utilised in section \ref{4d compact str} and chapter \ref{ch4}. 
Essential to this formalism is the use of the supergroup $\sSU(2,2|\N)$, which has two different but related roles: (i) as the $\N$-extended superconformal group in four dimensions; and (ii) as the $\N$-extended anti-de Sitter supergroup in five dimensions.
We will begin by outlining our conventions for the group $\sSU(2,2)$, including specific details that were omitted in appendix \ref{spinor conventions}. 
For this purpose we make use of twistor space $\frak{T}$, which can be identified with $\mathbb{C}^{4}$.
Then we will extend these considerations to the supersymmetric case. 

\subsection{Twistors of $\sSU(2,2)$}

A twistor $T \in {\mathfrak T}$ is a column vector
\begin{align}
	T = (T_{\ah}) = \left(\begin{array}{c}
		f_{\a} \\
		\bar{g}^{\ad}
	\end{array}\right)\,, 
	\qquad \a = 1\,,2\,, \quad \ad= \dot{1} \,, \dot{2}\,.
\end{align}
The group $\sSU(2,2)$  naturally acts on $\mathfrak T$ by transformations 
\begin{align}
	T_{\ah} \longrightarrow g_{\ah}{}^{\bh}T_{\bh}\,, \qquad g &= \big(g_{\ah}{}^{\bh}\big) \in \sSU(2,2)~.
\end{align}
By definition, the group elements $g \in \sSU(2,2)$ satisfy the master equation 
\begin{align}
g^{\dag} \O g = \O \,, \qquad \det g = 1\,,
\end{align}
where $\O$ is defined as 
\begin{align}
	\O &= \big(\O^{\ah\bh}\big) 
	= \left(\begin{array}{cc} 
	~0~ & \id_{2} \\
	\id_{2} & ~0~\end{array}\right)
	\, .
\end{align}
A unique $\sSU(2,2)$ invariant inner product on $\mathfrak T$ is given by 
\begin{align}
	\braket{T}{S} := T^{\dag}\O S \equiv \bar{T} S\, ,
\end{align}
where we have defined
\begin{align}
	\bar{T}= T^\dagger \O = (\bar{T}^{\ah})\,, \qquad \bar{T}^{\ah} = (T_{\bh})^* \,\O^{\bh\ah}~.
\end{align}
We will refer to $\bar T$ as the dual of $T$. It transforms as 
\begin{align}
	\bar{T}^{\ah} \longrightarrow \bar{T}^{\bh}(g^{-1})_{\bh}{}^{\ah}\,. 
\end{align}
The space of dual twistors will be denoted $\bar {\mathfrak T}$. 

Of special importance to the analysis of chapter \ref{ch4} is a subgroup of $\sSU(2,2)$, denoted $\sUSp(2,2)$, consisting of those group elements which satisfy the additional condition 
\begin{align} \label{usp cond}
	g ^{\T} C g = C\,, \qquad C = (\ve^{\ah \bh} ) = \left(\begin{array}{cc}
		~\ve^{\a\b} ~ & 0 
		\\
		0 & -\ve_{\ad \bd}
	\end{array}\right)\,,
\end{align}
with $ \ve^{\ab}$ and $\ve_{\ad \bd}$ being defined in \eqref{epsilon def}. 
The Lie algebras of $\sSU(2,2)$ and $\sUSp(2,2)$ consist of elements of the form \eqref{su(2,2) generic algebra element} and \eqref{so(4,1) generic algebra element}, respectively, and are denoted $\mathfrak{su}(2,2)$ and $\mathfrak{usp}(2,2)$.

\subsection{Supertwistors of $\sSU(2,2|\N)$} \label{su(2,2|n) sec}

We will now extend the above considerations to the supersymmetric case.
The supergroup $\sSU(2,2|\N)$
naturally acts on the space of {\it even} supertwistors and on
the space of {\it odd} supertwistors.
An arbitrary supertwistor $\bm{T}$ is a column vector
\bea
\bm{T} = \left(\bm{T}_{A}\right) = \left(\begin{array}{c}
	\bm{T}_{\ah} \\
	\hline \hline
	\bm{T}_{i}
\end{array}\right)\,, \qquad i = 1\,, \ldots\,, \N\,. 
\eea
In the case of even supertwistors, $ \bm{T}_\hal$ is bosonic
and $\bm{T}_i$ is fermionic.
In the case of odd supertwistors, $\bm{T}_\hal$ is fermionic while $\bm{T}_i$ is bosonic.
The even and odd supertwistors are called pure.
We introduce the parity function $\e ( \bm{T} )$ defined as:
$\e ( \bm{T} ) = 0$ if $ \bm{T}$ is even, and $\e ( \bm{T} ) =1$ if $\bm T $ is odd.
If we define
\begin{align} \label{parity convention}
	\e_A= \bigg\{\begin{array}{ccc}
		0 ~,& ~ & A = \ah \\
		1 ~,& ~ & A = i
	\end{array}
\end{align}
then the components $\bm{T}_A$ of a pure supertwistor
 have the following  Grassmann parities
\bea
\e ( \bm{T}_A) = \e ( \bm{T} ) + \e_A \quad (\mbox{mod 2})~.
\eea
The space of even supertwistors is naturally identified with
${\mathbb C}^{4|\cN}$,
while the space of odd supertwistors may be identified with
${\mathbb C}^{\cN |4}$.

Supertwistor space is equipped with the inner product 
\begin{align} \label{big o def}
	\braket{\bm{T}}{\bm{S}} = \bm{T}^{\dag}\bm{\O S}\,, \qquad \bm{\O} = \left(\begin{array}{ccc}
		0 & \id_{2} & 0 
		\\
		\id_{2} & 0 & 0 
		\\
		0 & 0 & -\id_{\N}
	\end{array}\right) = \left(\begin{array}{cc}
	\O & 0 \\
	0 & -\id_{\N}
\end{array}\right)\,. 
\end{align}
The inner product is invariant under the supergroup $\sSU(2,2|\N)$
spanned by supermatrices of the form 
\begin{align} \label{susy master eq}
	g = (g_{A}{}^{B}) \in \sSL(4|\N)\,, \qquad g^{\dag}\bm{\O}g &= \bm{\O}\,.
\end{align}

Associated with a supertwistor $\bm{T}$ is its dual
\begin{align} \label{bm dual supertwistor}
	\bar{\bm{T}} := \bm{T}^{\dag}\bm{\O} = (\bar{\bm{T}}{}^{A}) = (\bar{\bm{T}}{}^{\ah}\,, -\bar{\bm{T}}{}^{i})\,, \qquad \bar{\bm{T}}{}^{i} := (\bm{T}_{i})^{*}\,. 
\end{align}
$\sSU(2,2|\N)$ acts on supertwistors and their duals as 
\begin{align}
	\bm{T}_{A} \rightarrow g_{A}{}^{B}\bm{T}_{B}\,, \qquad \bar{\bm{T}}{}^{A} \rightarrow \bar{\bm{T}}{}^{B}(g^{-1})_{B}{}^{A}\,. 
\end{align}

The superalgebra $\mathfrak{su}(2,2|\N)$ is spanned by elements of the form
\bsubeq \label{su22n element}
\begin{align} 
	\mathfrak{g} = \left(\begin{array}{c|c||c}
		\o_{\a}{}^{\b} - (\frac{1}{2}\o^{45} - \frac{\ri\N}{\N-4}\t) \d_{\a}{}^{\b} & - \frac{\ri}{2}(\o^{a4} + \o^{a5})(\s_{a})_{\a\bd} & 2\eta_{\a}{}^{j} \\
		\hline 
		\frac{\ri}{2}(\o^{a4} - \o^{a5})(\tilde{\s}_{a})^{\ad\b} & 
		-\bar{\o}^{\ad}{}_{\bd} + (\frac{1}{2}\o^{45} + \frac{\ri\N}{\N-4}\t)\d^{\ad}{}_{\bd} & 2\bar{\e}^{\ad j}
		\\
		\hline \hline 
		2\e_{i}{}^{\b} & 2\bar{\eta}_{i\bd} & \frac{4\ri}{\N-4}\t\d_{i}{}^{j} + \L_{i}{}^{j}
	\end{array}\right)\,,
\end{align}
with 
\begin{align}
	\o_{\a}{}^{\b} = \frac{1}{2}\o^{ab}(\s_{ab})_{\a}{}^{\b}\,, \qquad  
	\o^{\ua\ub} = - \o^{\ub\ua} \,, \t \in \mathbb{R}\,, \qquad \L = (\L_i{}^j) \in \mathfrak{su}(\N)\,. 
\end{align}
\esubeq
The elements of $\mathfrak{su}(2,2|\N)$ satisfy the conditions
\begin{align}
	\text{str} \, \mathfrak{g} = 0\,, \qquad \mathfrak{g}^{\dag}\bm{\O} + \bm{\O}\mathfrak{g}= 0\,. 
\end{align}
More details on the AdS superalgebra in five dimensions will be elaborated in section \ref{5d geometry}.

\section{The supergroup $\sOSp(\mathcal{N}|4;\mathbb{R})$ and supertwistors} \label{Supertwistors}

This appendix is dedicated to outlining our conventions regarding the supergroup $\sOSp(\N|4;\mathbb{R})$. 
Similarly to what is described in appendix \ref{5d appendix}, this supergroup has two origins in supersymmetric field theory: 
(i) as the $\N$-extended superconformal group in three dimensions; and (ii) as the $\N$-extended anti-de Sitter supergroup in four dimensions. 
Its role in the $\N$-extended superconformal field theory in three dimensions is spelt out in \cite{Park:1999cw}.
Despite the obvious similarites of these `3D' and `4D' supertwistors (with the nomenclature stemming from their superconformal roots), there are a number of crucial differences, which we will elucidate below.

Our starting point is to consider the supergroup $\sOSp(\N|4;\mathbb{C})$. 
This supergroup naturally acts on the space of {\it  even} supertwistors and on
the space of {\it  odd} supertwistors.
An arbitrary supertwistor is a column vector
\bea
\bm{T} = (\bm{T}_A) =\left(
\begin{array}{c}
\bm{T}_\hal \\
\hline \hline
 \bm{T}_I
\end{array}
\right)~, 
\qquad \hat \a = 1,2, 3,4 ~, \quad I = 1, \dots, \cN ~.
\eea
Pure supertwistors, and their Grassmann parities, are defined in complete analogy with appendix \ref{5d appendix}.
It is also useful to define
\bea
 \e_A = \left\{
\begin{array}{c}
 0 \qquad A=\hal \\
 1 \qquad A=I
\end{array}
\right.{}~.
\non
\eea
Then the components $\bm{T}_A$ of a pure supertwistor
 have the following  Grassmann parities
\bea
\e ( \bm{T}_A) = \e ( \bm{T} ) + \e_A \quad (\mbox{mod 2})~.
\eea
The space of  even supertwistors is naturally identified with
${\mathbb C}^{4|\cN}$,
while the space of  odd supertwistors may be identified with
${\mathbb C}^{\cN |4}$. 
These definitions are analogous to those of appendix \ref{5d appendix}.

Supertwistor space is equipped with the symplectic inner product
\bea
\langle \bm{T} | \bm{S} \rangle  : = \bm{T}^{\rm sT} {\mathbb J} \, \bm{S}
~,
\label{innerp}
\eea
where the row vector  $T^{\rm sT} $ is defined by
\bea
{ \bm{T}}^{\rm sT} := \big( \bm{T}_\hal , - (-1)^{\e(\bm{T})}  \bm{T}_I \big)
= (  \bm{T}_A (-1)^{\e(\bm{T})\e_A +\e_A} )\,,
\eea
and the supermatrix $\mathbb{J}$ is defined by 
\bea
{\mathbb J} = ({\mathbb J}^{AB}) = \left(
\begin{array}{c ||c}
J ~&~ 0 \\\hline \hline
0 ~& ~{\rm i} \,\id_\cN
\end{array} \right) ~, \qquad
J
=\big(J^{\hat \a \hat \b} \big)
=\left(
\begin{array}{cc}
0  & \id_2\\
 -\id_2  &    0
\end{array}
\right) ~.
\label{supermetric}
\eea
The inner product \eqref{innerp} is graded antisymmetric, 
\bea
\langle \bm{T} |  \bm{S}  \rangle
= -(-1)^{ \e (\bm{T}) \e(\bm{S}) }  \langle \bm{S} | \bm{T}  \rangle ~.
\eea

We can now define the supergroup   $\sOSp(\cN|4; {\mathbb C})$ to 
consist of those even $(4|\cN) \times (4|\cN)$ supermatrices
\bea
g = (g_A{}^B) ~, \qquad \e(g_A{}^B) = \e_A + \e_B ~,
\eea
that preserve the inner product \eqref{innerp} under the action
\bea
\bm{T} =(\bm{T}_A) ~\to ~ g\bm{T} = (g_A{}^B \bm{T}_B)~. \label{2.9}
\eea
This implies the following master equation for the group elements $g$
\bea
g^{\rm sT} {\mathbb J}\, g = {\mathbb J} ~, \qquad
(g^{\rm sT})^A{}_B := (-1)^{\e_A \e_B + \e_B} g_B{}^A~.
\label{groupcond}
\eea
Finally, writing the supermatrix $g$ in block form, we have
\begin{align} \label{originalosp}
	g =  \left(
	\begin{array}{c||c}
		A & B\\
		\hline \hline
		C & D
	\end{array}
	\right)\,,
	\qquad g^{\sT} =\left(
	\begin{array}{c||c}
		A^\T & -C^\T\\
		\hline \hline
		B^{\T} & D^\T
	\end{array}
	\right)\,.
\end{align}

We will now extend the above definitions to the supergroup $\sOSp(\N|4;\mathbb{R})$. 
For this purpose it useful to first introduce the notion of a real supertwistor. 
A pure supertwistor is said to be real if its components obey the reality condition\footnote{Here we point out that $\overline{\bm{T}_{A}}$ denotes the complex conjugate of $\bm{T}_{A}$, rather than the components of the dual supertwistor which was introduced for $\sSU(2,2|\N)$.}
\bea
\overline{\bm{T}_A} = (-1)^{\e(\bm{T}) \e_A + \e_A} \bm{T}_A~.
\label{initrealitycond}
\eea
This reality condition is not preserved under the action of $\sOSp(\N|4;\mathbb{C})$.
Indeed, the supergroup $\sOSp(\N|4;\mathbb{R})$ is defined to consist of those supermatrices in $\sOSp(\N|4;\mathbb{C})$ that preserve the reality condition \eqref{initrealitycond}.
That is, 
\begin{subequations}
\bea
\overline{\bm{T}_A} = (-1)^{\e(\bm{T}) \e_A + \e_A} \bm{T}_A \quad \longrightarrow \quad
\overline{(g\bm{T})_A} = (-1)^{\e(\bm{T}) \e_A + \e_A} (g\bm{T})_A~.
\eea
This implies the supermatrices $g$ obey the reality condition
\bea
\overline{ g_A{}^B} = (-1)^{\e_A \e_B + \e_A} g_A{}^B \quad
\Longleftrightarrow \quad g^\dagger = g^{\rm sT}~, \qquad
\forall g \in \sOSp(\cN|4; {\mathbb R}) ~.
\label{A.13b}
\eea
Then, the supergroup $\sOSp(\N|4;\mathbb{R})$ can be defined to consist of those even $(4|\N)\times(4|\N)$ supermatrices which obey the group conditions
\bea
g^{\sT} \mathbb{J} g &= \mathbb{J}\,,
\\
g^\dagger {\mathbb J}\, g &= {\mathbb J} ~.
\label{A.13c}
\eea
\end{subequations}

Finally, we will discuss involution for supertwistors. 
Let us consider an arbitrary complex supertwistor $\bm{T}$. 
Then, the following involution can be defined 
\bsubeq
\begin{align}
\bm{T} = (\bm{T}_A)  &\rightarrow ~*\bm{T} =\big( (* \bm{T})_A\big) ~, 
\\
(* \bm{T})_A &:=  (-1)^{\e(\bm{T}) \e_A + \e_A} \overline{\bm{T}_A}~, \qquad *(*\bm{T})=\bm{T}~.
\label{A.14}
\end{align}
\esubeq
This involution is defined such that $*\bm{T}$ transforms as a supertwistor with respect to the supergroup $\sOSp(\N|4;\mathbb{R})$, 
\bea
g (*\bm{T} )= *(g\bm{T})~, \qquad \forall g \in \sOSp(\cN|4; {\mathbb R})~.
\eea
It follows that
\bea
(*\bm{T})^{\rm sT} = \bm{T}^\dagger~.
\eea
We point out that a real supertwistors $\bm{T}$ satisfies $*\bm{T} =\bm{T}$. 
More details on the AdS superalgebra in four dimensions can be found in chapter \ref{ch3}. 
Further, useful alternative realisations of $\sOSp(\N|4;\mathbb{R})$ and its connected component $\sOSp_)(\N|4;\mathbb{R})$ are outlined in appendices \ref{adsgroup} and \ref{another ads group}.


\end{subappendices}
 

\chapter{Embedding formalism for $\N$-extended AdS superspace in four dimensions} \label{ch3}
\thispagestyle{boef1}

This chapter is dedicated to the study of the $\N$-extended anti-de Sitter superspace in four dimensions, AdS$^{4|4\N}$. 
The group theoretic definition of AdS$^{4|4\N}$ has long been known in the literature as
\begin{align} \label{4d ads susy definition}
	\text{AdS}^{4|4\N} = \frac{\sOSp(\N|4;\mathbb{R})}{\sSL(2\,,\mathbb{C})\times\sO(\N)}\,,
\end{align} 
see, e.g., \cite{Castellani:1991et}. 
Its systematic study has, until recently, only been carried out in the $\N=1$ and $\N=2$ cases.
These are reviewed in the background material, see section \ref{bm super ads}.

Regarding the $\N$-extended AdS superspace, only the coset realisation eq. \eqref{4d ads susy definition} existed in the literature for many years. 
However, a global realisation of AdS$^{4|4\N}$ making use of (bi-)supertwistor techniques was recently developed \cite{KTM}. 
This is reviewed in the background material, see section \ref{ads4 bst}, and is the supersymmetric generalisation of the standard embedding formalism for AdS$_{4}$, see section \ref{bm ads sec}. 
Previously, (super)twistor techniques had been used to describe (super)particles propagating in AdS (super)spaces, see \cite{CGKRZ, CRZ, CKR, BLPS, Z, Cm, Cm2, ABGT, ABGT2, U, U2, Adamo:2016rtr}. 

An important aspect of AdS$^{4|4\N}$ is that it is conformally flat. 
This was first established for AdS$^{4|4}$ in \cite{IS}, and was later re-derived in the textbooks \cite{Gates:1983nr, BK} in the supergravity setting. 
The fact that AdS$^{4|8}$ is conformally flat was demonstrated in \cite{Kuzenko:2008qw,Butter:2011ym} making use of the off-shell $\N=2$ supergravity framework. 
Alternative conformally flat realisations for AdS$^{4|4}$ and AdS$^{4|8}$ were derived in \cite{Butter:2012jj}, based on the use of Poincar\'e coordinates. 
The conformal flatness of AdS$^{4|4\N}$, for arbitrary $\N$, was in fact demonstrated earlier in \cite{BILS} as part of their general study of AdS supergeometries of the form AdS$_{m} \times S^{n}$.

This chapter is based on the publications \cite{KKR} and \cite{KKR2} and partial results from \cite{KK3}.
It is organised as follows. 
In section \ref{4d n extended}, we introduce the $\N$-extended anti-de Sitter supergeometry in four dimensions, AdS$^{4|4\N}$, from the supergravity-inspired approach. 
Specifically, we introduce the constraints required to realise this supergeometry within the $\sSU(\N)$ superspace framework.  
This allows us to derive explicit expressions for the covariant derivatives and supervielbein one-forms in terms of those of the $\N$-extended Minkowski superspace $\mathbb{M}^{4|4\N}$. 
We then describe how the specific properties of the torsion tensor imply that there exists a gauge choice in which the $R$-symmetry group is reduced to $\sO(\N)\,.$ 
Finally, we derive two explicit conformally flat realisations of AdS$^{4|4\N}$ based on the use of stereographic and Poincar\'e coordinates, respectively, and determine the Killing supervectors of AdS$^{4|4\N}$. 
Then, in section \ref{4d coset} we describe the AdS supergeometry from the perspective of the embedding formalism. 
We introduce two coordinate charts that are a natural generalisation of stereographic coordinates and form an atlas for AdS$^{4|4\N}$. 
In section \ref{4d geometry} we develop aspects of the AdS supergeometry, first in a way which is coordinate system-agnostic and then specialising to the case of the two charts introduced. 
Finally, we show the precise correspondence between the two approaches to AdS$^{4|4\N}$. 
In section \ref{4d superparticle discussion} we describe applications of the analysis developed in the previous sections. 
Specifically, we introduce a deformed AdS supersymmetric interval and make use of it to derive new superparticle models. 
We show how such models arise in both the supergravity-inspired approach and the embedding formalism, and show how they are related.  
The main body of this chapter is accompanied by several appendices. 
In appendix \ref{adsgroup}, we derive an isomorphic realisation of the AdS supergroup $\sOSp(\N|4;\mathbb{R})$ which proves to be very useful for the analysis of section \ref{4d coset}. 
Appendix \ref{4d n ext coset} discusses implications of the expressions derived in the coset construction for $\N>1$. 
Appendices \ref{another ads group} and \ref{ads and mink} derive another realisation of the AdS supergroup and use it to demonstrate the precise relationship between the supertwistor realisations of AdS$^{4|4\N}$ and $\overline{\mathbb{M}}{}^{4|4\N}$. 
Finally, appendix \ref{cflat geo} reviews the $\sSU(\N)$ superspace formalism of \cite{KKR} and \cite{KKR2} and details the necessary information to realise AdS$^{4|4\N}$ in this framework.  


\section{$\cN$-extended AdS superspace}
\label{4d n extended}

In this section we describe how the $\N$-extended AdS supergeometry in four dimensions in realised in the $\sSU(\N)$ superspace setting, which is outlined in appendix \ref{cflat geo}. 
Our starting point is a four-dimensional curved superspace, parametrised by local bosonic ($x^{m}$) and fermionic ($\q_{\imath}^{\m}\,, \bar{\q}^{\imath}_{\dmu}$) coordinates 
\begin{align}
	z^{M} = (x^{m}\,, \q_{\imath}^{\m}\,, \bar{\q}^{\imath}_{\dmu})\,, \qquad m = 0\,,1\,,2\,,3\,, ~~ \m = 1\,,2\,, ~~ \dmu = \dot{1}\,,\dot{2}\,, ~~ \imath = \underline{1}\,, \ldots \,, \underline{\N}\,.
\end{align}
The superspace geometry is controlled by a number of constrained superfields. They are 
\bsubeq \label{n ext controlling fields}
\begin{align}
	\N &=1: \qquad \m\,, ~~ G_{\a\ad}\,,
	\\
	\N &>1: \qquad S^{ij}\,, ~~ Y_{\ab}^{ij}\,, ~~ G_{\a\ad}\,,
\end{align}
\esubeq
satisfying the Bianchi identities eq. \eqref{n=1 bianchi} and \eqref{sun bianchi}. 
Within this framework, the AdS$^{4|4\N}$ supergeometry is singled out by the requirement that these superfields are: (i) Lorentz-invariant; and (ii) covariantly constant. 
Imposing these conditions on the superfields \eqref{n ext controlling fields} implies the following 
\begin{subequations} \label{torsion conditions}
	\begin{align}
		\cN=1:& \qquad G_\aa = 0~, \qquad \cD_A \m = 0~. \\
		\label{4d n>1 conditions} \cN>1:& \qquad Y_{\a \b}^{ij} = 0 ~, \qquad G_\aa = 0~,\qquad \cD_A S^{jk} = 0\,,
	\end{align}
\end{subequations}
where $\cD_{A} = (\cD_{a}\,, \cD^{i}_{\a}\,, \bar{\cD}_{i}^{\ad})$ are the AdS covariant derivatives.

In the $\sSU(\N)$ superspace setting, the structure group is $\sSL(2\,,\mathbb{C}) \times \sSU(\N)$, and so the AdS covariant derivatives take the form 
\begin{align} \label{4d ads cd def}
	\cD_{A} = E_{A}{}^{M} \partial_{M} + \frac{1}{2}\O_{A}{}^{cd}M_{cd} + \F_{A}{}^{i}{}_{j}\mathbb{J}^{j}{}_{i}\,,
\end{align}
see eq. \eqref{su(n) cd}. Here, $M_{cd}$ are the Lorentz generators and $\mathbb{J}^{i}{}_{j}$ are the $\sSU(\N)$ generators. 
The superfield 
\begin{align}
	\O_{A} = \frac{1}{2}\O_{A}{}^{bc}M_{bc}
\end{align}
denotes the Lorentz connection; and 
\begin{align}
	\F_{A} = \F_{A}{}^{i}{}_{j}\mathbb{J}^{j}{}_{i} 
\end{align}
denotes the $\sSU(\N)$ connection. Their inclusion in the expression \eqref{4d ads cd def} ensures that the covariant derivative of a tensor superfield transforms as a tensor superfield.  
A powerful application of working within such a framework is the construction of an explicit realisation of the AdS covariant derivatives by applying a super-Weyl transformation, eq. \eqref{superweylGWZ} and \eqref{SU(N)sW}, to those of the $\N$-extended Minkowski superspace in four dimensions. 
In such a frame, the covariant derivatives are given by the following expressions\footnote{We emphasise that the relations \eqref{AdSBoost} are not the most general form of the covariant derivatives in AdS superspace. The most general form is given by \eqref{4d ads cd def}.} 
\begin{subequations}
	\label{AdSBoost}
	\begin{align}
	\cD_\a^{i}&= \re^{\frac{\cN-2}{2\cN} \s + \frac 1 \cN \bar{\s}} \Big( D_\a^i+ D^{\b i}\s M_{\a \b} + D_{\a}^j\s \mathbb{J}^{i}{}_j \Big) ~, 
	\\ 
	\bar{\cD}_{i}^{\ad}&=\re^{\frac{1}{\cN} \s + \frac{\cN-2}{2\cN} \bar{\s}} \Big( \bar{D}^\ad_i-\bar{D}_{ \bd i} \bar{\s} \bar{M}^{\ad \bd} - \bar{D}^{\ad}_j \bar{\s} \mathbb{J}^{j}{}_i \Big)~,
	\\
	\cD_\aa &= \re^{\hf \s + \hf \bar{\s}} \Big(\partial_\aa + \frac{\rm i}{2} D^i_{\a} \s \bar{D}_{\ad i} + \frac{\rm i}{2} \bar{D}_{\ad i} \bar{\s} D_{\a}^i + \hf \Big( \partial^\b{}_\ad (\s + \bar \s ) - \frac{\ri}{2} D^{\b i} \s \bar{D}_{\ad i} \bar{\s} \Big) M_{\a \b} \non \\ & + \hf \Big( \partial_{\a}{}^\bd (\s + \bar{\s}) + \frac{\ri}{2} D_{\a}^i \s \bar{D}^{\bd}_i \bar{\s} \Big) { \bar M}_{\ad \bd} 
	- \frac \ri 2 D_\a^i \s \bar{D}_{\ad j} \bar{\s} \mathbb{J}^j{}_i \Big) ~, 
	\end{align}
\end{subequations}
where $D_{A} = (\partial_{a}\,, D_{\a}^{i}\,, \bar{D}_{i}^{\ad})$ are the flat $\N$-extended covariant derivatives, see eq. \eqref{flat mink cd}, and $\s$ is a chiral superfield, $\bar{D}^{\ad}_{i}\s = 0$, whose constraints are determined by the requirements eq. \eqref{torsion conditions}. 
These constraints, and their solutions, are described in subsections \ref{cf1} and \ref{cf2}. 
Further, it is possible to extract the components of the supervielbein in a coordinate system in which the relations \eqref{AdSBoost} hold.
They are given by 
\bsubeq \label{n extend cf vierbein}
\begin{align}
	E^{a} &= \re^{-\frac{1}{2}(\s + \bar{\s})} \P^{a}\,, \qquad \P^{a} = \text{d}x^{a} + \ri(\q_{i}\s^{a}\text{d}\bar{\q}^{i} - \text{d}\q_{i}\s^{a}\bar{\q}^{i})\,,
	\label{cf vec vierbein}
	\\
	E_{i}^{\a} &= \re^{-(\frac{\N-2}{2\N}\s + \frac{1}{\N}\bar{\s})}\Big(\text{d}\q_{i}^{\a} + \frac{\ri}{4}\bar{D}_{\ad i}\bar{\s} \P^{\ad\a}\Big)\,, \label{cf dq}
	\\
	\bar{E}_{\ad}^{i} &= \re^{-(\frac{1}{\N}\s + \frac{\N-2}{2\N}\bar{\s})}\Big(\text{d}\bar{\q}_{\ad}^{i} - \frac{\ri}{4}D_{\a}^{i}\s \P_{\ad}{}^{\a}\Big) \, , \label{cf dqb}
\end{align}
\esubeq
where $\P^{a}$ is the $\N$-extended version of the Volkov-Akulov one-form. 
Any frame in which the relations eq. \eqref{AdSBoost} and \eqref{n extend cf vierbein} hold is said to be conformally flat. 

Finally, the constraints \eqref{torsion conditions} imply that the covariant derivatives satisfy the graded commutation relations
%
\begin{subequations} 
	\label{4d cflat ads algebra}
	\bea
	\{ \cD_\a^i , \cD_\b^j \}
	&=&
	4 S^{ij}  M_{\a\b} 	- 4 \ve_{\a\b}S^{k[i}  \mathbb{J}^{j]}{}_{k}
	~,
	\\
	\{ \bar{\cD}^\ad_i , \bar{\cD}^\bd_j \}
	&=&
	- 4 \bar{S}_{ij}  \bar{M}^{\ad\bd} 	+ 4 \ve^{\ad \bd} \bar{S}_{k[i}  \mathbb{J}^{k}{}_{j]}
	~,
	\\
	\{ \cD_\a^i , \bar{\cD}^\bd_j \}
	&=&
	- 2 \ri \d_j^i\cD_\a{}^\bd
	~,
	\\
	{[} \cD_\a^i, \cD_{\bb}{]}&=& 
	- \ri \ve_{\a \b} S^{ij} \bar{\cD}_{\bd j}
	~,
	\qquad
	{[} \bar{\cD}^\ad_i, \cD_{\bb}{]}= 
	\ri \d_\bd^\ad \bar{S}_{ij} \cD_\b^j
	~, 
	\\
	\big [ \cD_\aa , \cD_\bb \big] &=& - 2|S|^2
	(\ve_{\a \b} \bar{M}_{\ad \bd} + \ve_{\ad \bd} M_{\a \b})~,\qquad |S|^2: = \frac{1}{\cN}S^{ij} \bar{S}_{ij}>0~.
	\eea
\end{subequations}
In the above, $|S|$ is a constant parameter of the superspace. 
For $\N=1$, the $\sSU(\N)$ generators $\mathbb{J}^{i}{}_{j}$ are not present and, provided one uses the identification $\bar{\m} = -S$, we recover the known case for AdS$^{4|4}$, see, e.g., \cite{BK} or section \ref{bm n=1 ads sec}. 
For $\N=2$, the algebra coincides with that described in \cite{Kuzenko:2008qw, Kuzenko:2008ep} and reviewed in section \ref{bm n=2 ads sec}. 
Below we will discuss implications for the $\N$-extended case.

\subsection{Diagonal frame} \label{4d solving}
	
Associated with the constraints \eqref{4d n>1 conditions} are integrability conditions. 
In particular, the constraint $\cD_A S^{jk} = 0$ implies the following condition on the torsion $S^{ij}$:
\begin{subequations}\label{4d unitary S condition}
\begin{align}	
	\d_{(k}^{[i} S^{j]m} \bar{S}_{l)m} = 0 \quad \implies \quad 
	S^{ik} \bar{S}_{jk} =  |S|^2\d^i_j
	~.
\end{align}
It is instructive to rewrite the properties of $S^{ij}$ in matrix notation. 
With the definitions 
\begin{align}
	\hat{S} := \left(S^{ij}\right)\,, \qquad 
	{\mathbb S}:= 
|S|^{-1}
\hat{S}~,
\end{align}
the properties of the matrix $\mathbb{S}$ can be recast in the form:
\bea
{\mathbb S}^{\rm T} = {\mathbb S}~, \qquad {\mathbb S}^\dagger {\mathbb S}={\mathbbm 1}_{\cN}~.
\eea
\end{subequations}
Now, we can make use of a simple lemma, proven by Zumino in \cite{Zumino:1962smg}, to decompose the matrix $\mathbb{S}$ as follows
\bea
{\mathbb S} = U U^{\rm T} ~, \qquad U \in \sU(\cN)~,
\label{Zum}
\eea
where the unitary matrix $U$ is defined modulo the equivalence relation
\bea 
U ~\sim ~ U \cO ~, \qquad \cO \in \sO(\cN)~.
\eea

The above analysis plays a critical role in relating the supergravity-inspired description of AdS$^{4|4\N}$ to its description as a homogeneous space, which we will develop in subsequent sections. 
In particular, the decomposition \eqref{Zum} implies that there exists a gauge in which the torsion tensor $S^{ij}$ takes the following form
\begin{align}
	\label{4d diag gauge}
	S^{ij} = \d^{ij} S ~. 
\end{align}
This gauge can be reached by performing a local $\sU(\N)$ transformation.\footnote{Strictly speaking, this should be performed in the $\sU(\N)$ superspace setting \cite{Howe:1980sy,Howe:1981gz}, see, e.g., \cite{KKR2}. However, for our purposes it suffices to introduce a flat $\sU(1)$ connection. }
Further, in this gauge, the algebra of covariant derivatives \eqref{4d cflat ads algebra} contains the $\sSU(\N)$ generators only in the combination 
\begin{align}
	\mathcal{J}^{ij} := - 2 \d^{k[i} \mathbb{J}^{j]}{}_k = - \cJ^{ji} 
~,
\end{align}
which can be shown to act on an isospinor $\psi^k$ as follows
\begin{align}
	\cJ^{ij} \psi^k = 2 \d^{k[i} \psi^{j]}~
\end{align}
and thus leave $\d^{ij}$ invariant.
This means that $\cJ^{IJ}$ are generators of $\sSO(\N)$, and the $R$-symmetry group is reduced to $\sO(\N)$. 
In such a frame, isospinor indices are raised and lowered according to the rule
\begin{align}
	\psi^i = \d^{ij} \psi_j ~, \qquad \psi_i = \d_{ij} \psi^j~.
\end{align}
The result is an algebra of covariant derivatives of the form
\begin{subequations} 
	\label{4d ads coset algebra}
	\bea
	\{ \cD_\a^i , \cD_\b^j \}
	&=&
	4 S \d^{ij}  M_{\a\b} + 2 \ve_{\a\b} S \cJ^{ij}
	~,
	\\
	\{ \bar{\cD}^\ad_i , \bar{\cD}^\bd_j \}
	&=&
	- 4 \bar{S} \d_{ij}  \bar{M}^{\ad\bd} - 2 \ve^{\ad\bd} \bar{S}  \cJ_{ij}
	~,
	\\
	\{ \cD_\a^i , \bar{\cD}^\bd_j \}
	&=&
	- 2 \ri \d_j^i\cD_\a{}^\bd
	~,
	\\
	{[} \cD_\a^i, \cD_{\bb}{]}&=& 
	- \ri \ve_{\a \b} S \bar{\cD}_{\bd}^i
	~,
	\qquad
	{[} \bar{\cD}^\ad_i, \cD_{\bb}{]}= 
	\ri \d_\bd^\ad \bar{S} \cD_{\b j}
		~, 
		\\
		\big [ \cD_\aa , \cD_\bb \big] &=& - 2 |S|^{2} (\ve_{\a \b} \bar{M}_{\ad \bd} + \ve_{\ad \bd} M_{\a \b})~.
		\eea
\end{subequations}
This algebra coincides with that which is derived from the coset construction, eq. \eqref{NowarAlgebra}, as we will see below.

\subsection{Conformally flat realisation I: stereographic coordinates} \label{cf1}

It is often useful to relax the gauge condition \eqref{4d diag gauge} and instead work in a frame in which the covariant derivatives feature an $\sSU(\N)$ connection and are related to the flat ones by the rule \eqref{AdSBoost}. 
As described above, an advantage of doing so is that it is possible to find an explicit conformally flat realisation of AdS$^{4|4\N}$. 
These transformations are determined by a chiral superfield $\s$ subject to constraints imposed by the conditions \eqref{torsion conditions}. 
The goal of this section is to determine the chiral paramter $\s$ corresponding to the $\N$-extended AdS superspace. 

For the covariant derivatives \eqref{AdSBoost},
the curvature $S^{ij}$ takes the form
\begin{subequations}
	\label{AdSCurvatures}
	\begin{align}
		S^{ij }= - \frac 1 4 \re^{\frac{\cN-2}{\cN} \s + \frac 2 \cN \bar{\s}} \Big(D^{ij} \s - D^{\a i} \s D_\a^{j} \s \Big)
=  \frac 1 4 \re^{\frac{2}{\cN} [\bar \s + (\cN -1)\s ]} D^{ij} \re^{-\s}	
 ~.  \label{ads constraints 0}
	\end{align}
Imposing the conditions 
\eqref{torsion conditions} leads to the following constraints on the chiral parameter $\s$:	\begin{align} \label{ads constraints 1}
		Y_{\ab}^{ij} &= 0 \quad \implies \quad D_{(\a}^{[i} D_{\b)}^{j]} \re^\s = 0~,
		\\
		G_{\a\ad} &= 0 \quad \implies \quad [D_\a^i,\bar{D}_{\ad i}] \re^{\frac{\cN}{2}(\s + \bar{\s})} = 0 \label{ads constraints 2}
		~.
	\end{align}
In the $\cN=1$ case, the constraint \eqref{ads constraints 1} is absent, and the condition of covariant constancy of $S^{ij} $ becomes
\bea
\bar \m = - \frac 1 4 \re^{2 \bar \s } D^2 \re^{-\s}=\text{const}~.
\label{ads constraints 3}
\eea
\end{subequations}
If  the constraints \eqref{ads constraints 1} and \eqref{ads constraints 2} are satisfied, 
the tensor $S^{ij}$ defined by \eqref{ads constraints 0} proves to be covariantly constant. 
 
 In the $\cN=1$ case, the constraints \eqref{ads constraints 2} and 
 \eqref{ads constraints 3} were solved in section 6.5.4 of \cite{BK}. 
For $\N=2$, the constraints \eqref{ads constraints 1} and \eqref{ads constraints 2} reduce to those given in \cite{Kuzenko:2008qw}, where they were also solved. 
We will now solve them for arbitrary $\N$. 
As $\s$ is chiral, $\bar{D}_{i}^{\ad}\s = 0\,,$ it is a function of the coordinates $x_{+}^{a} = x^{a} + \ri\q_{i}\s^{a}\bar{\q}^{i}$ and $\q_{i}^{\a}\,.$
The most general Lorentz and $\sSU(\N)$ invariant ansatz is then given by 
\begin{align} \label{general ansatz}
	\re^{\s} = \sum_{n=0}^{\N}A_{n}^{i_{1}\ldots i_{2n}}(x_{+}^{2})\q_{i_{1}i_{2}}\ldots\q_{i_{2n-1}i_{2n}}\,, 
\end{align}
where $\q_{ij} = \q^{\a}_{i}\q_{\a j} = \q_{ji}$ and the coefficients $A_{n}^{i_{1}\ldots i_{2n}}(x_{+}^{2})$ in general furnish reducible representations of $\sSU(\N)\,.$ 
Inserting \eqref{general ansatz} into \eqref{ads constraints 1} yields the following 
\begin{align}
	\re^{\s} = a + bx_{+}^{2} + s^{ij}\q_{ij}\,, 
\end{align}
where $a,b$ and $s^{ij} = s^{ji}$ are constant. 
The constraint \eqref{ads constraints 2} then implies 
\bsubeq
\begin{align}
	s^{ik}\bar{s}_{kj} &= -4a\bar{b}\d^{i}_{j}\,, \qquad \bar{s}_{ij} = \overline{s^{ij}}\,, \label{c cbar}
	\\
	a\bar{b} &= \bar{a}b\,.
\end{align}
\esubeq
Since the left-hand side of \eqref{c cbar} is positive-definite, we find that $a\bar{b} < 0\,.$
Choosing the constant $a$ to be $a=1$, the above relations lead to 
\begin{align}
	b = -\frac{1}{4\N}s^{ij}\bar{s}_{ij}\,. 
\end{align}
The solution to \eqref{ads constraints 1} and \eqref{ads constraints 2} is then
\begin{align} \label{ads soln}
	\re^{\s} = 1 - \frac{1}{4\N}s^{ij}\bar{s}_{ij}x_{+}^{2} + s^{ij}\q_{ij}\,.
\end{align}

Evaluating the torsion superfield $S^{ij}$ yields 
\begin{align} \label{stereographic s}
	S^{ij} = \frac{1}{4}\re^{\frac{2(\N-1)}{\N}\s + \frac{2}{\N}\bar{\s}}D^{ij}\re^{-\s} = s^{ij} + \cO(\q)\,. 
\end{align}
We point out that 
\bea
S^{ij}(z) \bar S_{ij} (z)  \equiv \cN |S|^2 = s^{ij} \bar s_{ij}=\text{const} ~.
\eea

Finally, given the form of the conformally flat supervielbein \eqref{n extend cf vierbein}, it follows that the spacetime metric is 
\begin{align} \label{stereo metric}
	\text{d}s^{2} = \eta_{ab}E^{a}E^{b}|_{\q=0} = \frac{\eta_{ab}\text{d}x^{a}\text{d}x^{b}}{(1-\frac{x^{2}}{4\ell^{2}})^{2}}\,, 
\end{align}
with $\ell^{-2} = |S|^2$.
It should be pointed out that, while the constraints \eqref{ads constraints 1} and \eqref{ads constraints 2} appeared in \cite{KKR} for the first time, the super-Weyl parameter \eqref{ads soln} was constructed in ref. \cite{BILS} 
by making use of an alternative approach, 
although an explicit solution of the form \eqref{AdSBoost}  was not derived in \cite{BILS}.

\subsection{The Killing supervectors of AdS$^{4|4\N}$}

Now we turn to determining the Killing supervectors of AdS$^{4|4\N}$. 
An infinitesimal isometry of AdS$^{4|4\N}$ is generated by a Killing supervector $\bm{\x}^{A} E_{A}$ which is defined to satisfy the property 
\begin{align} \label{Ads killing def}
[\bm{\x}^{A}\cD_{A} + \frac{1}{2}\L^{cd}M_{cd} + \L^{i}{}_{j}\mathbb{J}^{j}{}_{i}\,, \cD_{B}] = 0\,, 
\end{align}
for a real antisymmetric tensor $\L^{cd}(z)$.
Since AdS$^{4|4\N}$ is conformally related to $\N$-extended Minkowski superspace $\mathbb{M}^{4|4\N}$, see the relations \eqref{AdSBoost}, 
the supervector $\bm{\x}^{A}E_{A}$ can be decomposed with respect to the AdS basis $\{E_{A}\}$ or the flat basis $\{D_{A}\}\,,$
as 
\begin{align} \label{killing decomp}
\x = \bm{\x}^{A}E_{A} = \x^{A}D_{A}\,. 
\end{align}
Here, $\x^{A}$ are the components of a conformal Killing supervector, which generates infinitesimal superconformal transformations in $\mathbb{M}^{4|4\N}$
\begin{align}
z^{A} \longrightarrow z^{A} + \x\,z^{A}\,, 
\end{align}
and is defined to satisfy the constraint 
\begin{align}
[\x\,, D_{\a}^{i}] \propto D_{\b}^{j}\,,
\end{align}
see, e.g., \cite{Kuzenko:2006mv,Kuzenko:1999pi,Park:1999pd}, for more details. 
With respect to the basis $\{D_{A}\}$, the components of $\x$ are 
\bsubeq \label{killing scf trf parameters}
\begin{align}
\tilde{\x}_{+} &= (\x_{+}^{\ad\a}) = \tilde{a} + \frac{1}{2}(\D + \bar{\D})\tilde{x}_{+} - \bar{\o}\tilde{x}_{+} - \tilde{x}_{+}\o + \tilde{x}_{+}b\tilde{x}_{+} + 4\ri\bar{\e}\q - 4\tilde{x}_{+}\eta\q
\,, \\
(\x_{i}^{\a}) &= \e + \frac{1}{2\N}\big((\N-2)\D + 2\bar{\D})\big) \q - \q\o + \L\q + \q b\tilde{x}_{+} - \ri\bar{\eta}\tilde{x}_{+} - 4\q\eta\q
\,, \\
\x^{a} &= \frac{1}{2}(\x_{+}^{a} + \x_{-}^{a}) + \ri\left( \q_i\s^{a}\bar{\x}^i - \x_i \s^{a}\bar{\q}^i \right)\,, \qquad \x^{a}_{+} = -\frac{1}{2}\x_{+}^{\ad\a}(\s^{a})_{\a\ad} = \overline{\x_{-}^{a}}\,,
\end{align}
\esubeq
where the parameters $\{a\,, b\,, \o\,, \bar{\o}\,, \D\,, \bar{\D}\,, \e\,, \bar{\e}\,, \eta\,, \bar{\eta}\,, \L \}$ are identified with those in \eqref{scf trf bm}. 
The general solution to the above was given for $\N=1$ in \cite{BK} and $\N>1$ in \cite{Park:1998nra,Park:1999pd}.

Now the solution \eqref{ads soln} plays the role of the compensator for AdS$^{4|4\N}$, see, e.g., \cite{BK,Kuzenko:2008qw},
\begin{align} \label{ads compensator}
W = \re^{-\s}\,.
\end{align}
Given a superconformal transformation, $W$ transforms as 
\bsubeq \label{comp var}
\begin{align}
\d W &= \x W +  \s[\x] W\,, 
\\
 \s[\x] &= \frac{1}{\N(\N-4)}\left( (\N-2)D_{\a}^{i}\x_{i}^{\a} -  2 \bar{D}_{i}^{\ad}\bar{\x}_{\ad}^{i} \right) \,. 
\end{align}
\esubeq
Then, the problem of determining the AdS Killing supervectors proves to be equivalent to determining those conformal Killing supervectors which do not change the compensator \eqref{ads compensator}, 
\begin{align} \label{comp var = 0}
\d W = 0\,.
\end{align}
The $\N=1$ and $\N=2$ cases were worked out in \cite{BK} and \cite{Kuzenko:2008qw}, respectively.
It can be shown that eq. \eqref{comp var = 0} imposes the following constraints on the transformation parameters in \eqref{killing scf trf parameters}
\bsubeq \label{ads killing}
\begin{align}
b^{a} &= \frac{s^{ij}\bar{s}_{ij}}{4\N}a^{a} \,, 
\\
\eta_{\a}^{i} &= \frac{1}{2}s^{ij}\e_{\a j}\,, 
\\
s^{k(i}\L_{k}{}^{j)} &= 0\,, \label{s lambda}
\\
\D &= 0\,.
\end{align}
\esubeq
In particular, eq. \eqref{s lambda} implies 
\begin{align}
\hat{s} \L + \L^{\T} \hat{s} = 0\,, \qquad \hat{s} := (s^{ij})\,. 
\end{align}
Then, making use of \eqref{killing decomp}, one can read off the components of the AdS Killing supervector $\bm{\x}^{A}\,.$ 
Appendix \ref{ads and mink} outlines another derivation for the relations \eqref{ads killing}.

\subsection{Conformally flat realisation II: Poincar\'e coordinates} \label{cf2}

As can be seen from \eqref{stereo metric}, the above realisation makes use of stereographic coordinates, in which the spacetime metric is manifestly invariant under the group of four-dimensional Lorentz transformations, $\sO(3,1)\,.$ 
One can also make use of Poincar\'e coordinates, in which the spacetime metric takes the form 
\begin{align} \label{pp metric}
	\text{d}s^{2} = \left(\frac{1}{sz}\right)^{2}\left(\eta_{\ha\hb}\text{d}x^{\ha}\text{d}x^{\hb} + \text{d}z^{2}\right)\,, \qquad \ha = 0\,,1\,,3\,, 
\end{align}
and is manifestly invariant under the group of three-dimensional Poincar\'e transformations, $\sIO(2,1)\,.$
The reason for this index convention will become clear below.
This coordinate system was utilised in \cite{Butter:2012jj} to derive an alternative conformally flat realisation for the $\N=1$ and $\N=2$ AdS superspaces, and we will now extend this analysis to the case of arbitrary $\N\,.$

As described above, the relations \eqref{AdSBoost} hold for a conformally flat realisation of AdS$^{4|4\N}$. 
Our point of departure from the analysis of section \ref{cf1} is in the realisation of the flat superspace covariant derivatives. 
For this purpose, it is convenient to introduce a $3+1$ splitting of the 4D vector indices as follows.\footnote{Our notation and conventions coincide with those in \cite{Butter:2012jj}.} 
We first delete the sigma-matrix with vector index $a=2$,
\bsubeq \label{3d gm}
\begin{align}
	(\s_{a})_{\a\bd}&\longrightarrow (\g_{\ha})_{\ab} = (\g_{\ha})_{\b\a}\,, \qquad \left((\g_{\ha})_{\ab}\right) = (\id_{2}\,, \s_{1}\,, \s_{3})\,, 
	\\
	(\tilde{\s}_{a})^{\ad\b} &\longrightarrow (\g_{\ha})^{\ab} = (\g_{\ha})^{\b\a} = \ve^{\a\g}\ve^{\b\d}(\g_{\ha})_{\g\d}\,. 
\end{align}
\esubeq
Spinor indices are raised and lowered using the $\sSL(2,\mathbb{R})$ invariant tensors $\ve^{\ab}$ and $\ve_{\ab}$, satisfying
\begin{align}
	\ve^{\ab} = -\ve^{\b\a}\,, \qquad \ve^{12} = -\ve_{12} = 1\,,
\end{align}
by the rule 
\begin{align}
	\j^{\a} = \ve^{\ab}\j_{\b}\,, \qquad \j_{\a} = \ve_{\a\b}\j^{\b}\,.
\end{align}
A four-vector $V_{a}$ can then be written as 
\bsubeq
\begin{align}
	V_{\a\bd} &= V^{a}(\s_{a})_{\a\bd} \longrightarrow V_{\ab} + \ri\ve_{\ab}V_{z}\,, \qquad V_{\ab} := V_{\ha}(\g^{\ha})_{\ab}\,, 
	\\
	V^{\ad\b} &= V^{a}(\tilde{\s}_{a})^{\ad\b} \longrightarrow V^{\ab} + \ri\ve^{\ab}V_{z}\,, \qquad V^{\ab} := V_{\ha}(\g^{\ha})^{\ab}\,. 
\end{align}
\esubeq
The flat spinor covariant derivatives then take the form 
\bsubeq
\begin{align}
	D_{\a}^{i} &= \frac{\partial}{\partial\q_{i}^{\a}} + \ri(\g^{\hm})_{\ab}\bar{\q}^{\b i}\partial_{\hm} - \bar{\q}^{i}_{\a}\partial_{z}\,,
	\\
	\bar{D}_{\a i} &= - \frac{\partial}{\partial \bar{\q}^{\a i}} - \ri(\g^{\hm})_{\ab}\q_{i}^{\b}\partial_{\hm} - \q_{\a i}\partial_{z}\,,
\end{align}
and they satisfy the anticommutation relations 
\begin{align}
	\{D_{\a}^{i}\,, D_{\b}^{j}\} = \{\bar{D}_{\a i}\,, \bar{D}_{\b j}\} = 0\,, \qquad \{D_{\a}^{i}\,, \bar{D}_{\b j}\} = -2\ri\d^{i}_{j}(\g^{\hm})_{\ab}\partial_{\hm} + 2\d^{i}_{j}\ve_{\ab}\partial_{z}\,. 
\end{align}
\esubeq
Finally, let us introduce the chiral coordinate $z_{L} := z - \q_{k}^{\a}\bar{\q}^{k}_{\a}\,, ~ \bar{D}_{\a i}z_{L} = 0\,.$ 

Given the form of the spacetime metric \eqref{pp metric}, we seek a solution to the constraints  \eqref{ads constraints 1} and \eqref{ads constraints 2} that is a function of the chiral coordinates $z_{L}$ and $\q_{i}^{\a}\,.$ 
The most general ansatz is then 
\begin{align} \label{pp ansatz}
	\re^{\s} = \sum_{n=0}^{\N}A_{n}^{i_{1} i_2 \ldots i_{2n-1}i_{2n}}(z_{L})\q_{i_{1}i_{2}}\ldots\q_{i_{2n-1}i_{2n}}\,.
\end{align}
The constraints prove to be solved by the following
\begin{align} \label{pp soln}
	\re^{ \s} = |s| z_{L} + s^{ij}\q_{ij}\,, \qquad 
	|s|^2 := \frac{1}{\N}s^{ij}\bar{s}_{ij} >0~,
\end{align}
for a constant symmetric tensor $s^{ij}$. The most general solution to the constraints proves to be at most quadratic in $\q$'s.

Similar to the stereographic case, evaluating the torsion superfield $S^{ij}$ yields
\begin{align}
	S^{ij} = s^{ij} + \cO(\q)\,.
\end{align}



%
\section{Coset construction} \label{4d coset}

The analysis of the previous sections has been devoted to studying the AdS supergeometry from the perspective of the supergravity-inspired approach. 
We will now turn to developing its supergeometry via the (bi-)supertwistor description of AdS$^{4|4\N}$ derived in \cite{KTM} and reviewed in section \ref{ads4 bst}.
For this purpose, we will make use of the coset construction of AdS$^{4|4\N}$, for which the alternative realisation of the AdS supergroup described in appendix \ref{adsgroup} is ideal.


As described in section \ref{ads4 bst}, AdS$^{4|4\N}$ can be realised in terms of constrained (bi-)supertwistors.
Making use of the equivalence transformations \eqref{equiv}, the conditions \eqref{newcon} on these supertwistors can be fine-tuned to\footnote{We emphasise that the inner products in \eqref{normalised conditions} are defined by rules \eqref{innerproducts}.}
\bsubeq \label{normalised conditions}
\begin{align} 
	\braket{\underline{\bm{T}}^\mu}{\underline{\bm{T}}^\nu}_\mathbb{K} &= \ell\ve^{\mu\nu}\,,
\qquad \ell = \bar \ell >0~,  \label{normalised conditions.a}	\\
	\braket{\underline{\bm{T}}^\mu}{\underline{\bm{T}}^\nu}_\mathbb{J} &= 0\,,
	 \label{normalised conditions.b}
\end{align}
\esubeq
for a fixed positive parameter $\ell$. Such a frame is said to be normalised.
The normalised conditions \eqref{normalised conditions.a} are preserved under the `reduced' equivalence relation
\begin{align} \label{equivalence relation}
	\underline{\bm{T}}^\m \sim \underline{\bm{T}}^\n N_\n{}^\m
	\,, \qquad N \in \sSL(2,\mathbb{C})\,,
\end{align}
which is the remnant of \eqref{equiv}. 
The space of such normalised frames will be denoted ${\mathfrak F}_\cN^{(\ell)}$.

The conditions \eqref{normalised conditions} can be recast in terms of the two-plane
\begin{align} \label{two plane def}
	\underline{\cP} = ( \underline{\bm{T}}^\mu )
	= \left( \begin{array}{c}
		F \\
		G \\
		\hline \hline
		\ri\Q
	\end{array} \right) 
\end{align}
and imply the following constraints:
\bsubeq \label{detcon}
\begin{align}
	\det F &+ \det G = \ell + \frac{1}{2}\text{tr}\left(\Q\ve^{-1} \Q^\T  \right)\,,  \label{detcon1}
	\\
	F^\dag G &- G^\dag F + \ri\Q^\dag \Q = 0\, . \label{detcon2}
\end{align}
\esubeq
An important implication of the relation \eqref{detcon1} is that at least one of the $2\times 2$ matrices $F$ and $G$ is nonsingular.

Along with the definition \eqref{3.11} given earlier,
the $\cN$-extended AdS superspace can equivalently be defined as 
\begin{align} \label{new ads4 def}
	\text{AdS}^{4|4\mathcal{N}} = \frak{F}_\mathcal{N}^{(\ell)}/ \!\! \sim\,,
\end{align}
where the equivalence relation is given by \eqref{equivalence relation}.


\subsection{Stabiliser}

In order to prove \eqref{new ads4 def}, it suffices to choose a base (also known as marked, or preferred) point and determine its stabiliser.
For the remainder of this section, we will set $\ell =1$.\footnote{Reinstating the length parameter amounts to a simple rescaling.} 
The most convenient choice of two-plane satisfying the conditions \eqref{normalised conditions} is then
\begin{align} \label{prefpoint}
	\underline{\cP}^{(0)} = \left( \begin{array}{c}
		{\mathbbm 1}_2\\
		0\\
		\hline \hline
		0
	\end{array} \right)\,.
\end{align}
The stabiliser $H$ of $\underline{\cP}^{(0)}$ consists of those elements $h$ of the AdS supergroup $\sOSp(\mathcal{N}|4;\mathbb{R})_{C}$
which satisfy the conditions
\begin{align}
	h \underline{\cP}^{(0)} = \left( \begin{array}{c}
		M  \\
		0\\
		\hline \hline
		0
	\end{array} \right)\,,
	\qquad M \in \sGL(2,\mathbb{C})\,.
\end{align}
These conditions imply that 
\begin{align}
	h = \left( \begin{array}{c||c}
		\begin{array}{cc}
			N & 0 \\
			0 & (N^\dagger)^{-1}
		\end{array}
		& 0 \\
		\hline \hline
		0 & R 
	\end{array} \right)\,, 
	\qquad N \in \sSL(2,\mathbb{C})\,, \quad R \in \sO(\mathcal{N})\,.
\end{align}
Thus the stability subgroup $H$ is isomorphic to 
\begin{align} \label{stabgroup}
	\sSL(2,\mathbb{C}) \times \sO(\mathcal{N})\,,
\end{align}
which proves \eqref{new ads4 def}.

The bi-supertwistors \eqref{new bi-super} corresponding to the preferred point $\underline{\cP}^{(0)}$ take the form
\begin{align} \label{pref bi-super}
	\underline{X}^{(0)} = \left(\begin{array}{c|c||c}
		-\ri\ve^{-1} & 0 & 0
		\\
		\hline
		0 & ~0~ & ~0~ 
		\\ 
		\hline \hline
		0 & 0 & 0
	\end{array}\right) \,, \qquad \underline{\bar{X}}^{(0)} = \left(\begin{array}{c|c||c}
	0 & 0 & 0
	\\
	\hline 
	~0~ & ~\ri\ve~ & ~0~ 
	\\
	\hline \hline 
	0 & 0 & 0
\end{array}\right) \,.
\end{align}

\subsection{Generalised coset representative}

A key role in the study of AdS$^{4|4\N}$ as a homogeneous space is played by a local coset representative, $S$, which is an injective map $S: U \rightarrow \sOSp(\N|4;\mathbb{R})_{C}$, defined for every chart
$U$ 
of the atlas on AdS$^{4|4\N}$ chosen, with the property $\p \circ S = {\rm id}_U$, where $\p$ denotes the natural (canonical)  projection 
$\p: \sOSp(\N|4;\mathbb{R})_{C} \to \text{AdS}^{4|4\N} =\mathsf{OSp}({\cal N}|4;\mathbb{R})_{C} /\big[ \mathsf{SL}(2, \mathbb{C}) \times \mathsf{O}({\cal N}) \big]$. 
In this subsection we will lay the groundwork for developing the coset representative for AdS$^{4|4\N}$.

Associated with the normalised two-plane  $\underline{\cP} $ is the following group element
\bsubeq \label{4d cosetrep}
\begin{align}
	S(\underline{\cP}) 
&= \left( \begin{array}{c||c}
		A & B \\
		\hline
		\hline
		C & D
	\end{array} \right)\,,	\\
	A &= \left( \begin{array}{c|c} \label{ablock}
		F ~& \ve^{-1} \bar{G} \ve^{-1} \\ \hline
		G ~& \ve \bar{F} \ve^{-1} 
	\end{array}
	\right) 
	\,,
	\\
	C &= \Big(\begin{array}{c|c}
		~\ri\Q~ &~ -\bar{\Q}\ve^{-1}
	\end{array} \Big) 
	\,,
	\\
	D &= \Big( {\mathbbm 1}_\mathcal{N} -  C [A^\T \me A]^{-1}C^\T\Big)^{-\frac{1}{2}}\,,
	\label{dblockdef} \\
	B&= \me^{-1}(A^{-1})^\T C^\T D \label{bblockdef}
	\,.
\end{align}
\esubeq
The fundamental property of $S(\underline{\cP})$ is that 
\begin{align} \label{4d coset rep fundamental}
	S(\underline{\cP})\underline{\cP}^{(0)} = \underline{\cP}\,,
\end{align}
for any normalised two-plane 
$\underline{\cP} \in {\mathfrak F}_\cN^{(\ell)}$.
%
The functional forms of the matrices $A$ and $C$ are fixed through the condition  \eqref{4d coset rep fundamental} and 
the reality conditions \eqref{areal} and \eqref{creal}.
 The remaining blocks are then fixed by the group requirements \eqref{groupreq1} and \eqref{groupreq2}.
 We point out that $D$ is symmetric, $D=D^\T$. 
It is possible to obtain alternate expressions for the blocks $D$ and $B$, which may be more suited to performing particular calculations. They take the following form 
\bsubeq
\begin{align}
	D &= \Big({\mathbbm 1}_{\N} + C\me^{-1}C^{\T}\Big)^{\frac{1}{2}} \,, \label{new d block def}
	\\
	B &= A\me^{-1}C^{\T}D^{-1} \,.
\end{align}
\esubeq
These expressions can be seen to coincide with \eqref{dblockdef} and \eqref{bblockdef} by using the group requirements and the general form for the inverse of a supermatrix, see, e.g. \cite{BK} for more details.

Due to the particular properties of the group elements of $\sOSp(\N|4;\mathbb{R})_{C}$, we have been able to parametrise them in a way that relates directly to the two-planes \eqref{two plane def} without making use of the gauge freedom \eqref{equivalence relation}. 
This is echoed in the fact that the element $S(\underline{\cP})$ introduced above satisfies the following relation
\begin{align} \label{rep equiv}
	S(\underline{\cP}N) = S(\underline{\cP})\mathfrak{N}\,, \quad 
	\mathfrak{N} = \left(\begin{array}{cc||c}
		N & 0 & 0 \\
		0 & (N^\dagger)^{-1} & 0 \\
		\hline \hline 
		0 & 0 & {\mathbbm 1}_{\N}
	\end{array} \right)\,,
\end{align}
with $N \in \sSL(2,\mathbb{C})$. 
What this means is that $S(\underline{\cP})$ is, in and of itself, not a genuine coset representative, but a group element that will allow us to obtain a coset representative if we pick a single two-plane in each equivalence class. 
This is readily achieved in coordinate charts for AdS$^{4|4\N}$.
An advantage of having the element $S(\underline{\cP})$ defined in such a way is that we can develop certain aspects of the AdS supergeometry prior to imposing a particular gauge, as we will see in subsequent sections. 
We then have the freedom to impose a gauge and read off geometric objects in a coordinate system of choice. 

%


\subsection{AdS space ($\cN=0$)}

We will begin by outlining the above setup in the non-supersymmetric ($\cN = 0$) case. 
As noted above, at least one of the $2\times 2$ matrices $F$ and $G$, see eq. \eqref{two plane def}, is nonsingular.
This naturally leads to the introduction of two coordinate charts for ${\mathfrak F}_\cN^{(\ell)}$, which provide an atlas.
We define the north chart to consist of those two-planes with nonsingular $F$, that is, $\det F \neq 0\,.$
Analogously, the south chart is defined to consist of those two-planes with nonsingular $G$, $\det G \neq 0\,.$

In the north chart, since $F$ is nonsingular, we can make use of the equivalence relation \eqref{equivalence relation} to choose
$F \propto {\mathbbm 1}_2$.
The conditions \eqref{normalised conditions} then imply that the two-planes in the north chart take the form
\begin{align} \label{north chart general}
	\underline{\cP} 
	= \l\left( \begin{array}{c}
		\ell {\mathbbm 1}_2 \\
		- \tilde{x}
	\end{array}\right)
\,, \qquad 
\tilde{x} = x^{m}\tilde{\s}_{m}\,, \qquad \tilde{\s}_m = ({\mathbbm 1}_2, - \vec \s )~,
\end{align}
where
$\l \neq 0$ is a scalar field, and $\vec \s$ are the Pauli matrices. The constraints \eqref{detcon1} and \eqref{detcon2} give, respectively, 
\begin{subequations}
\bea 
\l^2 ( \ell^2 + \det \tilde{x} ) &=& \ell ~~\implies ~~ \bar \l^2 ( \ell^2 + \det \tilde{x}^\dagger ) = \ell~, \label{4.12a} \\
\tilde{x}^\dagger &=& \tilde{x}~.
\eea
\end{subequations}
It follows that the coordinates $x^m$ are real, $x^2 := \eta_{mn} x^m x^n \neq \ell^2$, and $\l= \bar \l$.
We can fix $\l>0$ by making use of the remnant of the equivalence relation 
\eqref{equivalence relation}, 
$	\underline{T}^\m \sim - \underline{T}^\m$.
Then we observe that the coordinate chart is specified by 
\bea
x^2< \ell^2~, 
\eea
and the parameter $\l$ is given by 
\begin{align}
\l = \sqrt{ \frac{\ell}{\ell^{2}-x^2} }\,.
\end{align}
The real coordinates $x^{m}$ parametrise AdS$_{4}$ in the north chart. 
Direct calculation of the two-point function \eqref{ads two point function} in this chart yields
\begin{align} \label{4d bosonic interval}
	\o(\underline{\cP},\underline{\cP}+\text{d}\underline{\cP}) =  \frac{\ell^{4} \eta_{mn}\text{d}x^{m}\text{d}x^{n}}{(\ell^{2}-x^{2})^{2}}\,,
\end{align}
which is related to the induced metric in stereographic coordinates eq. \eqref{north chart induced metric appendix} as 
\begin{align}
	\o(\underline{\cP},\underline{\cP}+\text{d}\underline{\cP}) = \frac{1}{4}\rd s_{\rm{north}}^{2}\,.
\end{align}

In the south chart,  the gauge freedom \eqref{equivalence relation} can be used
to choose
$G \propto {\mathbbm 1}_2$.
Analogously, the two-plane conditions \eqref{normalised conditions} then yield
\begin{align} \label{south chart general}
	\underline{\cP} 
	= \g \left( \begin{array}{c}
		{y} \\
		\ell {\mathbbm 1}_2
	\end{array}\right)
\,, \qquad 
{y} = y^{m}{\s}_{m}\,, \qquad \s_m = ({\mathbbm 1}_2, \vec{\s})~,
\end{align}
for some parameter $\g \neq 0$. Now, repeating the north-chart analysis 
tells us that the local coordinates $y^m$ are real, and the following relations hold:
\bea
y^2 <\ell^2~, \qquad \g = \sqrt{ \frac{\ell}{\ell^{2}-y^2} }\,.
\eea
The two-point function \eqref{ads two point function} in the south chart is
\begin{align}
	\o(\underline{\cP},\underline{\cP}+\text{d}\underline{\cP}) =  \frac{\ell^{4} \eta_{mn}\text{d}y^{m}\text{d}y^{n}}{(\ell^{2}-y^{2})^{2}}\,. 
\end{align}

In the intersection of the two charts, the transition functions are
\bea
{y} = - \ell^2  \tilde{x}^{-1} ~~ \Longleftrightarrow ~~y^m =  \frac{\ell^2}{x^2} x^m~~\implies ~ ~y^2 x^2 = \ell^4~.
\label{4.19}
\eea
It follows that $x^2 <0 \Longleftrightarrow y^2 <0$.
Comparing the above relations with those described in appendix \ref{bm ads sec},
we find complete agreement except for the sign difference \eqref{4.19} and \eqref{B.9}.


%
%
\subsection{$\N\neq0$} \label{n s cosets}
The analysis of the previous subsection was for the non-supersymmetric case. We will now extend these considerations to the $\N>0$ situation.
%
We begin by considering the north chart, in which the matrix $F$ in \eqref{two plane def}
is nonsingular. 
The equivalence relation \eqref{equivalence relation}
can once again be used to 
choose $F \propto {\mathbbm 1}_2$.
Similarly to in the non-supersymmetric case, the two-plane constraints \eqref{normalised conditions} imply the general form for two-planes in the north chart is given by
\begin{align} \label{chiral chart}
	\underline{\cP} 
	=\l (x_+, \q) \left(\begin{array}{c}
		\ell {\mathbbm 1}_2 \\
		-\tilde{\bm x}_{+} \\
		\hline \hline 
		2\ri \sqrt{\ell} \q
	\end{array}\right)
\,,\qquad
	 \tilde{\bm x}_+ = x^{m}_+\tilde{\s}_{m}\,,
\end{align}
where the following relations hold
\bsubeq \label{chiral chart2}
\begin{align}
	&\tilde{x}_{+} - \tilde{x}_{-} = 4\ri\q^{\dag}\q\,, 
	\qquad \tilde{x}_{-} := (\tilde{x}_{+})^{\dag}\,, 	\\
	&\l = \sqrt{ \frac{\ell}{\ell^{2} - x^2_{+} - 2\ell\q^2}}\,, \qquad \q^{2} := 
	\text{tr}\left(\q\ve^{-1} \q^\T  \right)
	\,. 
\end{align}
\esubeq
The former is solved by 
\begin{align}
	\tilde{x}_{\pm} = x^{m}\Tilde{\s}_{m} \pm 2\ri\q^{\dag}\q\,.
\end{align}
It is useful to introduce the first-order differential operators 
\bsubeq \label{4d SO(N) derivatives}
\begin{align}
	D_{\m I} &= \partial_{\m I} + \ri(\s^{\m})_{\m\dmu}\bar{\q}_{I}{}^{\dmu}\partial_{m}\,,
	\\
	\bar{D}_{\dmu I} &= -\bar{\partial}_{\dmu I} - \ri \q_{I}{}^{\m}(\s^{m})_{\m\dmu}\partial_{m}\,.
\end{align}
\esubeq
The differential operators \eqref{4d SO(N) derivatives} mimic the flat superspace covariant derivatives but carry an $\sSO(\N)$ index as opposed to an $\sSU(\N)$ index. 
Then, the superfield $\l$ is chiral in the sense that 
\begin{align}
	\bar{D}_{\dmu I} \l = 0 \Longleftrightarrow D_{\m I} \bar{\l} = 0\,.
\end{align}
Further, we can see that the chiral superfield $\l$ is related to that derived from the supergravity-inspired approach eq. \eqref{ads soln} as $\l = \re^{-\frac{1}{2}\s}\,,$ provided one fixes $\ell =1$ and $s^{ij} = -2\d^{ij}$ as described in section \ref{4d solving}. 
Finally, we point out that the two-planes \eqref{chiral chart} are parametrised by the chiral coordinates $x_{+}^{m}$ and $\q_{I}{}^{\m}$, with $x_{+}^{m} = x^{m} + \ri\q_I \s^{m}\bar{\q}_I$. 

As described above, imposing a particular gauge on the group element $S(\underline{\cP})$, eq. \eqref{4d cosetrep}, allows us to read off the coset representative in a coordinate system of choice. 
The coset representative in the north chart is given by
\bsubeq
\begin{align}
	S(\underline{\cP})_{\text{north}} &= 
	\left(
	\begin{array}{c||c}
		A_{\text{n}} & B_{\text{n}} \\
		\hline\hline
		C_{\text{n}} & D_{\text{n}}
	\end{array}
	\right) \label{north chart coset rep}\,, 
	\\
	A_{\text{n}} &= 
	\left(
	\begin{array}{c|c}
		\l {\mathbbm 1}_{2} & -\bar{\l}\ve\tilde{x}_{-}^{\T}\ve \\ 
		\hline
		-\l \tilde{x}_{+} &  \bar{\l} {\mathbbm 1}_{2}
	\end{array}
	\right)
	\,,
	\\
	C_{\text{n}} &= 
	\left(
	\begin{array}{c|c}
		2\ri\l\q & ~2\bar{\l}\bar{\q}\ve
	\end{array}
	\right)
	\,,
	\\
	D_{\text{n}} &= \Big({\mathbbm 1}_{\N} + C_{\text{n}}\me^{-1}{C_{\text{n}}}^{\T}\Big)^{\frac{1}{2}} \,,
	\\
	B_{\text{n}} &= A_{\text{n}}\me^{-1}{C_{\text{n}}}^{\T}{D_{\text{n}}}^{-1}\,.
\end{align}
\esubeq
The two-point function \eqref{ads two point function} computed in the north chart yields 
\begin{align} \label{susy chiral interval}
	\o(\underline{\cP},\underline{\cP}+\text{d}\underline{\cP}) =  \frac{\ell^{4}\eta_{mn}\P^{m}\P^{n}}{(\ell^{2}-x_{+}^{2}-2\ell\q^{2})(\ell^{2}-x_{-}^{2}-2\ell\bar{\q}^{2})}\,,
\end{align}
where
\begin{align} 
	\P^{m} = \text{d}x^{m} + \ri(\q\s^{m}\text{d}\bar{\q}
	- \text{d}\q\s^{m}\bar{\q})\,, \qquad x_{\pm}^{2} = \eta_{mn}x_{\pm}^{m}x_{\pm}^{n} \,.
	\label{4.27}
\end{align}
In the non-supersymmetric case, $\N=0$, this reduces to
\eqref{4d bosonic interval}.

In the south chart, the gauge freedom \eqref{equivalence relation} can be used to fix $G \propto {\mathbbm 1}_2$. Repeating the analysis of the north chart leads to
\begin{align} \label{south chart two planes}
	\underline{\cP} = \g (y_+, \x) 
	\left(
	\begin{array}{c}
		{y}_{+} \\
		\ell {\mathbbm 1}_{2} \\
		\hline \hline
		2\ri\sqrt{\ell}\xi
	\end{array}
	\right) 
	\,,\qquad 
	{y}_+ = y^{m}_+{\s}_{m}\,,
\end{align}
with
\bsubeq
\begin{align}
	&{y}_{+} - {y}_{-} = 4\ri\xi^{\dag}\xi\,, 
	\qquad {y}_{-} := ({y}_{+})^{\dag}\,,	\\
	&\g = \sqrt{ \frac{\ell}{\ell^{2} - y_{+}^{2} - 2\ell\xi^2}}\,, \qquad \xi^{2} := \text{tr}\left(\xi\ve^{-1} \xi^\T  \right)\,. 
\end{align}
\esubeq
The former is solved by 
\begin{align}
	{y_{\pm}} = y^{m}\s_{m} \pm 2\ri\xi^{\dag}\xi\,.
\end{align}
We see that the two-planes in the south chart \eqref{south chart two planes} are parametrised by the chiral coordinates $y_{+}^{m}$ and $\xi_{I}{}^{\m}$, with $y_{+}^{m} = y^{m} + \ri\xi_I \tilde{\s}^{m}\bar{\xi}_I$. 

The coset representative in the south chart is given by
\bsubeq
\begin{align}
	S(\underline{\cP})_{\text{south}} &= 
	\left(
	\begin{array}{c||c}
		A_{\text{s}} & B_{\text{s}} \\
		\hline\hline
		C_{\text{s}} & D_{\text{s}}
	\end{array}
	\right) \label{south chart coset rep}\,, 
	\\
	A_{\text{s}} &= 
	\left(
	\begin{array}{c|c}
		\g y_{+} & -\bar{\g}{\mathbbm 1}_{2} \\ 
		\hline
		\g{\mathbbm 1}_{2} & - \bar{\g} \ve y_{-}^{\T} \ve
	\end{array}
	\right)
	\,,
	\\
	C_{\text{s}} &= 
	\left(
	\begin{array}{c|c}
		2\ri\g\xi & ~2\bar{\g}\bar{\xi}\ve
	\end{array}
	\right)
	\,,
	\\
	D_{\text{s}} &= \Big({\mathbbm 1}_{\N} + C_{\text{s}}\me^{-1}{C_{\text{s}}}^{\T}\Big)^{\frac{1}{2}} \,,
	\\
	B_{\text{s}} &= A_{\text{s}}\me^{-1}{C_{\text{s}}}^{\T}{D_{\text{s}}}^{-1}\,.
\end{align}
\esubeq
The two-point function \eqref{ads two point function} computed in the south chart yields 
\begin{align} \label{susy south interval}
	\o(\underline{\cP},\underline{\cP}+\text{d}\underline{\cP}) =  \frac{\ell^{4} \eta_{mn}\P'^{m}\P'^{n}}{(\ell^{2}-y_{+}^{2}-2\ell\xi^{2})(\ell^{2}-y_{-}^{2}-2\ell\bar{\xi}^{2})}\,,
\end{align}
where
\begin{align} 
	\P'^{m} = \text{d}y^{m} + \ri(\xi\s^{m}\text{d}\bar{\xi}
	- \text{d}\xi\s^{m}\bar{\xi})\,, 
	\qquad y_{\pm}^{2} = \eta_{mn}y_{\pm}^{m}y_{\pm}^{n}~.
\end{align}
In the intersection of the two charts, the transition functions are given by
\bsubeq \label{susy transition functions}
\begin{align}
	{y}_{+} &= -\ell^{2}\tilde{x}_{+}^{-1} ~~ \Longleftrightarrow ~~ y_{+}^{m} = \frac{\ell^{2}}{x_{+}^{2}}x_{+}^{m} ~~ \implies ~ y_{+}^{2}x_{+}^{2} = \ell^{4}\,,
	\\
	\xi &= -\q\tilde{x}_{+}^{-1} \,.
\end{align}
\esubeq
In addition, the two coset representatives \eqref{north chart coset rep} and \eqref{south chart coset rep} are related in the intersection by the point-dependent little group transformation
\begin{align} \label{s coset to n coset}
	S(\underline{\cP})_{\text{north}} &= S(\underline{\cP})_{\text{south}} h^{-1}(x_{+},\q) \,, \qquad h^{-1} \in H\,. 
\end{align}
Explicitly, $h^{-1}$ is given by
\bsubeq
\begin{align}
	h^{-1} &= \left(\begin{array}{c|c||c} \label{coset reps trf}
		n^{-1} & 0 & 0 \\
		\hline 
		0 & n^{\dag} & 0 \\
		\hline \hline
		0 & 0 & {\mathbbm 1}_{\N}
	\end{array}\right)\,, 
\\
n^{-1} &= -\l\g^{-1}\tilde{x}_{+}\,. \label{n inverse block} 
\end{align}
\esubeq
We see that $n^{-1}$ is chiral, through the transition functions \eqref{susy transition functions}.

Finally, the analysis of this section has focused specifically on the general forms and use of the two-planes in the north and south charts.
It is also instructive and useful to consider the bi-supertwistors \eqref{new bi-super} in an explicit coordinate system, for applications described later.
In the north chart they take the form
\begin{align}
	\left(\underline{\bm{X}}_{AB}\right) &= -\ri \l^{2} (x_+, \q) 
	\left(\begin{array}{c|c||c}
		\ell^{2}\ve_{\a\b} & - \ell {x}_{+ \a}{}^{\bd} & 2\ri\ell^{\frac{3}{2}}\q_{\a J}
		\\
		\hline 
		\ell {{x}}_{+}{}^{\ad}{}_{\b} & \ve^{\ad\bd}x_{+}^{2} & -2\ri\ell^{\frac{1}{2}}{{x}}_{+}{}^{\ad\m}\q_{\m J}
		\\
		\hline \hline
		-2\ri\ell^{\frac{3}{2}}\q_{I\b} & 2\ri\ell^{\frac{1}{2}}\q_{I\m}{{x}}_{+}{}^{\bd\m} & -4\ell\q_{I}{}^{\m}\q_{\m J}
	\end{array}\right)\,.
\end{align}
It is of interest to compare this supermatrix with a similar result for compactified 
$\cN$-extended Minkowski superspace, see eq. \eqref{bm mink bst}. 


\section{Superspace geometry} \label{4d geometry}

In this section we will develop certain aspects of the AdS supergeometry, making use of the coset construction described above. 
In particular, we will derive explicit expressions for the supervielbein and connection, as well as torsion and curvature tensors. 
This analysis will be based on the use of the `generalised coset representative', eq. \eqref{4d cosetrep}, which will then allow us to construct useful expressions in the coordinate systems derived above.


\subsection{Geometric structures in AdS$^{4|4\mathcal{N}}$} \label{geometric objects}
We will begin by recalling pertinent details of the AdS superalgebra. 
Let us denote by $\mathcal{G}$ the superalgebra of the AdS supergroup $\sOSp(\mathcal{N}|4;\mathbb{R})_{C}$, and by $\mathcal{H}$ the algebra of the stability group \eqref{stabgroup}. 
Let $\mathcal{K}$ be the complement of $\mathcal{H}$ in $\mathcal{G}$, $\mathcal{G}=\mathcal{H}\oplus\mathcal{K}$. 
The superalgebra $\mathcal{G}$ consists of even supermatrices 
\begin{align}
	\underline{\Omega} = 
	\left(\begin{array}{c|c||c}
		u_{\a}{}^{\b} & v_{\a\bd}&\ri \psi_{\a J}\\
		\hline
		-{v}^{\ad\b} \phantom{\Big|}& -(u^{\dag})^{\ad}{}_{\bd} & -\bar{\psi}^{\ad}{}_{J}\\
		\hline \hline 
		\ri\psi_{I}{}^{\b} & \bar{\psi}_{I\bd} & \Lambda_{IJ} 
	\end{array}\right)
			 \in \mathcal{G}\,,\quad u \in \mathfrak{sl} (2,{\mathbb C}) \,, 
			 \quad \L \in \mathfrak{so} (\cN)~,
\end{align}
		with 
		$v=v^{\dag}$.
		The elements $\underline{h} \in \mathcal{H}$ take the form
\begin{align} \label{4dstabform}
			\underline{h} = \left(\begin{array}{c|c||c}
				n_{\a}{}^{\b} & 0 &0 \\
				\hline
				0 & -(n^{\dag})^{\ad}{}_{\bd} & 0\\
				\hline \hline 
				0 & 0 & r_{IJ} 
			\end{array}\right)\,, \quad n \in \mathfrak{sl}(2,\mathbb{C})\,, \quad r \in \mathfrak{so}(\mathcal{N})\,.
\end{align}
		Additionally, the elements $\underline{k} \in \mathcal{K}$ take the form
\begin{align} \label{4dcompel}
			\underline{k} = \left(\begin{array}{c|c||c}
				0 & v_{\a\bd}&\ri \psi_{\a J}\\
				\hline
				-{v}^{\ad\b} &0 & -\bar{\psi}^{\ad}{}_{J}\\
				\hline \hline 
				\ri\psi_{I}{}^{\b} & \bar{\psi}_{I\bd} &0
			\end{array}\right) \,.
\end{align}
		It can be useful to introduce the following row-vector definition
\begin{align}
			\left(\varphi_{I}{}^{\hat{\a}}\right) := \left(\begin{array}{c|c}
				\ri\psi_{I}{}^{\a} & \bar{\psi}_{I \ad}
			\end{array}\right)\,,
\end{align}
		with which the elements \eqref{4dcompel} then take the form
\begin{align} \label{compform}
			\underline{k} = \left(\begin{array}{c||c}
				\begin{array}{c|c}
					0 & v_{\a\bd}  \\
					\hline
					-{v}^{\ad\b} & 0
				\end{array}&  \varphi_{\hat{\a} J} \\
				\hline \hline
				\varphi_{I}{}^{\hat{\b}} & 0 
			\end{array} \right) \,.
\end{align}
		It is straightforward to verify $[\mathcal{K},\mathcal{H}] \subset \mathcal{K}$.

		Using the coset representative one can introduce the Maurer-Cartan one-form, 
		\begin{align}
			\o = S^{-1}\rd S\,,
		\end{align}
		which proves to encode all of the geometric information of our superspace. 
We may uniquely decompose the Maurer-Cartan one-form as the sum 
\begin{align}
	\o &= \bf{E} + \bm{\O}\,, 
\end{align}				
where $\bf{E} = \o|_{\cK}$ and $\bm{\O} = \o|_{\cH}$ are the supervielbein and connection, respectively. 
The explicit form of $\o$ is given by
\begin{align} \label{mc1f}
			\omega = \left(\begin{array}{c||c}
				\o_{\text{Sp}(4)}  & \me^{-1} \cE^{\T} \\
				\hline \hline 
				\cE & \O_{\text{SO}(\N)}
			\end{array}\right)\,,
\end{align}
		where the blocks are
\bsubeq
\begin{align}
			\o_{\text{Sp}(4)} &= \left(\begin{array}{c|c}
				\ve^{-1} F^{\T}\ve\text{d}F - \ve^{-1} G^{\T}\ve^{-1} \text{d}G  & \ve^{-1} F^{\T}\text{d}\bar{G}\ve^{-1}  - \ve^{-1} G^{\T}\text{d}\bar{F}\ve^{-1}  \\
				+ \ve^{-1} \Q^{\T}\text{d}\Q
				& + \ri\ve^{-1} \Q^{\T}\text{d}\bar{\Q}\ve^{-1} \\
				\hline
				F^{\dag}\text{d}G - G^{\dag}\text{d}F &  F^{\dag}\ve\text{d}\bar{F}\ve^{-1}  - G^{\dag}\ve^{-1} \text{d}\bar{G}\ve^{-1} 
				\\
				+ \ri \Q^{\dag}\text{d}\Q & - \Q^{\dag}\text{d}\bar{\Q}\ve^{-1} 
			\end{array}\right)\,,
			\\
			\cE &= D\text{d}C - DC(A^{-1})\text{d}A\,, \label{vierbeinoddblock}
			\\
			\O_{\text{SO}(\N)} &= D^{-1}\text{d}D 
			-  DC[(A^{-1})\me^{-1}\text{d}(A^{-1})^{\T}C^{\T}
			+ (A^{-1})\me^{-1}(A^{-1})^{\T}\text{d}C^{\T}]D\,.
\end{align}
\esubeq
One can make use of the group conditions \eqref{groupreq1} to recast $\cE$ in an equivalent form
\begin{align} \label{new spinor vierbein}
			\cE &= D\text{d}C - D^{-1}C\me^{-1}A^{\T}\me\text{d}A\,,
			\\ 
			\notag 
			&= \left(\begin{array}{c|c}
				2 \ri (\textbf{E}_{\Q})_{I}{}^{\a} & 2(\bar{\textbf{E}}_{\bar{\Q}})_{I\ad}
			\end{array}\right) \,. 
\end{align}
In the above, $\textbf{E}_{\Q}$ is given by
		\begin{align} \label{e theta block general}
			\textbf{E}_{\Q} &= \frac{1}{2}D\text{d}\Q - \frac{1}{2} D^{-1}\Q\ve^{-1}(F^{\T}\ve\text{d}F - G^{\T}\ve^{-1}\text{d}G)
			\\
			\notag
			& \quad - \frac{1}{2} \ri D^{-1}\bar{\Q}\ve^{-1}(F^{\dag}\text{d}G-G^{\dag}\text{d}F)\,. 
		\end{align}

We can make use of the relations \eqref{4dstabform} and \eqref{compform} to decompose the Maurer-Cartan form into the supervielbein and connection. 		
The connection is 
		\bsubeq
\begin{align}
			\bm{\O} = 
			\left(\begin{array}{c||c}
				\hat{\O}_{\hat{\a}}{}^{\hat{\b}} &  0 \\ 
				\hline \hline 
				0 & \O_{\text{SO}(\mathcal{N})}{}_{IJ}
			\end{array}\right)
		=
		\left(\begin{array}{c|c||c}
			\O_{\a}{}^{\b} & 0 &0 \\
			\hline
			0 & -\bar{\O}^{\ad}{}_{\bd} & 0\\
			\hline \hline 
			0 & 0 & \O_{\text{SO}(\mathcal{N})}{}_{IJ}
		\end{array}\right)\,,
\end{align}
		where 
\begin{align} \label{lorentz connection gauge free}
			\O &= \ve^{-1} F^{\T}\ve\text{d}F - \ve^{-1} G^{\T}\ve^{-1} \text{d}G + \ve^{-1} \Q^{\T}\text{d}\Q\,.
\end{align}
\esubeq
It is possible that these expressions may be simplified by using an explicit form for $A^{-1}$, with $A$ given by \eqref{ablock}, however the above expressions appear most convenient for proving the required properties
\begin{align} \label{connection conditions}
			\text{tr}\,\O = 0\,, \qquad 
			\Omega_{\text{SO}(\mathcal{N})}^{\T} = -\Omega_{\text{SO}(\mathcal{N})}\,. 
\end{align}
The supervielbein is 
\begin{align} \label{vierbein}
			\textbf{E} = \left(\begin{array}{c||c}
				\hat{\textbf{E}}_{\hat{\a}}{}^{\hat{\b}} &  \cE_{\hat{\a} J} \\ 
				\hline \hline 
				\cE_{I}{}^{\hat{\b}} & 0
			\end{array}\right)
		= \left( \begin{array}{c|c||c}
			0 & \textbf{E}_{\a\bd} & 2\ri(\textbf{E}_{\Q})_{\a J} \\
			\hline
			- {\textbf{E}}^{\ad\b} & 0 & -2(\bar{\textbf{E}}_{\bar{\Q}})^{\ad}{}_{J} \\
			\hline \hline 
			2\ri (\textbf{E}_{\Q})_{I}{}^{\b} & 2(\bar{\textbf{E}}_{\bar{\Q}})_{I\bd} & 0
		\end{array} \right)
		 \,,
\end{align}
		where 
\begin{align} \label{vierbeinblock}
			\tilde{\textbf{E}} &= (\textbf{E}^{\ad \b})=-F^{\dag}\text{d}G + G^{\dag}\text{d}F - \ri\Q^{\dag}\text{d}\Q \,,
\end{align}
and $\cE$ is defined as in \eqref{vierbeinoddblock} or \eqref{new spinor vierbein}. 
It is straightforward to show that \eqref{vierbeinblock} is Hermitian, using \eqref{detcon2}. 
Making use of the definitions \eqref{lorentz connection gauge free} and \eqref{vierbeinblock}, as well as useful form for $D_{IJ}$ \eqref{new d block def}, one can prove the identity 
\begin{align} \label{d block spinor identity}
	D_{IJ} ({\bf{E}}_{\Q})_{J}{}^{\a} &= \frac{1}{2}\rd\Q_{I}{}^{\a} - \frac{1}{2}\Q_{I}{}^{\b}\O_{\b}{}^{\a} - \frac{1}{2}\ri\bar{\Q}_{I\ad}{\bf{E}}^{\ad\a}\,.
\end{align}		
		This identity will be useful for our later analysis in the $\N>1$ case. 

Using the above expressions we can now compute the torsion $\mathcal{T}$ and curvature $\mathcal{R}$ for AdS$^{4|4\N}$. 
In accordance with the coset construction, they are defined as follows:
\begin{align} \label{torscurvdef}
			\mathcal{T} = \text{d}\textbf{E} - \bm{\O} \wedge \textbf{E} - \textbf{E}\wedge\bm{\O}\,, \qquad \mathcal{R} = \text{d}\bm{\O} - \bm{\O} \wedge \bm{\O}\,,
\end{align}
and we will provide these expressions below. 
As shown in \cite{Kuzenko:2007aj}, however, it is possible to express these purely in terms of the supervielbein, which is useful for certain applications. 
Their expressions are then given by 
		%
		%
		%
		\begin{align} \label{ewedgee}
			\mathcal{T} = (\textbf{E}\wedge\textbf{E})|_{\mathcal{K}}\,, \qquad \mathcal{R} = (\textbf{E}\wedge\textbf{E})|_\mathcal{H}\,. 
		\end{align}

For completeness, we will first provide the explicit expressions for \eqref{torscurvdef}. 
The torsion takes the form
\begin{align} \label{torsion}
			\mathcal{T} = \left(\begin{array}{c||c}
				\mathcal{T}_1 & \mathcal{T}_3 \\
				\hline \hline
				\mathcal{T}_2 & 0 
			\end{array}\right)\,,
\end{align}
		where 
\bsubeq
\begin{align}
			\mathcal{T}_1 &= \left(\begin{array}{c|c} \label{t1}
				0 & \mathcal{T}_R \\
				\hline
				\mathcal{T}_L & 0  
			\end{array}\right)\,, \\
			\mathcal{T}_L &= - \text{d}F^{\dag}\text{d}G + \text{d}G^{\dag}\text{d}F - \ri\text{d}\Q^{\dag}\text{d}\Q 
			\\
			\notag 
			& \quad 
			- (F^{\dag}\text{d}G - G^{\dag}\text{d}F + \ri\Q^{\dag}\text{d}\Q)\ve^{-1}(F^{\T}\ve\text{d}F - G^{\T}\ve^{-1}\text{d}G + \Q^{\T}\text{d}\Q)
			\\
			\notag 
			& \quad 
			- (F^{\dag}\ve\text{d}\bar{F} - G^{\dag}\ve^{-1}\text{d}\bar{G}-\
			\Q^{\dag}\text{d}\bar{\Q})\ve^{-1}(F^{\dag}\text{d}G-G^{\dag}\text{d}F+\ri\Q^{\dag}\text{d}\Q)\,, 
			\\
			\mathcal{T}_R &= \ve^{-1} \bar{\mathcal{T}}_L\ve^{-1} \,, \label{tr}
			\\
			\mathcal{T}_2 &= -\text{d}D\text{d}C + \text{d}\big(DC(A^{-1})\big)\text{d}A - D\text{d}C\hat{\O} + DC(A^{-1})\text{d}A\hat{\O}
			\\
			\notag 
			& \quad
			- D^{-1}\text{d}D\big(D\text{d}C - DC(A^{-1})\text{d}A\big) 
			\\
			\notag 
			& \quad
			+ DC(A^{-1})\me^{-1}\text{d}\big((A^{-1})^{\T}C^{\T}\big)D\big(D\text{d}C - DC(A^{-1})\text{d}A\big)\,,
			\\
			\mathcal{T}_3 &= \me^{-1} \mathcal{T}_2^{\T}\,. 
\end{align}
\esubeq

The curvature is given by
\begin{align}
			\mathcal{R} = \left( \begin{array}{c|c||c}
				\mathcal{R}_{1} & 0 & 0 \\
				\hline 
				0 & \mathcal{R}_2 & 0 \\
				\hline \hline 
				0 & 0 & \mathcal{R}_{\sSO(\N)}
			\end{array} \right) \,,
\end{align}
		where 
\bsubeq
\begin{align}
			\mathcal{R}_1 &= \ve^{-1} \text{d}G^{\T}\ve^{-1} \text{d}G - \ve^{-1} \text{d}F^{\T}\ve\text{d}F - \ve^{-1} \text{d}\Q^{\T}\text{d}\Q 
			\\
			\notag 
			& \quad 
			-\ve^{-1} F^{\T}\ve\text{d}F\ve^{-1} F^{\T}\ve\text{d}F + \ve^{-1} F^{\T}\ve\text{d}F\ve^{-1} G^{\T}\ve^{-1} \text{d}G
			\\
			\notag 
			& \quad 
			-\ve^{-1} F^{\T}\ve\text{d}F\ve^{-1} \Q^{\T}\text{d}\Q + \ve^{-1} G^{\T}\ve^{-1} \text{d}G\ve^{-1} F^{\T}\ve\text{d}F
			\\
			\notag 
			& \quad 
			- \ve^{-1} G^{\T}\ve^{-1} \text{d}G\ve^{-1} G^{\T}\ve^{-1} \text{d}G + \ve^{-1} G^{\T}\ve^{-1} \text{d}G\ve^{-1} \Q^{\T}\text{d}\Q
			\\
			\notag 
			& \quad 
			-\ve^{-1} \Q^{\T}\text{d}\Q \ve^{-1} F^{\T} \ve\text{d}F + \ve^{-1} \Q^{\T}\text{d}\Q\ve^{-1} G^{\T}\ve^{-1} \text{d}G 
			\\
			\notag 
			& \quad 
			-\ve^{-1} \Q^{\T}\text{d}\Q\ve^{-1} \Q^{\T}\text{d}\Q\,,
			\\
			\mathcal{R}_2 &= \ve\bar{\mathcal{R}}_1\ve^{-1} \,,
			\\
			\mathcal{R}_{\sSO(\N)} &= 
			 D\text{d}C(A^{-1})\me^{-1}\text{d}(A^{-1})^{\T}C^{\T}D 
			+ \ DC\text{d}(A^{-1})\me^{-1}\text{d}(A^{-1})^{\T}C^{\T}D 
			\\
			\notag 
			& \quad 
			+ D\text{d}C(A^{-1})\me^{-1}(A^{-1})^{\T}\text{d}C^{\T}D
			+ DC\text{d}(A^{-1})\me^{-1}(A^{-1})^{\T}\text{d}C^{\T}D 
			\\
			\notag 
			& \quad 
			+ D^{-1}\text{d}C\me^{-1}C^{\T}C(A^{-1})\me^{-1}\text{d}((A^{-1})^{\T}C^{\T})D
			\\
			\notag
			&\quad 
			+ D^{-1}C\me^{-1}\text{d}C^{\T}C(A^{-1})\me^{-1}\text{d}((A^{-1})^{\T}C^{\T})D
			\\
			\notag 
			& \quad
			+ \big(DC(A^{-1})\me^{-1}\text{d}((A^{-1})^{\T}C^{\T})D\big)^2\,.
\end{align}
\esubeq
		%
%
The above expressions may be used to extract the components of the curvature and torsion tensors for the AdS superspace. 
		%
		%
\subsection{Covariant derivatives} \label{covariant derivatives}
The supervielbein and connection (as well as curvature and torsion) can be decomposed into the bases corresponding to the superalgebra $\mathcal{K}$ and the algebra $\mathcal{H}$. 
Accordingly, we must introduce a basis $K_A = (P_a, q_{I\a}, \bar{q}_{I}{}^{\ad})$ for the superalgebra $\mathcal{K}$ and a basis $H_{\hat{I}} = (M_{ab}, \mathcal{J}_{IJ})$ for the algebra $\mathcal{H}$. 
The elements $\underline{h}$ of $\mathcal{H}$ and $\underline{k}$ of $\mathcal{K}$, given by \eqref{4dstabform} and \eqref{4dcompel}, may be written as a linear combination of generators
\bsubeq
\begin{align} 
	\underline{h} &= \frac{1}{2}n^{ab}M_{ab} + \frac{1}{2}r^{IJ} \cJ_{IJ}\,, \label{stab decomp}
	\\
	\underline{k} &= v^{a}P_{a} + \ri( \psi^{I\a}q_{I\a} + \bar{\psi}_{I\ad}\bar{q}_{I}{}^{\ad})\,. \label{comp decomp}
\end{align}
\esubeq
Making use of \eqref{4dstabform} and \eqref{4dcompel}, one can read off the graded commutation relations for the generators of $\cH$ and $\cK$
\bsubeq \label{superalgebra}
\begin{align} 
	[M_{ab},M_{cd}] &= \eta_{ad}M_{bc} - \eta_{ac}M_{bd} + \eta_{bc}M_{ad} - \eta_{bd}M_{ac}
	\,,
	\\
	[P_{a},P_{b}] &= 4 M_{ab}
	\,,
	\\
	[M_{ab},P_{c}] &= \eta_{cb}P_{a}-\eta_{ca}P_{b}
	\,,
	\\
	[M_{ab},q_{I\a}] &= -(\s_{ab})_{\a}{}^{\b}q_{I\b}
	\,,
	\\
	[M_{ab},\bar{q}_{I}{}^{\ad}] &= -(\Tilde{\s}_{ab})^{\ad}{}_{\bd}\bar{q}_{I}{}^{\bd}
	\,,
	\\
	[P_{a},q_{I\a} ] &= -\ri(\s_a)_{\a\ad}\bar{q}_{I}{}^{\ad}
	\,,
	\\
	[P_{a},\bar{q}_{I}{}^{\ad}] &= -\ri(\Tilde{\s}_{a})^{\ad\a}q_{I\a}
	\,,
	\\
	[\mathcal{J}_{IJ},\mathcal{J}_{KL}] &= \d_{JK}\mathcal{J}_{IL} - \d_{IK}\mathcal{J}_{JL} - \d_{JL}\mathcal{J}_{IK} + \d_{IL}\mathcal{J}_{JK} 
	\,,
	\\
	[M_{ab},\mathcal{J}_{IJ}] &= 0
	\,,
	\\
	[P_a,\mathcal{J}_{IJ} ] &= 0
	\,,
	\\
	\{q_{I\a},q_{J\b}\} &= -8\d_{IJ}M_{\a\b} -4  \ve_{\a\b}\mathcal{J}_{IJ}
	\,,
	\\
	\{\bar{q}_{I}{}^{\ad},\bar{q}_{J}{}^{\bd}\} &= 8\d_{IJ}\bar{M}^{\ad\bd} +4  \ve^{\ad\bd}\mathcal{J}_{IJ}
	\,,
	\\
	\{q_{I\a},\bar{q}_{J}{}^{\ad}\} &= 2\ri\d_{IJ}(\s^{a})_{\a}{}^{\ad}P_a
	\,,
	\\
	[\mathcal{J}_{IJ}, q_{K\a} ] &= \d_{JK}q_{I\a} - \d_{IK}q_{J\a}
	\,.
\end{align}
\esubeq
These relations constitute the $\mathcal{N}$-extended AdS superalgebra
$ \mathfrak{osp}(\mathcal{N}|4;\mathbb{R})_{C}$.

Given that the supervielbein one-form and the torsion two-form are elements of $\cK$, they can be decomposed with respect to the basis $K_{A}$ as 
\bsubeq
\begin{align}
	\textbf{E} &= \textbf{E}^a P_a + \ri (\textbf{E}^{I\a} q_{I\a} + \bar{\textbf{E}}_{I\ad}\bar{q}^{I\ad})\,,
	\\
	\mathcal{T} &= \mathcal{T}^a P_a + \ri( \mathcal{T}^{I\a}q_{I\a}+\mathcal{T}_{I\ad}\bar{q}^{I\ad}) \,,
\end{align}
\esubeq
to obtain the one-forms $\textbf{E}^{A} = (\textbf{E}^{a}, \textbf{E}^{I\a},\bar{\textbf{E}}_{I\ad})$ and the torsion $\mathcal{T}^{A} = (\mathcal{T}^{a}, \mathcal{T}^{I\a}, \mathcal{T}_{I\ad})$.
A similar procedure follows for the connection and curvature.
The supervielbein one-forms ${\bf{E}}^{A}$ form a basis, and we can then further decompose the torsion and curvature as
\bsubeq
\begin{align}
	\mathcal{T}^{A} &= \frac{1}{2}\textbf{E}^B\wedge\textbf{E}^C\mathcal{T}_{CB}{}^{A}\,,
	\\
	\mathcal{R} &= \frac{1}{2}\textbf{E}^B\wedge\textbf{E}^{C}\left(\frac{1}{2}\mathcal{R}_{CB}{}^{de}M_{de} + \frac{1}{2}\mathcal{R}_{CB}{}^{MN}\mathcal{J}_{MN}\right)\,. 
\end{align}
\esubeq

Building on the approach used in \cite{Kuzenko:2007aj}, we can use the above definitions in conjunction with \eqref{ewedgee} and the graded commutation relations \eqref{superalgebra} to determine the non-vanishing components of the torsion and curvature. 
They are given by
\bsubeq \label{nonvanishingcomps}
\begin{align} 
	\mathcal{T}_{I\a}{}^{J\bd a} &= 2\ri\d_{I}{}^{J}(\s^{a})_{\a}{}^{\bd}
	\,,
	\\
	\mathcal{T}_{a I\a J\bd} &= \ri\d_{IJ}(\s_{a})_{\a\bd}
	\,,
	\\
	\mathcal{T}_{a}^{I\ad J\b} &= \ri\d^{IJ}(\Tilde{\s}_{a})^{\ad\b}
	\,,
	\\
	\mathcal{R}_{ab}{}^{cd} &= -4(\d_{a}{}^{c}\d_{b}{}^{d}-\d_{a}{}^{d}\d_{b}{}^{c})\label{4d ads curvature}
	\,,
	\\
	\mathcal{R}_{I\a J\b}{}^{\g\d} &= -4\d_{IJ}(\d_{\a}{}^{\g}\d_{\b}{}^{\d}+\d_{\a}{}^{\d}\d_{\b}{}^{\g})
	\,,
	\\
	\mathcal{R}_{I\a J\b}{}^{KL} &= -4 \ve_{\a\b}(\d_{I}{}^{K}\d_{J}{}^{L}-\d_{J}{}^{K}\d_{I}{}^{L})
	\,,
	\\
	\mathcal{R}^{I\ad J\bd}{}_{\gd\dd} &=  4\d^{IJ}(\d^{\ad}{}_{\gd}\d^{\bd}{}_{\dd}+\d^{\ad}{}_{\dd}\d^{\bd}{}_{\gd})
	\,,
	\\
	\mathcal{R}^{I\ad J\bd KL} &= 4\ve^{\ad\bd}(\d^{IK}\d^{JL}-\d^{JK}\d^{IL})
	\,.
\end{align}
\esubeq
We can see that the component \eqref{4d ads curvature} recovers the familiar AdS$_{4}$ curvature tensor. 

At this stage it is useful to introduce a generic local coordinate system $z^{M} = (x^{m}\,, \q_{\imath}{}^{\m}\,, \bar{\q}_{\imath \dmu})$ in order to outline the analysis that follows. 
We will fix this analysis to the north chart in section \ref{chiralgeometry}.  
Associated with the local coordinates $z^{M}$ are the one-forms $\rd z^{M} = (\rd x^{m}\,, \rd\q_{\imath}{}^{\m}\,, \rd\bar{\q}_{\imath\dmu})$ which are related to the supervielbein one-forms ${\bf{E}}^{A}$ via the relations 
\bsubeq
\begin{align}
	\rd z^{M} &= {\bf{E}}^{A} E_{A}{}^{M}
	\\
	{\bf{E}}^{A} &= \rd z^{M} E_{M}{}^{A}\,.
\end{align}
\esubeq
In the above, the vielbein supermatrix $E_{M}{}^{A}$ satisfies 
\begin{align}
	E_{M}{}^{A}E_{A}{}^{N} = \d_{M}{}^{N} \Longleftrightarrow E_{A}{}^{M}E_{M}{}^{B} = \d_{A}{}^{B}\,.
\end{align}
Further, we can introduce the vector fields $E_{A}$ as follows 
\begin{align}
	E_{A} = E_{A}{}^{M}\partial_{M}\,, \qquad \partial_{M} = \left(\partial_{m}\,, \frac{\partial}{\partial \q_{\imath}{}^{\m}}\,, \frac{\partial}{\partial \bar{\q}_{\imath \dmu}}\right)\,. 
\end{align}
Finally, making use of \eqref{stab decomp}, the connection $\bm{\O}$ can be decomposed into the following 
\begin{align}
	{\bm{\O}} = \frac{1}{2}{\bf{E}}{}^{A}\O_{A}{}^{ab}M_{ab} + \frac{1}{2}{\bf{E}}^{A}\O_{A}{}^{IJ}\cJ_{IJ}
\end{align}
to extract the superfields $\O_{A}{}^{ab}$ and $\O_{A}{}^{IJ}\,.$ 

We can now introduce the covariant derivatives 
$\cD_{A} = (\cD_{a}\,, \cD_{I \a}\,, \bar{\cD}_{I}^{\ad})$, which take the form 
\begin{align}
	\cD_{A} = E_{A} + \frac{1}{2}\O_{A}{}^{bc}M_{bc} + \frac{1}{2}\O_{A}{}^{JK}\cJ_{JK}\,.
\end{align}
Their graded commutation relations can be constructed from the components \eqref{nonvanishingcomps} as
\begin{align} \label{covdiv}
	[\mathcal{D}_A,\mathcal{D}_B\} &= - \mathcal{T}_{AB}{}^{C}\mathcal{D}_C + \frac{1}{2}\mathcal{R}_{AB}{}^{cd}M_{cd} + \frac{1}{2}\mathcal{R}_{AB}{}^{KL}\mathcal{J}_{KL}\,.
\end{align}
These relations are then given by
\bsubeq
\label{NowarAlgebra}
\begin{align}
	[\mathcal{D}_a,\mathcal{D}_b] &= - 4 M_{ab}
	\,,
	\\
	[\mathcal{D}_a,\mathcal{D}_{I\a}] &= -\ri(\s_{a})_{\a\bd}\bar{\mathcal{D}}_{I}{}^{\bd}
	\,,
	\\
	[\mathcal{D}_a,\bar{\mathcal{D}}_{I}{}^{\dot{\a}}] &= -\ri(\Tilde{\s}_a)^{\dot{\a}\b}\mathcal{D}_{I\b}
	\,,
	\\
	\{\mathcal{D}_{I\a},\mathcal{D}_{J\b}\} &= -8\d_{IJ}M_{\ab}  - 4\ve_{\a\b}\mathcal{J}_{IJ}
	\,,
	\\
	\{\bar{\mathcal{D}}_{I}{}^{\dot{\a}},\bar{\mathcal{D}}_{J}{}^{\dot{\b}}\} &= 
	8\d_{IJ}\bar{M}^{\ad\bd} + 4\ve^{\dot{\a}\dot{\b}}\mathcal{J}_{IJ}
	\,,
	\\
	\{\mathcal{D}_{I\a},\bar{\mathcal{D}}_{J}{}^{\dot{\b}}\} &= 
	-2\ri\d_{IJ}(\s^{a})_{\a}{}^{\dot{\b}}\mathcal{D}_{a}
	\,,
\end{align}
\esubeq
which coincide with the relations \eqref{4d ads coset algebra} derived from the supergravity-inspired approach in the diagonal gauge, with $S=-2$. 
\subsection{$\N=1$ AdS superspace} \label{n=1 geometric objects}
		
		Many of the explicit expressions derived in subsection \ref{geometric objects} contain terms involving the matrices $A^{-1}$ and $D$. These are, in principle, expressible in terms of $F$, $G$, and $\Q$. These expressions are, however, $\N$-dependent. 
		Below, we will discuss both of these in the $\N=1$ case. 
		
		Using the group requirements \eqref{aka con} we can rearrange for $(A^{-1})^{\T}$
		\begin{align}
			(A^{-1})^{\T} = \me A \me^{-1}\left({\mathbbm 1} +  C^{\T}C\me^{-1}\right)^{-1}\,,
		\end{align}
		which in the $\N=1$ case yields the following expression
		\begin{align}
			(A^{-1})^{\T}  
			= \left(
			\begin{array}{c|c}
				\Tilde{F} & \ve\Bar{\Tilde{G}}\ve \\
				\hline
				\Tilde{G} & \ve^{-1}\Bar{\Tilde{F}}\ve
			\end{array}\right)
			\,,
		\end{align}
		where
		\bsubeq
		\begin{align}
			\Tilde{F} &= \ve F\ve^{-1} + \ve F \ve^{-1} \q^{\T}\q\ve^{-1} -\ve F \ve^{-1} \Q^{\T}\bar{\Q}\ve^{-1}\Q^{\dag}\Q\ve^{-1}
			\\
			\notag 
			& \quad + \ri \bar{G}\ve^{-1}\Q^{\dag}\Q\ve^{-1}
			\,,
			\\
			\Tilde{G} &= 
			-\ve^{-1} G \ve^{-1} - \ve^{-1} G \ve^{-1} \Q^{\T}\Q\ve^{-1}
			+ \ve^{-1}G\ve^{-1}\Q^{\T}\bar{\Q}\ve^{-1}\Q^{\dag}\Q\ve^{-1}
			\\
			\notag 
			& \quad 
			-\ri \bar{F} \ve^{-1}\Q^{\dag}\Q\ve^{-1}
			\,.
		\end{align}
		\esubeq
		Furthermore, $D$ has the explicit solution 
		\begin{align}
			D &= 1 - \frac{1}{2}\Q^2 - \frac{1}{2}\bar{\Q}^2 - \frac{1}{4}\Q^2\bar{\Q}^2 \,.
		\end{align}
		We can use these expressions to compute $\cE$ from the vierbein \eqref{vierbeinoddblock}. For $\N=1$ it is
		\begin{align} \label{n=1vbodd}
			\cE = \left(\begin{array}{c|c}
				2\ri(\textbf{E}_{\Q})^{\a}  & 2(\bar{\textbf{E}}_{\bar{\Q}})_{\ad}
			\end{array}\right)\,,
		\end{align}
		where 
		\begin{align} \label{n=1spinorvierbein}
			\textbf{E}_{\Q} 
			&= 
			\frac{1}{2} D \text{d} \Q
			- \frac{1}{2} (1 - \frac{1}{2}\bar{\Q}\ve^{-1}\bar{\Q})\Q\ve^{-1}(F^{\T}\ve\text{d}F-G^{\T}\ve^{-1}\text{d}G)
			\\
			\notag 
			& \quad 
			- \frac{1}{2} \ri(1 + \frac{1}{2}\Q\ve^{-1}\Q)\bar{\Q}\ve^{-1}(F^{\dag}\text{d}G - G^{\dag}\text{d}F)
			\,.
		\end{align}
		This expression coincides with \eqref{e theta block general} when considered in the $\N=1$ case. 
\subsection{North and south charts}\label{chiralgeometry}
The results of subsections \ref{geometric objects}, \ref{covariant derivatives} and \ref{n=1 geometric objects} did not make use of the freedom eq. \eqref{equivalence relation} to fix a coordinate system. In this section we will use these results to describe the geometry in the $\N=1$ case for the north and south charts, given by two-planes of the form \eqref{chiral chart} and \eqref{south chart two planes}.  

In the north chart, the vierbein \eqref{vierbein} reads 
\begin{align} \label{chiral vierbein supermatrix}
	\textbf{E}_{\text{north}} = \left(\begin{array}{c|c||c}
		0 & \l\bar{\l}\P_{\a\bd} &  
		2\ri (\eta_{\q})_{\a}\\
		\hline 
		-\l\bar{\l}{\P}^{\ad\b} & 0 & - 2(\bar{\eta}_{\bar{\q}})^{\ad} \\
		 \hline \hline
		  2\ri (\eta_{\q})^{\b} &  2(\bar{\eta}_{\bar{\q}})_{\bd} & 0
	\end{array}\right)\,,
\end{align}

where $\eta_{\q}$ is computed using \eqref{n=1spinorvierbein} as
\begin{align} \label{chiralspinorvierbein}
	\eta_{\q} &= \left(\l - 2\l^{3}\q^{2} + 2\l\bar{\l}^{2}\bar{\q}^{2} + 4\l^{3}\bar{\l}^{2}\q^{2}\bar{\q}^{2} - 4\ri\l^{3}(\q x_{+} \bar{\q})\right)\text{d}\q 
	\\
	\notag
	& \quad + \left(\ri\l\bar{\l}^{2}(1+2\l^{2}\q^{2})\bar{\q}\ve^{-1} - \l^{3}(1+2\bar{\l}^{2}\bar{\q}^{2})(\q x_{+})\right)\Tilde{\P}\,.  
\end{align}
In the above, $\text{d}\q$ and $\tilde{\P} = \P^{a}(\tilde{\s}_{a})$ 
are the flat $\N=1$ superspace vielbeins. 
The general forms for the supervielbeins of an $\N=1$ superspace with superconformally flat geometry are 
\bsubeq \label{scfvierbeins}
\begin{align}
	\textbf{E}^{a} &= \re^{-\frac{1}{2}(\s+\bar{\s})}\P^{a}\,,
	\\
	(\textbf{E}_{\q})^{\a} &= \re^{\frac{1}{2}\s-\bar{\s}}\left(\text{d}\q^{\a} + \frac{\ri}{4}(\bar{D}_{\ad}\bar{\s})(\Tilde{\s}_{b})^{\ad\a}\P^{b}\right) \,,
\end{align}
\esubeq
where $\s \,\, (\bar{\s})$ is chiral (antichiral), see, e.g., eq. \eqref{n extend cf vierbein} and \cite{BK}. 
In our case it is straightforward to compute the coefficients in \eqref{scfvierbeins}, which yields the following expression
\begin{align}
	\label{5.37}
	\l &= \re^{-\frac{1}{2}\s}\,,
\end{align}
as described in section \ref{n s cosets}.
Indeed, direct calculation shows that \eqref{chiral vierbein supermatrix} is equal to 
\begin{align} \label{chiralvbsm2}
	\textbf{E}_{\text{north}} = \left(\begin{array}{c|c||c}
		0 & \re^{-\frac{1}{2}(\s+\bar{\s})}\P_{\a\bd} &  2\ri (\textbf{E}_{\q})_{\a} \\
		\hline 
		-\re^{-\frac{1}{2}(\s+\bar{\s})}\P^{\ad\b} & 0 & - 2(\bar{\textbf{E}}_{\bar{\q}})^{\ad} \\
		\hline \hline
		2\ri (\textbf{E}_{\q})^{\b} & 2(\bar{\textbf{E}}_{\bar{\q}})_{\bd} & 0
	\end{array}\right)\,,
\end{align}
with 
$(\textbf{E}_{\q})$ given by \eqref{scfvierbeins}. The connection is given by
\begin{align}
	\bm{\O}_{\text{north}} = \left(\begin{array}{c|c||c}
		\O_{\a}{}^{\b} & 0 & 0 \\
		\hline 
		0 & -\bar{\O}^{\ad}{}_{\bd} & 0 \\
		\hline \hline
		0 & 0 & ~0~
	\end{array}\right)\,,
\end{align}
where the components of the connection read
\begin{align} \label{connection components}
	\O_{\a}{}^{\b} &= 
	\P^{m}\big(\l^{-1}\d_{\a}{}^{\b}\partial_{m}\l + \l^{-1}(\tilde{\s}_{m})^{\bd\b}\partial_{\a\bd}\l\big)
	+ 
	\text{d}\q^{\g}\big(\l^{-1}\d_{\a}{}^{\b}D_{\g}\l - 2\l^{-1}\d_{\g}{}^{\b}D_{\a}\l\big)\,. 
\end{align}

At this stage it is useful to introduce the inverse $E_{A}{}^{M}$ of the vielbein supermatrix $E_{M}{}^{A}$, 
\bsubeq \label{inverse vierbein}
\begin{align} 
\ve^{M} &= (\P^{m}\,,\text{d}\q^{\m}\,, \text{d}\bar{\q}_{\dmu}) = \textbf{E}^{A}E_{A}{}^{M}\,, 
\\
\textbf{E}^{A} &= (\textbf{E}^{a}\,,(\textbf{E}_{\q})^{\a}\,, (\bar{\textbf{E}}_{\bar{\q}})_{\ad}) = \ve^{M}E_{M}{}^{A}\,,
\end{align}
\esubeq 
and the vector fields 
\begin{align}
	E_{A} &=
	\big(
	E_{a}\,, 
	E_{\a}\,,
	\bar{E}^{\ad}
	\big) =
	 E_{A}{}^{M}D_{M}\,.
\end{align}
 Here $D_M :=(\partial_{m}\,, D_{\m}\,, \bar{D}^{\dot \mu})$ are the $\N=1$ flat superspace covariant derivatives.\footnote{The analysis provided here differs slightly from that in subsection \ref{covariant derivatives} in that we have chosen a different basis for the flat one-forms. Here, it is more convenient to use the flat supersymmetric one-forms $\ve^{M}$ rather than the basis $\rd z^{M}$.} 
 We find 
 \bsubeq
 \begin{align}
 	E_{a} &= \l^{-1}\bar{\l}^{-1}\partial_{a}  - \frac{\ri}{4}\l^{-1}\bar{\l}^{-1}(\bar{D}_{\ad}\bar{\s})(\tilde{\s}_{a})^{\ad\a}D_{\a}
 	\\
 	\notag 
 	& \quad
 	- \frac{\ri}{4}\l^{-1}\bar{\l}^{-1}(D^{a}\s)(\s_{a})_{\a\ad}\bar{D}^{\ad}\,,
 	\\
 	E_{\a} &= \l\bar{\l}^{-2}D_{\a} \,,
 	\\
 	\bar{E}^{\ad} &= \bar{\l}\l^{-2}\bar{D}^{\ad}\,. 
 \end{align}
 \esubeq

 The components of the connection $\bm{\O}_{\text{north}}$ were given with respect to the basis $\{\ve^{M}\}$ in \eqref{connection components}. 
 Using the inverse supervielbein defined by \eqref{inverse vierbein}, the connection can be decomposed into the basis $\{\textbf{E}^{A}\}$,  with which we can then construct explicit expressions for the covariant derivatives
\begin{align}
	\mathcal{D}_{A} = E_{A} + \frac{1}{2}\O_{A}{}^{bc}M_{bc}
	= \big(\mathcal{D}_{a}\,, \mathcal{D}_{\a}\,, \bar{\mathcal{D}}^{\ad}\big)\,.
\end{align}
They take the following form
\bsubeq \label{north chart covariant derivatives}
\begin{align}
	\mathcal{D}_{a} &=- \frac{\rm i}{4} (\tilde{\s}_a)^{\bd \b} \big\{ \cD_\b, \bar \cD_\bd \big\} \,,
	\\
	\mathcal{D}_{\a} &= \l\bar{\l}^{-2}D_{\a} -2\bar{\l}^{-2}\big(D^{\b}\l\big)M_{\a\b} \,,
	\label{spinor derivative} \\
	\bar{\mathcal{D}}^{\ad} &= \bar{\l}\l^{-2}\bar{D}^{\ad} + 2\l^{-2}\big(\bar{D}_{\bd}\bar{\l}\big)\bar{M}^{\ad\bd} \,.
	\label{conjugate spinor derivative}
\end{align}
\esubeq
The expressions \eqref{north chart covariant derivatives} can be seen to coincide with the general form for the covariant derivatives of a conformally flat superspace, see, e.g., eq. \eqref{AdSBoost} and \cite{BK}. 


To conclude this section, we will briefly outline the supergeometry in the south chart. 
The supervielbein \eqref{vierbein} is given by
\begin{align} \label{south vierbein}
	\textbf{E}_{\text{south}} &= 
	\left(\begin{array}{c|c||c}
	0 & \g\bar{\g}\P'_{\a\bd} & 2\ri({\eta}_{{\xi}})_{\a} \\
	\hline 
	-\g \bar{\g} {\P}'^{\ad\b} & 0 & -2(\bar{\eta}_{\bar{\xi}})^{\ad} \\
	\hline \hline
	2\ri (\eta_{\xi})^{\b} & 2(\bar{\eta}_{\bar{\xi}})_{\bd} & 0	
	\end{array}
\right)\,,
\end{align}
where $\eta_{\xi}$ is 
\begin{align}
	\eta_{\xi} &= 
	\left(\g - 2\g^{2}\xi^{2} + 2\g\bar{\g}^{2}\bar{\xi}^{2} + 4\g^{2}\bar{\g}^{2}\xi^{2}\bar{\xi}^{2} -4\ri\g^{3}(\xi y_{+} \bar{\xi})\right)\text{d}\xi 
	\\
	\notag 
	& \quad 
	+ \left(\ri \g \bar{\g}^{2}(1+2\g^{2}\xi^{2})\bar{\xi}\ve^{-1} - \g^{3}(1+2\bar{\g}^{2}\bar{\xi}^{2})(\xi\ve^{-1}y_{+}^{\T})\right)\tilde{\P}' 
	\,.
\end{align}
We showed in section \ref{n s cosets} that the coset representatives in the north and south charts were related by a little group transformation, see \eqref{s coset to n coset}. 
Under such a transformation, the supervielbein and connection transform as follows
\bsubeq
\begin{align}
	\textbf{E} &\rightarrow \textbf{E}' = h\textbf{E}h^{-1} \,, 
	\\
	{\bm{\O}} &\rightarrow {\bm{\O}}' = h {\bm{\O}} h^{-1} - \text{d} h h^{-1} \,. 
\end{align}
\esubeq
We can see then that the supervielbein in the north chart is related to that in the south chart by 
\begin{align}
	\textbf{E}_{\text{north}} = h \textbf{E}_{\text{south}} h^{-1}\,,
\end{align}
which yields
\bsubeq
\begin{align}
	(\textbf{E}_{\text{north}})^{a} &=  (\textbf{E}_{\text{south}})^{b} \L(n)_{b}{}^{a}
	\,, \qquad \L(n)_{b}{}^{a} = -\frac{1}{2}\tr(n \s_{b} n^{\dag} \tilde{\s}{}^{a})\,,
	\\
	(\eta_{\q})^{\a} &= (\eta_{\xi})^{\b}(n^{-1})_{\b}{}^{\a} \,,
\end{align}
\esubeq
with $n^{-1}$ given by \eqref{n inverse block}. 
The vector fields $E_{A}$ are also related in the intersection of the two charts. We find
\bsubeq
\begin{align}
(E_{\text{north}})_{a} &= (\L(n)^{-1})_{a}{}^{b}(E_{\text{south}})_{b} \,,
\\
(E_{\text{north}})_{\a} &= n_{\a}{}^{\b}(E_{\text{south}})_{\b}\,. 
\end{align}
\esubeq

\section{New superparticle models in $\text{AdS}^{4|4{\cal N}} $} \label{4d superparticle discussion}

In this section we will collate various results of the two formalisms detailed above to describe example applications of both. 
In particular we will utilise the specific features of the $\N$-extended AdS supergeometry to construct new superparticle models in AdS$^{4|4\N}\,.$
We will first outline this from the supergravity-inspired approach. 

For a given supergravity background in four dimensions, the standard supersymmetric interval takes the form 
\begin{align} \label{4d standard interval}
	\rd s^{2} = \eta_{ab}{E}^{a}{E}^{b}\,,
\end{align}
where $E^{a}$ is the vector component of the supervielbein.
In particular, making use of the AdS supervielbein, given by eq. \eqref{n extend cf vierbein} in a conformally flat frame, the above expression yields the AdS supersymmetric interval. 
The analysis provided above shows us that we can introduce a deformation to this interval, making use of the specific properties of the AdS supergeometry. 
This deformation takes the form
\bea \label{interval def}
\rd s^2 =
\eta_{ab} {E}^a {E}^b + \frac{1}{|S|^2} \Big( \o \ve_{\a \b} S^{ij} {E}^\a_i {E}^\b_j 
+ \bar \o \ve^{\ad \bd} \bar S_{ij} \bar{E}_\ad^i \bar{E}_\bd^j \Big)~,
\eea
with $\o$ a dimensionless complex parameter. 
It is constructed using both the vector $({E}^{a})$ and spinor $({E}^{\a}_{i}\,, \bar{E}_{\ad}^{i})$ components of the supervielbein, as well as the torsion tensor $S^{ij}$. 

It can be useful to introduce a compact notation for such an interval. 
Indeed, it can be written as
\begin{align}
	\rd s^2 = 
	{E}^{A}\eta_{AB}{E}^{B}\,, 
\end{align}
where $\eta = (\eta_{AB})$ is a supermatrix defined as 
\begin{align} \label{supermetric def}
	\left(\eta_{AB}\right) = \left(
	\begin{array}{c||c|c}
		~\eta_{ab}~ & 0 & 0 \\
		\hline \hline 
		0 & \frac{\o}{|S|^{2}}\ve_{\ab}S^{ij} & 0 \\
		\hline 
		0 & 0 & \frac{\bar{\o}}{|S|^{2}}\ve^{\ad\bd}\bar{S}_{ij}
	\end{array}
	\right)\,,
	\qquad \text{Ber}(\eta) = -\left(\frac{|S|^{2}}{|\o|^{2}}\right)^{\N}\,.
\end{align}
The existence of such a supermetric means that there is a 
superparticle model
\begin{align} \label{ads model}
	S = \frac{1}{2}\int\text{d}\t\frak{e}^{-1}\left\{
	\dt{E}{}^{A}
	\eta_{AB}\dt{E}{}^{B} - (\frak{e}m)^{2} \right\}\,, 
	\qquad \dt{E}{}^{A} = \frac{\rd z^{M}}{\rd\t}E_{M}{}^{A}\,,
\end{align}
where $\mathfrak{e}$ is the einbein and $m$ the mass.\footnote{The concept of mass differs between flat and AdS spaces. This is elaborated on in chapter \ref{ch5}.} 
In a conformally flat frame, eq. \eqref{n extend cf vierbein}, we have
\begin{align}
	\dt{E}{}^{A}\eta_{AB}\dt{E}{}^{B} &= \re^{-(\s + \bar{\s})}\dt{\P}{}^{a}\dt{\P}{}^{b}\eta_{ab} 
	\notag \\
	& \quad ~  ~ ~ + \bigg ( \frac{\o S^{ij}}{|S|^{2}}\re^{-(\frac{\N-2}{\N}\s + \frac{2}{\N}\bar{\s})}\left(\dot{\q}_{ij}
	- \frac{\ri}{2}\dot{\q}_{i}^{\a}\bar{D}_{j}^{\ad}\bar{\s}\dt{\P}_{\a\ad}
	- \frac{1}{16}\bar{D}_{\ad i}\bar{\s}\bar{D}_{j}^{\ad}\bar{\s}\dt{\P}{}^{2}
	\right) + \text{c.c.} \bigg )\,, ~
\end{align}
with $\dt{\P}{}^{a} := \dot{x}^{a} + \ri(\q_{i}\s^{a}\dot{\bar{\q}}^{i} - \dot{\q}_{i}\s^{a}\bar{\q}^{i})$ and $\dot{\q}_{ij} := \dot{\q}^{\a}_{i}\dot{\q}_{\a j}\,.$
It is evident that for $\o = 0$ we recover the standard superparticle model which is discussed in the following subsection.\footnote{Note that the expression for $\text{Ber}(\eta)$ in \eqref{supermetric def} is not well-defined for $\o = 0$, see, e.g., \cite{BK} for details.} 

It is of interest to see how such a model can be derived from the embedding formalism, and we will now turn to performing this task. 
The most general superparticle model quadratic in derivatives of the evolution parameter $\t$ is given by the following
\begin{align} \label{embed model}
	S = - \frac{1}{2}\int \text{d}\t \frak{e}^{-1} \left\{\a \text{Str}(\dot{\bar{\underline{\bm{X}}}}\dot{\underline{\bm{X}}}) + \b\text{Str}(\dot{\underline{\bm{X}}}\dot{\underline{\bm{X}}}) +\bar{\b} \text{Str}(\dot{\bar{\underline{\bm{X}}}}\dot{\bar{\underline{\bm{X}}}})  + (\frak{e}m)^{2} \right\}\,,
\end{align} 
where the parameters $\a \in \mathbb{R}$ and $\b \in \mathbb{C}$ are constrained as
\begin{align}
	\a - (\b + \bar{\b}) = \frac{1}{4}\,.
\end{align}
The above parameters are constrained such that: (i) the model coincides with the bosonic one when the Grassmann variables are switched off; and (ii) the $\b=0$ choice recovers the standard superparticle propagating in AdS superspace.
Indeed, making use of the results of section \ref{n s cosets}, we can see that the kinetic term 
$\text{Str}(\dot{\bar{\underline{\bm{X}}}}\dot{\underline{\bm{X}}}) = -4\eta_{ab}\dot{E}^{a}\dot{E}^{b}$, see specifically eq. \eqref{susy chiral interval} and \eqref{new trace functions}. 
Furthermore, the special choice $\b = \ri\m$ ensures that the additional structures (the $\m$-dependent terms) generate no purely bosonic contributions.
 
In the north chart, the structures present in \eqref{embed model} take the form 
\bsubeq \label{str structures}
\begin{align}
	\text{Str}(\dot{\bar{\underline{\bm{X}}}}\dot{\underline{\bm{X}}}) &= -4 \re^{-(\s + \bar{\s})}\dt{\P}{}^{2}
	\,,
	\\
	\text{Str}(\dot{\underline{\bm{X}}}\dot{\underline{\bm{X}}}) 
	&= 4\re^{-(\s+\bar{\s})}\dt{\P}{}^{2}(1 - x_{-}^{2} - 2\bar{\q}^{2} + \re^{\bar{\s}-\s}x_{+}^{2})
	\notag \\
	& \quad
	+ 16\re^{-\s}\dot{\q}_{I \a}\P^{\ad \a}
	\Big(\re^{-\s}x_{+}^{a}(\q_{I}\s_{a})_{\ad}
	+ \ri\bar{\q}_{I \ad}(1+ \re^{-\s}x_{+}^{2}) \Big)
	\notag \\
	& \quad + 8\re^{-\s}\dot{\q}_{IJ}\Big(
	\d_{IJ} -4\re^{-\s}\q_{IJ} + 4\bar{\q}_{IJ}(1+\re^{-\s}x_{+}^{2})
	-8\ri\re^{-\s}x_{+}^{a}(\q_{I}\s_{a}\bar{\q}_{J})
	\Big)\,,
\end{align}
\esubeq
with $\dot{\q}_{IJ} := \dot{\q}_{I}{}^{\a}\dot{\q}_{J\a}\,.$
Making use of the expressions \eqref{str structures} and the results of appendix \ref{4d n ext coset}, actions \eqref{ads model} and \eqref{embed model} can be shown to coincide to leading order in the north chart provided one fixes 
\begin{align} \label{4d beta omega rel}
	\b = \frac{\o}{4|S|^{2}}\,. 
\end{align}

\subsection{$\k$-symmetry of the superparticle} \label{4d kappa discussion}
Any conformally flat frame for $\text{AdS}^{4|4{\cal N}} $ has
 applications to massless superparticle models, in the spirit of \cite{BILS}. 
In a  supergravity background, the model for a massless superparticle is
\begin{align} \label{general superparticle}
	S = \frac{1}{2}\int\text{d}\t \,\frak{e}^{-1}\dt{E}{}^{a}\dt{E}{}^{b}\eta_{ab}\,,
\end{align}
where $\dt{E}{}^{a}$ can be read off from \eqref{ads model}.
In a conformally flat frame the massless superparticle model takes the form 
\begin{align} \label{cflat sp}
	S = \frac{1}{2}\int \text{d}\t \,\frak{e}^{-1} \re^{-(\s + \bar{\s})}\dt{\P}{}^{a} \dt{\P}{}^{b}\eta_{ab}\,.
\end{align}
As pointed out in \cite{BILS}, such a model is classically equivalent to the massless superparticle model 
in ${\mathbb M}^{4|4{\cal N}} $ 
by a simple redefinition of the einbein $\frak{e} \rightarrow \tilde{\frak{e}} = \re^{(\s + \bar{\s})}\frak{e}\,,$ therefore this model is invariant under $\cN$-extended superconformal transformations which scale the flat-superspace interval $ \eta_{ab} \P^a \P^b $ \cite{Sohnius:1976pa}.
However, retaining the conformal factor explicitly has the advantage of keeping the symmetries of the background manifest. 
In particular, making use of the realisations \eqref{ads soln} or \eqref{pp soln}, one obtains the massless superparticle in AdS$^{4|4\N}$. 

A well-known feature of massless superparticle models is the presence of $\k$-symmetry, which was introduced in \cite{kappa1}. It was generalised to the superstring in \cite{kappa2}, see, e.g., \cite{Szg} for a review and  references. 
For the flat superparticle, these local transformations take the form 
\bsubeq \label{4d kappa trf def}
\begin{align}
	\d\q_{i}^{\a} &= -\ri\bar{\k}_{\ad i}\dt{\P}{}^{\ad\a}\,, \qquad \d\bar{\q}^{i\ad} = \ri \dt{\P}{}^{\ad\a}\k^{i}_{\a}\,, \label{4d kappa flat q trf}
	\\
	\d x^{m} &= \left(\q_{i}^{\a}(\s^{m})_{\a\ad}\dt{\P}{}^{\ad\b}\k^{i}_{\b} + \bar{\k}_{\ad i} \dt{\P}{}^{\ad\a}(\s^{m})_{\a\bd}\bar{\q}^{i\bd}\right)\,,
	\\
	\d \mathfrak{e} &= -4\frak{e}\left(\dot{\q}_{i}^{\a}\k^{i}_{\a} + \dot{\bar{\q}}^{i}_{\ad}\bar{\k}_{i}^{\ad}\right)\,, \label{flat e trf}
\end{align}
\esubeq
where $\k_{\a}^{i} = \k_{\a}^{i}(\t)$ is Grassmann-odd. 
This symmetry can be extended to conformally flat backgrounds by deforming the transformation \eqref{flat e trf} to the following 
\begin{align} \label{4d kappa trf def 2}
	\d \frak{e} = -4\frak{e}\left(\dot{\q}_{i}^{\a}\k_{\a}^{i} + \dot{\bar{\q}}_{\ad}^{i}\bar{\k}_{i}^{\ad} - \frac{\ri}{4}\bar{\k}_{\ad i} \dt{\P}{}^{\ad\b}D_{\b}^{i}\s + \frac{\ri}{4}\bar{D}_{\ad i}\bar{\s}\dt{\P}{}^{\ad\b}\k_{\b}^{i}
	\right)\,.
\end{align}
As a result, the models \eqref{ads model} and \eqref{embed model} are $\k$-symmetric in the $m = \o = 0$ case. 
It should be emphasised that, from the perspective of the model \eqref{embed model}, only the $\a$ term is $\k$-symmetric. 

Let us make some brief comments about the role of $\k$-symmetry. 
First of all, we point out that the equation of motion for the einbein $\mathfrak{e}$ enforces the mass-shell constraint $p^{2} = 0$. 
It follows that, for certain choices of transformation parameter $\k_\a^i$, the transformation \eqref{4d kappa flat q trf} vanishes on shell. 
This implies that not all of the components of $\k_\a^i$ are independent. 
As it turns out, only half of the components of $\k_\a^i$ are independent, see, e.g., \cite{DS}. 
Indeed, an important aspect of $\k$-symmetry is that it can be used to gauge away half of the fermionic degrees of freedom. 

\section{Discussion}
\label{4d discussion}

In this chapter we have systematically developed the $\N$-extended AdS supergeometry in four dimensions from both the supergravity-inspired approach and the embedding formalism. 
In the supergravity-inspired approach, we have realised AdS$^{4|4\N}$ as a curved supergeometry with local $\sSL(2\,,\mathbb{C})\times\sSU(\N)$ symmetry, possessing Lorentz-invariant and covariantly constant torsion tensor $S^{ij}$. 
We provided explicit expressions for the covariant derivatives, supervielbein one-forms, and $S^{ij}$ in a conformally flat frame, parametrised in terms of a chiral super-Weyl parameter, $\s$. 
We then solved the AdS constraints for $\s$, eq. \eqref{ads constraints 1} and \eqref{ads constraints 2}, in both stereographic and Poincar\'e coordinates, extending the results of \cite{IS,Kuzenko:2008qw,Butter:2012jj} to arbitrary $\N$. 
We also determined the Killing supervectors for AdS$^{4|4\N}$ for arbitrary $\N$.
Further, we showed how the properties of $S^{ij}$ imply that there is a gauge choice in which it has the diagonal form $S^{ij} = S \d^{ij}$, which reduces the structure group to $\sSL(2\,,\mathbb{C}) \times \sO(\N)$. 
This proves to be the starting point for determining the correspondence between the supergravity-inspired approach and the coset construction.

In the group-theoretic setting, we developed the supergeometry making use of the coset construction. 
We derived two coordinate charts, the north and south charts of section \ref{n s cosets}, which provide an atlas for AdS$^{4|4\N}$. 
We then developed certain aspects of the AdS supergeometry, in particular deriving the algebra \eqref{NowarAlgebra}, which coincides with that derived from the supergravity-inspired approach, eq. \eqref{4d ads coset algebra}, when the diagonal gauge choice \eqref{4d diag gauge} is applied. 
In doing so, we have demonstrated the precise correspondence between the two approaches. 

An important implication of the above analysis is that any frame derived from the embedding formalism is not conformally flat for $\N>1$. 
We will provide a concise explanation for this below. 
A cornerstone of the embedding formalism developed above is that the (bi-)supertwistors are variables that parametrise the coset superspace eq. \eqref{4d ads susy definition}.
This means that, by construction, the local coordinate systems that we can introduce for them have, at most, $\sSL(2\,,\mathbb{C})\times \sO(\N)$ as the structure group. 
On the other hand, from the supergravity-inspired approach, we have shown the specific parametrisation required for a frame to be conformally flat. 
For the case of AdS$^{4|4\N}$, these were the relations eq. \eqref{AdSBoost}. 
In particular, the local structure group in such a frame is $\sSL(2\,,\mathbb{C}) \times \sSU(\N)\,.$
Furthermore, we have shown that it is possible to diagonalise the torsion tensor $S^{ij}$ to reduce the $R$-symmetry group to $\sO(\N)$, as required to reach the frame derived from the embedding formalism. 
As described in section \ref{4d solving}, however, this means that the covariant derivatives are no longer related to those of Minkowski superspace by the super-Weyl transformations \eqref{AdSBoost}.
It immediately follows that the covariant derivatives no longer satisfy the conformally flat definition, hence the frame derived from the embedding formalism is not conformally flat. 
Finally, in appendix \ref{4d n ext coset}, we proved this was true for the north chart, and then generalised to an arbitrary coordinate system derived from the embedding formalism. 

It is important to emphasise that the above analysis does not mean the superspace AdS$^{4|4\N}$ is not conformally flat. 
Indeed, as mentioned at the beginning of this chapter, AdS$^{4|4\N}$ was proven to be conformally flat in \cite{BILS}.
In the $\N=1$ case, they made use of the Maurer-Cartan equations to derive a conformally flat realisation. 
For $\N>1$, they developed a new method based on embedding the AdS supergroup $\sOSp (\N |4; \mathbb{R} )$ into the superconformal group, $\sSU(2,2|\N)$. 
Our analysis provides an alternative proof of the conformal flatness of AdS$^{4|4\N}.$

Finally, we have provided an example application of the described formalisms in the form of new superparticle models in AdS$^{4|4\N}$. 
To do so, we introduced a deformed AdS supersymmetric interval making use of features specific to the AdS supergeometry in both formalisms. 
Specifically, in the supergravity-inspired approach, the deformation was introduced via terms containing the torsion tensor $S^{ij}$ and the spinor components of the supervielbein. 
On the other hand, in the embedding formalism, the deformation was derived making use of the three unique two-point invariants under the AdS supergroup, $\sOSp(\N|4;\mathbb{R})\,.$ 
We also showed how the two models were related.

\begin{subappendices}

\section{Isomorphic realisation of the AdS supergroup} \label{adsgroup}

In this appendix we will develop an alternative realisation for the AdS supergroup $\sOSp(\N|4;\mathbb{R})$ which proves to be useful for the applications described above.
To motivate such a realisation, it suffices to consider the twistor realisation for AdS space (i.e. the $\N=0$ case of section \ref{ads4 bst}). 
The conditions \eqref{planecon.a} tell us that for the two-plane
\begin{align} 
	{\cP}
	= \left( \begin{array}{c}
		F \\
		G 	
		\end{array} \right) \in {\mathfrak F}_0~, \qquad F, G \in {\sMat} (2, {\mathbb C}) ~,
\end{align}
the  $2\times 2$ matrices $F$ and $G$ are non-zero. 
With these conditions it is possible to parametrise the two-planes in Poincar\'e coordinates, following \cite{KTM}.
However, for the purposes of the analysis presented in section \ref{ch3}, it would be useful to deal with a realisation of $\sSp(4,\mathbb{R})$ in which one of the two matrices $F$ or $G$ can be zero.
This is inspired by the parametrisation used in the (super)twistor realisation of Minkowski (super)space in, e.g., \cite{K-compactified12}, see also section \ref{4d compact str}.


Let us consider a supergroup, denoted  $\sOSp(\mathcal{N}|4;\mathbb{R})_{C}$,
 consisting of 
all even $(4|\cN)\times(4|\cN)$ supermatrices $\underline{g}$ subject to the following constraints:
\bsubeq  \label{group constraints}
\begin{align}
	\underline{g}^{\sT} \mathbb{K} \underline{g} &=\mathbb{K}\,, \label{gkg}\\
	\underline{g}^\dag \mathbb{J} \underline{g} &= \mathbb{J} \label{gdagjg}\,.
\end{align}
\esubeq
Here we have introduced the graded antisymmetric $(4|\cN)\times (4|\cN)$ supermatrix
\begin{align} \label{kdef}
	\mathbb{K} = \left(
	\begin{array}{c||c}
		\hat{\ve} ~& ~0\\
		\hline \hline
		0 ~&~ {\mathbbm 1}_\mathcal{N}
	\end{array}
	\right)\,, 
	\qquad 
	\hat{\ve} = \left(
	\begin{array}{cc}
		\ve & ~~0\\
		0 &~ -\ve^{-1}
	\end{array}
	\right)\,,
\end{align}
where we have used the notation $\hat{\ve}$ to distinguish between the charge conjugation matrix \eqref{3,2 cc matrix} and the blocks of \eqref{4d cosetrep}.

 The supergroup $\sOSp(\mathcal{N}|4;\mathbb{R})_{C}$ 
proves to be isomorphic to $\sOSp(\mathcal{N}|4;\mathbb{R})$. 
The proof is based on considering the following supermatrix correspondence:
\bsubeq
\begin{align}
	g &\rightarrow \underline{g} := \mathbb{U}g\mathbb{U}^{-1}\,, \qquad 
	\forall g \in \sOSp(\mathcal{N}|4;\mathbb{R})\,,
\end{align} 
in conjunction with the supertwistor transformation
\begin{align}
	\bm{T} &\rightarrow \underline{\bm{T}} := \mathbb{U}\bm{T}\,, \label{transformed t}
\end{align}
\esubeq
for every supertwistor $\bm{T}$. Here the supermatrix $\mathbb{U}$ is defined as
\begin{align}
	\mathbb{U} = \left(
	\begin{array}{c||c}
		t_s & 0\\
		\hline \hline
		0 & {\mathbbm 1}_\mathcal{N}
	\end{array}
	\right)\,, 
	\qquad 
	t_s = \frac{1}{\sqrt{2}} \left(
	\begin{array}{cc}
		{\mathbbm 1}_2 & -\ri\ve^{-1}\\
		-\ri\ve & {\mathbbm 1}_2
	\end{array}
	\right)\,~.
\end{align}
It obeys the useful properties:
\begin{align} \label{ujk properties 1}
	\bar{\mathbb{U}}  = \mathbb{U}^{-1} \,, \qquad
	\mathbb{U}^{\sT} = 
	(\mathbb{U}^{-1})^\dag\,,
\end{align}
and
\bsubeq \label{ujk properties 2}
\begin{align} 
	(\mathbb{U}^{-1})^{\sT} \mathbb{J}\mathbb{U}^{-1} &= \ri\mathbb{K}\,, \label{ujuk} \\
	\mathbb{U}^{\sT} \mathbb{J} \mathbb{U}^{-1} &= \mathbb{J}\,. \label{ujuj}
\end{align}
\esubeq
These conditions imply that 
\begin{align} \label{uunitary}
	\mathbb{U}^\dag \mathbb{J} \mathbb{U} = \mathbb{J}\,.
\end{align}
Associated with $\sOSp(\mathcal{N}|4;\mathbb{R})_{C}$ are two invariant inner products defined as
\bsubeq \label{innerproducts}
\begin{align}
	\braket{\underline{\bm{T}}}{\underline{\bm{S}}}_\mathbb{K} &:= \underline{\bm{T}}^{\sT}\mathbb{K}\underline{\bm{S}}\,, \label{tkt} \\ 
	\braket{\underline{\bm{T}}}{\underline{\bm{S}}}_\mathbb{J} &:= \underline{\bm{T}}^\dag \mathbb{J} \underline{\bm{S}}\,, \label{tjs}
\end{align}
\esubeq
for arbitrary pure supertwistors $\underline{\bm{T}}$ and $\underline{\bm{S}}$. 
The conditions \eqref{group constraints} impose restrictions on the blocks of $\underline{g}$. For
\begin{align}
	\underline{g} &= \left( \begin{array}{c||c}
		A & B \\
		\hline
		\hline
		C & D
	\end{array} \right)\,,
\end{align}
these are:
\bsubeq \label{groupreq1}
\begin{align}
	A^\T \hat{\ve} A -  C^\T C &= \hat{\ve}\,, \label{aka con}
	\\
	B^\T \hat{\ve} B +  D^\T D &= {\mathbbm 1}_N\,, \label{bkb con}
	\\
	A^\T \me B -  C^\T D &= 0\,, \label{akb con}
\end{align}
\esubeq
and
\bsubeq \label{groupreq2}
\begin{align}
	A^\dag J A + \ri C^\dag C = J\,,
	\\
	B^\dag J B + \ri D^\dag D = \ri{\mathbbm 1}_N\,,
	\\
	A^\dag J B + \ri C^\dag D = 0\,.
\end{align}
\esubeq
In the original realisation of $\sOSp(\mathcal{N}|4;\mathbb{R})$ the reality condition could be realised as the coincidence of the supertranspose and the Hermitian conjugate, eq. \eqref{A.13b}.
For our new realisation of the supergroup, 
 \eqref{A.13b} is replaced with the following condition
\begin{align}
	\underline{g}^{\dag} = (\mathbb{U}^{\text{sT}})^{2}\underline{g}^{\text{sT}}(\mathbb{U}^{\dag})^{2}\,.
\end{align}
From this we have the following conditions
\bsubeq
\begin{align}
	\bar{A} &= t_s{}^{-2} A t_s{}^2\,, \label{areal}
	\\
	\bar{B} &= t_s{}^{-2} B\,, \label{breal}
	\\
	\bar{C} &= -C t_s{}^2\,, \label{creal}
	\\
	\bar{D} &= D\,. \label{dreal}
\end{align}
\esubeq

We will now discuss involution for the supertwistors $\underline{\bm{T}}$. Since the transformation \eqref{transformed t} applies to every supertwistor $\bm{T}$, we can also consider it applied to $*\bm{T}$. We have
\begin{align}
	*\bm{T} \rightarrow \underline{*\bm{T}} = \mathbb{U}(*\bm{T})\,. 
\end{align} 
This acts explicitly on a supertwistor $\underline{\bm{T}}$ as
\begin{align} \label{davidstar}
	\underline{\bm{T}} = \left(\begin{array}{c}
		f \\
		g \\
		\hline\hline
		\ri\psi
	\end{array} \right) ~\rightarrow~
\underline{*\bm{T}} = -\ri  \left(\begin{array}{c}
	\ve^{-1}\bar{g} \\
	 \ve\bar{f} \\
	\hline\hline
	(-1)^{1+\e({\underline{\bm{T}}})}\bar{\psi}
\end{array}\right)\,.
\end{align}
The components of $\underline{*\bm{T}}$ are given by
\begin{align}
	\underline{*\bm{T}}_{A} = (-1)^{\e({\underline{\bm{T}}})\e_{A}+\e_{A}}\overline{(\mathbb{U}^{-2}\underline{\bm{T}})_{A}}\,.
\end{align}

Let us introduce a new operation, denoted by $\star$, by removing the factor of $-\ri$ in \eqref{davidstar}:
\begin{align}
	\underline{\bm{T}} = \left(\begin{array}{c}
		f \\
		g \\
		\hline\hline
		\ri\psi
	\end{array} \right) \rightarrow
	\star\underline{\bm{T}} = \left(\begin{array}{c}
		\ve^{-1}\bar{g} \\
		\ve\bar{f} \\
		\hline \hline 
		(-1)^{1+\e({\underline{\bm{T}}})}\bar{\psi}
		\end{array}\right)\,.
\end{align}
The components of $\star\underline{\bm{T}}$ are given by
\begin{align}
	(\star\underline{\bm{T}})_{A} = (-1)^{\e({\underline{\bm{T}}})\e_{A}+\e_{A}}\ri\overline{(\mathbb{U}^{-2}\underline{\bm{T}})_{A}}\,.
\end{align}
We therefore have the following reality condition with respect to the map $\star$
\begin{align}
	\overline{\underline{\bm{T}}_A} = (-1)^{\e(\underline{\bm{T}})\e_{A}+\e_{A}}(-\ri)(\mathbb{U}^{-2}\underline{\bm{T}})_{A}\,.
\end{align}
The map $\star$ is an involution, since it satisfies the property
\begin{align}
	\star(\star\underline{\bm{T}}) = \underline{\bm{T}}\,. 
\end{align}
We also observe that 
\begin{align}
	(\star\underline{\bm{T}})^{\sT} = \ri\underline{\bm{T}}^{\dag}(\mathbb{U}^{\sT})^{2}\,,
\end{align}
which, in conjunction with the properties \eqref{ujk properties 2}, yields the following
\begin{align}
	\braket{\star\underline{\bm{T}}}{\underline{\bm{S}}}_\mathbb{K} = (\star\underline{\bm{T}})^{\sT}\mathbb{K}\underline{\bm{S}} = \underline{\bm{T}}^{\dag}\mathbb{J}\underline{\bm{S}} = \braket{\underline{\bm{T}}}{\underline{\bm{S}}}_\mathbb{J}\,.
\end{align}

It is useful to express the constraints \eqref{planecon}, the two-point functions \eqref{two-point}, and the bi-supertwistors \eqref{bi-super} in terms of the new realisation of the supergroup. 
The constraints can be expressed as
\bsubeq \label{newcon}
\begin{align}
	\ve_{\m\n} \braket{\underline{\bm{T}}^{\m}}{\underline{\bm{T}}^{\n}}_{\mathbb{K}} &\neq 0 \,,
	\\
	\braket{\underline{\bm{T}}^{\m}}{\underline{\bm{T}}^{\n}}_{\mathbb{J}} &= 0\,. 
\end{align}
\esubeq
For the two-point functions we find
\bsubeq \label{new two-point}
\begin{align} \label{ads two point function}
	\frac{1}{\ell^2} \o(\underline{\cP},\underline{\widetilde{\cP}}) &= -2 \frac{\braket{\underline{\bm{T}}^\mu}{\underline{\widetilde{\bm{T}}}{}^{\nu}}_\mathbb{J}\braket{\underline{\bm{T}}_\mu}{\underline{\widetilde{\bm{T}}}_{\nu}}_\mathbb{J}}{\braket{\star\underline{\bm{T}}{}^{\dot{\s}}}{\star\underline{\bm{T}}{}_{\dot{\s}}}_\mathbb{K}
		\braket{\underline{\widetilde{\bm{T}}}{}^{\s}}{\underline{\widetilde{\bm{T}}}{}_{\s}}_\mathbb{K}}\,,
	\\
	\frac{1}{\ell^2} \o_{(+)}(\underline{\cP},\underline{\widetilde{\cP}}) &= 2 \frac{\braket{\underline{\bm{T}}^\mu}{\underline{\widetilde{\bm{T}}}{}^{\nu}}_\mathbb{K}\braket{\underline{\bm{T}}_\mu}{\underline{\widetilde{\bm{T}}}_{\nu}}_\mathbb{K}}{\braket{\underline{\bm{T}}{}^{\s}}{\underline{\bm{T}}{}_{\s}}_\mathbb{K}
		\braket{\underline{\widetilde{\bm{T}}}{}^{\r}}{\underline{\widetilde{\bm{T}}}{}_{\r}}_\mathbb{K}}-1\,,
	\\
	\frac{1}{\ell^2} \o_{(-)}(\underline{\cP},\underline{\widetilde{\cP}}) &= 2 \frac{\braket{\star\underline{\bm{T}}^{\dot{\mu}}}{\star\underline{\widetilde{\bm{T}}}{}^{\dot{\nu}}}_\mathbb{K}\braket{\star\underline{\bm{T}}_{\dot{\mu}}}{\star\underline{\widetilde{\bm{T}}}_{{\dot{\nu}}}}_\mathbb{K}}{\braket{\star\underline{\bm{T}}{}^{\dot{\s}}}{\star\underline{\bm{T}}{}_{\dot{\s}}}_\mathbb{K}
		\braket{\star\underline{\widetilde{\bm{T}}}{}^{{\dot{\r}}}}{\star\underline{\widetilde{\bm{T}}}{}_{{\dot{\r}}}}_\mathbb{K}} -1\,.
\end{align}
\esubeq
The bi-supertwistors \eqref{bi-super} can be expressed in terms of transformed supertwistors $\underline{\bm{T}}$ as follows
\bsubeq \label{new bi-super}
\begin{align}
	\underline{\bm{X}}_{AB} &= \mathbb{U}_{A}{}^{C}\bm{X}_{CD}(\mathbb{U}^{\sT})^{D}{}_{B} \,,
	\\
	\underline{\bar{\bm{X}}}_{AB} &= \mathbb{U}_{A}{}^{C}\bar{\bm{X}}_{CD}(\mathbb{U}^{\sT})^{D}{}_{B}\,. 
\end{align}
\esubeq
They satisfy the following properties
\bsubeq
\begin{align}
	\underline{\bm{X}}_{[AB}\underline{\bm{X}}_{CD\}} &= 0\,,
	\\
	(-1)^{\e_{B}} \underline{\bm{X}}_{AB}\mathbb{K}^{BC}\underline{\bm{X}}_{CD} &= -\ri\ell\underline{\bm{X}}_{AD}\,,
	\\
	\mathbb{K}^{BA}\underline{\bm{X}}_{AB} &= -2\ri\ell\,,
	\\
	(-1)^{\e_{B}}\underline{\bm{X}}_{AB}\mathbb{K}^{BC}\underline{\bar{\bm{X}}}_{CD} &= 0\,. 
\end{align}
\esubeq
For the case $\N=0$, the $\me$-traceless parts of the bi-supertwistors take the form
\begin{align}
	\underline{X}_{\langle\hat{\a}\hat{\b}\rangle} = \underline{X}_{\hat{\a}\hat{\b}} + \frac{\ri}{2}\ell\me_{\hat{\a}\hat{\b}}\,, \qquad  \underline{\bar{X}}_{\langle\hat{\a}\hat{\b}\rangle} = \underline{\bar{X}}_{\hat{\a}\hat{\b}} + \frac{\ri}{2} \ell\me_{\hat{\a}\hat{\b}}\,.
\end{align}
As before, we can express the two-point functions \eqref{new two-point} in terms of the supermatrices $\underline{\bm{X}}_{A}{}^{B}$ and $\underline{\bar{\bm{X}}}_{A}{}^{B}$ defined by
\begin{align} \label{new raised Xs}
	\underline{\bm{X}}_{A}{}^{B} = (-1)^{\e_{C}}\underline{\bm{X}}_{AC}\mathbb{K}^{CB}\,, \qquad \underline{\bar{\bm{X}}}_{A}{}^{B} = (-1)^{\e_{C}}\underline{\bar{\bm{X}}}_{AC}\mathbb{K}^{CB}\,. 
\end{align}
They then take the form
\bsubeq \label{new trace functions}
\begin{align}
	\o(\underline{\cP},\underline{\widetilde{\cP}}) &= -\frac{1}{2}{\rm Str}\big(\underline{\bar{\bm{X}}}\underline{\widetilde{\bm{X}}}\big)\,,
	\\
	\o_{(+)}(\underline{\cP},\underline{\widetilde{\cP}}) &= \frac{1}{2}{\rm Str}\big(\underline{\bm{X}} \underline{\widetilde{\bm{X}}}\big) - \ell^{2}
	\,,
	\\
	\o_{(-)}(\underline{\cP},\underline{\widetilde{\cP}}) &= \frac{1}{2}{\rm Str}\big(\underline{\bar{\bm{X}}} \underline{\widetilde{\bar{\bm{X}}}}\big) - \ell^{2}\,.
\end{align}
\esubeq


\section{Coset construction for $\N>1$} \label{4d n ext coset}

In this appendix we will extend elements of the above analysis of the coset construction to $\N>1$, with particular focus on the north chart. 
Specifically, in the main body, the spinor component of the supervielbein was only explicitly determined for the $\N=1$ case. Here we will elaborate on the story for $\N >1\,,$ making use of the useful identity \eqref{d block spinor identity}.

We will now show the components of the supervielbein in the north chart for $\N>1$. 
For the vector component, we have 
\begin{align} \label{vec vierbein}
	\bm{E}^{a} &= \re^{-\frac{1}{2}(\s+\bar{\s})}\P^{a}\,,
\end{align}
which has the same form as that computed for $\N=1.$
The spinor component is given by the following expression
\begin{align} 
\label{4d n ext spinor vierbein}
	\bm{E}_{I}{}^{\a} &= \re^{-\frac{1}{2}\s}(D^{-1})_{IJ}
	\bigg(\text{d}\q_{K}{}^{\a}\big(\d_{JK}+\q_{J}{}^{\n}D_{\n K}\s \big) 
	+ \bm{\P}^{\dnu \a}
	(-\re^{-\s}\q_{J}{}^{\n}(x_{+})_{\n\dnu} 
	- \ri\re^{-\bar{\s}}\bar{\q}_{J\dnu})
	\bigg)~,
\end{align}
where $D = (D_{IJ}) = D^{\rm T}$ takes the form
\begin{align}
	D = \sqrt{\id_{\N} - 4\J}\,, \qquad \J = (\J_{IJ}) = (\l^{2}\q_{I}{}^{\a}\q_{\a J} + \bar{\l}^{2}\bar{\q}_{I \ad}\bar{\q}^{\ad}{}_{J}) = \J^{\rm T} \,.
\end{align}
In the above, we have made the identification $\l = \re^{-\frac{1}{2}\s}$ as described in sections \ref{n s cosets} and \ref{4d solving}. 
Further, it should be noted that in the above expressions we have defined the matrix $D$ as the square root of another matrix, whose body coincides with the unit matrix. 
Due to the fact that $\Psi$ is bodiless, this expression can be solved as a series expansion in $\Psi$ around the unit matrix, which is guaranteed to terminate at finite order. The order at which the expansion terminates grows with $\N$. 
In finding expressions for arbitrary $\N$, the lack of closed-form expression for $D_{IJ}$ is a serious computational roadblock. For the sake of the analysis in this section, however, it suffices to consider $\N=2$, as we will see below. 

The goal of this section is to prove that the frame constructed from the embedding formalism, corresponding to the covariant derivative algebras \eqref{4d ads coset algebra} and \eqref{NowarAlgebra}, is not conformally flat for $\N>1$. 
This is expected due to the analysis and discussion of section \ref{4d solving}. 
In particular, for $\N>1$, the local $R$-symmetry groups differ between the conformally flat frame and the `coset frame'.  
For this purpose, let us compare the spinor component of the supervielbein derived from the coset construction, eq. \eqref{4d n ext spinor vierbein}, with that from the supergravity-inspired approach, eq. \eqref{n extend cf vierbein}.

The spinor component of the conformally flat supervielbein, which may be extracted from \eqref{n extend cf vierbein}, takes the form
\begin{align} \label{cflat n=2}
	E_{i}^{\a} = \re^{-\frac{1}{2}\bar{\s}}\Big(\text{d}\q_{i}{}^{\a} + \frac{\ri}{4}\bar{D}_{\ad i}\bar{\s} \P^{\ad\a}\Big)\,. 
\end{align}
To compare the two expressions \eqref{4d n ext spinor vierbein} and \eqref{cflat n=2}, we first consider the $\text{d}\q$ terms. 
On the one-hand, the coefficient $\re^{-\frac{1}{2}\bar{\s}}$ in \eqref{cflat n=2} is antichiral.
On the other hand, expression \eqref{4d n ext spinor vierbein} has non-vanishing chiral contributions of the form 
\begin{align}
	\frak{Y}_{IJ}\text{d}\q_{J}{}^{\a}\,, \qquad \bar{D}_{\ad I}\frak{Y}_{JK} = 0\,,
\end{align}
where $\frak{Y} = (\frak{Y}_{IJ})$ is given by 
\begin{align}
	\frak{Y} = \l(\id_{2} - 2\l^{2}\hat{\q} - 2\l^{4}\hat{\q}^{2} - 4\l^{6}\hat{\q}^{3} - 10\l^{8}\hat{\q}^{4})\,, \qquad \hat{\q} := (\q_{I}{}^{\a}\q_{\a J})\,.
\end{align}
The two expressions therefore do not coincide, hence the frame corresponding to \eqref{vec vierbein} and \eqref{4d n ext spinor vierbein} is not conformally flat.

This analysis makes use of a particular local coset representative, corresponding to the north chart of AdS$^{4|8}\,.$ 
We will now show that it is independent of the choice of coset representative. 
Let us denote the north chart by $U_{N}$, and its coset representative $S_{N}$. 
Consider a generic coset representative $S'$, with corresponding coordinate chart $U'$ such that $U' \cap U_{N} = V \neq \emptyset\,.$ 
On $V$, the two coset representatives are related as follows 
\begin{align}
	S' = S_{N}h^{-1}\,, \qquad h \in \sSL(2\,,\mathbb{C}) \times \sO(2)\,.
\end{align}
The vielbein supermatrices then satisfy
\begin{align}
	\bm{E}' = h \bm{E} h^{-1}\,,
\end{align}
where $\bm{E}'$ is the vielbein supermatrix corresponding to $U'\,.$
Since the vector component of the supervielbein, \eqref{vec vierbein}, is already in conformally flat form, we restrict our attention to those coset representatives related to $S_{N}$ by an orthogonal transformation only, in which case the spinor components \eqref{4d n ext spinor vierbein} are related as 
\begin{align} \label{vierbein On}
\bm{E}'_{I}{}^{\a} = R_{IJ}\bm{E}_{J}{}^{\a}\,, \qquad R^{\T}R = \id_{2}\,, \quad R = \bar{R}\,. 
\end{align}
Now let us assume that $\bm{E}'_{I}{}^{\a}$ is in conformally flat form, \eqref{cflat n=2}. 
This means that the matrix $\tilde{R}$ relating \eqref{4d n ext spinor vierbein} and \eqref{cflat n=2} satisfies the conditions \eqref{vierbein On}.
However, direct comparison of the $\text{d}\q$ terms in both expressions shows that $\tilde{R}$ is necessarily complex, hence they are not related by the rule \eqref{vierbein On}.
It follows that $\bm{E}'_{I}{}^{\a}$ cannot be in conformally flat form. 

One can consider instead the following more general condition on the matrices $R$ in \eqref{vierbein On}
\begin{align}
	R^{\dag} R = \id_{2}\,,
\end{align}
as the conformally flat form of \eqref{vec vierbein} is unspoilt by such a transformation. 
It can be shown that the matrix $\tilde{R}$ satisfies 
\begin{align}
	\tilde{R}^{\dag}\tilde{R} = \id_{2} + \cO(\q)\,,
\end{align}
and hence is not unitary.



\section{Another realisation of the AdS supergroup} \label{another ads group}

In this appendix we will develop another useful realisation of the connected component of the AdS supergroup $\sOSp_0(\N|4;\mathbb{R})$.
We will explicitly realise it as a subgroup of the superconformal group in four dimensions $\sSU(2,2|\N)$, for which our conventions are outlined in appendix \ref{5d appendix}.

Let 
$\sOSp_0(\N|4;\mathbb{R})_{\mathfrak U}$
be 
the subgroup of $\sSU(2,2|\N)$
consisting of those  supermatrices $\hat{g} \in  \sSL(4|\N;{\mathbb C})$
which are singled out by the conditions
\bsubeq \label{new group con}
\begin{align}
\hat{g}^{\dag} \bm{\O} \hat{g} &= \bm{\O} \,,
\\ 
\hat{g}^{\sT} \mathfrak{J} \hat{g} &= \mathfrak{J}\,,
\end{align}
\esubeq
where $\mathfrak{J}$ denotes follows symplectic supermatrix
\begin{align}
	\mathfrak{J} = \left(\begin{array}{c|c||c}
		\ve & ~0~ & 0 \\
		\hline 
		~0~ & \ve^{-1} & 0 \\
		\hline \hline 
		0 & 0 & \ri \id_{\N}
	\end{array}\right)\,,
\end{align}
and $\bm{\O}$ is given by \eqref{big o def}.

The supergroup $\sOSp_0(\N|4;\mathbb{R})_{\mathfrak U}$ proves to be isomorphic to $\sOSp_0(\N|4;\mathbb{R})\,.$ 
The proof is based on the following supermatrix correspondence 
\begin{align} \label{g def}
	f \rightarrow g = {\mathfrak U}^{-1} f \mathfrak{U}\,, \qquad \forall f \in \sOSp_0(\N|4;\mathbb{R})\,,
\end{align}
in conjunction with the supertwistor transformation 
\begin{align}
\bm{T} \rightarrow \hat{\bm{T}} := \frak{U}^{-1}\bm{T}\,.
\end{align}
Here the supermatrix $\mathfrak U$ is defined as 
\begin{align} \label{u def}
	\mathfrak{U} &= \left(\begin{array}{c||c}
	\mathfrak{m} & ~0~ \\
	\hline \hline 
	~0~ &  \id_{\N}
\end{array}\right)\,, \qquad \mathfrak{m} = \frac{1}{2}\left(\begin{array}{c|c}
	\a \id_{2} + \bar{\a}\ve  &  \bar{\a} \id_{2} + \a \ve \\ \hline
	-\a \id_{2} + \bar{\a}\ve & \bar{\a}\id_{2} - \a \ve
\end{array}\right)\,, \qquad \a = \re^{\ri \p/4}
= \frac{1+\ri}{\sqrt{2}}\,. 
\end{align}
It obeys the useful relations 
\bsubeq \label{useful props}
\begin{align}
	\mathfrak{U}^{\dag} &= \mathfrak{U}^{-1}\,, 
	\\
	\mathfrak{U}^\dagger \mathbb{J}  \mathfrak{U}  &= -\ri \bm{\O}\,, 
	\\
	(\mathfrak{U})^{\sT}\mathbb{J} \mathfrak{U} &=  \mathfrak{J}\,.
\end{align}
\esubeq
It can be constructed making use of the alternative realisations for $\sOSp_0(\N|4;\mathbb{R})$ and $\sSU(2,2|\N)$ provided in \cite{Kuzenko:2006mv} and \cite{Buchbinder:2015qsa}. 
Specifically, 
\begin{align}
\frak{U}^{-1} = M \S \mathbb{V}\,,
\end{align}
with 
\bsubeq
\begin{align}
M &= \frac{1}{\sqrt{2}}\left(\begin{array}{c|c||c}
\id_{2} & -\ve & 0 \\ 
\hline 
-\ve & \id_{2} & 0 \\ \hline \hline 
~0~ & ~0~ & \sqrt{2}\id_{\N}
\end{array}\right)\,, 
\\
\S &= 
\frac{1}{\sqrt{2}}\left(\begin{array}{c|c||c}
\id_{2} & -\id_{2} & 0 \\ 
\hline 
\id_{2} & \id_{2} & 0 \\ \hline \hline 
~0~ & ~0~ & \sqrt{2}\id_{\N}
\end{array}\right)\,, 
\\
\mathbb{V} &= 
\frac{1}{\sqrt{2}}\left(\begin{array}{c|c||c}
\id_{2} & \ri\id_{2} & 0 \\ 
\hline 
\ri\id_{2} & \id_{2} & 0 \\ \hline \hline 
~0~ & ~0~ & \sqrt{2}\id_{\N}
\end{array}\right)\,. 
\end{align}
\esubeq
The relations \eqref{useful props} can be proven with the aid of the following properties 
\bsubeq
\begin{align}
\mathbb{V} \mathbb{J} \mathbb{V}^{\dag} &= -\ri \mathbb{I}\,, \qquad (\mathbb{V}^{-1})^{\sT}\mathbb{J} \mathbb{V}^{-1} = \mathbb{J}\,, 
\\
\S \mathbb{I} \S^{\dag} &= \bm{\O}\,, \qquad ~~(\S^{-1})^{\sT}\mathbb{J} \S^{-1} = \mathbb{J}\,, \\
M \bm{\O} M^{\dag} &= \bm{\O}\,, \qquad  (M^{-1})^{\sT} \mathbb{J} M^{-1} = \frak{J}\,,
\end{align}
\esubeq
where $\mathbb{I}$ is defined as 
\begin{align}
\mathbb{I} = \left(\begin{array}{c|c||c}
		\id_{2} & ~0~ & 0 \\
		\hline 
		~0~ & -\id_{2} & 0 \\
		\hline \hline 
		0 & 0 & - \id_{\N}
	\end{array}\right)\,. 
\end{align}
Further, the matrices $M$, $\S$ and $\mathbb{V}$ are unitary,
\begin{align}
M^{-1} = M^{\dag}\,, \qquad \S^{-1} = \S^{\dag}\,, \qquad \mathbb{V}^{-1} = \mathbb{V}^{\dag}\,. 
\end{align}
These properties imply that the supermatrix $g$ defined by eq. \eqref{g def} obeys the conditions \eqref{new group con}, and hence $g \in \sOSp_0(\N|4;\mathbb{R})_{\frak{U}}\,.$

Associated with $\sOSp_0(\N|4;\mathbb{R})_{\mathfrak U}$ are two invariant inner products 
\bsubeq
\begin{align}
\braket{\hat{\bm{T}}}{\hat{\bm{S}}}_{\bm{\O}} &:= \hat{\bm{T}}{}^{\dag} \bm{\O} \hat{\bm{S}}\,, 
\\
\braket{\hat{\bm{T}}}{\hat{\bm{S}}}_{\mathfrak{J}} &:= \hat{\bm{T}}{}^{\sT}\mathfrak{J}\hat{\bm{S}}\,,
\end{align}
\esubeq
for arbitrary pure supertwistors $\hat{\bm{T}}$ and $\hat{\bm{S}}$.

The supergroup elements $\hat{g}$ satisfy the reality condition 
\begin{align} \label{new ads reality}
	\hat{g}^{\dag} = \U^{-1}\hat{g}^{\sT}\U \,, \qquad \U = \frak{U}^{\sT}\frak{U} = \left(\begin{array}{c|c||c}
	~0~ & \ri\ve & ~0~ \\ 
	\hline 
	-\ri\ve & ~0~ & 0 \\ 
	\hline \hline 
	0 & 0 & \id_{\N}
	\end{array}\right)\,.
\end{align}
Then, making use of eq. \eqref{new ads reality}, one can introduce an involution operation $\star$ defined as 
\begin{align}
	\hat{\bm{T}} \rightarrow \star \hat{\bm{T}}\,, \qquad 	(\star \hat{\bm{T}})_{A} = (-1)^{\e(\hat{\bm{T}})\e_{A} + \e_{A}}\left( \U^{-1}\overline{\hat{\bm{T}}} \right)_{A}\,.  
\end{align}
Its key properties are 
\bsubeq
\begin{align}
	\star (\star \hat{\bm{T}}) &= \hat{\bm{T}}\,, \\
	\hat{g}(\star \hat{\bm{T}}) &= \star(\hat{g} \hat{\bm{T}})\,. 
\end{align}
\esubeq
In our new realisation of the AdS supergroup, a supertwistor $\hat{\bm{T}}$ is said to be real if it satisfies 
\begin{align}
\star \hat{\bm{T}} = \hat{\bm{T}}\,.
\end{align}
Further, we observe that 
\begin{align}
\star{\hat{\bm{T}}}{}^{\sT} = \hat{\bm{T}}{}^{\dag} \U^{-1}\,,
\end{align}
which, in conjunction with the relations \eqref{useful props}, yields 
\begin{align}
\braket{\star{\hat{\bm{T}}}}{\hat{\bm{S}}}_{\frak{J}} = - \ri\braket{\hat{\bm{T}}}{\hat{\bm{S}}}_{\bm{\O}}\,.
\end{align}

It turns out that there is another supermatrix which will enact the similarity transformation \eqref{g def}. 
Let us introduce the supermatrix $\frak{N}$ defined as 
\begin{align}
\frak{N} = \left(\begin{array}{c||c}
\frak{n} & 0 \\
\hline \hline 
0 & \id_{\N}
\end{array}\right)\,, \qquad 
\frak{n} = \frac{\re^{-\ri\p/4}}{\sqrt{2}} \left(\begin{array}{c|c}
\id_{2} & - \ve \\ 
\hline
\ri\ve & - \ri\id_{2} 
\end{array}\right)\,.
\end{align}
The supermatrix $\frak{N}$ enjoys the properties
\bsubeq
\begin{align}
\frak{N}^{\dag} &= \frak{N}^{-1}\,, 
\\
\frak{N}^{\dag}\mathbb{J}\frak{N} &= -\ri\bm{\O}\,,
\\
\frak{N}^{\sT}\mathbb{J}\frak{N} &= \frak{J}\,. 
\end{align}
\esubeq
For every $f \in \sOSp_{0}(\N|4;\mathbb{R})$, the supermatrix defined by 
\begin{align}
\hat{g} = \frak{N}^{-1}f\frak{N}
\end{align}
belongs to $\sOSp_0(\N|4;\mathbb{R})_{\frak{U}}\,.$

As the supermatrices $\frak{N}$ and $\frak{U}$ both take us to the realisation $\sOSp_0(\N|4;\mathbb{R})_{\frak{U}}$, they must be related to each other in the following way 
\bsubeq \label{u to n}
\begin{align} 
\frak{U} = \frak{N} \frak{S}\,,
\end{align}
where $\frak{S}$ satisfies the following properties 
\begin{align}
\frak{S}^{\dag} = \frak{S}^{-1}\,, \qquad \frak{S}^{\dag}\bm{\O} \frak{S} = \bm{\O}\,, \qquad \frak{S}^{\sT}\frak{J}\frak{S} = \frak{J}\,.
\end{align}
\esubeq
It can be shown that the solution to eq. \eqref{u to n} takes the form 
\begin{align}
\frak{S} = \left(\begin{array}{c|c}
\frak{s} & ~0~ \\
\hline \hline 
~0~ & \id_{\N}
\end{array}\right)\,, \qquad 
\frak{s} = \frac{1}{\sqrt{2}}\left(\begin{array}{c|c}
\ve & \ri \ve \\ 
\hline 
\ri \ve & \ve
\end{array}\right)\,.
\end{align}

\section{AdS superspace and compactified Minkowski superspace} \label{ads and mink}

In this appendix we will make use of the realisation $\sOSp_0(\N|4;\mathbb{R})_{\frak{U}}$ to show the relationship between AdS$^{4|4\N}$ and $\overline{\mathbb{M}}{}^{4|4\N}$, described in section \ref{4d compact str}.
To be precise, AdS$^{4|4\N}$ arises as an open domain of $\N$-extended compactified Minkowski superspace. 
We will illustrate this below. 
For the remainder of this section we will use the notation $\hat{\bm{T}} = \bm{T}$.

Let us begin with compactified Minkowski superspace. 
We recall that it is described by those null two-planes $\bm{T}^{\m}$ which satisfy the constraints \eqref{compact null} and are defined modulo the equivalence relation \eqref{compact equiv}.
These conditions are preserved under the action of the superconformal group $\sSU(2,2|\N)$.

Let us now restrict our attention to those two-planes for which the following property holds 
\begin{align} \label{new ads symp neq}
\braket{\bm{T}^{\m}}{\bm{T}^{\n}}_{\frak{J}} \neq 0\,.
\end{align}
Then, making use of the equivalence relation \eqref{compact equiv}, one can impose the normalisation condition 
\begin{align} \label{new ads normal symp}
\braket{\bm{T}^{\m}}{\bm{T}^{\n}}_{\frak{J}} = \ell \ve^{\m\n}\,,
\end{align}
for a constant $\ell > 0\,.$ The conditions \eqref{new ads normal symp} and \eqref{compact null} prove to be invariant under the action of the connected component of the AdS supergroup $\sOSp_0(\N|4;\mathbb{R})_{\frak{U}}$. 
They are preserved under equivalence transformations of the form 
\begin{align} \label{new ads appendix equiv}
\bm{T}^{\m} \sim \bm{T}^{\n} N_{\n}{}^{\m}\,, \qquad N = (N_{\n}{}^{\m}) \in \sSL(2\,,\mathbb{C})\,.
\end{align}
As before, such a pair of supertwistors is referred to as a frame, and the space of such frames is denoted $\frak{F}^{(\frak{U})}_{\N}$.
The supergroup $\sOSp_0(\N|4;\mathbb{R})_{\frak{U}}$ acts on $\frak{F}^{(\frak{U})}_{\N}$ as 
\begin{align}
\bm{T}^{\m} \rightarrow g\bm{T}^{\m}\,, \qquad g \in \sOSp_0(\N|4;\mathbb{R})_{\frak{U}}\,.
\end{align}
This action is naturally extended to the quotient space $\frak{F}^{(\frak{U})}_{\N}/\sim\,,$ which can be identified with AdS superspace, 
\begin{align}
\text{AdS}^{4|4\N} = \frak{F}^{(\frak{U})}_{\N}/\sim\,,
\end{align}
where the equivalence relation is given by \eqref{new ads appendix equiv}.

Now, one can repeat the analysis of section \ref{n s cosets} to find, for our new realisation, the two-planes $\cP = (\bm{T}_{A}{}^{\m})$ in the north chart take the form 
\begin{align} \label{new nc 2p}
\cP  = \l \left(\begin{array}{c}
\id_{2} \\ 
-\ri \tilde{x}_{+} \\ \hline \hline 
2\q
\end{array}\right)\,,
\end{align}
with $\l$ given by 
\begin{align}
\l = (1 - x_{+}^{2} + 2\ri\q^{2})^{-\frac{1}{2}}\,. 
\end{align}
It follows that, in this realisation, $s^{ij}$ is given by 
\begin{align} \label{appendix s def}
s^{ij} = 2\ri\d^{ij}\,. 
\end{align}
Further, we find that our two-planes in the north chart are related to those two-planes in the Minkowski chart of $\overline{\mathbb{M}}{}^{4|4\N}$ \eqref{mink 2plane} by multiplication by the scale parameter $\l\,.$

In this realisation it is straightforward to read off the superconformal transformation rules for AdS$^{4|4\N}$.
They are given by the relations \eqref{scf trf bm}. 
The AdS transformations can then be singled out as those superconformal transformations which preserve the AdS condition \eqref{new ads normal symp}. 
This requirement proves to impose the following constraints on the parameters in \eqref{scf trf bm}
\bsubeq \label{ads translation con}
\begin{align}
	a^{a} &= b^{a}\,, 
	\\
	\eta_{\a}{}^{i} &= \ri\d^{ij}\e_{\a j}\,,
	\\
	\D &= 0\,.
\end{align}
\esubeq
Further, only the antisymmetric component of $\L$ remains
\begin{align} \label{ads so(n) con}
\L = -\L^{\T}\,. 
\end{align}
Then, making use of the definition of $s^{ij}$ in this realisation, eq. \eqref{appendix s def}, we find complete agreement with the relations \eqref{ads killing} for the components of the AdS Killing supervectors.


\section{$\sSU(\cN)$ superspace}
\label{cflat geo}

This appendix is devoted to a review of the pertinent details of the $\N$-extended $\sSU(\N)$ superspace geometry described in \cite{KKR, KKR2}. 
The goal of this appendix is to outline the formalism required to formulate the $\N$-extended anti-de Sitter supergeometry described in section \ref{4d n extended}. 
As such, specific discussion of the conformal superspace starting point, as well as the degauging procedure via $\sU(\N)$ superspace which was introduced in \cite{Howe:1980sy,Howe:1981gz}, is not included. 
The reader is referred to section 6 of \cite{KKR} and section 2 of \cite{KKR2} for more details.

For the purposes of this appendix, our starting point is a curved superspace $\cM^{4|4\N}$ parametrised by local coordinates 
\begin{align}
	z^{M} = (x^{m}\,, \q_{\imath}^{\m}\,, \bar{\q}^{\imath}_{\dmu})\,, \qquad m = 0\,,1\,,2\,,3\,, ~~ \m = 1\,,2\,, ~~ \dmu = \dot{1}\,,\dot{2}\,, ~~ \imath = \underline{1}\,, \ldots \,, \underline{\N}\,.
\end{align}
Its structure group is $\sSL(2\,,\mathbb{C}) \times \sSU(\N)$, and so the covariant derivatives take the form 
\begin{align} \label{su(n) cd}
	\cD_{A} = (\cD_{a}\,, \cD_{\a}^{i}, \cDB_{i}^{\ad}) = E_{A} + \O_{A} + \F_{A}\,. 
\end{align}
Here, $E_{A} = E_{A}{}^{M}\partial_{M}$ denotes the frame field, with $E_{A}{}^{M}$ being the inverse supervielbein. 
The superfield 
\begin{align}
	\O_{A} = \frac{1}{2}\O_{A}{}^{bc}M_{bc}\,, \qquad M_{bc} = - M_{cb}\,,
\end{align}
denotes the Lorentz connection; and 
\begin{align}
	\F_{A} = \F_{A}{}^{i}{}_{j}\mathbb{J}^{j}{}_{i} 
\end{align}
denotes the $\sSU(\N)$ connection. 

The structure of the algebra of covariant derivatives differs between the $\N=1$ and $\N>1$ cases, and so we will describe them separately below. 

\subsubsection{$\cN = 1$ case}

For $\N=1$, the algebra of covariant derivatives is given by 
\begin{subequations} \label{GWZalgebra}
	\bea
	\{ \cD_{\a}, \cD_{\b} \} &=& -4{\bar R} M_{\a \b}~, \qquad
	\{\cDB_{\ad}, \cDB_{\bd} \} =  4R {\bar M}_{\ad \bd}~, \\
	&& {} \qquad \{ \cD_{\a} , \cDB_{\ad} \} = -2{\rm i} \cD_{\a \ad} ~, 
	\\
	\big[ \cD_{\a} , \cD_{ \b \bd } \big]
	& = &
	{\rm i}
	{\ve}_{\a \b}
	\Big({\bar R}\,\cDB_\bd + G^\g{}_\bd \cD_\g
	- (\cD^\g G^\d{}_\bd)  M_{\g \d}
	\Big)
	+ {\rm i} (\cDB_{\bd} {\bar R})  M_{\a \b}~,\\
	\big[ {\bar \cD}_{\ad} , \cD_{\b\bd} \big]
	& = &
	- {\rm i}
	\ve_{\ad\bd}
	\Big({R}\,\cD_{\b} + G_\b{}^\gd \cDB_\gd
	- (\cDB^{\gd} G_{\b}{}^{\dd})  \bar M_{\gd \dd}
	\Big) 
	- {\rm i} (\cD_\b R)  {\bar M}_{\ad \bd}~,
	\eea
\end{subequations}
which describes a conformally flat GWZ geometry \cite{Grimm:1977kp, Grimm:1978ch}.
This geometry is described by the superfields $R$ and $G_{a} = \overline{G_{a}}$ subject to the constraints 
\begin{align} \label{n=1 bianchi}
	\cDB_{\ad}R = 0\,, \qquad \cD_{\a}R = \cDB^{\ad}G_{\a\ad}\,, \qquad \cD_{\a}{}^{(\g}G_{\b)\gd} = 0\,. 
\end{align}

The structure of the above algebra is invariant under the super-Weyl transformations
\begin{subequations} 
	\label{superweylGWZ}
	\bea
	\cD'_\a &=& \re^{ {\bar \s} - \hf \s} \Big(  \cD_\a + \cD^\b \s \, M_{\a \b} \Big) ~, \\
	\bar \cD'_\ad & = & \re^{  \s -  \hf {\bar \s}} \Big(\bar \cD_\ad +  \bar \cD^\bd  {\bar \s}  {\bar M}_{\ad \bd} \Big) ~,\\
	\cD'_{\a\ad} &=& \re^{\hf \s + \hf \bar \s} \Big( \cD_{\a\ad} 
	+\frac{\ri}{2} \bar \cD_\ad \bar \s \cD_\a + \frac{\ri}{2} \cD_\a  \s \bar \cD_\ad  
	+ \Big( \cD^\b{}_\ad \s + \frac{\ri}{2} \bar{\cD}_\ad \bar{\s} \cD^\b \s \Big) M_{\a\b} \non \\
	&& \qquad + \Big( \cD_\a{}^\bd \bar \s + \frac{\ri}{2} \cD_\a \s \bar{\cD}^\bd \bar{\s} \Big) \bar M_{\ad \bd} \Big)~, \\
	R' &=& \re^{2\s - \bar{\s}} \Big( R +\frac{1}{4} \bar{\cD}^2 \bar{\s} - \frac{1}{4} (\bar{\cD} \bar{\s})^2 \Big) ~, \\
	G_{\a\ad}'= &=& \re^{\hf \s + \hf \bar \s} \Big( G_{\a\ad} +\ri \cD_{\a\ad} ( \s - \bar \s) + \hf \cD_\a \s \bar{\cD}_\ad \bar{\s} \Big) ~,
	\eea
\end{subequations} 
where the super-Weyl parameter $\s$ is chiral, $\cDB_{\ad}\s = 0\,.$ 
In the infinitesimal limit, these transformations are a special case of those given in \cite{HT}, see also \cite{Siegel:1978fc}.

\subsubsection{$\cN > 1$ case}

We will now detail the $\N>1$ case. 
Up to dimension-1, the covariant derivatives obey the algebra
\begin{subequations} 
	\label{sun cov algebra}
	\bea
	\{\cD_\a^i,\cD_\b^j\}&=&
	4 S^{ij}  M_{\a\b} 
	+4\ve_{\a\b} Y^{ij}_{\g\d}  M^{\g\d}  
	-4\ve_{\a \b} S^{k[i} \mathbb{J}^{j]}{}_k
	+ 8{Y}_{\a\b}^{k(i}  \mathbb{J}^{j)}{}_k~,
	\label{acr1} \\
	\{\cD_\a^i,\cDB^\bd_j\}&=&
	-2\ri\d^i_j \cD_\a{}^\bd
	+4\d^{i}_{j}G^{\d\bd}M_{\a\d}
	+4\d^{i}_{j}G_{\a\gd}\bar{M}^{\gd\bd}
	+8 G_\a{}^\bd \mathbb{J}^{i}{}_{j}~.~~~~~~~~~
	\eea
\end{subequations}
This supergeometry is described by the complex superfields $S^{ij} = S^{(ij)}$ and $Y_{\ab}^{ij} = Y_{(\ab)}^{[ij]}$, as well as the real superfield $G_{a} = \overline{G_{a}}$, subject to the Bianchi identities 
\bsubeq \label{sun bianchi}
\begin{align}
	\cD_{\a}^{(i}S^{jk)} &= 0\,, \qquad \cDB_{i}^{\ad}S^{jk} = -4\d_{i}^{(j}\cD_{\a}^{k)}G^{\a\ad}\,, 
	\\
	\cD_{(\a}^{(i}Y_{\b\g)}^{j)k} &= 0\,, \qquad \cD^{\b k}Y_{\ab}^{ij} = - \cD_{\a}^{[i}S^{j]k} \,, \qquad \cDB_{j}^{\bd}Y_{\ab}^{ij} = 2\cD_{(\a}^{i}G_{\b)}{}^{\bd}\,. 
\end{align}
\esubeq

In the $\cN=2$ case this algebra of covariant derivatives coincides with the conformally flat limit of the one derived by Grimm \cite{Grimm1980}. 

The structure of the algebra \eqref{sun cov algebra} is invariant under the following super-Weyl transformations 
\begin{subequations}
	\label{SU(N)sW}
	\begin{align}
	\cD_\a^{'i}&= \re^{\frac{\cN-2}{2\cN} \s + \frac 1 \cN \bar{\s}} \Big( \cD_\a^i+ \cD^{\b i}\s M_{\a \b} + \cD_{\a}^j\s \mathbb{J}^{i}{}_j \Big) ~, 
	\\ 
	\bar{\cD}_{i}^{' \ad}&=\re^{\frac{1}{\cN} \s + \frac{\cN-2}{2\cN} \bar{\s}} \Big( \bar{\cD}^\ad_i-\bar{\cD}_{ \bd i} \bar{\s} \bar{M}^{\ad \bd} - \bar{\cD}^{\ad}_j \bar{\s} \mathbb{J}^{j}{}_i \Big)~,
	\\
	\cD_\aa' &= \re^{\hf \s + \hf \bar{\s}} \Big(\cD_\aa + \frac{\rm i}{2} \cD^i_{\a} \s \cDB_{\ad i} + \frac{\rm i}{2} \cDB_{\ad i} \bar{\s} \cD_{\a}^i + \hf \Big( \cD^\b{}_\ad (\s + \bar \s ) - \frac{\ri}{2} \cD^{\b i} \s \cDB_{\ad i} \bar{\s} \Big) M_{\a \b} \non \\ & + \hf \Big( \cD_{\a}{}^\bd (\s + \bar{\s}) + \frac{\ri}{2} \cD_{\a}^i \s \cDB^{\bd}_i \bar{\s} \Big) { \bar M}_{\ad \bd} 
	- \frac \ri 2 \cD_\a^i \s \cDB_{\ad j} \bar{\s} \mathbb{J}^j{}_i \Big) ~, \\
	S^{' ij}&= \re^{\frac{\cN-2}{\cN} \s + \frac 2 \cN \bar{\s}} \Big( S^{ij}-{\frac14}\cD^{ij} \s + \frac 1 4 \cD^{\a (i} \s \cD_\a^{j)} \s \Big)~, 
	\label{super-Weyl-S} \\
	Y^{' ij}_{\a\b}&= \re^{\frac{\cN-2}{\cN} \s + \frac 2 \cN \bar{\s}} \Big( Y_{\a\b}^{ij}-{\frac14}\cD_\a^{[i} \cD_\b^{j]}\s - \frac14 \cD_\a^{[i} \s \cD_\b^{j]} \s \Big)~,
	\label{super-Weyl-Y} \\
	G_{\aa}' &=
	\re^{\frac12 \s + \frac12 \bar{\s}} \Big(
	G_{\a\bd} -{\frac{\ri}4}
	\cD_{\a \ad} (\s-\bar{\s})
	-\frac{1}{8} \cD_\a^i \s \bar{\cD}_{\ad i} \bar{\s}
	\Big)
	~,
	\label{super-Weyl-G}
	\end{align}
\end{subequations}
where the super-Weyl paramter $\s$ is chiral, $\cDB_{i}^{\ad}\s = 0\,.$ 
For $\cN=2$, these transformations are a special case of those given in \cite{Kuzenko:2008ep, Kuzenko:2009zu}.
For $\N=3$, they are a special case of those given in \cite{Kuzenko:2023qkg}. 
It is important to point out that for $\cN=4$ the super-Weyl parameter appears only in the real combination $\s + \bar{\s}$.

A specific application of the above construction is that one can take $\cD_A = D_A = (\pa_a , D_\a^i , \bar D^\ad_i)$ to be the covariant derivatives of $\cN$-extended Minkowski superspace ${\mathbb M}^{4|4\cN}$ to obtain a conformally flat realisation of an arbitrary conformally flat $\sSU(\N)$ superspace. 
Indeed, the relations \eqref{SU(N)sW} are used in the main body to define a conformally flat frame for AdS$^{4|4\N}\,.$


\end{subappendices}

\chapter{Embedding formalism for $\N$-extended AdS superspace in five dimensions} \label{ch4}
\thispagestyle{boef2}

In this chapter we develop the $\N$-extended anti-de Sitter supergeometry in five dimensions.
This supergeometry can be defined from a group-theoretic point of view as 
\begin{align} \label{5d n ext ads def}
	\text{AdS}^{5|8\N} = \frac{\sSU(2,2|\N)}{\sUSp(2,2)\times\sU(\N)}\,,
\end{align}
see, e.g., \cite{BILS}. 
This definition makes use of one of the key underpinnings of the AdS/CFT duality, being the dual role played by the supergroup $\sSU(2,2|\N)$.
On the one hand, this supergroup is the $\N$-extended superconformal group in four dimensions. 
On the other hand, it is the supersymmetric extension of the AdS$_{5}$ group, $\sSU(2,2)$, see section \ref{bm ads sec} and appendix \ref{spinor conventions}. 
Indeed, the body of the coset superspace \eqref{5d n ext ads def} can be identified as 
\begin{align}
	\text{AdS}_{5} = \frac{\sSU(2,2)}{\sUSp(2,2)} = \frac{\sSpin(4,2)}{\sSpin(4,1)}\,.
\end{align}

Regarding the coset \eqref{5d n ext ads def}, the $\N$=1 case is exceptional in that it is the only case which is conformally flat, as argued in \cite{BILS}. 
This is due to the existence of the unique superconformal algebra in five dimensions \cite{Nahm:1977tg}, known as the $F(4)$ superalgebra \cite{Kac:1977em}.\footnote{It should be emphasised that the $F(4)$ superalgebra is associated with superconformal symmetry in five dimensions, and as such is identified with the AdS superalgebra in six dimensions.} 
Further, there is only one type of superconformal tensor calculus in five dimensions, developed independently by Fujita, Kugo, and Ohashi \cite{Kugo:2000hn,Kugo:2000af,Fujita:2001kv,Kugo:2002vc}, and Bergshoeff et al. \cite{Bergshoeff:2001hc,Bergshoeff:2002qk,Bergshoeff:2004kh}. 
Finally, there is only one type of conformal superspace in five dimensions \cite{Butter:2014xxa}.
An explicit conformally flat realisation of AdS$^{5|8}$ was derived in \cite{KTM2} in the supergravity setting, making use of the construction of \cite{Kuzenko:2008wr}. 
It is also known that AdS$^{5|8}$ is a maximally symmetric background of the minimal $\N=1$ supergravity geometry sketched by Howe \cite{Howe:1981ev}, see also \cite{Howe:1981nz}. 
This was fully developed in \cite{Kuzenko:2007cj,Kuzenko:2007hu}. 

Nevertheless, the study of the $\N>1$ AdS superspaces in five dimensions remains an interesting problem, in particular because of their relevance in higher-dimensional supergravity theories with solutions of the form AdS$_{m}\times S^{n}$ \cite{Kallosh:1998qs}. 
As we have pointed out in chapter \ref{ch3}, the conformal flatness of these superspaces is analysed in \cite{BILS}. 
An example of these is the famous AdS$_{5}\times S^{5}$ superspace, which can be defined as\footnote{The AdS$_{5}\times S^{5}$ supergeometry is sometimes defined as a homogeneous space of $\sPSU(2,2|4)$, see, e.g., \cite{Roiban:2000yy}. We have chosen to retain the $\sU(1)$ factor, see \cite{Lee:2004jx}.} 
\begin{align}
	\text{AdS}_{5}\times S^{5} = \frac{\sSU(2,2|4)}{\sUSp(2,2)\times\sUSp(4)\times\sU(1)}\,,
\end{align}
see, e.g., \cite{Metsaev:1998it, Claus:2000ka}. 
Supertwistor techniques have been used extensively to describe superparticles \cite{U,U2} and superstrings \cite{CGKRZ, Roiban:2000yy} in AdS$_{5}\times S^{5}$ superspace, however, until recently it remained an open problem to derive a global realisation of the `pure' AdS$^{5|8\N}$ cosets \eqref{5d n ext ads def}.
The goal of this chapter is to derive such an embedding formalism for AdS$^{5|8\N}$. 

This chapter is based on the publication \cite{KK}, and is organised as follows. 
In section \ref{boson ads5}, we introduce the twistor realisation for AdS$_{5}$. We also recall the standard bi-twistor realisation and prove their equivalence. 
In section \ref{susy ads5}, we introduce the supertwistor and bi-supertwistor realisations for the $\N$-extended anti-de Sitter supergeometry in five dimensions. 
We make use of these realisations to construct two-point functions on AdS$^{5|8\N}$, and introduce a superparticle model. 
Then, in section \ref{5d coset rep section} we introduce coset representatives for AdS$^{5|8\N}$ and derive their corresponding local coordinate systems. 
In the $\N=0$ case, we derive the six-vector $X^{\ua}$ satisfying the AdS$_{5}$ condition $X^{\ua}X_{\ua} = -\ell^{2}$ for the two coordinate systems introduced. 
For $\N>0$, we relate the bi-supertwistor in Poincar\'e-like coordinates to that of the four-dimensional compactified Minkowski superspace, see \cite{K-compactified12}.
In section \ref{5d geometry}, we develop aspects of the supergeometry of AdS$^{5|8\N}$, and show how they are related to the known $\N=1$ case in the literature. 
Section \ref{5d projective} is devoted to developing (bi-)supertwistor realisations of AdS$^{5|8\N}$ superspaces with additional bosonic degrees of freedom, such as the AdS$_{5}\times S^{5}$ superspace.
Finally, in section \ref{5d superparticle section} we derive the deformed AdS supersymmetric interval shown in chapter \ref{ch3} for the five dimensional case. 
We also show how the corresponding superparticle model is related to that introduced via the (bi-)supertwistor realisation.

\section{AdS$_5$} \label{boson ads5}

Earlier in this thesis, in section \ref{st realisations} and chapter \ref{ch3}, we discussed (bi-)supertwistor realisations for compactified Minkowski superspace and AdS superspace in four dimensions. 
These techniques all made use of (super)twistors, which were defined as column vectors transforming in the fundamental representation of the (super)conformal (or AdS) group. 
For the purposes of this chapter, twistor space is a four-dimensional complex vector space, $\mathfrak{T} = {\mathbb C}^4$, endowed with the linear action of the group $\sSU(2,2)$, 
\bea
\sSU(2,2) = \left\{ g \in \sSL(4,\mathbb{C})\,, \quad g^{\dag}\O g = \O \, ,\quad 
\O =  \left(\begin{array}{cc}
	0 & \id_{2} \\
	\id_{2} & 0
\end{array}\right)
\right\}~.
\label{su(2,2) master}
\eea
As we will see below, these twistors can be used to describe AdS$_{5}$.


An important role in the analysis of this section is played by the group $\sUSp(2,2)$, the conventions for which are detailed in appendix \ref{4,1 conv}.
In appendix \ref{4,1 conv}, the group $\sUSp(2,2)$ is realised as a subgroup of $\sSU(2,2)$, singled out by imposing the symplectic condition 
\begin{align} \label{5d chapter usp symp}
g^{\T} C g = C\,, \qquad C = \left(\begin{array}{cc}
\ve & ~0~\\
~0~ & -\ve^{-1} 
\end{array}\right)\,,
\end{align}
where $C$ is the charge conjugation matrix in five dimensions. 
For the purposes of this chapter, however, where we will have a proliferation of indices transforming under representations of different groups, it is necessary to modify the presentation of appendix \ref{4,1 conv}. These modifications are specified below. 
 
The group $\sUSp(2,2)$ is spanned by matrices 
\begin{align}
h = (h_{\tmu}{}^{\tnu}) \in \sSL(4,\mathbb{C})\,, \qquad \tmu\,, \tnu = 1\,,2\,,3\,,4\,,
\end{align}
satisfying the master equations 
\bsubeq
\begin{align}
h^{\dag} \O h = \O\,, \qquad \O = (\O^{\tmu\tnu}) = \left(\begin{array}{cc}
~0~ & \id_{2}
\\
\id_{2} & ~0~
\end{array}\right)\,,
\end{align}
and 
\begin{align}
h^{\T}\frak{C}h = \frak{C}\,, \qquad 
	\mathfrak{C} = (\mathfrak{C}^{\tmu\tnu}) = \left(\begin{array}{cc}
		\ve & ~0~ 
		\\
		~0~ & -\ve^{-1}
	\end{array}\right)\,.
\end{align}
\esubeq
Of course, the matrix $\O$ is identified with $\g_0$ as defined in appendix \ref{4,1 conv}, and the matrix $\frak{C}$ is identified with the charge conjugation matrix in five dimensions. 
However, the charge conjugation matrix specifically will appear throughout this chapter in different forms.
When possessing indices transforming with respect to $\sSU(2,2)$, it is denoted $C$. 
When possessing indices transforming with respect to $\sUSp(2,2)$, it is denoted $\frak{C}$.
It is necessary to distinguish between the two cases, which is the reason for the different notation introduced above.

\subsection{Twistor realisation}

Here we will introduce the twistor realisation for AdS$_{5}$. 
Let us introduce a set of four linearly independent twistors 
\begin{align}
	T^\tmu = \big( T_{\ah}{}^{\tmu} \big)\,, \qquad \tmu = 1\,,2\,,3\,,4\,,
\end{align}
and restrict our attention to those quartets $T^\tmu$ satisfying the constraints 
\bsubeq \label{5d group cond with I}
\begin{align} 
	(T_{\ah}{}^\tmu )^* \O^{\ah\bh} T_{\bh}{}^{\tnu} &= \O^{\tmu\tnu}\,, 
\\
 \det \left(T_{\ah}{}^{\tmu}\right) &= 1\,. \label{det const}
\end{align}
\esubeq
The space of all such quartets will be denoted $\mathfrak{F}$.

On the space $\mathfrak{F}$, we introduce the equivalence relation
\begin{align} \label{usp equiv}
	T_{\ah}{}^{\tmu} &\sim T_{\ah}{}^{\tnu} \l_{\tnu}{}^{\tmu}\,, \qquad \l \in \sUSp(2,2)\,.
\end{align}
We emphasise that the constraints \eqref{5d group cond with I} are invariant under the equivalence transformations \eqref{usp equiv}. 
The group $\sSU(2,2)$ acts on the space $\mathfrak{F}$ as 
\begin{align}
	T_{\ah}{}^{\tmu} \rightarrow g_{\ah}{}^{\bh}T_{\bh}{}^{\tmu}\,, \qquad g \in \sSU(2,2)\,. 
\end{align}
This action is naturally extended to the quotient space $\mathfrak{F} /\sim$. 
The latter proves to be a homogeneous space for the group $\sSU(2,2)$. 
It turns out that 
\begin{align} \label{5d ads twistor def}
	\text{AdS}_{5} = \mathfrak{F}/\sim\,. 
\end{align}

To prove \eqref{5d ads twistor def}, we introduce the bi-twistor 
\begin{align} \label{5d ads bitwistor}
	X_{\ah\bh} := \ell \mathfrak{C}_{\tmu\tnu}T_{\ah}{}^{\tmu}T_{\bh}{}^{\tnu} = -X_{\bh\ah}\,.
\end{align}
The bi-twistor $X_{\ah\bh}$ is invariant under the equivalence relation \eqref{usp equiv}, and as such can be used to parametrise the space \eqref{5d ads twistor def}.
As shown in appendix \ref{spinor conventions}, a one-to-one correspondence 
exists between complex vectors $V^{\ua}$ in $4+2$ dimensions and antisymmetric bi-twistors $V_{\ah\bh} = -V_{\bh\ah}$. It is defined by the rule \eqref{o-to-o}, 
\begin{align}
	V^{\ua} ~\to~ V_{\ah\bh}  = V^{\ua}(\S_{\ua})_{\ah\bh}\,,
\end{align}
where the matrices $\S_{\ua}$ are given in \eqref{bt basis def}.
The six-vector corresponding to \eqref{5d ads bitwistor}, 
\begin{align}
	X^{\ua} = \frac{1}{4}(\tilde{\S}^{\ua})^{\ah\bh}X_{\ah\bh}\,,
\end{align}
is real, because the bi-twistor $X_{\ah\bh}$ satisfies the reality condition, 
\begin{align} \label{5d ads real bt}
	\bar{X}^{\ah\bh} := \O^{\ah\gh}(X^{\dag})_{\gh\dhat}\O^{\dhat\bh} = \frac{1}{2}\ve^{\ah\bh\gh\dhat}X_{\gh\dhat}\,,
\end{align}
see \eqref{index reality}.
In order to prove \eqref{5d ads real bt} we use the identities 
\bsubeq
\begin{align}
	T_{\ah}{}^{\tmu}T_{\bh}{}^{\tnu}T_{\gh}{}^{\tro}T_{\dhat}{}^{\tsi}\ve^{\ah\bh\gh\dhat} &= \ve^{\tmu\tnu\tro\tsi}\,,
	\\
	\mathfrak{C}^{\tmu\tnu}\mathfrak{C}^{\tro\tsi} + \mathfrak{C}^{\tmu \tro}\mathfrak{C}^{\tsi \tnu} + \mathfrak{C}^{\tmu \tsi}\mathfrak{C}^{\tnu \tro} &= - \ve^{\tmu\tnu\tro\tsi}\,,
\end{align}
\esubeq
the first of which is equivalent to $\det (T_{\ah}{}^{\tmu}) = 1$.
Making use of the completeness relation
\begin{align}
	(\tilde{\S}^{\ha})^{\ah\bh}(\tilde{\S}_{\ha})^{\gh\dhat} &= 2\ve^{\ah\bh\gh\dhat}\,,
\end{align}
see \eqref{completeness relations}, we obtain 
\begin{align}
	X^{\ua}X_{\ua} = -\ell^{2}\,. 
\end{align}


\subsection{Bi-twistor realisation} \label{5d bt realisation}

A bi-twistor realisation for (complexified) AdS$_5$ exists in the literature, see \cite{Sinkovics:2004fm, Adamo:2016rtr}.
In this section we will prove its equivalence with the twistor realisation described above. 

Let  $\mathcal{L}$ be the set of bi-twistors, $X_{\ah\bh} = -X_{\bh\ah}$, which obey the following constraints:
\bsubeq \label{5d bt properties}
\begin{align} 
	\bar{X}^{\ah\bh} &= \O^{\ah\gh}(X^{\dag})_{\gh\dhat}\O^{\dhat\bh} = \frac{1}{2}\ve^{\ah\bh\gh\dhat}X_{\gh\dhat}\,, \label{bt tilde bar con}
	\\
	\bar{X}^{\ah\gh}X_{\gh\bh} &= \ell^{2} \d^{\ah}{}_{\bh}\,, \label{delta con}
	\\
	X_{[\ah\bh}X_{\gh\dhat]} &= -\frac{1}{3} \ell^{2}\ve_{\ah\bh\gh\dhat}\
	\quad \Longleftrightarrow \quad \ve^{\ah\bh\gh\dhat}X_{\ah\bh}X_{\gh\dhat} = -8\ell^{2} \, .
	 \label{XX epsilon}
\end{align}
\esubeq
Our aim is to show that the space $\mathcal{L}$ can be identified with $\mathfrak{F} / \sim$. 
First of all, it is not difficult to see that $\mathcal L$ provides an equivalent realisation of AdS$_5$.
Making use of the correspondence \eqref{o-to-o}, we associate a unique six-vector $X^{\ua}$ with each $X_{\ah\bh} \in \cL\,.$
The condition \eqref{bt tilde bar con} means that the six-vector $X^{\ua}$ is real, whereas \eqref{delta con} recreates the AdS condition $X^{2} = -\ell^{2}$. 
We therefore can identify AdS$_5$ with $\cL$. 
It remains to show that every $\hat{X} = (X_{\ah\bh} ) \in \cL $ can be written in the form \eqref{5d ads bitwistor}, 
where the twistors $T^\tmu$ are constrained as in \eqref{5d group cond with I}.

An $\sSU(2,2)$ transformation acts on $\hat{X} = (X_{\ah\bh} ) $ as 
\begin{align}
	\hat{X} &\longrightarrow g \hat{X} g^{\T} = X'^{\ua} (\S_{\ua})\,, \qquad g \in \sSU(2,2)\,,
\end{align}
with $X'^{\ua}$ given by 
\begin{align}
	X'^{\ua} = \L^{\ua}{}_{\ub} X^{\ub}\,, \qquad \L \in \sSO_{0}(4,2)\,.
\end{align}
This transformation preserves the conditions \eqref{5d bt properties}. 
Now we make use of the fact that AdS$_5$ is a homogeneous space of $\sSO_{0}(4,2)$, and has a global coset representative $s(X)$, see, e.g., \cite{Kuzenko:1995aq}. That is, for any $X^{\ua} \in$ AdS$_5$ it holds that 
\begin{align}
	X^{\ua} = s(X)^{\ua}{}_{\ub}X_{(0)}^{\ub}\,, \qquad s(X) \in \sSO_{0}(4,2)\,,
\end{align}
for some base point $X_{(0)}^{\ua}$. 
We can choose the base point to be 
\begin{align}
	X_{(0)}^{\ua} = (0\,,0\,,0\,,0\,,0\,,-\ell)\,,
\end{align}
which corresponds to the bi-twistor 
\begin{align}
	X_{(0)} = X_{(0)}^{\ua}(\S_{\ua}) = \ell C^{-1}\,. 
\end{align}
Then, any bi-twistor corresponding to a point in AdS$_5$ can be written as 
\begin{align} \label{x bt general}
	\hat{X} = \ell g C^{-1} g^{\T}\,, \qquad g \in \sSU(2,2)\,. 
\end{align}
The group element $g$, however, is not uniquely defined. It is defined modulo the equivalence relation
\begin{align}
	g \sim g \l\,, \qquad \l \in \sUSp(2,2)\,. 
\end{align}
Finally, we identify the columns of $g$ with the quartet $T^{\tmu}$.
This completes the proof.

\section{$\N$-extended AdS superspace in five dimensions} \label{susy ads5}

In this section we will extend the above description to the supersymmetric case. 
The goal of this section is thus to develop a (bi-)supertwistor realisation for the $\N$-extended anti-de Sitter superspace in five dimensions, eq. \eqref{5d n ext ads def}.
For this purpose, we will make use of supertwistors transforming in the fundamental representation of the AdS supergroup, $\sSU(2,2|\N)$, 
and a particular parametrisation that is suited to the case for AdS$^{5|8\N}$. 
Technical details regarding $\sSU(2,2|\N)$ and supertwistors are located in appendix \ref{su(2,2|n) sec}.

\subsection{Supertwistor realisation}

In complete analogy with the bosonic construction described in section \ref{boson ads5}, consider four {\it even} supertwistors with linearly independent bodies\footnote{Here $A$ is a supertwistor index, see appendix \ref{su(2,2|n) sec}, and $\e_A$ is defined as in \eqref{parity convention}.}
\begin{align}
\bm{T}^{\tmu}	=(\bm{T}_{A}{}^{\tmu})\,, \qquad \e (\bm{T}_{A}{}^{\tmu}) = \e_A\, , \qquad
\tmu = 1\,,2\,,3\,,4\,.
\end{align}
We restrict our attention to those quartets satisfying the constraint 
\begin{align} \label{5d susy const}
	\bar{\bm{T}}{}^{\tmu A} \bm{T}_{A}{}^{\tnu} = \O^{\tmu\tnu}\,, \qquad 
	\bar{\bm{T}}{}^{\tmu A} :=  (\bm{T}_{B}{}^\tmu)^* \,\bm{\O}^{BA}\,.
\end{align}
The space of such quartets will be denoted
$\mathfrak{F}_{\N}$.
On $\mathfrak{F}_{\N}$, we introduce the equivalence relation
\bsubeq \label{5d susy equiv}
\begin{align} 
	\bm{T}_{A}{}^{\tmu} &\sim \bm{T}_{A}{}^{\tnu} \l_{\tnu}{}^{\tmu} \,, \qquad \l \in \sUSp(2,2)\,, \label{susy usp}
	\\
	\bm{T}_{A}{}^{\tmu} &\sim \re^{\ri\vf}\bm{T}_{A}{}^{\tmu} \,, \qquad \vf \in \mathbb{R}\,. \label{susy u1}
\end{align}
\esubeq
We emphasise that the condition \eqref{5d susy const} is invariant under the equivalence transformations \eqref{5d susy equiv}. 
For $\N > 0$, the constraint \eqref{det const} is substituted with 
\eqref{susy u1}.
We note that there is no $\sU(\N)$ factor present in the relations \eqref{5d susy equiv}, despite the definition of AdS$^{5|8\N}$ as \eqref{5d n ext ads def}. 
This is due to the fact that, under an arbitrary $\sSU(2,2|\N)$ transformation, only a $\sU(1)$ subgroup of $\sU(\N)$ acts on the body of the even supertwistors $\bm{T}_{A}{}^{\tmu}$. Indeed, as we will see below, the equivalence relations \eqref{5d susy equiv} lead to the correct definition of the five dimensional AdS superspace. 

The supergroup $\sSU(2,2|\N)$ acts on $\mathfrak{F}_{\N}$ as 
\begin{align}
	\bm{T}_{A}{}^{\tmu} \rightarrow g_{A}{}^{B}\bm{T}_{B}{}^{\tmu}\,, \qquad g \in \sSU(2,2|\N)\,. 
\end{align}
This action is naturally extended to the quotient space $\mathfrak{F}_{\N} / \sim$, which proves to be a homogeneous space of $\sSU(2,2|\N)$. 
To show this, we note that the quartets $\bm{T}_{A}{}^{\tmu}$ can be embedded into a group element 
\bea \label{embed t in g}
g = ( \bm{T}_{A}{}^{\tmu} , \bm{\X}_{A}{}^{\ti})  \in \sSU(2,2|\N)~, \qquad  \ti =1,\dots, \cN\, , 
\eea
where $\bm{\X}_{A}{}^{\ti}$ are $\N$ linearly independent odd supertwistors, and $\sSU(2,2|\N)$ acts transitively on itself from the left. 
It turns out that 
\begin{align} \label{5d ads susy def}
	\mathfrak{F}_{\N} / \sim~ =~ \text{AdS}^{5|8\N}\,. 
\end{align}

Let us discuss some implications of eq. \eqref{embed t in g}. 
Specifically, the embedding of the supertwistors $\bm{T}_{A}{}^{\tmu}$ into the group element $g$ is essentially the inverse of the process advocated for $\mathbb{M}^{4|4\N}$ and $\mathbb{M}^{3|2\N}$ in \cite{Park:1997bq,Park:1998nra,Park:1999pd,Park:1999cw}.
In these references, one begins with the coset representative, and then `reduces' it to consider only those columns necessary to parametrise it. 
Here, our starting point is precisely the `reduced' coset representative, which contains the information about the superspace \eqref{5d n ext ads def}. 
In this sense, our formalism can be viewed as the AdS$^{5|8\N}$ extension of that developed in \cite{Park:1997bq,Park:1998nra,Park:1999pd,Park:1999cw}.

\subsection{Stabiliser}

To prove \eqref{5d ads susy def}, it suffices to choose a base point and determine its stabiliser. 
As a base point, $\bm{T}^{(0)}$, we choose 
\begin{align} \label{5d origin}
	\bm{T}^{(0)} = \left(\begin{array}{c}
		\id_{4} 
		\\
		\hline 
		\hline 
		0
	\end{array}\right)\,.
\end{align}
This is the simplest quartet satisfying the constraint \eqref{5d susy const}.
By definition, the stabiliser $H$ of $\bm{T}^{(0)}$ consists of those elements $h \in \sSU(2,2|\N)$ that satisfy the condition 
\begin{align}
	h \bm{T}^{(0)} = \left(\begin{array}{c}
		\re^{\ri\vf}\l 
		\\
		\hline 
		\hline 
		0
	\end{array}\right)\,, \qquad \vf \in \mathbb{R}\,, \quad \l \in \sUSp(2,2)\,. 
\end{align}
That is, it consists of those group elements that map $\bm{T}^{(0)}$ to an equivalent point with respect to \eqref{5d susy equiv}. 
These conditions imply that the group elements $h$ take the form
\bsubeq
\begin{align} \label{little group element}
	h &=
	 \left(\begin{array}{c||c}
		~\re^{\ri\vf} \id_{4}~ & 0
		\\
		\hline \hline 
		0 & ~\re^{\frac{4}{\N}\ri\vf}\id_{\N}~
	\end{array}\right) 
\left(
\begin{array}{c||c}
	~\l~ & 0 
	\\
	\hline \hline 
	0 & ~\id_{\N}~
\end{array}
\right)
\left(
\begin{array}{c||c}
	~\id_{4}~ & 0 
	\\
	\hline \hline 
	0 & ~U~
\end{array}
\right)\,,
\end{align}
with 
\begin{align}
	\vf \in \mathbb{R}\,, \qquad \l & \in \sUSp(2,2)\,, \qquad U \in \sSU(\N)\,. 
\end{align}
\esubeq
We therefore have $H$ is isomoprhic to
\begin{align}
	\sUSp(2,2)\times\sU(\N)\,. 
\end{align}

\subsection{Bi-supertwistor realisation}

In this subsection we will describe a bi-supertwistor realisation of the AdS superspace described above. 
Given a point in $\mathfrak{F}_{\N}$, we associate with it the following graded antisymmetric supermatrix $\bm{X} = (\bm{X}_{AB})$,
\begin{align} \label{sbt def}
	\bm{X}_{AB} := \ell \mathfrak{C}_{\tmu \tnu} \bm{T}_{A}{}^{\tmu} \bm{T}_{B}{}^{\tnu} = - (-1)^{\e_A \e_B}\bm{X}_{BA}\,. 
\end{align}
These supermatrices are invariant under arbitrary equivalence transformations of the form \eqref{susy usp}. 
In order to parametrise the space $\mathfrak{F}_{\N} / \sim$, they must be defined modulo the equivalence relation
\begin{align} \label{sbt u1}
	\bm{X}_{AB} \sim \re^{\ri\vf}\bm{X}_{AB}\,, \qquad \vf \in \mathbb{R}\,,
\end{align}
due to eq. \eqref{susy u1}.
Using the dual supertwistors we can introduce the following supermatrix 
\begin{align} \label{dual sbt def}
	\bm{\bar{X}} := \bm{\O}\bm{X}^{\dag}\bm{\O} = (\bm{\bar{X}}{}^{AB}) \,, \qquad \bm{\bar{X}}{}^{AB} = -\ell \mathfrak{C}_{\tmu \tnu}  \bm{\bar{T}}{}^{\tmu A}\bm{\bar{T}}{}^{\tnu B}\,. 
\end{align}
The important properties of the supermatrices $\bm{X}$ and $\bm{\bar{X}}$ are
\bsubeq \label{s5d bt properties}
\begin{align}
	\bm{X}_{[AB}\bm{X}_{CD}\bm{X}_{E\}F} &= 0 \implies \bm{X}_{[AB}\bm{X}_{CD}\bm{X}_{EF\}} = 0\,, \label{graded antisym sbt}
	\\
	\bm{\bar{X}}{}^{AB}\bm{X}_{BA} &= 4\ell^{2}\,, \label{sbt l squared}
	\\
	(-1)^{\e_{C}}\bm{X}_{AC}\bar{\bm{X}}{}^{CD}\bm{X}_{DB} &= \ell^{2}\bm{X}_{AB}\,, \label{triple product sbt}
\end{align}
\esubeq
where $[\ldots\}$ denotes the graded antisymmetrisation of indices.
Property \eqref{graded antisym sbt} follows from the fact that $\bm{X}_{AB}$ is constructed from four supertwistors. 
An important implication of this property is that the body of the bosonic block $\bm{X}_{ij}$ defined by
\begin{align}
	\bm{X} = \left(\begin{array}{c||c}
		\bm{X}_{\ah\bh} & \bm{X}_{\ah j}
		\\
		\hline \hline
		\bm{X}_{i\bh} & \bm{X}_{ij}
	\end{array}\right)\,,
\end{align}
vanishes. Indeed, let us consider the case 
\begin{align}
	\bm{X}_{(ij}\bm{X}_{kl}\bm{X}_{mn)} = 0\,. 
\end{align}
Since $\bm{X}_{ij}$ is symmetric, the only non-zero solution to the above is a symmetric, bodiless matrix $\bm{X}_{ij}$.

Now, we would like to describe the superspace AdS$^{5|8\N}$ solely in terms of bi-supertwistors, without any reference to the supertwistor realisation above. 
Let us consider the space of graded antisymmetric supermatrices $\bm{X}_{AB}$, 
\begin{subequations}
\begin{align}
	\bm{X}_{AB} &= -(-1)^{\e_A\e_B}\bm{X}_{BA}\, , 
	\\
	\e(\bm{X}_{AB}) &= \e_{A}+ \e_{B}\,. 
\end{align}
\end{subequations}
In this space, we consider the surface $\mathfrak{L}$ consisting of those supermatrices which obey the constraints \eqref{s5d bt properties}.
We then introduce the quotient space $\mathfrak{L}/\sim$, where
the equivalence relation is given by \eqref{sbt u1}. 
Our goal is to show that $\mathfrak{L}/\sim$ can be identified with AdS$^{5|8\N}$. 

We begin by considering all the elements of $\mathfrak{L}$ with vanishing soul.
These elements take the form 
\begin{align} \label{5d soulless sbt}
	\bm{X}| := X = \left(\begin{array}{c||c}
		X_{\ah\bh} & 0
		\\
		\hline\hline
		0 & ~0~
	\end{array}\right)\,, \qquad X_{\ah\bh} = -X_{\bh\ah} \in {\mathbb C}\,, 
\end{align}
where $\bm{X}|$ denotes the soulless component of $\bm{X}\,.$
The space of such elements will be denoted $\mathfrak{L}|$. 

It turns out that $\mathfrak{L}|/\sim $ is a homogeneous space of the subgroup of $\sSU(2,2|\N)$ spanned by matrices of the form
\begin{align} \label{reduced supergroup element}
	v = \left(\begin{array}{c||c}
	\re^{\ri\vf} \S & ~0~ \\
	\hline \hline 
	0 & \re^{\frac{4\ri\vf}{\N}} U
\end{array}\right)\,, \qquad \vf \in \mathbb{R}\,, \quad \S \in \sSU(2,2)\,, \quad  U \in \sSU(\N)\,. 
\end{align} 
The proof is analogous to that described in subsection \ref{5d bt realisation} and is described below. 
As a base point we can choose
\begin{align} \label{pref Y}
	X_{(0)} = \left(\begin{array}{c||c}
		\ell C^{-1} & ~0~ \\
		\hline \hline 
		0 & 0
	\end{array}\right)\,. 
\end{align}
Then, any point in $\mathfrak{L}|/\sim $ can be reached by a group transformation of the form 
\begin{align} \label{l y l}
	X = v X_{(0)} v^{\T}\,,
\end{align}
with $v$ given by \eqref{reduced supergroup element}. 
As in the bosonic case, the transformation $v$ is not uniquely defined. 
The stabiliser of the bi-supertwistor $X_{(0)}$ consists precisely of those group elements that take the form \eqref{little group element}. 
We therefore have that $\mathfrak{L}| / \sim$ coincides with the body of the coset superspace $\sSU(2,2|\N) / \left(\sUSp(2,2) \times \sU(\N)\right)\,.$

To prove \eqref{l y l}, we note that  $X_{\ah\bh}$ in \eqref{5d soulless sbt} can be written as
\begin{align}
	X_{\ah\bh} = X^{\ua}(\S_{\ua})_{\ah\bh}\,. 
\end{align}
The matrix $\hat{X} = (X_{\ah\bh})$ is non-singular. 
Because of property \eqref{triple product sbt}, we have 
\begin{align} \label{ybary}
	\hat{X}^{\dag} \O \hat{X} = \ell^{2}\O \implies \bar{X}^{\ah\gh}X_{\gh\bh} = \ell^{2}\d^{\ah}{}_{\bh}\,. 
\end{align}
This property has two implications: 
\begin{enumerate}
\item
the mutually conjugate six-vectors $\bar{X}^{\ua}$ and $X^{\ua}$ satisfy the following
\bsubeq
\begin{align} 
	\bar{X}^{\ua}X_{\ua} = -\ell^{2}\,,
	\label{i}
\end{align}
\item
$\bar{X}^{\ua}$ and $X^{\ua}$  are linearly dependent, which follows from
\begin{align}
	\bar{X}^{\ua}X^{\ub}(\tilde{\S}_{\ua\ub})^{\ah}{}_{\bh} = 0\,,
	\label{ii}
\end{align}
\esubeq
\end{enumerate}
where we have used \eqref{i}. 
Property \eqref{i} can be seen by taking the trace of \eqref{ybary}.
To prove property \eqref{ii} it is instructive to write out the above expression explicitly,
\begin{align}
	\bar{X}^{\ua}X^{\ub}(\tilde{\S}_{\ua\ub})^{\ah}{}_{\bh} 
	&= 
	-\frac{1}{4}\bar{X}^{\ua}X^{\ub}\left(\tilde{\S}_{\ua}\S_{\ub} - \tilde{\S}_{\ub}\S_{\ua}\right)^{\ah}{}_{\bh}\,,
\end{align}
and then make use of \eqref{s stilde clifford}. 
Then we have 
\begin{align}
	X^{\ua} = \re^{\ri\vf}Y^{\ua}\,, \qquad \vf \in \mathbb{R}\,, \quad Y^{\ua} \in \text{AdS}_{5}\,. 
\end{align}
Now we can repeat the argument from the non-supersymmetric case.
Specifically, we have 
\begin{align}
	X^{\ua} = \re^{\ri\vf}s(Y)^{\ua}{}_{\ub}Y_{(0)}^{\ub}\,, \qquad s(Y) \in \sSO_{0}(4,2)\,,
\end{align}
for a base point $Y_{(0)}^{\ua}$. 
Expression \eqref{l y l} follows. 

We now note that both the full supergroup, $\sSU(2,2|\N)$, and a generic bi-supertwistor $\bm{X} \in \mathfrak{L} / \sim$ have $8\N$ real Grassmann odd degrees of freedom. 
We conclude that, by replacing $v$ in \eqref{l y l} with some $g \in \sSU(2,2|\N)$, we can reach an arbitrary $\bm{X} \in \mathfrak{L} / \sim$ by a transformation acting on $X_{(0)}$. 
That is, the space $\mathfrak{L} / \sim$ is itself a homogeneous space of $\sSU(2,2|\N)$, which shows the equivalence of the bi-supertwistor and supertwistor realisations of AdS$^{5|8\N}$. 

\subsection{Projection-like operator}

The specific properties of the (bi-)supertwistors described above allow us to introduce an additional structure. 
Let us introduce the supermatrix
\begin{align}
\bm{Y} = (\bm{Y}_{A}{}^{B})\, , \qquad
	\bm{Y}_{A}{}^{B} := \bm{T}_{A}{}^{\tmu}\O_{\tmu\tnu}\bar{\bm{T}}{}^{\tnu B}~,
\end{align}
with the following transformation law under the AdS supergroup 
\begin{align}
	\bm{Y} \rightarrow g \bm{Y} g^{-1}\,, \qquad 
	g \in \sSU(2,2|\N)\,. 
\end{align}
This structure is unique to the supersymmetric case in the sense that for $\N=0$ it coincides with the unit matrix, 
\bea
\cN=0 \,: \qquad \bm{Y} = {\mathbbm 1}_4\, .
\eea
It is invariant under the equivalence relation \eqref{5d susy equiv}, and satisfies the following properties: 
\bsubeq
\begin{align}
	\bm{Y}_{A}{}^{C}\bm{Y}_{C}{}^{B} &= \bm{Y}_{A}{}^{B}\,, 
	\\
	\bm{Y}_{A}{}^{B}\bm{T}_{B}{}^{\tmu} &= \bm{T}_{A}{}^{\tmu}
	\\
	\bar{\bm{T}}{}^{\tmu B} \bm{Y}_{B}{}^{A}&= \bar{\bm{T}}{}^{\tmu A} \,,
	\\
	\bm{Y}_{A}{}^{C}\bm{X}_{CB} &= \bm{X}_{AB}\,, 
	\\
	\bar{\bm{X}}^{AC}	\bm{Y}_{C}{}^{B} &= \bar{\bm{X}}^{AB}\,, 
	\\
	(-1)^{\e_{A}}\bm{Y}_{A}{}^{A} &= 4\,. 
\end{align}
\esubeq
The operator $\bm{Y}_{A}{}^{B}$ can be related to the bi-supertwistors $\bm{X}_{AB}$ as 
\begin{align}
	\bm{Y}_{A}{}^{B} = \frac{(-1)^{\e_{C}}}{\ell^{2}}\bm{X}_{AC}\bar{\bm{X}}{}^{CB}\,. 
\end{align}

A similar structure is available in the case of $3$D $(p,q)$ AdS superspaces, see \cite{KT}.

\subsection{$\sSU(2,2|\N)$-invariant two-point function}

An important application of the formalism described above is the construction of manifestly invariant two-point functions. 
Indeed, let $\cF= (\bm{T}^{\tmu})$ and $\cF' = (\bm{T}'^{\tmu})$ be arbitrary points of $\mathfrak{F}_{\N}$. 
The following two-point function 
\begin{align} \label{5d tp function}
	\o(\cF,\cF') := -\frac{1}{4}\ell^{2}\mathfrak{C}_{\tmu\tnu}\mathfrak{C}_{\tro\tsi}\braket{\bm{T}^{\tmu}}{\bm{T}'^{\tro}}\braket{\bm{T}^{\tnu}}{\bm{T}'^{\tsi}}
\end{align}
is $\sSU(2,2|\N)$-invariant. It is also invariant under the equivalence transformations \eqref{5d susy equiv}, and is therefore defined on the quotient space \eqref{5d ads susy def}.

The two-point function \eqref{5d tp function} can be expressed in terms of the bi-supertwistors $\bm{X}$ and $\bar{\bm{X}}$. 
Let us denote the bi-supertwistors corresponding to $\cF$ and $\cF'$ as $\bm{X}$ and $\bm{X}'$. Then the two-point function takes the form 
\begin{align} \label{5d sbt tp function}
	\o(\bm{X},\bm{X}') = - \frac{1}{4}\bar{\bm{X}}^{AB}\bm{X}'_{BA}\,. 
\end{align}
Making use of the operator $\bm{Y}$ introduced above, we can also consider the following $n$-point function 
\begin{align} \label{y n point}
	\x(\bm{Y}_{1}\,, \ldots\,, \bm{Y}_{n}) := \frac{1}{4}\text{Str}\left(\bm{Y}_{1}\bm{Y}_{2}\ldots\bm{Y}_{n}\right)\,. 
\end{align}
In the non-supersymmetric case, $\N = 0$, \eqref{5d tp function} and \eqref{5d sbt tp function} coincide with the AdS$_5$ two-point function $X^{\ua}X'_{\ua}$, and \eqref{y n point} is constant, $\x(Y_{1}\,, \ldots Y_{n}) = 1\,.$ 

In chapter \ref{ch3}, we used a similar construction to describe new models for superparticles propagating in AdS$^{4|4\N}$. 
In the present context, such models take the form
\begin{align} \label{5d superparticle model}
	S = \frac{1}{2}\int \text{d}\t \mathfrak{e}^{-1}\left\{\o(\dot{\bm{X}},\dot{\bm{X}}) + \a \,\x(\dot{\bm{Y}},\dot{\bm{Y}}) - 
	(\mathfrak{e}m)^{2} \right\}\,, 
\end{align}
for a real parameter $\a$. 
It can be shown that the $\a$-term vanishes when the Grassmann variables are switched off, and the resulting model coincides with the bosonic one. 
This means that the $\a$ term generates purely fermionic contributions. This will be elaborated on later, in section \ref{5d superparticle section}. 

\section{Coset representative} \label{5d coset rep section}

This section is devoted to the derivation of a coset representative and its associated local coordinate system. 
As illustrated in chapter \ref{ch3}, such a coset representative plays a key role in the study of the geometry of a homogeneous space. 
A global coset representative exists for $d$-dimensional AdS in the context of the standard embedding formalism, see \cite{Kuzenko:1995aq}. 
For the embedding formalism developed in this chapter, suitable for AdS superspace, an exponential coset parametrisation is used and we therefore find a local coset representative (and local coordinate system).
These are detailed below. 

\subsection{AdS space ($\N = 0$)}

The Lie algebras $\mathfrak{su}(2,2)$ and $\mathfrak{usp}(2,2)$ are given by expressions \eqref{su(2,2) generic algebra element} and  \eqref{so(4,1) generic algebra element}, respectively.
Let $\mathfrak{K}$ be the complement of $\mathfrak{usp}(2,2)$ in $\mathfrak{su}(2,2)$.
We then have
\begin{align} \label{su22 decomp}
	\mathfrak{su}(2,2) = \mathfrak{usp}(2,2) \oplus \mathfrak{K}\,.
\end{align}
The elements of $\mathfrak{K}$ take the form 
\begin{align} \label{compel}
	\mathfrak{t} = \left(\begin{array}{c|c}
		- \frac{1}{2}\o^{45}\d_{\a}{}^{\b} & - \frac{\ri}{2}(\o^{a 5})(\s_{a})_{\a\bd}
		\\
		\hline 
		- \frac{\ri}{2}(\o^{a5})(\tilde{\s}^{a})^{\ad\b} & \frac{1}{2}\o^{45}\d^{\ad}{}_{\bd}
	\end{array}\right) \in \mathfrak{K}\,, \qquad \o^{\ha 5} \in \mathbb{R}\,.
\end{align}
Then, for a group element $g \in \sSU(2,2)$ in a neighbourhood of $\id_{4}$, it holds that 
\begin{align}
	g = \re^{\mathfrak{t} + \mathfrak{h}}\,, \qquad \mathfrak{h} \in \mathfrak{usp}(2,2)\,. 
\end{align}
We seek to (locally) factorise $g$ in the following way 
\begin{align}
	g = \cS(y) h\,, \qquad \cS(y) \in \sSU(2,2)\, ,
	\qquad h \in \sUSp(2,2)\,,
\end{align}
where $\cS(y) $ is the coset representative, and $y$ are local coordinates on AdS$_5$. 
We can make the following ansatz for $\cS(y)$:
\begin{align}\label{5d bosonic coset rep}
	\cS(y) = \left(
	\begin{array}{c|c}
		~\x\r \id_{2}~ & -\ri x \\
		\hline
		-\ri \tilde{x} & \x\r^{-1} \id_{2}
	\end{array}
	\right) = \re^{\mathfrak t}
	\,, 
\end{align}
with 
\bsubeq
\begin{align}
	x &:= x^{a}(\s_{a}) = x^{\dag}\,, \qquad \tilde{x}:= x^{a}(\tilde{\s}_{a}) = \tilde{x}^{\dag}\,,
	\\
	\x &:= (1+x^{2})^{\frac{1}{2}}\,, \qquad \,~~x^{2} = x^{a}x^{b}\eta_{ab}\,,
	\\
	\quad \qquad \r &= \bar{\r}\,.  
\end{align}
\esubeq
It follows that the local coordinates $x^{a}$ and $\r$ are real, and the coordinate chart is specified by 
\begin{align}
	x^{2} > -1\,, \qquad \r \neq 0\,. 
\end{align}
The bi-twistor corresponding to \eqref{5d bosonic coset rep} is given by 
\bsubeq
\begin{align} \label{cr bt}
	(X_{\ah\bh}) &= \ell \left( \mathfrak{C}_{\tmu\tnu}\cS(y)_{\ah}{}^{\tmu}\cS(y)_{\bh}{}^{\tnu} \right) = 
	\ell \left(
	\begin{array}{c|c}
		\big(\x^{2}\r^{2} + x^{2}\big)\ve_{\a\b} & - \ri\x(\r + \r^{-1})x_{\a}{}^{\bd}
		\\
		\hline 
		\ri\x(\r+\r^{-1})x^{\ad}{}_{\b}	& - \big(\x^{2}\r^{-2} + x^{2}\big)\ve^{\ad\bd} 
	\end{array}
	\right)\,,
\end{align}
where
\begin{align}
	\cS(y)_{\ah}{}^{\tmu} &:= \cS(y)_{\ah}{}^{\bh}T^{(0)}_{\bh}{}^{\tmu} \,, \qquad 
	T^{(0)} = \id_{4}\,. 
\end{align}
\esubeq
Associated with $X_{\ah\bh}$ is the following point in AdS$_5$
\begin{align}
	X^{\ua} = \left(
	\ell \x (\r + \r^{-1}) x^{a} \,, - \frac{1}{2}\ell \x^{2} (\r^{2} - \r^{-2})\,,
	-\ell\big(x^{2} + \frac{1}{2}\x^{2}(\r^{2} + \r^{-2})\big)
	\right)\,. 
\end{align}
Instead of the coset parametrisation used above, we could choose a coset representative corresponding to Poincar\'e coordinates in AdS$_5$. 
This coset representative, used in \cite{CRZ, Kuzenko:2007aj}, is given by
\begin{align}
	\cS(y) = \left(
	\begin{array}{c|c}
		\r^{\frac{1}{2}}\id_{2} & ~0~ 
		\\
		\hline 
		-\ri\r^{\frac{1}{2}}\tilde{x} & \r^{-\frac{1}{2}}\id_{2}
	\end{array}
	\right)\,.
\end{align}
Its corresponding bi-twistor is 
\begin{align} \label{pp bt}
	(X_{\ah\bh}) = \frac{1}{z}\left(
	\begin{array}{c|c}
		~\ve_{\a\b}~ & -\ri x_{\a}{}^{\bd}
		\\
		\hline
		\ri x^{\ad}{}_{\b} & -\big((\ell z)^{2} + x^{2}\big)\ve^{\ad\bd}
	\end{array}
	\right)\,, \qquad z:= (\ell \r)^{-1}\,. 
\end{align}
The real coordinates $z>0$ and $x^{a}$ parametrise AdS$_5$ in the Poincar\'e patch.
They are related to the embedding coordinates $X^{\ua}$ as follows
\begin{align}
	X^{\ua} = \frac{1}{z}\left(
	x^{a} \,, -\frac{1}{2}(1 - x^{2} - (\ell z)^{2})\,,
	-\frac{1}{2}(1 + x^{2} + (\ell z)^{2} )
	\right)\,. 
\end{align}

\subsection{$\N \neq 0$} \label{5d n>0 cr section}

The analysis of the previous section can be extended to the supersymmetric case in similar fashion. 
Coset representatives for the $\N = 1$ superspaces AdS$^{5|8} \times S^{1}$ and AdS$^{5|8}$ are given in \cite{Kuzenko:2001ag,Kuzenko:2007aj}. 
Building on the approach of these references, again denoting the local coordinates by $y$, we choose the coset representative in a supersymmetric generalisation of Poincar\'e coordinates as
\begin{align} \label{pp coset rep}
	\cS(y) &= g(\bm{z})\cdot g_{S} \cdot g_{D}
	\\ 
	\notag
	&= 
	\left(
	\begin{array}{c|c||c}
		\id_{2} & ~0~ & ~0~ \\
		\hline
		-\ri \tilde{x}_{+} & \id_{2} & 2\bar{\q} \\
		\hline \hline
		2 \q & 0 & \id_{\N}
	\end{array}
	\right)
	\left(
	\begin{array}{c|c||c}
		\id_{2} & 2\eta \bar{\eta} & 2 \eta
		\\ \hline 
		0 & \id_{2} & 0 
		\\
		\hline \hline
		~0~ & 2\bar{\eta} & \id_{\N}
	\end{array}
	\right)
	\left(
	\begin{array}{c|c||c}
		\r^{\frac{1}{2}}\id_{2}& 0 & 0 \\
		\hline 
		0 & \r^{-\frac{1}{2}}\id_{2} & 0 \\
		\hline \hline 
		0 & 0 & \id_{\N}
	\end{array}
	\right)
	\\
	\notag 
	&= 
	\left(
	\begin{array}{c|c||c}
		\r^{\frac{1}{2}}\d_{\a}{}^{\b} 
		& 
		2 \r^{-\frac{1}{2}}\eta_{\a}{}^{k}\bar{\eta}_{k \bd} 
		& 
		2 \eta_{\a}{}^{j}
		\\
		\hline 
		~-\ri \r^{\frac{1}{2}}x_{+}^{\ad\b} ~
		&
		\r^{-\frac{1}{2}}\big(\d^{\ad}{}_{\bd} - 2\ri x_{+}^{\ad\g}\eta_{\g}{}^{k}\bar{\eta}_{k \bd} + 4\bar{\q}^{\ad k}\bar{\eta}_{k \bd}\big)
		&
		2 \big( \bar{\q}^{\ad j} - \ri x_{+}^{\ad\g}\eta_{\g}{}^{j} \big)
		\\
		\hline \hline
		2\r^{\frac{1}{2}}\q_{i}{}^{\b} 
		& 
		2\r^{-\frac{1}{2}}\big(\bar{\eta}_{i\bd} + 2 \q_{i}{}^{\g}\eta_{\g}{}^{k}\bar{\eta}_{k\bd} \big) 
		&
		\d_{i}{}^{j} + 4\q_{i}{}^{\g}\eta_{\g}{}^{j} 
	\end{array}
	\right)\,,
\end{align}
where $x_{\pm}^{a} = x^{a} \pm \ri\q_{i}\s^{a}\bar{\q}^{i}$.
The coset representative $g(\bm{z})$ corresponds to that of $4$D $\N$-extended Minkowski superspace, see eq. \eqref{smink global crep}. 
This coordinate system will be referred to as Poincar\'e-like. 

The action of $\cS(y)$ on the base point $\bm{T}^{(0)}$, see \eqref{5d origin}, yields the general form of the quartet $\bm{T}_{A}{}^{\tmu}$ in Poincar\'e-like coordinates 
\begin{align}
	\left(\bm{T}_{A}{}^{\tmu}\right) = 
	\left(
	\begin{array}{c|c}
		~\r^{\frac{1}{2}}\d_{\a}{}^{\b} ~
		& 
		2 \r^{-\frac{1}{2}}\eta_{\a}{}^{k}\bar{\eta}_{k \bd} 
		\\
		\hline 
		~-\ri \r^{\frac{1}{2}}x_{+}^{\ad\b} ~
		&
		\r^{-\frac{1}{2}}\big(\d^{\ad}{}_{\bd} - 2\ri x_{+}^{\ad\g}\eta_{\g}{}^{k}\bar{\eta}_{k \bd} + 4\bar{\q}^{\ad k}\bar{\eta}_{k \bd}\big)
		\\
		\hline \hline
		2\r^{\frac{1}{2}}\q_{i}{}^{\b} 
		& 
		2\r^{-\frac{1}{2}}\big(\bar{\eta}_{i\bd} + 2 \q_{i}{}^{\g}\eta_{\g}{}^{k}\bar{\eta}_{k\bd} \big) 
	\end{array}
	\right)\,.
\end{align}
Furthermore, the bi-supertwistor $\bm{X}_{AB}$ is 
\begin{align} \label{pp bst}
	\bm{X} = \left(\begin{array}{c|c||c}
		\bm{X}_{\ab} & \bm{X}_{\a}{}^{\bd} & \bm{X}_{\a j} \\
		\hline
		\bm{X}^{\ad}{}_{\b} & \bm{X}^{\ad\bd} & \bm{X}^{\ad}{}_{j} \\
		\hline \hline
		\bm{X}_{i\b} & \bm{X}_{i}{}^{\bd} & \bm{X}_{ij}
	\end{array}\right)\,,
\end{align}
where the components are given by 
\bsubeq
\begin{align}
	\bm{X}_{\ab} &= \frac{1}{z} \big(\e_{\a\b} + 4(\ell z)^{2}\eta_{\a}{}^{k}\eta_{\b}{}^{l}\bar{\eta}_{k \gd}\bar{\eta}_{l}{}^{\gd} \big)\,, 
	\\
	\bm{X}_{\a}{}^{\bd} &= \frac{1}{z} \big( 
	-\ri  x_{+}{}_{\a}{}^{\bd}
	+2(\ell z)^{2}\eta_{\a}{}^{k}\bar{\eta}_{k}{}^{\gd} 
	\big( \d^{\bd}{}_{\gd} - 2\ri x_{+}^{\bd \g}\eta_{\g}{}^{l}\bar{\eta}_{l \gd} + 4\bar{\q}^{\bd l}\bar{\eta}_{l\gd}
	\big)
	\big)
	\,,
	\\
	\bm{X}_{\a j} &= 
	\frac{1}{z} \big(2 \q_{\a j} - 4(\ell z)^{2}\eta_{\a}{}^{k}\bar{\eta}_{k\gd}\big(\bar{\eta}_{j}{}^{\gd} + 2\q_{j}{}^{\g}\eta_{\g}{}^{l}\bar{\eta}_{l}{}^{\gd}
	\big) \big) \,,
	\\
	\bm{X}^{\ad\bd} &= \frac{1}{z}\big(
	- (\ell z)^{2}\big(\d^{\ad}{}_{\gd} - 2\ri x_{+}^{\ad\g}\eta_{\g}{}^{k}\bar{\eta}_{\k \gd} + 4\bar{\q}^{\ad k}\bar{\eta}_{k \gd} \big)
	\big(\ve^{\gd\bd} - 2\ri x_{+}^{\bd \s} \eta_{\s}{}^{l}\bar{\eta}_{l}{}^{\gd} + 4\bar{\q}^{\bd l}\bar{\eta}_{l}{}^{\gd}
	\big)
	\\
	\notag 
	& \quad 
	- x_{+}^{2}\ve^{\ad\bd} \big)\,,
	\\
	\bm{X}^{\ad}{}_{j} &= \frac{1}{z} \big(
	-2(\ell z)^{2}
	\big(
	\d^{\ad}{}_{\gd} - 2\ri x_{+}^{\ad\g}\eta_{\g}{}^{k}\bar{\eta}_{k\gd} + 4\bar{\q}^{\ad k}\bar{\eta}_{k \gd}
	\big)
	\big(
	\bar{\eta}_{j}{}^{\gd} + 2\q_{j}{}^{\g}\eta_{\g}{}^{l}\bar{\eta}_{l}{}^{\gd}
	\big)
	\\
	\notag 
	& \quad - 2\ri x_{+}^{\ad\g}\q_{\g j} \big)\,,
	\\
	\bm{X}_{ij} &= \frac{1}{z} \big( 
	4 \q_{i}{}^{\a}\q_{\a j} 
	- 4(\ell z)^{2}\big(
	\bar{\eta}_{\i \ad} + 2\q_{i}{}^{\g}\eta_{\g}{}^{k}\bar{\eta}_{k \ad}
	\big)
	\big(
	\bar{\eta}_{j}{}^{\ad} + 2\q_{j}{}^{\d}\eta_{\d}{}^{l}\bar{\eta}_{l}{}^{\ad}
	\big)
	\big)\,.
\end{align}
\esubeq
The coordinates $z$ and $\r$ are related as in \eqref{pp bt}. 

Using the above formulation we can readily describe the conformal boundary of AdS superspace.
To do so, we should switch from $\bm{X}_{AB} $ to $ z\bm{X}_{AB}$ and take the limit $z \rightarrow 0$.  
In this limit, the Grassmann variables $\eta$ disappear and the resulting bi-supertwistor is given by 
\begin{align}
	\left(\bm{X}^{\text{\tiny{boundary}}}_{AB}\right) = 
	\left(
	\begin{array}{c|c||c}
		\ve_{\ab} 
		& 
		-\ri  x_{+}{}_{\a}{}^{\bd} 
		& 
		2 \q_{\a j}
		\\
		\hline 
		\ri  x_{+}{}^{\ad}{}_{\b}
		&
		- x_{+}^{2} \ve^{\ad\bd} 
		&
		-2 \ri x_{+}{}^{\ad \g}\q_{\g j} 
		\\
		\hline \hline 
		-2 \q_{i \b}
		&
		2 \ri \q_{i \g} x_{+}{}^{\g \bd} 
		&
		4  \q_{i}{}^{\a}\q_{\a j}
	\end{array}
	\right)\,. 
\end{align}
This expression coincides with eq. $(3.17)$ in \cite{K-compactified12} and eq. \eqref{bm mink bst}.
Making use of the results of \cite{K-compactified12}, $\bm{X}^{\text{\tiny{boundary}}}_{AB}$ can be decomposed into a pair of supertwistors 
\begin{align}
	\left(\bm{T}_{A}{}^{\b}\right) &= \left(\begin{array}{c}
		\d_{\a}{}^{\b}
		\\
		-\ri x_{+}^{\ad\b} 
		\\
		\hline\hline
		2\q_{i}{}^{\b}
	\end{array}\right)\,, \qquad \bm{T}_{A}{}^{\b} \sim \bm{T}_{A}{}^{\g} M_{\g}{}^{\b}\,, \qquad M \in \sGL(2\,,\mathbb{C})\,. 
\end{align}
We then arrive at the situation described in section \ref{4d compact str}.
The local bosonic $(x_{+}^{a})$ and fermionic $(\q_{i}{}^{\a})$ coordinates can be interpreted as coordinates in the chiral subspace of a four-dimensional $\N$-extended Minkowski superspace, $\mathbb{M}^{4|4\N}$.
Indeed, this two-plane coincides with that described in the background material, see eq. \eqref{mink 2plane}.
This is supported by the action of the AdS supergroup, $\sSU(2,2|\N)$, as superconformal transformations on $\mathbb{M}^{4|4\N}$, see the relations \eqref{scf trf bm}.

\section{Superspace geometry} \label{5d geometry}

In this section we will elaborate on details of the supergeometry of the $\N$-extended AdS superspace introduced above. 
For this purpose it is useful to introduce local bosonic $(x)$ and fermionic $(\q)$ coordinates on AdS superspace as 
\begin{align}
	y^{\hat{M}} = (x^{\hm}\,, \q_{\imath}{}^{\mh}\,, \bar{\q}^{\imath}{}_{\mh})\,, \qquad  \mh = 1\,,2\,,3\,,4\,,
	\qquad \imath = 1\,, \ldots \N\,.  
\end{align}

\subsection{Geometric structures in AdS$^{5|8\N}$}

We will begin by formalising some of the earlier discussion about the AdS superalgebra. 
Let us denote the following 
\begin{align}
	\cH := \mathfrak{usp}(2,2) \oplus \mathfrak{u}(1) \oplus \mathfrak{su}(\N)\,. 
\end{align}
Denote the complement of $\cH$ in $\mathfrak{su}(2,2|\N)$ by $\cK$,
\begin{align}
	\mathfrak{su}(2,2|\N) = \cK \oplus \cH\,.
\end{align}
The elements $k \in \cK$ are given by
\begin{align} 
	k = 
	\left(\begin{array}{c|c||c}
		- \frac{1}{2}\o^{45} \d_{\a}{}^{\b} 
		& - \frac{\ri}{2}(\o^{a5})(\s_{a})_{\a\bd}
		& ~2\eta_{\a}{}^{j}~ \\
		\hline 
		-\frac{\ri}{2}(\o^{a5})(\tilde{\s}_{a})^{\ad\b} 
		&  \frac{1}{2}\o^{45}\d^{\ad}{}_{\bd} 
		& 2\bar{\e}^{\ad j}
		\\
		\hline \hline 
		2\e_{i}{}^{\b} 
		& 2\bar{\eta}_{i\bd} 
		& ~0~
	\end{array}\right)\,,
	\qquad \o^{\ha 5} \in \mathbb{R}
	\,. 
\end{align}
The elements $m \in \cH$ are given by  
\bsubeq
\begin{align} \label{stabel}
	m = \left(\begin{array}{c|c||c}
		\frac{1}{2}\o^{ab}(\s_{ab})_{\a}{}^{\b} + \frac{\ri \N}{\N-4}\t \d_{\a}{}^{\b} & - \frac{\ri}{2}(\o^{a4})(\s_{a})_{\a\bd} & 0 \\
		\hline 
		\frac{\ri}{2}(\o^{a4})(\tilde{\s}_{a})^{\ad\b} & \frac{1}{2}\o^{ab}(\tilde{\s}_{ab})^{\ad}{}_{\bd} + \frac{\ri\N}{\N-4}\t\d^{\ad}{}_{\bd} & 0
		\\
		\hline \hline 
		0 & 0 & \frac{4\ri}{\N-4}\t\d_{i}{}^{j} + \L_{i}{}^{j}
	\end{array}\right)\,, 
\end{align}
with 
\begin{align}
	\o^{\ha\hb}\,,\t \in \mathbb{R}\,, \qquad \L \in \mathfrak{su}(\N)\,.
\end{align}
\esubeq

Using the coset representative, one can introduce the Maurer-Cartan form, $\cS^{-1}\text{d}\cS$, which proves to encode all the information about the geometry of our superspace. 
The Maurer-Cartan form can be decomposed into the vielbein and the connection as 
\bsubeq
\begin{align}
	\cS^{-1}\text{d}\cS &= \tb{E} + \bm{\F}\,, 
	\\
	\tb{E} &= (\cS^{-1}\text{d}\cS) |_{\cK}\,,
	\\
	\bm{\F} &= (\cS^{-1}\text{d}\cS)|_{\cH}\,. 
\end{align}
\esubeq
Under a group transformation $g \in \sSU(2,2|\N)$, the coset representative transforms as
\begin{align} \label{group trf on coset rep}
	\cS(y) \rightarrow \cS(y') = g\,\cS(y)\,h^{-1}(g\,,y)\,, \qquad h \in H\,,
\end{align}
and the vielbein and connection transform as 
\bsubeq
\begin{align}
	\tb{E} &\rightarrow h \tb{E} h^{-1}\,,
	\\
	\bm{\F} & \rightarrow h\bm{\F}h^{-1} - \text{d}h\, h^{-1}\,. 
\end{align}
\esubeq

Since the vielbein and the connection are elements of the AdS superalgebra, it is useful to introduce a basis for the superalgebra as
\bsubeq
\begin{align}
	K_{\hA} &:= (P_{\ha}\,, q_{\ah}{}^{i}\,, \bar{q}{}^{\ah}{}_{i})\,, 
	\\
	H_{\hat{I}} &:= (M_{\ha\hb}\,, \mathbb{J}^{i}{}_{j}\,, \mathbb{Y})\,. 
\end{align}
\esubeq
The basis elements $K_{\hA}$ correspond to supertranslations, while $H_{\hat{I}}$ correspond to Lorentz, $\sSU(\N)$, and $\sU(1)$ transformations.  
They obey the following graded commutation relations 
\bsubeq \label{5d superalgebra}
\begin{align}
	[M_{\ha\hb}\,, M_{\hc\hd}] &= 
	2\eta_{\hd[\ha} M_{\hb] \hc}
	- 2\eta_{\hc[\ha} M_{\hb] \hd}
	\,, \label{5d lorentz comm}
	\\
	[M_{\ha\hb}\,, P_{\hc}] &= 2\eta_{\hc [\hb}P_{\ha]}
	\,, \label{5d lorentz tran comm}
	\\
	[P_{\ha}\,, P_{\hb}] &= M_{\ha\hb} \label{5d tran comm}
	\,,
	\\
	[\mathbb{J}^{i}{}_{j}\,, \mathbb{J}^{k}{}_{l}] &= 
	\d^{k}_{j}\mathbb{J}^{i}{}_{l} - \d^{i}_{l}\mathbb{J}^{k}{}_{j}
	\,,
	\\
	[M_{\ha\hb}\,, q_{\ah}{}^{i}] &=
	-(\S_{\ha\hb})_{\ah}{}^{\bh}q_{\bh}{}^{i} 
	\,,
	\\
	[P_{\ha}\,, q_{\ah}{}^{i}] &= 
	\frac{\ri}{2}(\g_{\ha})_{\ah}{}^{\bh}q_{\bh}{}^{i}
	\,,
	\\
	[P_{\ha}\,, \bar{q}^{\ah}{}_{i}] &= \frac{\ri}{2}(\g_{\ha})^{\ah}{}_{\bh}\bar{q}^{\bh}{}_{i}
	\,,
	\\
	[\mathbb{Y}\,,q_{\ah}{}^{i}] &= 
	- q_{\ah}{}^{i}
	\,,
	\\
	\{q_{\ah}{}^{i}\,, \bar{q}^{\bh}{}_{j}\} &= 
	2 \d^{i}_{j}M^{\ha\hb}(\S_{\ha\hb})_{\ah}{}^{\bh} 
	+2\ri \d^{i}_{j}P^{\ha}(\g_{\ha})_{\ah}{}^{\bh} 
	\\
	\notag
	& \quad 
	- 4\d_{\ah}{}^{\bh}\mathbb{J}^{i}{}_{j} 
	- \frac{\N-4}{\N}\,\d^{i}_{j}\d_{\ah}{}^{\bh} \mathbb{Y} 
	\,,
	\\
	[\mathbb{J}^{i}{}_{j}\,, q_{\ah}{}^{k}] &= 
	\d^{k}_{j}q_{\ah}{}^{i} - \frac{1}{\N}\,\d^{i}_{j}q_{\ah}{}^{k}
	\,,
\end{align} 
\esubeq
with $(\g_{\ha})_{\ah}{}^{\bh}$ defined as \eqref{5d gamma}.\footnote{The graded commutation relations \eqref{5d superalgebra} constitute the superalgebra $\frak{su}(2,2|\N)$. We have presented them in a basis that proves most suitable for analysis regarding AdS$^{5|8\N}$.
They reproduce those for the $\N$-extended superconformal algebra in four dimensions, see, e.g., \cite{Park:1999pd}, when recast in the basis appropriate for four-dimensional analysis.}

Any element $n$ of the AdS superalgebra, $n \in \mathfrak{su}(2,2|\N) \,,$ can be written as a linear combination of generators 
\begin{align}
	n &= \frac{1}{2}n^{\ha\hb}M_{\ha\hb} + n^{j}{}_{i}\mathbb{J}^{i}{}_{j} + \ri (n_{\sU(1)}) \mathbb{Y} 
	\\
	\notag 
	& \quad + n^{\ha}P_{\ha} + \ri\big(n_{i}{}^{\ah}q_{\ah}{}^{i} + \bar{n}^{i}{}_{\ah}\bar{q}^{\ah}{}_{i} \big)\,. 
\end{align}
This means that the vielbein can be decomposed as follows 
\begin{align}
	\textbf{E} &= \textbf{E}^{\hA}K_{\hA} = \textbf{E}^{\ha}P_{\ha} + \ri \big(\textbf{E}_{i}{}^{\ah} q_{\ah}{}^{i} + \bar{\textbf{E}}{}^{i}{}_{\ah}\bar{q}^{\ah}{}_{i}\big)\,,
\end{align}
and the connection can be decomposed as 
\begin{align}
	\bm{\F} &= \bm{\F}^{\hat{I}}H_{\hat{I}} = \frac{1}{2}\bm{\F}^{\ha\hb}M_{\ha\hb} + \bm{\F}^{j}{}_{i}\mathbb{J}^{i}{}_{j} + \ri (\bm{\F}_{\sU(1)})\mathbb{Y}\,. 
\end{align}
In the above expressions we have used the definition
\begin{align}
	\tb{E}^{\hA} = (\tb{E}^{\ha}\,, \tb{E}_{i}{}^{\ah}\,, \bar{\tb{E}}{}^{i}{}_{\ah})\,. 
\end{align}
The components of the connection, $\bm{\F}^{\hat{I}}\,,$ can be decomposed with respect to the basis $\tb{E}^{\hA}$ to yield the superfields $\F_{\hA}{}^{\hat{I}}$:
\begin{align}
	\bm{\F}^{\hat{I}} = \tb{E}^{\hA}\F_{\hA}{}^{\hat{I}}\,. 
\end{align}

In accordance with the coset construction, we can introduce the curvature and torsion two-forms as
\bsubeq \label{5d torsion and curvature}
\begin{align}
	\tb{R} := \text{d}\bm{\F} - \bm{\F} \wedge \bm{\F}\,, 
	\qquad
	\tb{T} := \text{d}\tb{E} - \tb{E} \wedge \bm{\F} - \bm{\F} \wedge \tb{E}\,. 
\end{align}
The curvature and torsion can alternatively be expressed as
\begin{align}
	\tb{R} = (\tb{E} \wedge \tb{E})|_{\cH}\,, 
	\qquad 
	\tb{T} = (\tb{E} \wedge \tb{E})|_{\cK}\,. 
\end{align}
\esubeq
Under group transformations, \eqref{group trf on coset rep}, they transform covariantly
\begin{align}
	\tb{R}' = h \tb{R} h^{-1}\,, \qquad \tb{T}' = h \tb{T} h^{-1}\,. 
\end{align}
The components of the torsion and curvature two-forms are defined via the relations
\bsubeq \label{5d components def}
\begin{align}
	\tb{T}&= \frac{1}{2}\tb{E}^{\hB}\wedge\tb{E}^{\hA}\big(
	\cT_{\hA\hB}{}^{\hc}P_{\hc} + \ri(\cT_{\hA\hB k}{}^{\gh}q_{\gh}{}^{k} + \cT_{\hA\hB}{}^{k}{}_{\gh}\bar{q}^{\gh}{}_{k})
	\big)
	\,,
	\\	
	\tb{R} &= \frac{1}{2}\tb{E}^{\hB}\wedge\tb{E}^{\hA}\big(
	\frac{1}{2}\cR_{\hA\hB}{}^{\hc\hd}M_{\hc\hd}
	+\cR_{\hA\hB}{}^{l}{}_{k}\mathbb{J}^{k}{}_{l}
	+\ri(\cR_{\sU(1)})_{\hA\hB}\mathbb{Y}
	\big)\,. 
\end{align}
\esubeq
Building on the approach utilised in \cite{Kuzenko:2007aj, KT} and outlined in chapter \ref{ch3}, using the definitions \eqref{5d components def} and the graded commutation relations of the generators, the non-vanishing components of the torsion and curvature can be determined. 
They are given by 
\bsubeq \label{5d ads torsions}
\begin{align}
	\cT^{i}{}_{\ah j}{}^{\bh \ha} &= 2\ri\d^{i}_{j}(\g^{\ha})_{\ah}{}^{\bh} 
	\,, \label{global tors}
	\\
	\cT_{\ha}{}^{i}{}_{\ah j}{}^{\bh} &= -\frac{\ri}{2}\d^{i}_{j}(\g_{\ha})_{\ah}{}^{\bh}
	\,,
	\\
	\cT_{\ha i}{}^{\ah j}{}_{\bh} &= - \frac{\ri}{2}\d^{j}_{i}(\g_{\ha})^{\ah}{}_{\bh}
	\,,
\end{align}
\esubeq
and
\bsubeq \label{5d ads curvatures}
\begin{align}
	\cR_{\ha\hb}{}^{\hc\hd} &= -2\d_{[\ha}{}^{\hc}\d_{\hb]}{}^{\hd} \,,
	\\
	\cR_{j}{}^{\bh i }{}_{\ah}{}^{\ha\hb} &= 4\d^{i}_{j}(\S^{\ha\hb})_{\ah}{}^{\bh} \,,
	\\
	\cR^{i}{}_{\ah j}{}^{\bh}{}^{l}{}_{k} &= 
	-4\d_{\ah}{}^{\bh}\d^{i}_{k}\d^{l}_{j}
	\,,
	\\
	(\cR_{\sU(1)})^{i}{}_{\ah j}{}^{\bh} &= \frac{\ri(\N-4)}{\N}\d^{i}_{j}\d_{\ah}{}^{\bh}\,. 
\end{align}
\esubeq

\subsection{Covariant derivatives}
In this subsection we will use the results of the previous subsection to detail the construction of covariant derivatives on AdS superspace and determine their algebra.

Associated with each generator $K_{\hA}$ is a vector field $E_{\hA}\,,$ and each generator $H_{\hat{I}}$ a connection $\bm{\F}^{\hat{I}}\,.$ 
To determine $E_{\hA}$ it is useful to introduce the vielbein supermatrix $E_{\hM}{}^{\hA}$ used implicitly above:
\bsubeq
\begin{align}
	\tb{E}^{\hA} &= \bm{\cE}^{\hM} E_{\hM}{}^{\hA} \,,
	\\
	\bm{\cE}^{\hM} &= \tb{E}^{\hA} E_{\hA}{}^{\hM} \,,
\end{align}
\esubeq
with $\bm{\cE}^{\hM}$ given by
\begin{align}
	\bm{\cE}^{\hM} &= (\text{d}x^{\hm}\,, \text{d}\q_{\imath}{}^{\mh}\,, \text{d}\bar{\q}^{\imath}{}_{\mh})\,.
\end{align}
We then have 
\begin{align}
	E_{\hA} = E_{\hA}{}^{\hM} \partial_{\hM}\,, \qquad 
	\partial_{\hM} = (\partial_{\hm}\,, \frac{\partial}{\partial \q_{\imath}{}^{\mh}}\,, \frac{\partial}{\partial \bar{\q}^{\imath}{}_{\mh}})\,. 
\end{align}

Now we define the covariant derivatives
\begin{align}
	\cD_{\hA} := (\cD_{\ha}\,, \cD_{\ah}{}^{i}\,, \bar{\cD}^{\ah}{}_{i})\,. 
\end{align}
They take the form 
\begin{align} \label{5d cd form}
	\cD_{\hA} = E_{\hA} + \frac{1}{2}\F_{\hA}{}^{\ha\hb}M_{\ha\hb} + \F_{\hA}{}^{j}{}_{i}\mathbb{J}^{i}{}_{j} + \ri (\F_{\sU(1)}) _{\hA} \mathbb{Y}\,,
\end{align}
and their algebra can be constructed making use of the components \eqref{5d ads torsions} and \eqref{5d ads curvatures}
\begin{align}
	[\cD_{\hA}\,, \cD_{\hB}\} = -\cT_{\hA\hB}{}^{\hC}\cD_{\hC} + \frac{1}{2}\cR_{\hA\hB}{}^{\hc\hd}M_{\ha\hb} + \cR_{\hA\hB}{}^{l}{}_{k}\mathbb{J}^{k}{}_{l} + \ri(\cR_{\sU(1)})_{\hA\hB}\mathbb{Y}\,. 
\end{align}
Explicitly, we have
\bsubeq \label{5d ads cd alg}
\begin{align}
	[\cD_{\ha}\,, \cD_{\hb}] &= 
	-M_{\ha\hb} 
	\,,
	\\
	[\cD_{\ha}\,, \cD_{\ah}{}^{i}] &= 
	\frac{\ri}{2}(\g_{\ha})_{\ah}{}^{\bh}\cD_{\bh}{}^{i} 
	\,,
	\\
	[\cD_{\ha}\,, \bar{\cD}^{\ah}{}_{i}] &= 
	\frac{\ri}{2}(\g_{\ha})^{\ah}{}_{\bh}\bar{\cD}^{\bh}{}_{i}
	\,,
	\\
	\{\cD_{\ah}{}^{i} \,, \bar{\cD}^{\bh}{}_{j} \} &= 
	-2\ri\d^{i}_{j}(\g^{\ha})_{\ah}{}^{\bh}\cD_{\ha}
	+ 4\d^{i}_{j} M_{\ah}{}^{\bh} 
	\\
	\notag 
	& \quad 
	- 4\d_{\ah}{}^{\bh}\mathbb{J}^{i}{}_{j} - \frac{(\N-4)}{\N}\d^{i}_{j}\d_{\ah}{}^{\bh}\mathbb{Y} 
	\,.
\end{align}
\esubeq

In the $\N=1$ case, we can present the above algebra in a different basis, in line with the conventional approach to $5$D superspaces in the literature. 
Let us combine the derivatives $\cD_{\ah}$ and $\bar{\cD}_{\ah}$ into the following $\sSU(2)$ doublet
\begin{align}
	\tilde{\cD}_{\ah}^{\underline{i}}\,, \qquad  \tilde{\cD}_{\ah}^{\underline{1}} = \cD_{\ah}\,, 
	\qquad \tilde{\cD}_{\ah}^{\underline{2}} = \bar{\cD}_{\ah}\,, \qquad \underline{i}\,, \underline{j} = \underline{1}\,, \underline{2}\,.  
\end{align}
The corresponding local coordinates $\q_{\ah}$ and $\bar{\q}_{\ah}$ are combined in a similar way and satisfy a pseudo-Majorana condition.
The $\sSU(2)$ indices are raised and lowered by $\ve^{\underline{ij}}$ and $\ve_{\underline{ij}}$, $\ve^{\underline{12}} = \ve_{\underline{21}} = 1$, as $\tilde{\cD}_{\ah \underline{i}} = \ve_{\underline{ij}}\tilde{\cD}_{\ah}^{\underline{j}}\,.$  
We then have the following relations
\bsubeq
\begin{align}
	[\cD_{\ha}\,, \tilde{\cD}^{\ui}_{\ah}] &= \frac{\ri}{2}(\s_{3})^{\ui}{}_{\uj}(\g_{\ha})_{\ah}{}^{\bh}\tilde{\cD}^{\uj}_{\bh}
	\,,
	\\
	\{\tilde{\cD}_{\ah}^{\underline{i}}\,, \tilde{\cD}_{\bh}^{\underline{j}} \} &= 
	-2\ri \ve^{\underline{ij}} (\g^{\ha})_{\ah \bh}\cD_{\ha} - 4(\s_{3})^{\underline{ij}} M_{\ah\bh} - 3\ve_{\ah\bh}\ve^{\underline{ij}} \mathbb{Y}\,,
\end{align}
\esubeq
and our algebra of covariant derivatives coincides with that of \cite{Kuzenko:2007aj}, see equations (2.21) and (7.49).


\section{AdS superspaces with internal bosonic dimensions} \label{5d projective}

The analysis of the previous sections made use of \textit{even} supertwistors. 
One can also consider \textit{odd} supertwistors,
\begin{align}
	\bm{\X} = (\bm{\X}_{A}) = \left(\begin{array}{c}
		\bm{\X}_{\ah} \\
		\hline\hline
		\bm{\X}_{i}
	\end{array}\right)\,,
\end{align}
with opposite Grassmann parities 
\begin{align}
	\e(\bm{\X}_{A}) = 1 + \e_{A} \qquad \text{(mod 2)}\,. 
\end{align}
The superspace AdS$^{5|8\N}$ can be extended to AdS$^{5|8\N} \times \mathbb{X}^{\N}_{m}$, where the internal space $\mathbb{X}^{\N}_{m}$ is realised in terms of $m \leq \N$ odd supertwistors. 
In the $\N$-extended case, there are many choices for the space $\mathbb{X}^{\N}_{m}$, corresponding to the various ways in which the $m$ odd supertwistors $\bm{\X}^{\ti}\,, \ti = 1\,,\ldots\,,m$ can be constrained.
In all cases, the odd supertwistors are required to be orthogonal to the quartet $\bm{T}^{\tmu}$,
which means that they contribute no Grassmann odd degrees of freedom.
The remaining constraints on the odd supertwistors should be chosen in such a way that the space 
AdS$^{5|8\N} \times \mathbb{X}^{\N}_{m}$ is a homogeneous space of $\sSU(2,2|\N)$. 
This implies that $\mathbb{X}^{\N}_{m}$ is a homogeneous space of $\sSU(\N)$, and can therefore be realised as a coset space 
\begin{align}
	\mathbb{X}^{\N}_{m} = \frac{\sSU(\N)}{\cG}\,, 
\end{align}
where $\cG$ is a subgroup of $\sSU(\N)$. 
We will outline a few possibilities below. 

Here we consider the superspaces AdS$^{5|8\N} \times \mathbb{X}_{1}^{\N}$. 
Let us accompany the quartet $\bm{T}^{\tmu}$ with a single odd supertwistor, $\bm{\X}$, subject to the orthonormality conditions 
\bsubeq
\begin{align}
	\braket{\bm{T}^{\tmu}}{\bm{\X}} &= 0\,, \label{even odd ip}
	\\
	\braket{\bm{\X}}{\bm{\X}} &= -1\,. \label{odd odd ip}
\end{align}
\esubeq
The second condition implies that the body of $\bm{\X}_{i}$ is non-zero. 
For $\N=1$, the internal space is given by $S^{1}$, whereas for $\N>2$ it can be shown that 
\begin{align} \label{x1n space}
	\mathbb{X}_{1}^{\N} = \frac{\sSU(\N)}{\sSU(\N-1)}\,, \qquad \N>2\,. 
\end{align}
As a consistency check, we can compare the dimension of the space \eqref{x1n space} with the degrees of freedom of the supertwistor $\bm{\X}$. 
The latter has $2\N - 1$ even degrees of freedom, because of the $2\N$ real components and the constraint \eqref{odd odd ip}. 
This coincides with the $\N^{2} - 1 - ((\N-1)^{2}-1) = 2\N-1$ dimensions of \eqref{x1n space}.  

For $\N=2$, we can modify the construction as follows. 
First, we relax the condition \eqref{odd odd ip}. We still require that the body of $\bm{\X}_{i}$ is non-zero. 
Then, we define the supertwistor $\bm{\X}$ modulo the equivalence relation 
\begin{align}
	\bm{\X} \sim d ~ \bm{\X} \,, \qquad d \in \mathbb{C} - \{0\}\,.  
\end{align}
The superspace obtained can be seen to be AdS$^{5|16} \times \mathbb{C}P^{1}$. It is a homogeneous space for $\sSU(2,2|2)\,.$


\subsection{The $\N=4\,, m=4$ case}

As outlined above, for a given $\N$, the maximal number of odd supertwistors we can consider is $\N$. 
With this in mind, for the $\N=4$ case let us introduce the quartet $\bm{\X}^{\ti}\,, ~ \ti = 1\,,2\,,3\,,4\,,$ subject to the orthonormality conditions
\bsubeq
\begin{align}
	\braket{\bm{T}^{\tmu}}{\bm{\X}^{\ti}} &= 0\,,
	\\
	\braket{\bm{\X}^{\ti}}{\bm{\X}^{\tj}} &= -\d^{\ti\tj}\,,
\end{align}
\esubeq
and defined modulo the eqiuvalence relation 
\begin{align} \label{s5 equiv}
	\bm{\X}^{\ti} \sim \re^{\ri\vf}\bm{\X}^{\tj} \g_{\tj}{}^{\ti}\,, \qquad \vf \in \mathbb{R}\,, \qquad \g \in\sUSp(4)\,. 
\end{align}
Making use of the equivalence relations \eqref{susy u1} and \eqref{s5 equiv}, the $\sU(1)$ freedom on the even and odd supertwistors can be fixed such that the supermatrix  formed by 
$(\bm{T}^{\tmu} \,, \bm{\X}^{\ti})$ has unit Berezinian.
This means that they can be represented as
\begin{align}
	(\bm{T}^{\tmu}\,, \bm{\X}^{\ti}) := \bm{\mathbb{T}} = \left(\begin{array}{c||c}
		\bm{T}_{\ah}{}^{\tmu} & \bm{\X}_{\ah}{}^{\ti} \\
		\hline \hline 
		\bm{T}_{i}{}^{\tmu} & \bm{\X}_{i}{}^{\ti}
	\end{array}\right) 
	\in \sSU(2,2|4)
	\,. 
\end{align}
Furthermore, the equivalence relations that preserve this choice can be written as 
\begin{align} \label{s5 big equiv}
	\bm{\mathbb{T}} \sim \re^{\ri\vf} \bm{\mathbb{T}} \left(\begin{array}{c||c}
		 \l & ~0~
		\\
		\hline \hline 
		~0~ & \g
	\end{array}\right)
	\,. 
\end{align}
It can be shown that the resulting superspace is
\begin{align} \label{ads times s5}
	\text{AdS}^{5|32}\times\mathbb{X}^{4}_{4} = \frac{\sSU(2,2|4)}{\sUSp(2,2)\times \sU(1) \times\sUSp(4)}\,. 
\end{align}
To prove \eqref{ads times s5}, we note that the simplest  base point is 
\begin{align}
	\bm{\mathbb{T}}^{(0)} &= \left(\begin{array}{c||c}
		\id_{4} & 0 
		\\
		\hline \hline 
		0 & \id_{4}
	\end{array}\right)\,. 
\end{align}
The stabiliser of $\bm{\mathbb{T}}^{(0)}$ consists of those supergroup elements that map $\bm{\mathbb{T}}^{(0)}$ to an equivalent point with respect to \eqref{s5 big equiv}. 
Since the stabiliser of $\bm{\mathbb{T}}^{(0)}$ is a subgroup of $\sSU(2,2|4)$, its elements have unit Berezinian and take the form 
\bsubeq
\begin{align}
	h = \left(\begin{array}{c||c}
		\re^{\ri\vf}\l & ~0~ \\
		\hline \hline 
		~0~ & \re^{\ri\vf} \g
	\end{array}\right)\,, 
\end{align}
with 
\begin{align}
	\l \in \sUSp(2,2)\,, \quad \vf \in \mathbb{R}\,, \quad \g \in \sUSp(4)\,. 
\end{align}
\esubeq
It is therefore isomorphic to 
\begin{align}
	\sUSp(2,2) \times \sU(1) \times \sUSp(4)\,,
\end{align}
and we arrive at \eqref{ads times s5}. 
In this situation, the internal space is given by
\begin{align}
	\mathbb{X}_{4}^{4} = \frac{\sSU(4)}{\sUSp(4)} = \frac{\sSO(6)}{\sSO(5)}
	\,,
\end{align}
where the latter relation may be established using arguments analogous to those given in subsection \ref{5d bt realisation}. Thus the space $\mathbb{X}_{4}^{4}$ is equivalent to $S^{5}$. 

We can also introduce bi-supertwistors for the space $\mathbb{X}^{4}_{4}$. 
Let us introduce the following supermatrix 
\begin{align} \label{s5 bst}
	\bm{\z}_{AB} = r\bm{\X}_{A}{}^{\ti}\bm{\X}_{B}{}^{\tj}C_{\ti\tj}\,,
\end{align}
where $C_{\ti\tj}$ is given by 
\begin{align}
	(C_{\ti\tj}) = \left(\begin{array}{cc}
		\ve & ~0~
		\\
		~0~ & \ve
	\end{array}\right)\,, 
\end{align} 
and $r$ is the $S^{5}$ radius.
The supermatrix $\bm{\z}_{AB}$ is defined modulo the equivalence relation
\begin{align}
	\bm{\z}_{AB} \sim \re^{\ri\vf}\bm{\z}_{AB}\,, \qquad \vf \in \mathbb{R}\,,
\end{align}
which is the remnant of the  
equivalence transformations  \eqref{s5 equiv}. 
Because the quartet $\bm{\X}^{\ti}$ has opposite Grassmann parity to $\bm{T}^{\tmu}$, the supermatrix $\bm{\z}_{AB}$ has the following graded symmetry property
\begin{align}
	\bm{\z}_{AB} = -(-1)^{(1+\e_A)(1+\e_B)}\bm{\z}_{BA}\,.
\end{align}
It also satisfies the conditions
\bsubeq
\begin{align}
	\bar{\bm{X}}{}^{AC}\bm{\z}_{CB} &= 0\,,
	\\
	\bar{\bm{\z}}{}^{AB}\bm{\z}_{BA} &= 4r^{2}\,,
\end{align}
\esubeq
where $\bar{\bm{\z}}{}^{AB}$ is defined in the same way as \eqref{dual sbt def}. 

The bi-supertwistors $\bm{X}_{AB}$ and $\bm{\z}_{AB}$ can be combined in the following way. 
Let us introduce the supermatrix 
\begin{align}
	\cX = \left(\begin{array}{c||c}
		\mathfrak{C}_{\tmu\tnu} & 0
		\\
		\hline\hline
		0 & C_{\ti\tj}
	\end{array}\right) \in \sSU(2,2|4)\,.
\end{align}
Then, we find the following 
\begin{align}
	\U &:= \bm{\mathbb{T}} \cX \bm{\mathbb{T}}^{\T} = \ell^{-1}\bm{X} + r^{-1}\bm{\z}\,, \qquad 
	\bm{X} = (\bm{X}_{AB})\, , \quad \bm{\z} = (\bm{\z}_{AB})\,.
\end{align}
Due to the $\sU(1)$ freedom enjoyed by $\bm{X}$ and $\bm{\z}$, the supermatrix $\U$ is defined modulo 
\begin{align}
	\U \sim \re^{\ri\vf}\U\,, \qquad \vf \in \mathbb{R}\,. 
\end{align}
Furthermore, it satisfies the following condition
\begin{align}
	\U^{\dag}\bm{\O}\U = \bm{\O}\,. 
\end{align}
%
\section{New superparticle models in AdS$^{5|8}$} \label{5d superparticle section}

One of the main applications of the formalism developed in this chapter is the new model \eqref{5d superparticle model} for a superparticle propagating in AdS$^{5|8\N}$. 
An important feature of this model is that it involves two independent two-derivative terms. 
This is indicative that it has an analogous structure to the deformed model introduced in four dimensions, see section \ref{4d superparticle discussion}. 
As we saw in chapter \ref{ch3}, we were able to relate the deformed structure of the model to a deformed interval constructed within the supergravity framework. 
It is therefore interesting to see how such two-derivative structures may originate within the supergravity setting in five dimensions, which is available in the $\N=1$ case. 
This section is devoted to re-deriving the superparticle model \eqref{5d superparticle model} within the $\N=1$ supergravity framework in five dimensions as reviewed in \cite{KTM2}.

Let $z^{\hM} = (x^{\hm}\,, \q_{\underline{\imath}}^{\mh})$, with $\underline{\imath} = \underline{1}\,, \underline{2}$ and $\overline{\q_{\underline{\imath}}^{\mh}} = \q^{\underline{\imath}}_{\mh}$, be local coordinates on the $\N=1$ superspace AdS$^{5|8}\,.$
The  covariant derivatives of AdS$^{5|8}$, $\cD_A = \big( \cD_{\ha} \, ,\cD_{\ah}^{\ui}\big)$, obey the graded commutation relations 
\bsubeq
\begin{align}
	\{\cD_{\ah}^{\ui}\,,\cD_{\bh}^{\uj}\} &= -2\ri\ve^{\underline{ij}}\cD_{\ah\bh} + 4\ri\cS^{\underline{ij}}M_{\ah\bh} 
	+ 3\ri\ve_{\ah\bh}\ve^{\underline{ij}}\cS^{\uk\ul}J_{\uk\ul}\,,
	\\
	[\cD_{\ha}\,, \cD_{\ah}^{\ui}] &= \frac{1}{2}(\g_{\ha})_{\ah}{}^{\bh}\cS^{\ui}{}_{\uj}\cD_{\bh}^{\uj}\,.
\end{align}
\esubeq
Here $M_{\ah\bh} = M_{\bh\ah} $ is the Lorentz generator, $J_{\uk\ul}\ = J_{\ul\uk}$ the $\sSU(2)$ generator,
and $\cS^{\underline{ij}}$ is a non-vanishing 
covariantly constant 
tensor superfield with the algebraic properties:
\begin{align} \label{s prop}
	\cS^{\underline{ij}} = \cS^{\underline{ji}}\,, \qquad \overline{\cS^{\underline{ij}}} = \cS_{\underline{ij}}
	= \ve_{\underline{ik}} \ve_{\underline{jl}}\cS^{\underline{kl}} 
	 ~\implies ~ \cS^{\underline{ik}}\cS_{\underline{kj}} = \cS^{2}\d^{\ui}{}_{\uj}\,, \quad \cS^{2} = \frac{1}{2}\cS^{\underline{ij}}\cS_{\underline{ij}}\,. 
\end{align}
In matrix notation, the properties of $\mathbb{S} = \cS^{-1}(\cS^{\underline{ij}})$ can be written as 
\begin{align} \label{MatrixConditions}
	\mathbb{S} = \mathbb{S}^{\T}\,, \qquad \mathbb{S}^{\dag}\mathbb{S} = \id_{2}\,, 
	\qquad \mathbb{S}^{\dag} = \ve \mathbb{S}\ve^{-1} \,.
\end{align}
These properties are analogous to those described in section \ref{4d solving} for the four dimensional case. 
It follows that $\mathbb{S} $ admits the representation\footnote{Had we omitted the reality condition in  
\eqref{MatrixConditions},
$\mathbb{S}^{\dag} = \ve \mathbb{S}\ve^{-1} $,  the general solution would be $\mathbb{S} = U U^{\T}$ with  $U \in \sU(2)$, 
 see, e.g., \cite{Zumino:1962smg}. }
\begin{align}
	\mathbb{S} = U U^{\T}\,, \qquad U \in \sSU(2)\,.
\end{align} 
In the above, the matrix $U$ is defined modulo the equivalence relation 
\begin{align}
	U \sim U\cO\,, \qquad \cO \in \sSO(2)\,.
\end{align}
This means that the space of solutions to \eqref{MatrixConditions} can be identified with the coset space 
\begin{align}
	\frac{\sSU(2)}{\sSO(2)} = S^{2}\,,
\end{align}
which is the unit two-sphere. 
In particular, applying an $\sSU(2)$ transformation allows one to choose $\cS^{\underline{ij}} = \d^{\underline{ij}}$. 
A different choice is employed in this chapter, however, allowing one to make direct correspondence with \cite{Kuzenko:2007aj}.
Making use of a unitary transformation, $\cS^{\underline{ij}}$ can be brought to the form 
\begin{align} \label{sigma 3}
	\cS^{\underline{ij}} = \ri \cS(\s^{3})^{\underline{ij}}\,,
\end{align}
and the resulting algebra of covariant derivatives coincides with \eqref{5d ads cd alg} provided one fixes $\cS = 1$. 
As outlined in chapter \ref{ch3}, the cost of making such a choice is the loss of a conformally flat frame.
Unless otherwise stated, we will not impose the gauge condition \eqref{sigma 3} below. 

We now turn to deriving the new superparticle model in the supergravity setting. 
Making use of $\cS^{\underline{ij}}$, we can introduce a one-parameter deformation of the supersymmetric interval $\eta_{\ha\hb}E^{\ha}E^{\hb}$ 
\begin{align}
	\rd s^{2} = \eta_{\ha\hb}E^{\ha}E^{\hb} + \frac{\ri\o}{\cS^{2}}\cS^{\underline{ij}}\ve_{\ah\bh}E_{\ui}^{\ah}E_{\uj}^{\bh}\,,
\end{align}
for a real dimensionless parameter $\o\,.$
This can be written as 
\begin{align}
	\rd s^{2} = E^{\hA}\eta_{\hA\hB}E^{\hB}\,,
\end{align}
where the supermatrix $\eta_{\hA\hB}$ is defined as 
\begin{align}
	(\eta_{\hA\hB}) = \left(
	\begin{array}{c||c}
		\eta_{\ha\hb} & 0 \\
		\hline \hline
		~0~ & \frac{\ri \o}{\cS^{2}}\cS^{\underline{ij}}\ve_{\ah\bh}
	\end{array}
	\right)\,.
\end{align}
This is very reminiscent to the four-dimensional case, see eq. \eqref{supermetric def}, and in complete analogy, it follows that there is a superparticle model 
\begin{align} \label{def model}
	S = \frac{1}{2}\int \rd \t \frak{e}^{-1} \left\{ \dt{E}{}^{\hA}\eta_{\hA\hB}\dt{E}{}^{\hB} - (\frak{e}m)^{2}\right\}\,, \qquad \dt{E}{}^{\hA} = \frac{\rd z^{\hM}}{\rd\t} E_{\hM}{}^{\hA}\,,
\end{align}
where $\frak{e}$ is the einbein and $m$ is the mass. 
For the case $\o = 0$ we recover the standard superparticle model in five dimensions.

In order to make contact with the model \eqref{5d superparticle model}, one must make use of the results of \cite{Kuzenko:2007aj}, in which the components of the supervielbein for the AdS$^{5|8}$ coset superspace are provided in the Poincar\'e-like coordinate system discussed in section \ref{5d n>0 cr section}. 
Then, in the gauge \eqref{sigma 3}, the deformation takes the form
\begin{align}
	\frac{\ri\o}{2\cS^{2}}\cS^{\underline{ij}}\ve_{\ah\bh}\dt{E}_{\ui}{}^{\ah}\dt{E}_{\uj}{}^{\bh} &= 
	\frac{\o}{\cS^{2}}\bigg(
	\dot{\q}^{\a}\dot{\eta}_{\a} + \dot{\bar{\q}}_{\ad}\dot{\bar{\eta}}^{\ad} + 2\dot{\q}^{2}\eta^{2} + 2\dot{\bar{\q}}^{2}\bar{\eta}^{2} + 4\dot{\q}^{\a}\eta_{\a}\bar{\eta}_{\ad}\dot{\bar{\q}}^{\ad} 
	- 2\eta^{2}\bar{\eta}^{2}\dt{\P}{}^{2}
	\\
	\notag & \qquad \qquad  + 4\ri\dt{\P}{}^{\ad\a}(\bar{\eta}^{2}\eta_{\a}\dot{\bar{\q}}_{\ad} - \eta^{2}\dot{\q}_{\a}\bar{\eta}_{\ad}) + \ri\dt{\P}{}^{\ad\a}(\bar{\eta}_{\ad}\dot{\eta}_{\a} - \dot{\bar{\eta}}_{\ad}\eta_{\a}) 
	 \bigg)\,, 
\end{align}
where $\dt{\P}{}^{a} = \dot{x}^{a} + \ri(\q\s^{a}\dot{\bar{\q}} - \dot{\q}\s^{a}\bar{\q})$.
This can be shown to coincide with the $\a$-term in \eqref{5d superparticle model} provided one fixes 
\begin{align}
	\a = \frac{\o}{4\cS^{2}}\,.
\end{align}

Finally, as we pointed out in the four-dimensional case, $\k$-symmetry is an important feature of models of this kind. 
It is most conveniently described in a conformally flat frame, which was derived for AdS$^{5|8}$ in \cite{KTM2}. 
In such a frame, the components of the supervielbein are
\bsubeq
\begin{align}
	E^{\ha} &= \re^{-2\s}\P^{\ha}\,, \qquad \P^{\ha} = \rd x^{\ha} + \ri \rd \q_{\ui}^{\ah}(\g^{\ha})_{\ah\bh}\q^{\bh\ui}\,,
	\\
	E_{\ui}^{\ah} &= \re^{-\s}\left\{\rd \q_{\ui}^{\ah} + \ri(D_{\bh \ui}\s)\P^{\bh\ah} \right\}\,, 
\end{align}
\esubeq
where $\P^{\ha}$ is the flat supersymmetric one-form in five dimensions, $\s$ is the super-Weyl parameter, and $D_{\ah}^{\ui} = \partial_{\ah}^{\ui} - \ri(\g^{\ha})_{\ah\bh}\q^{\bh \ui}\partial_{\ha}$ is the flat spinor covariant derivative.
Then, for $m = \o = 0$, the model \eqref{def model} takes the form 
\begin{align}
	S = \frac{1}{2}\int \rd\t\frak{e}^{-1}\re^{-4\s} \eta_{\ha\hb} \dt{\P}{}^{\ha}\dt{\P}{}^{\hb}\,, \qquad \dt{\P}{}^{\ha} = \dot{x}{}^{\ha} + \ri\dot{\q}_{\ui}^{\ah}(\g^{\ha})_{\ah\bh}\q^{\bh\ui}\,. 
\end{align} 
and is invariant under the following local $\k$-symmetry transformations 
\bsubeq \label{5d kappa trf def}
\begin{align}
	\d\q_{\ui}^{\ah} &= \ri\dt{\P}{}^{\ah\bh}\k_{\bh\ui}\,,
	\\
	\d x^{\ha} &= \ri \q_{\ui}^{\ah}(\g^{\ha})_{\ah\bh}\d\q^{\bh\ui}\,,
	\\
	\d\frak{e} &= -4\frak{e}\left\{ \dot{\q}_{\ui}^{\ah}\k_{\ah}^{\ui} + \d\q_{\ui}^{\ah}D_{\ah}^{\ui}\s  \right\}\,, 
\end{align}
\esubeq
where $\k_{\ah}^{\ui} = \k_{\ah}^{\ui}(\t)$ is Grassmann-odd. 
It is an interesting problem to extend these considerations beyond $\N=1$ and recast the local $\k$-symmetry in terms of supertwistors. 


\section{Discussion} \label{5d conclusion}

In this chapter we introduced (i) the twistor realisation for AdS$_{5}$, and showed its equivalence to the standard bi-twistor realisation; (ii) the supertwistor realisation for AdS$^{5|8\N}$; and (iii) the bi-supertwistor realisation for AdS$^{5|8\N},$ and showed its equivalence to the supertwistor realisation.
In particular, the (bi-)supertwistor variables derived parametrise the coset superspace eq. \eqref{5d n ext ads def}.
As an extension, we also considered superspaces of the form AdS$^{5|8\N} \times \mathbb{X}^{\N}_{m}$, where the internal space is a homogeneous space of $\sSU(\N)$. 
For the $\N=m=4$ case, we derived the necessary constraints for the (bi-)supertwistors to describe the AdS$_5 \times S^{5}$ superspace \eqref{ads times s5}.
These variables have numerous applications. 
For example, we presented the two-point functions and superparticle model discussed in section \ref{5d superparticle section}.
We extended the analysis of the four-dimensional case of chapter \ref{ch3} to five dimensions, and derived the deformed AdS$^{5|8}$ supersymmetric interval and its corresponding superparticle model in the supergravity framework. 
It would be interesting to extend these considerations to more general models, and to recast the type IIB superstring action in AdS$_5 \times S^{5}$ background \cite{Metsaev:1998it} in terms of the bi-supertwistors $\bm{X}_{AB}$ and $\bm{\x}_{AB}$. 
This will be discussed elsewhere. 

As pointed out at the beginning of this chapter, the analysis presented here is based on the published work \cite{KK}. 
To the best of our knowledge, this work contained the first systematic study of the $\N$-extended anti-de Sitter supergeometry in five dimensions. 
We showed that the results of section \ref{5d geometry} reduce to the known case for AdS$^{5|8}$ in the literature. 
Further, we presented an $\N$-extended version of the Poincar\'e-like coordinate system derived for $\N=1$ in \cite{Kuzenko:2007aj}, and computed the bi-supertwistor in this coordinate system. 
We also showed how this bi-supertwistor is related to that for the compactified Minkowski superspace in four dimensions, see \cite{K-compactified12}.

An interesting point that we did not explore in this chapter is the possible construction of $n$-point functions within this formalism. 
Due to the transformation properties of the bi-supertwistors, such a function necessarily requires an even total number of bi-supertwistors.
In particular, associated with $2n$-points of AdS$^{5|8\N}$ is the following $\sSU(2,2|\N)$ invariant function
\begin{align} \label{2n point function}
	(-1)^{\sum_{i=2}^{n}\e_{A_{i}}}\bar{\bm{X}}{}^{A_{1}B_{1}}(1)\bm{X}_{B_{1}A_{2}}(2)\bar{\bm{X}}{}^{A_{2}B_{2}}(3)\bm{X}_{B_{2}A_{3}}(4)\ldots\bar{\bm{X}}{}^{A_{n}B_{n}}(2n-1)\bm{X}_{B_{n}A_{1}}(n)\,.
\end{align}
The $2n$-point function \eqref{2n point function} can be represented in the compact form
\begin{align}
	(-1)^{\sum_{i=2}^{n}\e_{A_{i}}}\prod_{k=1}^{n} \bar{\bm{X}}{}^{A_{k}B_{k}}(2k-1)
	\prod_{l=1}^{n} \bm{X}_{B_{l}A_{l+1}}(2l)\,,
\end{align}
with the periodicity condition $A_{n+1} = A_{1}$.
Similar considerations can be extended to the case for $S^{5}$ making use of the bi-supertwistors \eqref{s5 bst}.

\chapter{New superparticle models in AdS superspaces in diverse dimensions}
\label{ch5}
\thispagestyle{boef3}

This chapter is dedicated to the derivation of new superparticle models in the three and two-dimensional anti-de Sitter supergeometries. 
In a generic D-dimensional supergravity background, parametrised by local coordinates $z^{M} = (x^{m}\,,\q^{\m})$, where the fermionic variables $\q^{\m}$ carry an appropriate spinor index, the standard Brink-Schwarz superparticle is described by an action of the form 
\begin{align} \label{bs sp}
	S = \frac{1}{2}\int\rd\t \frak{e}^{-1} \left\{\dt{E}{}^{a}\eta_{ab}\dt{E}{}^{b} - (\frak{e}m)^{2}\right\}\,, \qquad \dt{E}{}^{a} = \frac{\rd z^{M}}{\rd\t}E_{M}{}^{a}\,,
\end{align}
where $\eta_{ab}$ is the D-dimensional Minkowski metric and $E_{M}{}^{a}$ are components of the (inverse) supervielbein. 
These models were first introduced in the late 1970s by Casalbuoni \cite{C}, and were soon after reformulated by Brink and Schwarz \cite{BS}. 
There is a wealth of literature available on the subject, see, e.g., \cite{Szg} for a review and references, and they have by now become a textbook subject \cite{PW}. 
Further, they remain an active field of research, see \cite{BF1,BF2,Boffo:2024lwd} and references therein, in part due to their natural role as an arena to test concepts for more involved field and string models. 

In chapters \ref{ch3} and \ref{ch4}, we introduced new models for superparticles propagating in AdS$^{4|4\N}$ and AdS$^{5|8\N}$. 
The key feature of these models that distinguishes them from the standard superparticle, model \eqref{bs sp}, is the presence of additional two-derivative structures. 
As we can see from \eqref{bs sp}, the Brink-Schwarz superparticle is derived from an interval which makes use of only the vector component of the supervielbein.
In contrast to this, for AdS$^{4|4\N}$ the model \eqref{ads model} was derived from an interval containing both the vector $({E}^{a})$ and spinor $({E}_{i}^{\a}\,, \bar{E}^{i}_{\ad})$ components of the supervielbein.\footnote{See also eq. \eqref{def model} for the five-dimensional case.}  
Such an interval can be constructed in the supergravity setting with the help of a nowhere-vanishing, covariantly constant superfield $S^{ij}$. 
In the embedding formalism, these deformed intervals were derived making use of the unique two-point invariants of the AdS supergroup. 

In addition to the possibility of including additional two-derivative terms, models in AdS (super)space exhibit several other interesting features. One such feature is that, in contrast with the flat spacetime case, the mass-squared term of a free spin-0 particle propagating in AdS$_d$ may be negative, provided it does not violate the Breitenlohner-Freedman bound
\begin{align} \label{bf bound ch5}
m^{2} \geq -\frac{(d-1)^{2}}{4\ell^{2}}\,,
\end{align}
which was first derived in \cite{Breitenlohner:1982jf}.
This bound may also be derived from the perspective of worldline models, similar to those discussed in this chapter, see \cite{ABGT}.

Regarding the three-dimensional AdS supergeometry, the $\N$-extended conformally flat superspaces can be defined from a group-theoretic point of view as 
\begin{align} \label{sp 3d ads def}
	\text{AdS}_{(3|p\,,q)} = \frac{\sOSp(p|2;\mathbb{R})\times\sOSp(q|2;\mathbb{R})}{\sSL(2;\mathbb{R})\times\sSO(p)\times\sSO(q)}\,,
\end{align}
see, e.g., \cite{BILS}.
They were introduced in the supergravity setting in \cite{KLT-M12} as backgrounds of the $\N$-extended off-shell conformal supergravity in three dimensions \cite{HIPT, KLTM2}.
The conformal flatness of AdS$_{(3|p\,,q)}$ was argued in \cite{BILS} and conformally flat realisations were derived in \cite{KLT-M12} and \cite{KT}, based on the use of stereographic and Poincar\'e coordinates, respectively. 
A supertwistor realisation of AdS$_{(3|p\,,q)}$ was introduced in \cite{KTM} and its bi-supertwistor realisation was derived in \cite{KT}. 
The study of AdS$_{(3|p\,,q)}$ as a homogeneous space was also carried out in \cite{KT}. 
In the component setting, a Chern-Simons action for $(p\,,q)$ AdS supergravity theories in three dimensions was developed by Achuccaro and Townsend in \cite{Achucarro:1986uwr}, see also \cite{Achucarro:1989gm}. 

On the other hand, supersymmetric theories in AdS$_{2}$ have been attracting interest in recent years, see, e.g., \cite{BHT} and references therein. 
As an alternative to these studies based on a component framework, reference \cite{KR} developed the $\sSO(p)\times\sSO(q)$ superspace formulation for conformal supergravity in two dimensions and used it to realise the $\N$-extended AdS supergeometry. 
Essential to the approach of \cite{KR} is the use of the supergroup $\sOSp(\N|2;\mathbb{R})$, although there are other AdS supergroups in two dimensions, as demonstrated in \cite{GST}.

This chapter is based on the paper \cite{KRsp} and is organised as follows. 
In section \ref{3d ads sec} we introduce the deformed supersymmetric interval in the three-dimensional $(p\,,q)$ anti-de Sitter supergeometry, and use it to derive a new superparticle model.
This model is the three-dimensional analogue of those discussed in chapters \ref{ch3} and \ref{ch4}.
We then show how this model arrises in the embedding formalism for AdS$_{(3|p\,,q)}$ developed in \cite{KT}.  
Section \ref{2d ads sec} is devoted to showing how such a model can be derived for the two-dimensional anti-de Sitter supergeometry. 
This chapter is accompanied by appendix \ref{3d conventions appendix}, which reviews the technical details of the $\N$-extended conformal supergravity framework in three dimensions that are required for the analysis in this chapter.

\section{AdS superspaces in three dimensions} \label{3d ads sec}

In this section we introduce the deformed supersymmetric interval in the three-dimensional AdS supergeometry, first in the supergravity setting, and then from the embedding formalism.
Making use of this deformed interval, we derive the corresponding superparticle model. 
Our conventions for conformal supergravity in three dimensions are described in appendix \ref{3d conventions appendix}.

\subsection{$(p\,,q)$ AdS superspace}

We consider a three-dimensional curved superspace $\cM^{3|2\N}$, parametrised by local bosonic $(x^{m})$ and fermionic $(\q_{\tt I}^{\m})$ coordinates 
\begin{align}
    z^{M} = (x^{m}\,, \q_{\tt I}^{\m})\,, \qquad m = 0\,,1\,,2\,, \quad \m = 1\,,2\,, \quad {\tt I} = 1\,, \ldots\,, \N\,.
\end{align}
For the $\N=1$ case, the superspace geometry is controlled by the torsion superfields 
\begin{align}
    C_{\a\b\g} = C_{(\a\b\g)}\,, \qquad S\,,
\end{align}
whereas for $\N>1$ we have
\begin{align}
    X^{IJKL} = X^{[IJKL]}\,, \qquad C_{a}{}^{IJ} = C_{a}{}^{[IJ]}\,, \qquad S^{IJ} = S^{(IJ)}\,.
\end{align}
We emphasise that for $\N<4$ the super-Cotton tensor $X^{IJKL}$ vanishes identically, $X^{IJKL} = 0\,.$ 
As pointed out in the introduction, the AdS supergeometry can be singled out by the requirement that these torsion superfields are Lorentz-invariant and covariantly constant. 
This implies
\bsubeq \label{3d ads constraints}
\begin{align} 
\N &= 1: \qquad C_{\a\b\g} = 0\,, \qquad \cD_{A}S = 0\,,
\\
\N&>1: \qquad     C_{a}{}^{IJ} = 0\,, \qquad \cD_{A} S^{JK} = 0\,, \qquad \cD_{A}X^{JKLM} =0\,,  
\end{align}
\esubeq
where $\cD_{A} = (\cD_{a}\,, \cD_{\a}^{I})$ are the covariant derivatives.
Keeping these constraints in mind, the covariant derivatives obey the algebra
\bsubeq
\begin{align}
    \{\cD_{\a}^{I}\,,\cD_{\b}^{J}\} &= -2\ri\d^{IJ}\cD_{\ab} - 4\ri S^{IJ}\cM_{\ab} + \ri\ve_{\ab}(X^{IJKL}-4S^{K[I}\d^{J]L})\N_{KL}\,,
    \\
    [\cD_{a}\,,\cD_{\a}^{I}] &= S^{I}{}_{J}(\g_{a})_{\a}{}^{\b}\cD_{\b}^{J}\,,
    \\
    [\cD_{a}\,,\cD_{b}] &= -4S^{2}\cM_{ab}\,,
\end{align}
\esubeq
where $S^{2} := \frac{1}{\N}S^{IJ}S_{IJ}\,.$
This is the most general $\N$-extended AdS supergeometry of \cite{KLT-M12}. 

The superfield $S^{IJ}$ and super-Cotton tensor $X^{IJKL}$ satisfy non-trivial integrability conditions.
These are
\bsubeq
\begin{align} \label{s props}
    S^{IK}S_{K}{}^{J} = S^{2}\d^{IJ}\,,
\end{align}
and
\begin{align} \label{x const}
   X_{N}{}^{IJ[K}X^{LPQ]N} = 0 \implies X^{KLM[I}X^{J]}{}_{KLM} = 0\,. 
\end{align}
\esubeq
It follows from \eqref{s props} that, by a local $\sSO(\N)$ transformation, $S^{IJ}$ can be brought to the form 
\begin{align} \label{3d diag}
S^{IJ} = S ~ \text{diag}(\overbrace{+1\,, \cdots \,, +1}^{p} \,, \overbrace{-1\,, \cdots \,, -1}^{q = \N-p})\,, 
\end{align}
resulting in an unbroken local group $\sSO(p)\times \sSO(q)\,.$ 
The corresponding supergeometry is referred to as AdS$_{(3|p\,,q)}$.
Further, it was shown in \cite{KLT-M12} that a non-zero super-Cotton tensor $X^{IJKL}$ is only compatible with positive-definite $S^{IJ}\,.$ 
In other words, all three-dimensional $(p\,,q)$ AdS superspaces with $q >0$ have vanishing $X^{IJKL}\,.$

For the remainder of this section we will only consider the $q>0 \implies X^{IJKL} = 0$ case. This setup will be revisited in section \ref{sp discussion}.

\subsection{Conformally flat frame}

Bringing the superfield $S^{IJ}$ to the diagonal form \eqref{3d diag} is a special gauge choice which was made use of in \cite{KLT-M12}. 
It is often useful to relax this gauge condition, and unless otherwise stated we will assume that $S^{IJ}$ is not in the diagonal form \eqref{3d diag}. 
The advantage of doing so is that one can preserve local $\sSO(\N)$ freedom and, as a result, have a frame in which the covariant derivatives are related to those of the 3D Minkowski superspace by a super-Weyl transformation \eqref{cd trans}.\footnote{Indeed, as described in chapter \ref{ch3}, the gauge choice \eqref{3d diag} is not compatible with the frame \eqref{3d cflat derivatives}.}

In this case, the AdS covariant derivatives $\cD_{A}$ in three dimensions look like
\bsubeq \label{3d cflat derivatives}
\begin{align}
    \cD_{\a}^{I} &= \re^{\frac{1}{2}\s}\left\{ D_{\a}^{I} + (D^{\b I}\s)\cM_{\ab} + (D_{\a J}\s)\N^{IJ} \right\}\,,
    \\
    \cD_{a} &= \re^{\s}\bigg\{\partial_{a} + \frac{\ri}{2}(\g_{a})^{\ab}(D^{K}_{(\a}\s)D_{\b)K} + \ve_{abc}(\partial^{b}\s)\cM^{c} + \frac{\ri}{16}(\g_{a})^{\ab}([D^{[K}_{(\a}\,,D^{L]}_{\b)}]\s)\N_{KL}
    \\
    \notag 
    & ~~~~~~~\qquad - \frac{\ri}{8}(\g_{a})^{\ab}(D_{K}^{\r}\s)(D^{K}_{\r}\s)\cM_{\ab} + \frac{3\ri}{8}(\g_{a})^{\ab}(D^{[K}_{(\a}\s)(D_{\b)}^{L]}\s)\N_{KL}
    \bigg\}\,, 
\end{align}
\esubeq
where $D_{\a}^{I}$ is the flat 3D spinor derivative, defined as
\begin{align}
    D_{\a}^{I} = \partial_{\a}^{I} + \ri(\g^{c})_{\ab}\q^{\b I}\partial_{c}\,, \qquad \{D_{\a}^{I}\,,D_{\b}^{J}\} = 2\ri\d^{IJ}(\g^{c})_{\ab}\partial_{c}\,,
\end{align}
and $\s = \bar{\s}$ is the super-Weyl parameter, enjoying the constraint 
\bsubeq  \label{3d ads constraint}
\begin{align}
\N&=1: \qquad \partial_{(\ab}D_{\g)}\re^{\s} = 0\,,
\\
\N&>1: \qquad D_{(\a}^{[I}D_{\b )}^{J]}\re^{\s} = 0\,.    
\end{align}
\esubeq
Such a frame will be called conformally flat. 
We point out that the expressions \eqref{3d cflat derivatives} are also valid for $\N=1$. In this case, the terms containing the $\sSO(\N)$ generators $\N_{IJ}$ disappear, see eq. \eqref{n=1 sw cd}. 
Solutions to \eqref{3d ads constraint} were derived in \cite{KLT-M12, KT} making use of stereographic and Poincar\'e coordinates, respectively. 

It follows from \eqref{3d cflat derivatives} that the components of the supervielbein $E^{A} = (E^{a}\,, E^{\a}_{I})$ are given by
\bsubeq \label{cflat 3d vb}
\begin{align}
    E^{a} &= \re^{-\s}\P^{a}\,, \qquad \P^{a} = \rd x^{a} + \ri\q_{I}^{\a}(\g^{a})_{\ab}\rd\q_{I}^{\b}\,, \label{cflat 3d vector}
    \\
    E_{I}^{\a} &= \re^{-\frac{1}{2}\s}\left\{ \rd\q_{I}^{\a} - \frac{\ri}{2}(D_{\b I}\s)\P^{\ab} \right\}\,. 
\end{align}
\esubeq
Here, $\P^{a}$ is the three-dimensional analogue of the Volkov-Akulov supersymmetric one-form.

\subsection{Interval deformation for $p\geq q>0$}

We will now use the above considerations to derive a quadratic deformation to the supersymmetric interval in AdS$_{(3|p\,,q)}$.

The standard supersymmetric interval is constructed from the vector component of the supervielbein, and is given by
\begin{align} \label{3d interval}
    \rd s^{2} = \eta_{ab}E^{a}E^{b}\,, \qquad \eta_{ab} = \text{diag}(-1\,,+1\,,+1)\,.
\end{align}
Making use of the torsion superfield $S^{IJ}$ and the spinor components of the supervielbein $E^{\a}_{I}\,,$ we can introduce a one-parameter deformation as follows 
\begin{align} \label{3d def}
    \rd s^{2} = E^{A}\eta_{AB}E^{B} = \eta_{ab}E^{a}E^{b} + \frac{\ri\o}{S^{2}}S^{IJ}\ve_{\ab}E_{I}^{\a}E_{J}^{\b}\,,
\end{align}
for a real, dimensionless parameter $\o$.
The supermatrix $\eta_{AB}$ is defined as 
\begin{align} \label{supermetric}
    (\eta_{AB}) = \left(\begin{array}{c||c}
        ~\eta_{ab}~ & 0 \\
        \hline \hline
        ~0~ & \frac{\ri\o}{S^{2}}S^{IJ}\ve_{\ab}
    \end{array}\right)\,. 
\end{align}
We emphasise that the deformation \eqref{3d def} is invariant under the three-dimensional AdS superisometries.

Given the deformation \eqref{3d def}, we can introduce a new model for a superparticle propagating in AdS$_{(3|p\,,q)}$, defined by 
\begin{align} \label{3d model}
    S = \frac{1}{2}\int \rd\t \frak{e}^{-1}\left\{\dt{E}{}^{A}\eta_{AB}\dt{E}{}^{B} - (\frak{e}m)^{2}\right\}\,, \qquad \dt{E}{}^{A} := \frac{\rd z^{M}}{\rd \t} E_{M}{}^{A}\,,
\end{align}
where $\frak{e}$ is the einbein, $m$ is the mass, and $\t$ is the evolution parameter. 
In the conformally flat frame, where the components of the supervielbein are given by the expressions \eqref{cflat 3d vb}, we find 
\begin{align}
    \dt{E}{}^{A}\eta_{AB}\dt{E}{}^{B} &= \re^{-2\s}\dt{\P}{}^{2} + \frac{\ri\o}{S^{2}}\re^{-\s}S^{IJ}\left\{ \dot{\q}_{IJ} - \ri(D_{\a I}\s)\dt{\P}{}^{\ab}\dot{\q}_{\b J} + \frac{1}{4}(D^{\a}_{I}\s)(D_{\a J}\s)\dt{\P}{}^{2}
   \right\}\,, 
\end{align}
where $\dt{\P}{}^{a} = \dot{x}{}^{a} + \ri\q_{I}^{\a}(\g^{a})_{\ab}\dot{\q}{}_{I}^{\b}\,.$
Further, in the diagonal frame \eqref{3d diag}, we have
\begin{align}
    \dt{E}{}^{A}\eta_{AB}\dt{E}{}^{B} = \eta_{ab}\dt{E}{}^{a}\dt{E}{}^{b} + \frac{\ri\o}{S}\ve_{\ab}\left\{ \dt{E}{}_{\bar{I}}^{\a}\dt{E}{}_{\bar{I}}^{\b} - \dt{E}{}_{\uI}^{\a}\dt{E}{}_{\uI}^{\b} \right\}\,, 
\end{align}
with $\OI = 1\,, \ldots\,, p$ and $\UI = p+1\,,\ldots\,,\N\,.$
We point out that for $\o = 0$, we recover the standard superparticle propagating in AdS superspace.
As a result, the model \eqref{3d model} is the three-dimensional analogue of those proposed in \cite{KKR,KKR2,KK} and discussed in chapters \ref{ch3} and \ref{ch4}.

It is of interest to see how the above two-derivative structures arise in the embedding formalism for AdS$_{(3|p\,,q)}$ derived in \cite{KT}. 
A brief review of the bi-supertwistor realisation of AdS$_{(3|p\,,q)}$ is included in appendix \ref{bst 3d appendix}. 
We point out that it shares many features with those described for AdS$^{4|4\N}$ and AdS$^{5|8\N}$ in section \ref{ads4 bst} and chapters \ref{ch3} and \ref{ch4}. 
In particular, AdS$_{(3|p\,,q)}$ is described by three kinds of graded antisymmetric supermatrices (bi-supertwistors), denoted $\bm{X}\,, \bm{Y}\,,$ and $\bm{Z}$, which are defined following \eqref{3d smat}.  
Then, the most general superparticle model quadratic in derivatives of the evolution parameter $\t$ is given by the following 
\begin{align} \label{3d embed model}
    S = -\frac{1}{2}\int \rd \t \frak{e}^{-1} \left\{\frac{1}{2} \text{Str}(\dot{\bm{Z}}\dot{\bm{Z}}) + \bigg(\a- \frac{1}{4}\bigg)\big(\text{Str}(\dot{\bm{X}}\dot{\bm{X}}) + \text{Str}(\dot{\bm{Y}}\dot{\bm{Y}})\big) + (\frak{e}m^{2})\right\}\,, 
\end{align}
with $\a \in \mathbb{R}$.  
The coefficients of each structure are chosen to satisfy the following two conditions: (i) the model coincides with the bosonic one when the Grassmann variables are switched off; and (ii) the $\a = 0$ case recovers the non-deformed superparticle model. 
Indeed, it was shown in \cite{KT} that the following combination 
\begin{align} \label{3d canonical 2pf}
    \rd s^{2} = -\frac{1}{2}\text{Str}(\rd \bm{Z} \rd\bm{Z}) + \frac{1}{4}(\text{Str}(\rd\bm{X}\rd\bm{X}) + \text{Str}(\rd\bm{Y}\rd\bm{Y})),
\end{align}
yields the interval \eqref{3d interval}.  
Furthermore, in the Poincar\'e-like\footnote{Strictly speaking, ref. \cite{KT} introduced two coordinate systems for AdS$_{(3|p\,,q)}$ based on the use of Poincar\'e coordinates. One provided a conformally flat realisation, see eq. \eqref{3d cflat derivatives}, and the other corresponded to the diagonal frame \eqref{3d diag}. Following chapter \ref{ch4}, we refer to the latter as Poincar\'e-like coordinates.} coordinate system developed for the embedding formalism in \cite{KT}, which makes use of the gauge choice \eqref{3d diag}, the additional structures present in \eqref{3d embed model} take the form
\bsubeq
\begin{align}
    \text{Str}(\dot{\bm{X}}\dot{\bm{X}}) &= 4\ri z^{-2}\left(\dot{\q}^{+}_{\OI}\q^{-}_{\OI}(\dot{z}+ \ri\dot{\q}^{+}_{\OJ}\q^{-}_{\OJ}) + \dot{\q}^{-}_{\OI}\q^{-}_{\OI}(\dot{u}^{++} - \ri\dot{\q}^{+}_{\OJ}\q^{+}_{\OJ})
  \right) + 4\ri z^{-1}\dot{\q}^{-}_{\OI}\dot{\q}^{+}_{\OI}\,, 
  \\
  \text{Str}(\dot{\bm{Y}}\dot{\bm{Y}}) &= 4\ri z^{-2} \left( 
  -\dot{\q}^{-}_{\UI}\q^{+}_{\UI}(\dot{z} + \ri\dot{\q}^{-}_{\UJ}\q^{+}_{\UJ})
  -\dot{\q}^{+}_{\UI}\q^{+}_{\UI}(\dot{u}^{--} - \ri\dot{\q}^{-}_{\UJ}\q^{-}_{\UJ})
  \right) + 4\ri z^{-1}\dot{\q}^{-}_{\UI}\dot{\q}^{+}_{\UI}\,. 
\end{align}
\esubeq
As a result, the $\a$ term in \eqref{3d embed model} generates purely fermionic contributions. 
Making use of the above expressions and the components of the supervielbein derived in \cite{KT}, the models \eqref{3d model} and \eqref{3d embed model} can be shown to coincide to leading order provided one fixes 
\begin{align}
    \a = -\frac{\o}{8 S^{2}}\,. 
\end{align}

Finally, let us discuss $\kappa$-symmetry of the AdS superparticle in three dimensions. 
As in the four and five-dimensional case, such a symmetry is most conveniently formulated in the conformally flat frame, given by the relations \eqref{cflat 3d vb}. 
Then, it is straightforward to verify that, in the $m = \a = 0$ case, the model is invariant under the following $\k$-symmetry transformations 
\bsubeq
\begin{align}
\d \q_I^\a &= \ri \k_{\b I}\dt{\P}{}^{\b\a}\,,
\\
\d x^{a} &= \q_I^\a (\g^{a})_{\a\b}\dt{\P}{}^{\b\d}\k_{\d I}\,,
\\
\d \frak{e} &= -4\frak{e}\left( \dt{\q}{}^{\a}_{I}\k_{\a I} - \frac{1}{4}\d\q_I^\a D_{\a I}\s  \right)\,.
\end{align}
\esubeq
These transformations constitute the three-dimensional analogues of \eqref{4d kappa trf def}, \eqref{4d kappa trf def 2} and \eqref{5d kappa trf def}. We emphasise that it is only the combination \eqref{3d canonical 2pf} which leads to a $\kappa$-symmetric model. 

\section{New AdS superparticle in two dimensions} \label{2d ads sec}

In this section we will extend the above considerations to the two-dimensional $\N$-extended AdS supergeometry, which is realised in the conformal $(p\,,q)$ supergravity framework of \cite{KR}, with $p=q=\N\,.$ 
It should be pointed out that certain $(p\,,q)$ supergeometries in two dimensions were considered earlier in the literature, see, e.g., for $\N = (2\,,2)$ \cite{H2D,Grisaru:1994dm,Grisaru:1995dr,Gates:1995du}, $\N = (4\,,4)$ \cite{Bellucci:2000yx,TM2D},\footnote{Ref. \cite{Bellucci:2000yx} formulated $(4\,,4)$ matter-coupled supergravity in $\sSU(2)\times\sSU(2)$ harmonic superspace.} and for the $\N=(p\,,0)$ with $p \geq 2$ case \cite{Evans:1986ada,Brooks:1986uh,GOVINDARAJAN1992251}, as well as references therein.

\subsection{$\cN$-extended AdS superspace}

Let us introduce local coordinates for the two-dimensional AdS supergeometry as $z^M = (x^{\hat + \hat +},x^{\hat - \hat -},\q^{\hat{+}\overline{\cI}},\q^{\hat{-}\underline{\cI}})$ where $\overline{\cI} = 1, \dots , \N$ and $\underline{\cI} = 1, \dots , \N$.
The structure group is $\sSO(1,1)\times\sSO(\N)\times\sSO(\N)$, where the first $\sSO(\N)$ factor acts on the overlined $R$-symmetry indices, and the second $\sSO(\N)$ factor acts on the underlined $R$-symmetry indices.
These will be referred to as the left and right sectors. 
The AdS covariant derivatives $\cD_A = (\cD_{++}, \cD_{--}, \cD_{+}^{\overline{I}}, \cD_{-}^{\underline{I}})$ take the form 
\begin{align}
    \cD_A = E_A{}^{M}\partial_{M} + \O_A M +  \hf \F_A{}^{\overline{J} \overline{K}} \frak{L}^{\overline{J} \overline{K}} + \hf \F_A{}^{\underline{J} \underline{K}} \frak{R}^{\underline{J}\underline{K}}~,
\end{align}
where $\O_{A}$ is the Lorentz connection, $\F_{A}{}^{\OJ\OK}$ is the left $\sSO(\N)$ connection, and $\F_{A}{}^{\uJ\uK}$ is the right $\sSO(\N)$ connection. 
Finally, the Lorentz ($M$), left $\sSO(\N)$ ($\frak{L}^{\OI\OJ}$), and right $\sSO(\N)$ ($\frak{N}^{\uI\uJ}$) generators act on the derivatives and amongst themselves by the rules 
\bsubeq
    \begin{align}
        [M,\cD_{++}] &= \cD_{++} ~, \qquad [M,\cD_{+}^{\overline{I}}] = \hf \cD_{+}^{\overline{I}}~, \\
        [M,\cD_{--}] &= - \cD_{--} ~, \qquad [M,\cD_{-}^{\underline{I}}] = - \hf \cD_{-}^{\underline{I}}~, \\
        [\mathfrak{L}^{\OI \OJ} , \cD_{+}^{\OK}] &= 2 \d^{\OK [\OI} \cD_{+}^{\OJ]}~, \\
        [\mathfrak{R}^{\UI \UJ} , \cD_{-}^{\UK}] &= 2 \d^{\UK [\UI} \cD_{-}^{\UJ]}~, \\
        [\mathfrak{L}^{\OI \OJ} , \mathfrak{L}^{\OK \OL}] &= 2 \d^{\OK[\OI} \mathfrak{L}^{\OJ] \OL} - 2 \d^{\OL[\OI} \mathfrak{L}^{\OJ] \OK}~, \\
        [\mathfrak{R}^{\UI \UJ} , \mathfrak{R}^{\UK \UL}] &= 2 \d^{\UK[\UI} \mathfrak{R}^{\UJ] \UL} - 2 \d^{\UL[\UI} \mathfrak{R}^{\UJ] \UK}~.
    \end{align}
\esubeq

The covariant derivatives $\cD_{A}$ obey the algebra 
\begin{subequations} \label{2DAdSalgebra}
	\bea
	\{ \cD_{+}^{\OI}, \cD_{+}^{\OJ} \} &=& 2 \ri \d^{\OI \OJ} \cD_{++} ~, \qquad \{ \cD_{-}^{\UI}, \cD_{-}^{\UJ} \} = 2 \ri \d^{\UI \UJ} \cD_{--} ~, \\
	\{ \cD_{+}^{\OI}, \cD_{-}^{\UJ} \} &=& 4 \ri S^{\OI \UJ} M - 2 \ri S^{\OK \UJ} \mathfrak{L}^{\OK \OI} + 2 \ri S^{\OI \UK} \mathfrak{R}^{\UK \UJ} ~, 
	 \\
	\big[ \cD_{+}^{\OI} , \cD_{--} \big]
	& = & - 2 S^{\OI \UJ} \cD_{-}^{\UJ} ~, \qquad 
	\big[ \cD_{-}^{\UI} , \cD_{++} \big]
	= 2 S^{\OJ \UI} \cD_{+}^{\OJ} ~,
	\\
	\big[ \cD_{+}^{\OI} , \cD_{++} \big]
	& = & 0 ~, \qquad
	\big[ \cD_{-}^{\UI} , \cD_{--} \big] =  0  ~,
	\\
	\big[ \cD_{++} , \cD_{--} \big]
	& = & - 8  S^2 M
	~.
	\eea
\end{subequations}
The above algebra is controlled by the torsion superfield $S^{\OI\uJ}$, which is covariantly constant, 
\bsubeq
\begin{align}
	\cD_{A} S^{\OI\uJ}\,,
\end{align}
and satisfies 
\begin{align}
    S^{\OI \UK} S^{\OJ \UK} = \d^{\OI \OJ} S^2~, \qquad S^{\OK \UI} S^{\OK \UJ} = \d^{\UI \UJ} S^2~, \qquad S^2 := \frac{1}{\cN} S^{\OI \UJ} S^{\OI \UJ}~.
\end{align}
\esubeq
The torsion superfield $S^{\OI \UJ}$ is related to the scalar curvature by $\cR = - 16 S^2 < 0$.

The properties of $S^{\OI\uJ}$ are analogous to those for the non-vanishing component of the torsion in three-dimensional case of section \ref{3d ads sec}, the four-dimensional case of chapter \ref{ch3}, and the five-dimensional case of chapter \ref{ch4}. 
In particular, a local $\sSO(\N) \times \sSO(\N)$ transformation can be utilised to bring $S^{\OI\uJ}$ to the form 
\begin{align} \label{2d diag}
	S^{\OI\uJ} = S\d^{\OI\uJ}\,, \qquad \cD_{A} S = 0\,.
\end{align}
This special gauge choice was made use of in \cite{KR}. 
It should be emphasised that the gauge choice \eqref{2d diag} spoils the conformal flatness of the frame, as the local $\sSO(\N)\times\sSO(\N)$ freedom is broken to the diagonal subgroup. 
This is analogous to the three, four, and five-dimensional cases discussed earlier in this work, and implies that the covariant derivatives $\cD_{A}$ are no longer related to those of the flat $(\N\,,\N)$ superspace in two dimensions by a super-Weyl transformation, which are discussed in the following subsection.
The frame in which \eqref{2d diag} holds will be referred to as the diagonal frame.

\subsection{Conformally flat frame}
\label{2d conflat frame sec}

It can be advantageous to not impose the gauge condition \eqref{2d diag}, and instead remain in a frame with local $\sSO(\N)\times\sSO(\N)$ freedom. 
Unless otherwise stated, we will not impose \eqref{2d diag} below. 
It is useful to introduce the flat spinor covariant derivatives, $D_{+}^{\OI}$ and $D_{-}^{\UI}$, defined as
\bsubeq
\bea
D_{+}^{\OI}
:=
\frac{\pa}{\pa\q^{+ \OI}}
+ \ri \q^{+ \OI} \pa_{++}
~,\quad
D_{-}^{\UI}
:=
\frac{\pa}{\pa\q^{- \UI}}
+ \ri \q^{- \UI}\pa_{--}
~,
\eea
and satisfying the algebra
\begin{align}
 \{D_{+}^{\OI}\,,D_{+}^{\OJ}\} = 2\ri\d^{\OI\OJ}\partial_{++}\,, \qquad \{D_{-}^{\UI}\,,D_{-}^{\UJ}\} = 2\ri\d^{\UI\UJ}\partial_{--}\,, \qquad \{D_{+}^{\OI}\,,D_{-}^{\UJ}\} = 0\,.
\end{align}
\esubeq
Then, in such a frame, the AdS covariant derivatives $\cD_{A}$ are related to the flat ones $D_{A}$ by the rule
\begin{subequations} \label{ads frame}
\begin{align}
	{\cD}_+^\OI &= \re^{\hf \s} \Big( D_+^\OI + D_+^\OI \s M - D_+^\OJ \s \mathfrak{L}^{\OJ \OI} \Big) ~, \\
	\cD_-^\UI &= \re^{\hf \s} \Big( D_-^\UI - D_-^\UI \s M - D_-^\UJ \s \mathfrak{R}^{\UJ \UI} \Big) ~, \\
	\cal{D}_{++} &= \re^{\s} \Big( \partial_{++} - \ri D_+^\OI \s D_+^\OI + \partial_{++} \s M + \frac{\ri}{2} D_{+}^{\OI} \s D_{+}^{\OJ} \s \mathfrak{L}^{\OJ \OI} \Big) ~, \\
	\cal{D}_{--} &= \re^{\s} \Big( \partial_{--} - \ri D_-^\UI \s D_-^\UI - \partial_{--} \s M + \frac{\ri}{2} D_{-}^{\UI} \s D_{-}^{\UJ} \s \mathfrak{R}^{\UJ \UI} \Big) ~,
\end{align}
\end{subequations}
for a constrained superfield $\s = \bar{\s}$.\footnote{These transformations constitute super-Weyl transformations in two dimensions. In the $\N=(1\,,1)$ and $\N=(2\,,2)$ cases, they reduce to those given in \cite{Howe:1978ia} and \cite{H2D}, respectively. For the $\N=(4\,,4)$ case, it is expected that they are equivalent to those of \cite{TM2D}.} 
As shown in \cite{KRsp}, the super-Weyl parameter $\s$ is required to satisfy the constraints 
\begin{subequations}
    \begin{align}
    [D_+^{\OI}, D_{+}^{\OJ}] \re^{\s} &= 0 ~, \\
    [D_-^{\UI}, D_{-}^{\UJ}] \re^{\s} &= 0 ~.
    \end{align}
\end{subequations}

Finally, it follows from \eqref{ads frame} that the components of the supervielbein one forms $E^{A}$ are given by
\begin{subequations} 
    \begin{align} 
        {E}^{++} &= \re^{-\s} \P^{++}~, \qquad {E}^{+ \OI} = \re^{- \frac \s 2} \Big( \rd \q^{+ \OI} + \ri D_+^{\OI} \s \P^{++} \Big) ~, \\
        {E}^{--} &= \re^{-\s} \P^{--}~, \qquad {E}^{- \UI} = \re^{- \frac \s 2} \Big( \rd\q^{- \UI} + \ri D_-^{\UI} \s \P^{--} \Big) ~,
    \end{align}
\end{subequations}
where $\P^{++} = \rd x^{++} + \ri\q^{+ \OI}\rd\q^{+ \OI}$ and $\P^{--} = \rd x^{--} + \ri\q^{- \UI}\rd\q^{- \UI}$ are the two-dimensional analogues of the Volkov-Akulov supersymmetric one-form.

\subsection{Interval deformation}
In two dimensions, in the lightcone coordinates used above, the standard supersymmetric interval takes the form
\begin{align}
    \rd s^{2} = E^{++}E^{--}\,. 
\end{align}
We can introduce a deformation to this interval as 
\begin{align}
    \rd s^{2} = E^{++}E^{--} + \frac{\ri\o}{S^{2}}S^{\OI\UJ}E^{+\OI}E^{-\UJ}\,,
\end{align}
for a real, dimensionless parameter $\o$. 
It then follows that there is a corresponding superparticle model, given by 
\begin{align}
    S = \frac{1}{2}\int\rd\t\frak{e}^{-1}\left\{ \dt{E}{}^{++}\dt{E}{}^{--} + \frac{\ri\o}{S^{2}}S^{\OI\UJ}\dt{E}{}^{+\OI}\dt{E}{}^{-\UJ} - (\frak{e}m)^{2}\right\}\,.
\end{align}
 In a conformally flat frame \eqref{ads frame}, we have the following 
 \bsubeq
 \begin{align}
\dt{E}{}^{++}\dt{E}{}^{--} &= \re^{-2\s}\dt{\P}{}^{2}\,, \qquad \qquad \qquad \dt{\P}{}^{2}:= \dt{\P}{}^{++}\dt{\P}{}^{--}\,, 
\\
S^{\OI\UJ}\dt{E}{}^{+\OI}\dt{E}{}^{-\UJ} &= 
\re^{-\s}S^{\OI\UJ}\left\{ \dot{\q}^{+\OI}\dot{\q}^{-\UJ} + \ri D_{+}^{\OI}\s\dot{\q}^{-\UJ}\dt{\P}{}^{++} + \ri\dot{\q}^{+\OI}D_{-}^{\UJ}\s\dt{\P}{}^{--} 
     -D_{+}^{\OI}\s D_{-}^{\UJ}\s\dt{\P}{}^{2} \right\}\,,
 \end{align}
 \esubeq
where $\dt{\P}{}^{++} = \dot{x}^{++} + \ri\q^{+ \OI}\dot{\q}{}^{+ \OI}$ and $\dt{\P}{}^{--} = \dot{x}{}^{--} + \ri\q^{- \UI}\dot{\q}{}^{- \UI}\,.$

\section{Discussion}\label{sp discussion}

In this chapter we have extended the construction of new superparticle models in the four and five-dimensional $\N$-extended AdS superspaces, detailed in chapters \ref{ch3} and \ref{ch4}, to the three and two-dimensional AdS superspaces. 
For the three dimensional case, we also showed how such models arise within the embedding formalism, and related the construction of each approach. 
To the best of our knowledge, such an embedding formalism is yet to be developed for the $(\N\,,\N)$ AdS supergeometry in two dimensions, though this is an interesting avenue to explore.

The analysis of section \ref{3d ads sec} only holds for those AdS superspaces with vanishing super-Cotton tensor $X^{IJKL} = 0\,.$ 
As described in \cite{KLT-M12}, there are also superspaces of the form AdS$_{(3|\N,0)}$, with $\N \geq 4$, possessing non-zero $X^{IJKL}$. 
Given the homogeneous transformation rule of $X^{IJKL}$, eq. \eqref{t trans}, such superspaces are necessarily not conformally flat. 
The presence of such a dimensionful superfield, however, allows us to construct further deformations of the AdS supersymmetric interval.
We will briefly outline this below.

It follows from eq. \eqref{x const} that $X^{KLM(I}X^{J)}{}_{KLM}$ is unconstrained. 
With the following notation, 
\bsubeq
\begin{align}
    |X| &:= \sqrt{X^{IJKL}X_{IJKL}}\,\,,
    \\
    X^{IJ}& := X^{KLM(I}X^{J)}{}_{KLM}\,, \label{sym x}
\end{align}
\esubeq
we can then introduce an additional quadratic deformation, unique to the $(\N\,,0)$ case, as
\begin{align} \label{new def}
    \rd s^{2} = \eta_{ab}E^{a}E^{b} + \left\{\frac{\o}{S}\d^{IJ} 
 + \frac{\l}{|X|^{3}}X^{IJ}\right\}\ve_{\ab}E_{I}^{\a}E_{J}^{\b} \,,
\end{align}
where both $\o$ and $\l$ are real, dimensionless parameters.
For $\N=4$, $X^{IJKL} = X\ve^{IJKL}$ and so \eqref{sym x} is proportional to $\d^{IJ}$. 
The $\o$ and $\l$ terms then correspond to the same structure, and we return to the model \eqref{3d model}. 
The story is different for $\N \geq 5\,,$ however, for which eq. \eqref{new def} contains additional structure.
This is another interesting avenue for further exploration. 

The analysis of this chapter has been restricted to the case of superparticle models. These belong to a family of models of extended supersymmetric objects, including superstrings and super-$p$-branes, all of which make use of a supersymmetric interval in the target space. A key result of this thesis is the derivation of deformed supersymmetric intervals in AdS superspaces in diverse dimensions. It is an interesting problem to explore further new models, e.g., for superstrings, which are based on the use of the deformed supersymmetric interval, in similar fashion to the models described in this chapter. This is left for future work.

\begin{subappendices}

\section{Conformal supergravity in three dimensions} \label{3d conventions appendix}

Our conventions coincide with those used in \cite{KLT-M12,KLTM2}.
As a starting point, we consider a three-dimensional curved superspace $\cM^{3|2\N}$ which is parametrised by local bosonic $(x^{m})$ and fermionic $(\q_{\tt I}^{\m})$ coordinates 
\begin{align}
    z^{M} = (x^{m}\,, \q_{\tt I}^{\m})\,, \qquad m = 0\,,1\,,2\,, \quad \m = 1\,,2\,, \quad {\tt I} = 1\,, \ldots\,, \N\,.
\end{align}
Its structure group is $\sSL(2\,,\mathbb{R}) \times \sSO(\N)\,,$ and so the covariant derivatives take the form 
\begin{align}
    \cD_{A} \equiv (\cD_{a}\,, \cD_{\a}^{I}) = E_{A} + \O_{A} + \F_{A}\,. 
\end{align}
Here, $E_{A} = E_{A}{}^{M}\partial_{M}$ denotes the frame field, with $E_A{}^M$ being the inverse supervielbein.\footnote{The supermatrix $E_{A}{}^M$ is assumed to be non-singular, hence there exists a unique inverse $E_M{}^A$. It defines the supervielbein one forms $E^A = \rd z^M E_M{}^A$, which constitute a basis for the cotangent space at each point.}
The superfield 
\begin{align}
    \O_{A} = \frac{1}{2}\O_{A}{}^{bc}M_{bc} = - \O_{A}{}^{b}M_{b} =  \frac{1}{2}\O_{A}{}^{\g\d}M_{\g\d}\,, \qquad M_{ab} = - M_{ba}\,, \quad M_{\ab} = M_{\b\a}\,,
\end{align}
denotes the Lorentz connection; and 
\begin{align}
    \F_{A} = \frac{1}{2}\F_{A}{}^{KL}\N_{KL}\,, \qquad \N_{KL} = -\N_{LK}\,,
\end{align}
is the $\sSO(\N)$ connection. 

Our spinor conventions in three dimensions follow \cite{KLTM2}, and are compatible with the 4D two-component formalism used in \cite{WB,BK} as well as throughout the main body of this work.
In particular, two-component spinor indices are raised and lowered with the $\sSL(2;\mathbb{R})$ invariant tensors 
\begin{align}
    \left(\ve^{\ab}\right) = \left(\begin{array}{cc}
     ~0~    & 1 \\
       -1  & ~0~
    \end{array}\right)\,, \qquad \left(\ve_{\ab}\right) = \left(\begin{array}{cc}
  ~0~       & -1 \\
      1   & ~0~
    \end{array}\right)\,,
\end{align}
according to the rule 
\begin{align}
    \psi_{\a} = \ve_{\ab}\psi^{\b}\,, \qquad \psi^{\a} = \ve^{\ab}\psi_{\b}\,.
\end{align}
Furthermore, the Lorentz generators with spinor indices are related to those with vector indices by the rule 
\begin{align}
    M_{\ab} = \frac{1}{2}(\g^{a})_{\ab}\ve_{abc}M^{bc}\,,
\end{align}
where $\left((\g^{a})_{\ab}\right) = (\id_{2}\,, \s_{1}\,, \s_{3})$ are the gamma-matrices in three dimensions. 

The generators of $\sSL(2\,,{\mathbb{R}}) \times \sSO(\N)$ act on the covariant derivatives as follows:
%
\begin{align}
    [M_{\ab}\,, \cD_{\g}^{I}] &= \ve_{\g(\a}\cD_{\b)}^{I}\,, \quad [M_{ab}\,, \cD_{c}] = 2\eta_{c[a}\cD_{b]}\,, \quad 
    [\N_{KL}\,, \cD_{\a}^{I}] = 2\d^{I}_{[K}\cD_{\a L]}\,,
\end{align}
%
and amongst themselves as 
\bsubeq
\begin{align}
    [M_{ab}\,, M_{cd}] &= \eta_{ad}M_{bc} - \eta_{ac}M_{bd} + \eta_{bc}M_{ad} - \eta_{bd}M_{ac}\,,
    \\
    [M_{\ab}\,, M_{\g\d}] &= \frac{1}{2}\left( \ve_{\a\g}M_{\b\d}+\ve_{\a\d}M_{\b\g} + \ve_{\b\g}M_{\a\d} + \ve_{\b\d}M_{\a\g} \right) \,,
    \\
    [\N_{IJ}\,,\N_{KL}] &= \d_{IL}\N_{JK} - \d_{IK}\N_{JL} + \d_{JK}\N_{IL} - \d_{JL}\N_{IK} \,. 
\end{align}
\esubeq
All other commutators vanish. 

The supergravity gauge group is generated by local transformations of the form 
\begin{align}
    \d_{\cK}\cD_{A} = [\cK\,,\cD_{A}]\,, \qquad \cK = \x^{C}\cD_{C} + \frac{1}{2} K^{cd}M_{cd} + \frac{1}{2}\L^{PQ}\N_{PQ}\,,
\end{align}
with the parameters obeying natural reality conditions. 
These transformations act on tensor superfields (with indices suppressed) as 
\begin{align}
    \d_{\cK}\cU = \cK\cU\,. 
\end{align}

The structure of the algebra of covariant derivatives differs between the $\N=1$ and $\N>1$ cases.
We specify both cases below. 

\subsection{$\N=1$}

For $\N=1$, the algebra of covariant derivatives is given by 
\bsubeq\label{n=1 alg}
\begin{align}
\{\cD_{\a}\,,\cD_{\b}\} &= 
2\ri\cD_{\ab}-4\ri S M_{\ab}\,,
\\
[\cD_{\ab}\,,\cD_{\g}] &= 
-2S\ve_{\g(\a}\cD_{\b)} + 2\ve_{\g(\a}C_{\b)\d\r}M^{\d\r} + \frac{2}{3}\left( (\cD_{\g}S)M_{\ab} - 4(\cD_{(\a}S)M_{\b)\g} \right)\,,
\\
[\cD_{a}\,,\cD_{b}]&= 
\frac{1}{2}\ve_{abc}(\g^{c})^{\ab}\bigg\{ -\ri C_{\ab\g}\cD^{\g} - \frac{4\ri}{3}(\cD_{\a}S)\cD_{\b} + \ri\cD_{(\a}C_{\b\g\d)}M^{\g\d} 
\\
& \qquad \qquad \qquad ~~~~ -\left(\frac{2\ri}{3}(\cD^{2}S)+4S^{2}\right)M_{\ab}\bigg\}\,, 
\end{align}
\esubeq
see \cite{KLTM2}. 
It is expressed in terms of the dimension-1 scalar $S$ and a totally symmetric dimension-$3/2$ spinor $C_{\a\b\g} = C_{(\a\b\g)}\,,$ obeying the constraint 
\begin{align}
    \cD_{\a}C_{\b\g\d} = \cD_{(\a}C_{\b\g\d)} - \ri\ve_{\a(\b}\cD_{\g\d)}S\,. 
\end{align}

The structure of the algebra \eqref{n=1 alg} is invariant under the super-Weyl transformations 
\bsubeq \label{n=1 sw cd}
\begin{align}
  \hat{\cD}_{\a} &= \re^{\frac{1}{2}\s}\left\{ \cD_{\a} + (\cD^{\b}\s)\cM_{\ab} \right\}\,,
    \\
    \hat{\cD_{a}} &= \re^{\s}\bigg\{\cD_{a} + \frac{\ri}{2}(\g_{a})^{\ab}(\cD_{\a}\s)\cD_{\b} + \ve_{abc}(\cD^{b}\s)\cM^{c}  - \frac{\ri}{8}(\g_{a})^{\ab}(\cD^{\r}\s)(\cD_{\r}\s)\cM_{\ab} 
    \bigg\}\,. 
\end{align}
\esubeq
The corresponding transformations of the torsion superfields are 
\bsubeq
\begin{align}
    \hat{S} &= \frac{\ri}{2}\re^{\frac{3}{2}\s}\left\{\cD^{2} - 2\ri S\right\}\re^{-\frac{1}{2}\s}
    \,,
    \\
    \hat{C}_{\a\b\g} &= -\frac{1}{2}\re^{\frac{1}{2}\s}\left\{D_{(\ab}D_{\g)} -2C_{\a\b\g} \right\}\re^{\s}\,. 
\end{align}
\esubeq
The finite form of these super-Weyl transformations was given in \cite{KNTM}. 
It follows from \eqref{n=1 sw cd} that the transformations for the supervielbein one forms are given by
\bsubeq 
\begin{align}
    \hat{E}^{a} &= \re^{-\s}E^{a}\,, 
    \\
    \hat{E}^{\a} &= \re^{-\frac{1}{2}\s}\left\{ E^{\a} - \frac{\ri}{2}(\cD_{\b}\s)E^{\ab} \right\}\,. 
\end{align}
\esubeq

\subsection{$\N>1$}
We now detail the $\N>1$ case.
Up to dimension-1, the covariant derivatives $\cD_{A}$ obey the following algebra
\begin{align}
\label{3d alg}
    \{\cD_{\a}^{I}\,, \cD_{\b}^{J}\} &= 2\ri\d^{IJ}\cD_{\ab} - 2\ri\ve_{\ab}C^{\g\d IJ}M_{\g\d} - 4\ri S^{IJ}M_{\ab} \\
    \notag
     & \quad + \left(\ri \ve_{\ab} X^{IJKL} - 4\ri\ve_{\ab}S^{K[I}\d^{J]L} + \ri C_{\ab}{}^{KL}\d^{IJ} - 4\ri C_{\ab}{}^{K(I}\d^{J)L}   \right)\N_{KL}\,.
\end{align}
This algebra is given in terms of the real superfields $X^{IJKL}$, $S^{IJ}$ and $C_{a}{}^{IJ}$ which have the following symmetry properties 
\begin{align} \label{3d torsions}
    X^{IJKL} = X^{[IJKL]}\,, \qquad S^{IJ} = S^{(IJ)}\,, \qquad C_{a}{}^{IJ} = C_{a}{}^{[IJ]}\,.
\end{align}
Importantly, $C_{a}{}^{IJ} = 0$ for $\N=1$, and $X^{IJKL} = 0$ for $\N < 4\,.$

The structure of the algebra \eqref{3d alg} is invariant under super-Weyl transformations. The super-Weyl transformations of the covariant derivatives are given by 
\bsubeq \label{cd trans}
\begin{align}
    \hat{\cD}_{\a}^{I} &= \re^{\frac{1}{2}\s}\left\{ \cD_{\a}^{I} + (\cD^{\b I}\s)\cM_{\ab} + (\cD_{\a J}\s)\N^{IJ} \right\}\,,
    \\
    \hat{\cD_{a}} &= \re^{\s}\bigg\{\cD_{a} + \frac{\ri}{2}(\g_{a})^{\ab}(\cD^{K}_{\a}\s)\cD_{\b K} + \ve_{abc}(\cD^{b}\s)\cM^{c} + \frac{\ri}{16}(\g_{a})^{\ab}([\cD^{K}_{\a}\,,\cD^{L}_{\b}]\s)\N_{KL}
    \\
    \notag 
    & ~~~~~~~\qquad - \frac{\ri}{8}(\g_{a})^{\ab}(\cD_{K}^{\r}\s)(\cD^{K}_{\r}\s)\cM_{\ab} + \frac{3\ri}{8}(\g_{a})^{\ab}(\cD^{K}_{\a}\s)(\cD_{\b}^{L}\s)\N_{KL}
    \bigg\}\,. 
\end{align}
\esubeq
The dimension-1 torsion superfields transform as
\bsubeq \label{t trans}
\begin{align}
    \hat{S}^{IJ} &= \re^{\s}\left\{ S^{IJ} - \frac{\ri}{8}([\cD^{\r I}\,,\cD_{\r}^{J}]\s) + \frac{\ri}{4}(\cD^{\r I}\s)(\cD_{\r}^{J}\s) - \frac{\ri}{8}\d^{IJ}(\cD^{\r}_{K}\s)(\cD^{K}_{\r}\s)  \right\}\,,
    \\
    \hat{C}_{a}{}^{IJ} &= \re^{\s}\left\{  C_{a}{}^{IJ} - \frac{\ri}{8}(\g_{a})^{\ab}([\cD_{\a}^{I}\,,\cD_{\b}^{J}]\s) - \frac{\ri}{4}(\g_{a})^{\ab}(\cD_{\a}^{I}\s)(\cD_{\b}^{J}\s) \right\}\,,
    \\
    \hat{X}^{IJKL} &= \re^{\s}X^{IJKL}\,. 
\end{align}
\esubeq
It follows from \eqref{cd trans} that the corresponding transformations for the supervielbein one forms $E^{A}$ are 
\bsubeq \label{vb trans}
\begin{align}
    \hat{E}^{a} &= \re^{-\s}E^{a}\,, 
    \\
    \hat{E}_{I}^{\a} &= \re^{-\frac{1}{2}\s}\left\{ E_{I}^{\a} - \frac{\ri}{2}(\cD_{\b I}\s)E^{\ab} \right\}\,. 
\end{align}
\esubeq
The relations \eqref{cd trans}, \eqref{t trans}, and \eqref{vb trans} are used in the main body to define a conformally flat frame for AdS$_{(3|p\,,q)}\,.$

\section{Bi-supertwistor realisation of AdS$_{(3|p\,,q)}$} \label{bst 3d appendix}

In this appendix we outline the salient details of the bi-supertwistor realisation of AdS$_{(3|p\,,q)}$, derived in \cite{KT}, building on the supertwistor realisation derived in \cite{KTM}. 

As can be seen from \eqref{sp 3d ads def}, the isometry supergroup of AdS$_{(3|p\,,q)}$ is $\sOSp(p|2;\mathbb{R}) \times \sOSp(q|2;\mathbb{R})\,.$
For the purposes of the bi-supertwistor realisation of AdS$_{(3|p\,,q)}$, it is useful to use two different but equivalent realisations of the supergroup $\sOSp(n|2;\mathbb{R})$ denoted $\sOSp_{+}(n|2;\mathbb{R})$ and $\sOSp_{-}(n|2;\mathbb{R})$. We will briefly outline these below, following \cite{KT}. 

The supergroups $\sOSp_{+}(n|2;\mathbb{R})$ and $\sOSp_{-}(n|2;\mathbb{R})$ naturally act on the space of even and odd supertwistors. 
For our purposes, a supertwistor is a column vector 
\begin{align}
\bm{T} = \left(\bm{T}_{A}\right) = 
\left(\begin{array}{c}
\bm{T}_{\a} \\ 
\hline \hline 
\bm{T}_{I}
\end{array}\right)\,, \qquad \a = 1\,,2\,, ~~ I = 1\,,2\,,\ldots \,, n\,.
\end{align}
Even and odd supertwistors are defined in analogy with appendices \ref{5d appendix} and \ref{Supertwistors}.

Let us introduce the graded antisymmetric supermatrices $\mathbb{J}_{+}$ and $\mathbb{J}_{-}$
\begin{align}
    \mathbb{J}_{\pm} = \left(\mathbb{J}^{AB}\right) =  \left( \begin{array}{c||c}
      \ve   & ~0~ \\
      \hline \hline 
       ~0~  & \pm\ri \id_{n}
    \end{array}\right)\,.
\end{align}
By definition, the supergroup $\sOSp_{+}(n|2;\mathbb{R})$ consists of those even $(2|n) \times (2|n)$ supermatrices 
\begin{align}
    g = (g_{A}{}^{B})\,, \qquad \e(g_{A}{}^{B}) = \e_{A} + \e_{B}\,,
\end{align}
obeying the master equations
\bsubeq
\begin{align}
    g^{\sT} \mathbb{J}_{+} g &= \mathbb{J}_{+}\,, \\
    g^{\dag}\mathbb{J}_{+} g &= \mathbb{J}_{+}\,. 
\end{align}
\esubeq
In the above, the parity of the supermatrix components is defined in analogy with appendices \ref{5d appendix} and \ref{Supertwistors}, and the supertranspose of $g$ is defined as 
\begin{align}
    (g^{\sT})^{A}{}_{B} = (-1)^{\e_{A}\e_{B}+\e_{B}}g_{B}{}^{A}\,. 
\end{align}
Analogous definitions hold for $\sOSp_{-}(n|2;\mathbb{R})\,.$

Following \cite{KT}, the isometry supergroup\footnote{The implications of choosing other realisations are discussed in \cite{KT}.} of AdS$_{(3|p\,,q)}$ is taken as
\begin{align}
    G_{\pm} = \sOSp_{+}(p|2;\mathbb{R}) \times \sOSp_{-}(q|2;\mathbb{R})\,, 
\end{align}
in agreement with \cite{Achucarro:1986uwr,Achucarro:1989gm}. 
In what follows, indices transforming with respect to $\sOSp_{+}(p|2;\mathbb{R})$ are denoted $\oA\,, \oB$, and indices transforming with respect to $\sOSp_{-}(q|2;\mathbb{R})$ are denoted $\uA\,, \uB$.
Then, associated with a point $\cP$ of AdS$_{(3|p\,,q)}$ are three graded antisymmetric supermatrices, known as bi-supertwistors. 
They are denoted
\bsubeq
\begin{align}
    \bm{Z}_{\oA\uB} &= -(-1)^{\e_{\oA}\e_{\uB}}\bm{Z}_{\uB\oA}\,,
    \\ 
    \bm{X}_{\oA\oB} &= -(-1)^{\e_{\oA}\e_{\oB}}\bm{X}_{\oB\oA}\,,
    \\
    \bm{Y}_{\uA\uB} &= -(-1)^{\e_{\uA}\e_{\uB}}\bm{Y}_{\uB\uA}\,. 
\end{align}
\esubeq
In order to describe AdS$_{(3|p\,,q)}$, these bi-supertwistors must satisfy a number of constraints. 
Firstly, the obey 
\bsubeq
\begin{align}
\bm{X}_{\{\oA\oB}\bm{Z}_{\oC\}\uD} &= 0\,,
\\
\bm{X}_{\{\oA\oB}\bm{X}_{\oC\}\oD} &= 0\,,
\\
\bm{Y}_{\{\uA\uB}\bm{Z}_{\uC\}\oD} &= 0\,,
\\
\bm{Y}_{\{\uA\uB}\bm{Y}_{\uC\}\uD} &= 0\,.
\end{align}
\esubeq
Further, they satisfy the product relations 
\bsubeq
\begin{align}
    (-1)^{\e_{\uB}}\bm{Z}_{\oA\uB}\mathbb{J}^{\uB\uC}\bm{Z}_{\uC\oD} &= \bm{X}_{\oA\oD}\,,
    \\
    (-1)^{\e_{\oB}}\bm{Z}_{\uA\oB}\mathbb{J}^{\oB\oC}\bm{Z}_{\oC\uD} &= \bm{Y}_{\uA\uD}\,,
    \\
    (-1)^{\e_{\oB}}\bm{X}_{\oA\oB}\mathbb{J}^{\oB\oC}\bm{Z}_{\oC\uD} &= \bm{Z}_{\oA\uD}\,,
    \\
    (-1)^{\e_{\uB}}\bm{Y}_{\uA\uB}\mathbb{J}^{\uB\uC}\bm{Z}_{\uC\oD} &= \bm{Z}_{\uA\oD}\,. 
\end{align}
\esubeq
Finally, they satisfy the trace identities 
\begin{align}
    \mathbb{J}^{\oA\oB}\bm{X}_{\oB\oA} = \mathbb{J}^{\uA\uB}\bm{Y}_{\uB\uA} = 2\,.
\end{align}
These bi-supertwistors parametrise the $(p\,,q)$ AdS superspace in three dimensions, eq. \eqref{sp 3d ads def}.

We now introduce the supermatrices employed in the main body of this work. 
They are 
\bsubeq \label{3d smat}
\begin{align}
    \bm{Z} &= \left(\bm{Z}_{\oA}{}^{\uB}\right)\,, \qquad~~ \bm{Z}_{\oA}{}^{\uB} = (-1)^{\e_{\uC}}\bm{Z}_{\oA\uC}\mathbb{J}^{\uC\uB}\,,
    \\
    \bm{X} &= \left(\bm{X}_{\oA}{}^{\oB}\right)\,, \qquad \bm{X}_{\oA}{}^{\oB} = (-1)^{\e_{\oC}}\bm{X}_{\oA\oC}\mathbb{J}^{\oC\oB}\,, 
    \\
    \bm{Y} &= \left(\bm{Y}_{\uA}{}^{\uB}\right)\,, \qquad~~ \bm{Y}_{\uA}{}^{\uB} = (-1)^{\e_{\uC}}\bm{Y}_{\uA\uC}\mathbb{J}^{\uC\uB}\,. 
\end{align}
\esubeq
Under group transformations, we have
\begin{align}
    \bm{Z}_{\oA}{}^{\uB} &\longrightarrow g_{+\oA}{}^{\oB}\bm{Z}_{\oB}{}^{\uC}(g_{-}^{-1})_{\uC}{}^{\uB}\,,
    \\
    \bm{X}_{\oA}{}^{\oB} &\longrightarrow g_{+\oA}{}^{\oC}\bm{X}_{\oC}{}^{\oD}(g_{+}^{-1})_{\oD}{}^{\oB}\,, 
    \\
    \bm{Y}_{\uA}{}^{\uB} &\longrightarrow g_{-\uA}{}^{\uC}\bm{Y}_{\uC}{}^{\uD}(g_{-}^{-1})_{\uD}{}^{\uB}\,,
\end{align}
with $g_{+}\in\sOSp_{+}(p|2;\mathbb{R})$ and $g_{-} \in \sOSp_{-}(q|2;\mathbb{R})\,.$
Analogous definitions hold for $\bm{Z}_{\uA}{}^{\oB}\,.$

\end{subappendices}

\chapter{Conclusion}

This thesis has been dedicated to the study of the $\N$-extended anti-de Sitter supergeometries in four and five dimensions. 
Specifically, we have focused on the development of embedding formalisms for AdS$^{4|4\N}$ and AdS$^{5|8\N}$, as well as their applications. 
In this section, we will briefly recapitulate the main outcomes of our analysis as well as comment on possible avenues for future research. 

Let us first provide some general comments. 
As we have seen throughout this thesis, anti-de Sitter superspaces can be realised within a local framework, such as the supergravity setting; or globally, via a suitable embedding formalism. 
Both approaches are distinct and have natural advantages over their counterparts. 
For example, in the supergravity setting, one is able to derive elegant expressions for the superspace covariant derivatives in AdS, whereas in the embedding formalism such expressions can prove to be unwieldy.
On the other hand, the embedding formalism provides a simple way to construct models (of various kinds), and can provide insight into the allowed geometric structures in such models, which may not be obvious in the supergravity framework. 
These considerations have been applied in chapter \ref{ch3} to AdS$^{4|4\N}$; chapter \ref{ch4} to AdS$^{5|8\N}$; and chapter \ref{ch5} to superparticle models in three and two-dimensional AdS superspaces.

Supertwistor techniques have been utilised extensively in this thesis for developing the embedding formalisms for AdS$^{4|4\N}$ and AdS$^{5|8\N}$. 
As an example application of these formalisms, we derived new superparticle models in AdS superspace. 
In the literature, (super)twistor techniques have been applied to AdS (super)particles quite extensively, see, e.g., \cite{CGKRZ, CRZ, CKR, BLPS, Z, Cm, Cm2, ABGT, ABGT2, U, U2, Adamo:2016rtr}. 
In these approaches, one typically introduces (super)twistor variables by solving the mass-shell constraint, as described in, e.g., \cite{Mezincescu:2013nta}. 
Our approach differs in the sense that our models are derived from bi-supertwistors, which are the analogues of embedding coordinates for AdS superspace. 
In certain cases in the literature, e.g., \cite{U2}, the off-shell $\k$-symmetry of the superparticle has been expressed in terms of odd supertwistors. 
It is an interesting question to see how such symmetries originate within our setting. 

A natural extension of the above considerations is to the construction of models for superstrings propagating in AdS superspace. 
Such models can also be constructed within both the supergravity setting and the embedding formalism, and therefore possess deformations analogous to those we presented for superparticles. 
Supertwistor techniques have been applied to superstrings in, e.g., \cite{Uvarov:2007vs,Roiban:2000yy}. 
Of special interest is the extension of the interval deformation to AdS$_5 \times S^{5}$ superspace, which can be constructed in terms of the bi-supertwistor realisation detailed in chapter \ref{ch4}.

It is also interesting to consider AdS superspaces in other dimensions. In chapter \ref{ch5}, we extended our superparticle analysis to two dimensions to derive new models in the two-dimensional $(\N\,,\N)$ AdS supergeometry. 
We restricted our analysis to the realisation of $(\N\,,\N)$ AdS superspace within the supergravity setting, as, to the best of our knowledge, an embedding formalism for this AdS superspace is yet to be developed in the literature. 
The development of such an embedding formalism is left for future work.

It is well known that, in (super)conformal field theories, the structure of two and three-point functions are highly constrained by the (super)conformal symmetry, see, e.g., \cite{Park:1997bq,Park:1999pd,Poland:2018epd,Stone:2023ctp} for an incomplete list of references. 
Our formalism for AdS$^{4|4\N}$ and AdS$^{5|8\N}$ provides an interesting way to explore such considerations in an AdS/CFT context, similar in spirit to \cite{Binder:2020raz,Binder:2021euo}. 
In particular, one can naturally construct manifestly AdS-supersymmetric $n$-point functions in terms of the bi-supertwistors for AdS$^{4|4\N}$ and AdS$^{5|8\N}$.

In this thesis we have focused on anti-de Sitter supergeometries and corresponding superspace techniques. 
It is interesting to explore their analogues and applications to the (non-supersymmetric) de Sitter spacetimes. 
Surveys of quantum field theoretic techniques in de Sitter space can be found in, e.g., \cite{Gazeau:2006gq,Akhmedov:2013vka}. 
 Further, in recent years, there has been a surge of interest in generalised superalgebras and their applications to phenomenology, see, e.g., \cite{Aizawa:2023lhx,Kuznetsova:2021byk,Aizawa:2020ovt,Aizawa:2020xni,Toppan:2020rjz} and references therein.
 Actually, in the mid-1980s, Vasiliev already constructed a de Sitter supergravity theory with positive cosmological constant, based on precisely such a generalised superalgebra \cite{Vasiliev:1985vtv}.\footnote{See also \cite{Pilch:1984aw} and \cite{Bergshoeff:2015tra}, and references therein, for more on de Sitter supergravity.}  
 It is interesting to see whether an analogous embedding formalism can be developed in the framework of these generalised superalgebras.

 
\begin{footnotesize}
\bibliography{chapters/bibliography/bibliography}{}

@article{KK3,
    author = "Koning, Nowar E. and Kuzenko, Sergei M.",
    title = "{Anti{\textendash}de Sitter flag superspace}",
    eprint = "2512.14347",
    archivePrefix = "arXiv",
    primaryClass = "hep-th",
    doi = "10.1103/r56q-l5cx",
    journal = "Phys. Rev. D",
    volume = "113",
    number = "6",
    pages = "065019",
    year = "2026"
}

@article{KRsp,
    author = "Koning, Nowar E. and Raptakis, Emmanouil S. N.",
    title = "{New superparticle models in AdS superspaces}",
    eprint = "2506.17897",
    archivePrefix = "arXiv",
    primaryClass = "hep-th",
    doi = "10.1007/JHEP10(2025)099",
    journal = "JHEP",
    volume = "10",
    pages = "099",
    year = "2025"
}

@article{KR,
    author = "Kuzenko, Sergei M. and Raptakis, Emmanouil S. N.",
    title = "{Conformal (p, q) supergeometries in two dimensions}",
    eprint = "2211.16169",
    archivePrefix = "arXiv",
    primaryClass = "hep-th",
    doi = "10.1007/JHEP02(2023)166",
    journal = "JHEP",
    volume = "02",
    pages = "166",
    year = "2023"
}

@article{Keck,
    author = "Keck, B. W.",
    title = "{An Alternative Class of Supersymmetries}",
    reportNumber = "THEP 74/5-2",
    doi = "10.1088/0305-4470/8/11/018",
    journal = "J. Phys. A",
    volume = "8",
    pages = "1819--1827",
    year = "1975"
}

@article{Zumino,
    author = "Zumino, Bruno",
    title = "{Nonlinear Realization of Supersymmetry in de Sitter Space}",
    reportNumber = "CERN-TH-2316",
    doi = "10.1016/0550-3213(77)90211-5",
    journal = "Nucl. Phys. B",
    volume = "127",
    pages = "189--201",
    year = "1977"
}

@article{KTM,
    author = "Kuzenko, Sergei M. and Tartaglino-Mazzucchelli, Gabriele",
    title = "{Supertwistor realisations of AdS superspaces}",
    eprint = "2108.03907",
    archivePrefix = "arXiv",
    primaryClass = "hep-th",
    doi = "10.1140/epjc/s10052-022-10072-y",
    journal = "Eur. Phys. J. C",
    volume = "82",
    number = "2",
    pages = "146",
    year = "2022"
}

@article{KK,
    author = "Koning, Nowar E. and Kuzenko, Sergei M.",
    title = "{Embedding formalism for AdS superspaces in five dimensions}",
    eprint = "2406.10875",
    archivePrefix = "arXiv",
    primaryClass = "hep-th",
    doi = "10.1007/JHEP06(2025)016",
    journal = "JHEP",
    volume = "06",
    pages = "016",
    year = "2025"
}

@article{KT,
    author = "Kuzenko, Sergei M. and Turner, Kai",
    title = "{Embedding formalism for (p, q) AdS superspaces in three dimensions}",
    eprint = "2303.03082",
    archivePrefix = "arXiv",
    primaryClass = "hep-th",
    doi = "10.1007/JHEP06(2023)142",
    journal = "JHEP",
    volume = "06",
    pages = "142",
    year = "2023"
}

@article{KKR,
    author = "Koning, Nowar E. and Kuzenko, Sergei M. and Raptakis, Emmanouil S. N.",
    title = "{Embedding formalism for $ \mathcal{N} $-extended AdS superspace in four dimensions}",
    eprint = "2308.04135",
    archivePrefix = "arXiv",
    primaryClass = "hep-th",
    doi = "10.1007/JHEP11(2023)063",
    journal = "JHEP",
    volume = "11",
    pages = "063",
    year = "2023"
}

@article{KKR2,
    author = "Koning, Nowar E. and Kuzenko, Sergei M. and Raptakis, Emmanouil S. N.",
    title = "{The anti-de Sitter supergeometry revisited}",
    eprint = "2412.03172",
    archivePrefix = "arXiv",
    primaryClass = "hep-th",
    doi = "10.1007/JHEP02(2025)175",
    journal = "JHEP",
    volume = "02",
    pages = "175",
    year = "2025"
}

@article{KTM2,
    author = "Kuzenko, Sergei M. and Tartaglino-Mazzucchelli, Gabriele",
    title = "{Conformally flat supergeometry in five dimensions}",
    eprint = "0804.1219",
    archivePrefix = "arXiv",
    primaryClass = "hep-th",
    doi = "10.1088/1126-6708/2008/06/097",
    journal = "JHEP",
    volume = "06",
    pages = "097",
    year = "2008"
}

@book{PW,
    author = "West, Peter",
    title = "{Introduction to strings and branes}",
    isbn = "978-0-521-81747-9, 978-1-139-41529-3, 978-0-521-81747-9",
    publisher = "Cambridge University Press",
    month = "7",
    year = "2012"
}

@article{BS,
    author = "Brink, L. and Schwarz, J. H.",
    title = "{Quantum Superspace}",
    doi = "10.1016/0370-2693(81)90093-9",
    journal = "Phys. Lett. B",
    volume = "100",
    pages = "310--312",
    year = "1981"
}

@article{C,
    author = "Casalbuoni, R.",
    title = "{Relativity and Supersymmetries}",
    reportNumber = "Print-76-0053 (FLORENCE)",
    doi = "10.1016/0370-2693(76)90044-7",
    journal = "Phys. Lett. B",
    volume = "62",
    pages = "49--50",
    year = "1976"
}

@article{DS,
    author = "Sorokin, Dmitri P.",
    title = "{Superbranes and superembeddings}",
    eprint = "hep-th/9906142",
    archivePrefix = "arXiv",
    reportNumber = "HUB-EP-99-26, DFPD-99-TH-25",
    doi = "10.1016/S0370-1573(99)00104-0",
    journal = "Phys. Rept.",
    volume = "329",
    pages = "1--101",
    year = "2000"
}

@inbook{BDS,
    author = "Bandos, Igor A. and Sorokin, Dmitri P.",
    title = "{Superembedding Approach to Superstrings and Super-p-branes}",
    eprint = "2301.10668",
    archivePrefix = "arXiv",
    primaryClass = "hep-th",
    year = "2024"
}

@article{BHT,
    author = "Beccaria, Matteo and Jiang, Hongliang and Tseytlin, Arkady A.",
    title = "{Supersymmetric Liouville theory in AdS$_{2}$ and AdS/CFT}",
    eprint = "1909.10255",
    archivePrefix = "arXiv",
    primaryClass = "hep-th",
    reportNumber = "Imperial-TP-AT-2019-07",
    doi = "10.1007/JHEP11(2019)051",
    journal = "JHEP",
    volume = "11",
    pages = "051",
    year = "2019"
}

@book{WB,
    author = "Wess, J. and Bagger, J.",
    title = "{Supersymmetry and supergravity}",
    isbn = "978-0-691-02530-8",
    publisher = "Princeton University Press",
    address = "Princeton, NJ, USA",
    year = "1992"
}

@book{BK,
    author = "Buchbinder, I. L. and Kuzenko, S. M.",
    title = "{Ideas and Methods of Supersymmetry and Supergravity or A Walk Through Superspace: A Walk Through Superspace}",
    isbn = "978-1-4200-5051-6, 978-0-7503-0506-8, 978-0-367-80253-0",
    publisher = "IOP, Bristol",
    year = "1995, Revised Edition: 1998"
}

@article{KLTM2,
    author = "Kuzenko, Sergei M. and Lindstrom, Ulf and Tartaglino-Mazzucchelli, Gabriele",
    title = "{Off-shell supergravity-matter couplings in three dimensions}",
    eprint = "1101.4013",
    archivePrefix = "arXiv",
    primaryClass = "hep-th",
    reportNumber = "UUITP-39-10",
    doi = "10.1007/JHEP03(2011)120",
    journal = "JHEP",
    volume = "03",
    pages = "120",
    year = "2011"
}

@article{HIPT,
    author = "Howe, Paul S. and Izquierdo, J. M. and Papadopoulos, G. and Townsend, P. K.",
    title = "{New supergravities with central charges and Killing spinors in (2+1)-dimensions}",
    eprint = "hep-th/9505032",
    archivePrefix = "arXiv",
    reportNumber = "DAMTP-R-95-13",
    doi = "10.1016/0550-3213(96)00091-0",
    journal = "Nucl. Phys. B",
    volume = "467",
    pages = "183--214",
    year = "1996"
}

@article{BILS,
    author = "Bandos, Igor A. and Ivanov, Evgeny and Lukierski, Jerzy and Sorokin, Dmitri",
    title = "{On the superconformal flatness of AdS superspaces}",
    eprint = "hep-th/0205104",
    archivePrefix = "arXiv",
    reportNumber = "DFPD-02-TH-07, IFIC-02-16, FTUV-02-1904",
    doi = "10.1088/1126-6708/2002/06/040",
    journal = "JHEP",
    volume = "06",
    pages = "040",
    year = "2002"
}

@article{CGKRZ,
    author = "Claus, Piet and Gunaydin, Murat and Kallosh, Renata and Rahmfeld, J. and Zunger, Yonatan",
    title = "{Supertwistors as quarks of SU(2, 2|4)}",
    eprint = "hep-th/9905112",
    archivePrefix = "arXiv",
    reportNumber = "SU-ITP-99-22, KUL-TF-99-16, PSU-TH-208",
    doi = "10.1088/1126-6708/1999/05/019",
    journal = "JHEP",
    volume = "05",
    pages = "019",
    year = "1999"
}

@article{CRZ,
    author = "Claus, Piet and Rahmfeld, J. and Zunger, Yonatan",
    title = "{A Simple particle action from a twistor parametrization of AdS(5)}",
    eprint = "hep-th/9906118",
    archivePrefix = "arXiv",
    reportNumber = "SU-ITP-99-27, KUL-TF-99-21",
    doi = "10.1016/S0370-2693(99)01128-4",
    journal = "Phys. Lett. B",
    volume = "466",
    pages = "181--189",
    year = "1999"
}

@article{CKR,
    author = "Claus, Piet and Kallosh, Renata and Rahmfeld, J.",
    title = "{BRST quantization of a particle in AdS(5)}",
    eprint = "hep-th/9906195",
    archivePrefix = "arXiv",
    reportNumber = "SU-ITP-99-30, KUL-TF-99-25",
    doi = "10.1016/S0370-2693(99)00931-4",
    journal = "Phys. Lett. B",
    volume = "462",
    pages = "285--293",
    year = "1999"
}

@article{BLPS,
    author = "Bandos, Igor A. and Lukierski, Jerzy and Preitschopf, Christian and Sorokin, Dmitri P.",
    title = "{OSp supergroup manifolds, superparticles and supertwistors}",
    eprint = "hep-th/9907113",
    archivePrefix = "arXiv",
    reportNumber = "HUB-EP-99-32, TUW-99-15, MPI-PHT-99-28",
    doi = "10.1103/PhysRevD.61.065009",
    journal = "Phys. Rev. D",
    volume = "61",
    pages = "065009",
    year = "2000"
}

@article{Z,
    author = "Zunger, Yonatan",
    title = "{Twistors and actions on coset manifolds}",
    eprint = "hep-th/0001072",
    archivePrefix = "arXiv",
    reportNumber = "SU-ITP-00-01",
    doi = "10.1103/PhysRevD.62.024030",
    journal = "Phys. Rev. D",
    volume = "62",
    pages = "024030",
    year = "2000"
}

@article{Cm,
    author = "Cederwall, Martin",
    title = "{Geometric construction of AdS twistors}",
    eprint = "hep-th/0002216",
    archivePrefix = "arXiv",
    doi = "10.1016/S0370-2693(00)00552-9",
    journal = "Phys. Lett. B",
    volume = "483",
    pages = "257--263",
    year = "2000"
}

@article{Cm2,
    author = "Cederwall, Martin",
    editor = "Lukierski, Jerzy and Sorokin, Dimitri",
    title = "{AdS twistors for higher spin theory}",
    eprint = "hep-th/0412222",
    archivePrefix = "arXiv",
    doi = "10.1063/1.1923331",
    journal = "AIP Conf. Proc.",
    volume = "767",
    number = "1",
    pages = "96--105",
    year = "2005"
}

@article{ABGT,
    author = "Arvanitakis, Alex S. and Barns-Graham, Alec E. and Townsend, Paul K.",
    title = "{Anti\textendash{}de Sitter Particles and Manifest (Super)Isometries}",
    eprint = "1608.04380",
    archivePrefix = "arXiv",
    primaryClass = "hep-th",
    reportNumber = "DAMTP-2016-56",
    doi = "10.1103/PhysRevLett.118.141601",
    journal = "Phys. Rev. Lett.",
    volume = "118",
    number = "14",
    pages = "141601",
    year = "2017"
}

@article{ABGT2,
    author = "Arvanitakis, Alex S. and Barns-Graham, Alec E. and Townsend, Paul K.",
    title = "{Twistor description of spinning particles in AdS}",
    eprint = "1710.09557",
    archivePrefix = "arXiv",
    primaryClass = "hep-th",
    reportNumber = "DAMTP-2017-35, IMPERIAL-TP-2017-ASA-01",
    doi = "10.1007/JHEP01(2018)059",
    journal = "JHEP",
    volume = "01",
    pages = "059",
    year = "2018"
}

@article{U,
    author = "Uvarov, D. V.",
    title = "{Supertwistor formulation for massless superparticle in $AdS_5\times S^5$ superspace}",
    eprint = "1807.08318",
    archivePrefix = "arXiv",
    primaryClass = "hep-th",
    doi = "10.1016/j.nuclphysb.2018.10.006",
    journal = "Nucl. Phys. B",
    volume = "936",
    pages = "690--713",
    year = "2018"
}

@article{U2,
    author = "Uvarov, D. V.",
    title = "{Multitwistor mechanics of massless superparticle on AdS5$\times$S5 superbackground}",
    eprint = "1907.13613",
    archivePrefix = "arXiv",
    primaryClass = "hep-th",
    doi = "10.1016/j.nuclphysb.2019.114830",
    journal = "Nucl. Phys. B",
    volume = "950",
    pages = "114830",
    year = "2020"
}

@article{KNTM,
    author = "Kuzenko, Sergei M. and Novak, Joseph and Tartaglino-Mazzucchelli, Gabriele",
    title = "{Higher derivative couplings and massive supergravity in three dimensions}",
    eprint = "1506.09063",
    archivePrefix = "arXiv",
    primaryClass = "hep-th",
    doi = "10.1007/JHEP09(2015)081",
    journal = "JHEP",
    volume = "09",
    pages = "081",
    year = "2015"
}

@article{GST,
    author = "Gunaydin, M. and Sierra, G. and Townsend, P. K.",
    title = "{The Unitary Supermultiplets of $d=3$ Anti-de Sitter and $d=2$ Conformal Superalgebras}",
    reportNumber = "UCRL-93923",
    doi = "10.1016/0550-3213(86)90293-2",
    journal = "Nucl. Phys. B",
    volume = "274",
    pages = "429--447",
    year = "1986"
}

@article{Nahm:1977tg,
    author = "Nahm, W.",
    title = "{Supersymmetries and Their Representations}",
    reportNumber = "CERN-TH-2341",
    doi = "10.1201/9781482268737-2",
    journal = "Nucl. Phys. B",
    volume = "135",
    pages = "149",
    year = "1978"
}

@article{Kuzenko:2008wr,
    author = "Kuzenko, Sergei M. and Tartaglino-Mazzucchelli, Gabriele",
    title = "{Super-Weyl invariance in 5D supergravity}",
    eprint = "0802.3953",
    archivePrefix = "arXiv",
    primaryClass = "hep-th",
    doi = "10.1088/1126-6708/2008/04/032",
    journal = "JHEP",
    volume = "04",
    pages = "032",
    year = "2008"
}

@article{Kac:1977em,
    author = "Kac, V. G.",
    title = "{Lie Superalgebras}",
    doi = "10.1016/0001-8708(77)90017-2",
    journal = "Adv. Math.",
    volume = "26",
    pages = "8--96",
    year = "1977"
}

@article{Kugo:2000hn,
    author = "Kugo, Taichiro and Ohashi, Keisuke",
    title = "{Supergravity tensor calculus in 5-D from 6-D}",
    eprint = "hep-ph/0006231",
    archivePrefix = "arXiv",
    reportNumber = "KUNS-1672",
    doi = "10.1143/PTP.104.835",
    journal = "Prog. Theor. Phys.",
    volume = "104",
    pages = "835--865",
    year = "2000"
}

@article{Kugo:2000af,
    author = "Kugo, Taichiro and Ohashi, Keisuke",
    title = "{Off-shell D = 5 supergravity coupled to matter Yang-Mills system}",
    eprint = "hep-ph/0010288",
    archivePrefix = "arXiv",
    reportNumber = "KUNS-1693",
    doi = "10.1143/PTP.105.323",
    journal = "Prog. Theor. Phys.",
    volume = "105",
    pages = "323--353",
    year = "2001"
}

@article{Fujita:2001kv,
    author = "Fujita, Tomoyuki and Ohashi, Keisuke",
    title = "{Superconformal tensor calculus in five-dimensions}",
    eprint = "hep-th/0104130",
    archivePrefix = "arXiv",
    reportNumber = "KUNS-1716",
    doi = "10.1143/PTP.106.221",
    journal = "Prog. Theor. Phys.",
    volume = "106",
    pages = "221--247",
    year = "2001"
}

@article{Kugo:2002vc,
    author = "Kugo, Taichiro and Ohashi, Keisuke",
    title = "{Gauge and nongauge tensor multiplets in 5-D conformal supergravity}",
    eprint = "hep-th/0208082",
    archivePrefix = "arXiv",
    doi = "10.1143/PTP.108.1143",
    journal = "Prog. Theor. Phys.",
    volume = "108",
    pages = "1143--1164",
    year = "2003"
}

@article{Bergshoeff:2001hc,
    author = "Bergshoeff, Eric and de Wit, Tim and Halbersma, Rein and Cucu, Sorin and Derix, Martijn and Van Proeyen, Antoine",
    title = "{Weyl multiplets of N=2 conformal supergravity in five-dimensions}",
    eprint = "hep-th/0104113",
    archivePrefix = "arXiv",
    reportNumber = "UG-01-01, KUL-TF-01-11",
    doi = "10.1088/1126-6708/2001/06/051",
    journal = "JHEP",
    volume = "06",
    pages = "051",
    year = "2001"
}

@article{Bergshoeff:2002qk,
    author = "Bergshoeff, Eric and Cucu, Sorin and De Wit, Tim and Gheerardyn, Jos and Halbersma, Rein and Vandoren, Stefan and Van Proeyen, Antoine",
    title = "{Superconformal N=2, D = 5 matter with and without actions}",
    eprint = "hep-th/0205230",
    archivePrefix = "arXiv",
    reportNumber = "UG-02-37, KUL-TF-02-04, SPIN-2002-14, ITP-UU-02-22",
    doi = "10.1088/1126-6708/2002/10/045",
    journal = "JHEP",
    volume = "10",
    pages = "045",
    year = "2002"
}

@article{Bergshoeff:2004kh,
    author = "Bergshoeff, Eric and Cucu, Sorin and de Wit, Tim and Gheerardyn, Jos and Vandoren, Stefan and Van Proeyen, Antoine",
    title = "{N = 2 supergravity in five-dimensions revisited}",
    eprint = "hep-th/0403045",
    archivePrefix = "arXiv",
    reportNumber = "UG-03-08, KUL-TF-04-06, SPIN-03-39, ITP-UU-03-58",
    doi = "10.1088/0264-9381/23/23/C01",
    journal = "Class. Quant. Grav.",
    volume = "21",
    pages = "3015--3042",
    year = "2004"
}

@article{Butter:2014xxa,
    author = "Butter, Daniel and Kuzenko, Sergei M. and Novak, Joseph and Tartaglino-Mazzucchelli, Gabriele",
    title = "{Conformal supergravity in five dimensions: New approach and applications}",
    eprint = "1410.8682",
    archivePrefix = "arXiv",
    primaryClass = "hep-th",
    reportNumber = "NIKHEF-2014-046",
    doi = "10.1007/JHEP02(2015)111",
    journal = "JHEP",
    volume = "02",
    pages = "111",
    year = "2015"
}

@article{Howe:1981nz,
    author = "Howe, Paul S. and Lindstrom, Ulf",
    title = "{The Supercurrent in Five-dimensions}",
    reportNumber = "CERN-TH-3075",
    doi = "10.1016/0370-2693(81)90074-5",
    journal = "Phys. Lett. B",
    volume = "103",
    pages = "422--426",
    year = "1981"
}

@article{Kuzenko:2007cj,
    author = "Kuzenko, Sergei M. and Tartaglino-Mazzucchelli, Gabriele",
    title = "{Five-dimensional Superfield Supergravity}",
    eprint = "0710.3440",
    archivePrefix = "arXiv",
    primaryClass = "hep-th",
    doi = "10.1016/j.physletb.2008.01.055",
    journal = "Phys. Lett. B",
    volume = "661",
    pages = "42--51",
    year = "2008"
}

@article{Kuzenko:2007hu,
    author = "Kuzenko, Sergei M. and Tartaglino-Mazzucchelli, Gabriele",
    title = "{5D Supergravity and Projective Superspace}",
    eprint = "0712.3102",
    archivePrefix = "arXiv",
    primaryClass = "hep-th",
    doi = "10.1088/1126-6708/2008/02/004",
    journal = "JHEP",
    volume = "02",
    pages = "004",
    year = "2008"
}

@article{Kuzenko:2007aj,
    author = "Kuzenko, Sergei M. and Tartaglino-Mazzucchelli, Gabriele",
    title = "{Five-dimensional N = 1 AdS superspace: Geometry, off-shell multiplets and dynamics}",
    eprint = "0704.1185",
    archivePrefix = "arXiv",
    primaryClass = "hep-th",
    doi = "10.1016/j.nuclphysb.2007.06.014",
    journal = "Nucl. Phys. B",
    volume = "785",
    pages = "34--73",
    year = "2007"
}

@article{Sinkovics:2004fm,
    author = "Sinkovics, Annamaria and Verlinde, Erik P.",
    title = "{A Six dimensional view on twistors}",
    eprint = "hep-th/0410014",
    archivePrefix = "arXiv",
    reportNumber = "ITFA-2004-42",
    doi = "10.1016/j.physletb.2004.12.033",
    journal = "Phys. Lett. B",
    volume = "608",
    pages = "142--150",
    year = "2005"
}

@article{Adamo:2016rtr,
    author = "Adamo, Tim and Skinner, David and Williams, Jack",
    title = "{Twistor methods for AdS$_{5}$}",
    eprint = "1607.03763",
    archivePrefix = "arXiv",
    primaryClass = "hep-th",
    reportNumber = "DAMTP-2016-49",
    doi = "10.1007/JHEP08(2016)167",
    journal = "JHEP",
    volume = "08",
    pages = "167",
    year = "2016"
}

@article{Roiban:2000yy,
    author = "Roiban, R. and Siegel, W.",
    title = "{Superstrings on AdS(5) x S**5 supertwistor space}",
    eprint = "hep-th/0010104",
    archivePrefix = "arXiv",
    reportNumber = "YITP-SB-00-64",
    doi = "10.1088/1126-6708/2000/11/024",
    journal = "JHEP",
    volume = "11",
    pages = "024",
    year = "2000"
}

@article{Ferber:1977qx,
    author = "Ferber, Alan",
    title = "{Supertwistors and Conformal Supersymmetry}",
    reportNumber = "EFI 77/50-CHICAGO",
    doi = "10.1016/0550-3213(78)90257-2",
    journal = "Nucl. Phys. B",
    volume = "132",
    pages = "55--64",
    year = "1978"
}

@article{Kotrla:1984ky,
    author = "Kotrla, M. and Niederle, J.",
    title = "{Supertwistors and Superspace}",
    reportNumber = "IC/84/237",
    doi = "10.1007/BF01595531",
    journal = "Czech. J. Phys. B",
    volume = "35",
    pages = "602",
    year = "1985"
}

@article{Howe:1994ms,
    author = "Howe, Paul S. and Leeming, M. I.",
    title = "{Harmonic superspaces in low dimensions}",
    eprint = "hep-th/9408062",
    archivePrefix = "arXiv",
    reportNumber = "KCL-TH-94-15",
    doi = "10.1088/0264-9381/11/12/004",
    journal = "Class. Quant. Grav.",
    volume = "11",
    pages = "2843--2852",
    year = "1994"
}

@article{Kuzenko:2010rp,
    author = "Kuzenko, Sergei M. and Park, Jeong-Hyuck and Tartaglino-Mazzucchelli, Gabriele and Unge, Rikard",
    title = "{Off-shell superconformal nonlinear sigma-models in three dimensions}",
    eprint = "1011.5727",
    archivePrefix = "arXiv",
    primaryClass = "hep-th",
    reportNumber = "UUITP-36-10",
    doi = "10.1007/JHEP01(2011)146",
    journal = "JHEP",
    volume = "01",
    pages = "146",
    year = "2011"
}

@article{Rosly:1985nyf,
    author = "Rosly, A. A.",
    title = "{Gauge Fields in Superspace and Twistors}",
    doi = "10.1088/0264-9381/2/5/011",
    journal = "Class. Quant. Grav.",
    volume = "2",
    pages = "693--699",
    year = "1985"
}

@article{Lukierski:1988vw,
    author = "Lukierski, Jerzy and Nowicki, Anatol",
    title = "{General Superspaces From Supertwistors}",
    reportNumber = "UWR-88-702",
    doi = "10.1016/0370-2693(88)90903-3",
    journal = "Phys. Lett. B",
    volume = "211",
    pages = "276--280",
    year = "1988"
}

@article{Howe:1995md,
    author = "Howe, Paul S. and Hartwell, G. G.",
    title = "{A Superspace survey}",
    doi = "10.1088/0264-9381/12/8/005",
    journal = "Class. Quant. Grav.",
    volume = "12",
    pages = "1823--1880",
    year = "1995"
}

@article{Kuzenko:2006mv,
    author = "Kuzenko, Sergei M.",
    title = "{On compactified harmonic/projective superspace, 5-D superconformal theories, and all that}",
    eprint = "hep-th/0601177",
    archivePrefix = "arXiv",
    doi = "10.1016/j.nuclphysb.2006.03.019",
    journal = "Nucl. Phys. B",
    volume = "745",
    pages = "176--207",
    year = "2006"
}

@article{Buchbinder:2015qsa,
    author = "Buchbinder, Evgeny I. and Kuzenko, Sergei M. and Samsonov, Igor B.",
    title = "{Superconformal field theory in three dimensions: Correlation functions of conserved currents}",
    eprint = "1503.04961",
    archivePrefix = "arXiv",
    primaryClass = "hep-th",
    doi = "10.1007/JHEP06(2015)138",
    journal = "JHEP",
    volume = "06",
    pages = "138",
    year = "2015"
}

@article{Howe:2020xrg,
    author = {Howe, P. S. and Lindstr{\"o}m, U.},
    title = "{Local supertwistors and conformal supergravity in six dimensions}",
    eprint = "2008.10302",
    archivePrefix = "arXiv",
    primaryClass = "hep-th",
    reportNumber = "Uppsala University: UUITP-28/20, Imperial College:
  Imperial-TP-2020-UL-03",
    doi = "10.1098/rspa.2020.0683",
    journal = "Proc. Roy. Soc. Lond. A",
    volume = "476",
    number = "2243",
    pages = "20200683",
    year = "2020"
}

@article{Howe:2020hxi,
    author = {Howe, P. S. and Lindstr{\"o}m, U.},
    title = "{Superconformal geometries and local twistors}",
    eprint = "2012.03282",
    archivePrefix = "arXiv",
    primaryClass = "hep-th",
    reportNumber = "Uppsala University Theoretical Physics: UUITP-49/20",
    doi = "10.1007/JHEP04(2021)140",
    journal = "JHEP",
    volume = "04",
    pages = "140",
    year = "2021"
}

@article{Penrose:1967wn,
    author = "Penrose, R.",
    title = "{Twistor algebra}",
    doi = "10.1063/1.1705200",
    journal = "J. Math. Phys.",
    volume = "8",
    pages = "345",
    year = "1967"
}

@article{Penrose:1972ia,
    author = "Penrose, R. and MacCallum, Malcolm A. H.",
    title = "{Twistor theory: An Approach to the quantization of fields and space-time}",
    doi = "10.1016/0370-1573(73)90008-2",
    journal = "Phys. Rept.",
    volume = "6",
    pages = "241--316",
    year = "1972"
}

@article{Kuzenko:1995aq,
    author = "Kuzenko, S. M. and Lyakhovich, S. L. and Segal, A. Yu. and Sharapov, A. A.",
    title = "{Massive spinning particle on anti-de Sitter space}",
    eprint = "hep-th/9509062",
    archivePrefix = "arXiv",
    reportNumber = "ITP-UH-24-95, DESY-95-195",
    doi = "10.1142/S0217751X96001589",
    journal = "Int. J. Mod. Phys. A",
    volume = "11",
    pages = "3307--3330",
    year = "1996"
}

@article{Kuzenko:2001ag,
    author = "Kuzenko, S. M. and McArthur, Ian N.",
    title = "{Goldstone multiplet for partially broken superconformal symmetry}",
    eprint = "hep-th/0109183",
    archivePrefix = "arXiv",
    doi = "10.1016/S0370-2693(01)01294-1",
    journal = "Phys. Lett. B",
    volume = "522",
    pages = "320--326",
    year = "2001"
}

@article{Metsaev:1998it,
    author = "Metsaev, R. R. and Tseytlin, Arkady A.",
    title = "{Type IIB superstring action in AdS(5) x S**5 background}",
    eprint = "hep-th/9805028",
    archivePrefix = "arXiv",
    reportNumber = "FIAN-TD-98-21, IMPERIAL-TP-97-98-44, NSF-ITP-98-055",
    doi = "10.1016/S0550-3213(98)00570-7",
    journal = "Nucl. Phys. B",
    volume = "533",
    pages = "109--126",
    year = "1998"
}

@article{Zumino:1962smg,
    author = "Zumino, Bruno",
    title = "{Normal Forms of Complex Matrices}",
    doi = "10.1063/1.1724294",
    journal = "J. Math. Phys.",
    volume = "3",
    number = "5",
    pages = "1055",
    year = "1962"
}

@article{IS,
    author = "Ivanov, E. A. and Sorin, Alexander Savelievich",
    title = "{Superfield Formulation of OSp(1,4) Supersymmetry}",
    doi = "10.1088/0305-4470/13/4/013",
    journal = "J. Phys. A",
    volume = "13",
    pages = "1159--1188",
    year = "1980"
}

@article{WZ,
    author = "Wess, J. and Zumino, B.",
    title = "{Superfield Lagrangian for Supergravity}",
    reportNumber = "CERN-TH-2453",
    doi = "10.1016/0370-2693(78)90057-6",
    journal = "Phys. Lett. B",
    volume = "74",
    pages = "51--53",
    year = "1978"
}

@article{Stelle:1978ye,
    author = "Stelle, K. S. and West, Peter C.",
    title = "{Minimal Auxiliary Fields for Supergravity}",
    reportNumber = "ICTP/77-78/6",
    doi = "10.1016/0370-2693(78)90669-X",
    journal = "Phys. Lett. B",
    volume = "74",
    pages = "330--332",
    year = "1978"
}

@article{Ferrara:1978em,
    author = "Ferrara, S. and van Nieuwenhuizen, P.",
    title = "{The Auxiliary Fields of Supergravity}",
    reportNumber = "CERN-TH-2463",
    doi = "10.1016/0370-2693(78)90670-6",
    journal = "Phys. Lett. B",
    volume = "74",
    pages = "333",
    year = "1978"
}

@article{Townsend:1977qa,
    author = "Townsend, P. K.",
    title = "{Cosmological Constant in Supergravity}",
    doi = "10.1103/PhysRevD.15.2802",
    journal = "Phys. Rev. D",
    volume = "15",
    pages = "2802--2804",
    year = "1977"
}

@article{Kaku:1978ea,
    author = "Kaku, M. and Townsend, P. K.",
    title = "{Poincare Supergravity as Broken Superconformal Gravity}",
    reportNumber = "ITP-SB-78-6",
    doi = "10.1016/0370-2693(78)90098-9",
    journal = "Phys. Lett. B",
    volume = "76",
    pages = "54--58",
    year = "1978"
}

@book{Gates:1983nr,
    author = "Gates, S. J. and Grisaru, Marcus T. and Rocek, M. and Siegel, W.",
    title = "{Superspace Or One Thousand and One Lessons in Supersymmetry}",
    eprint = "hep-th/0108200",
    archivePrefix = "arXiv",
    reportNumber = "YITP-SB-01-53",
    isbn = "978-0-8053-3161-5",
    series = "Frontiers in Physics",
    volume = "58",
    year = "1983"
}

@article{Butter:2011vg,
    author = "Butter, Daniel and Kuzenko, Sergei M.",
    title = "{A dual formulation of supergravity-matter theories}",
    eprint = "1106.3038",
    archivePrefix = "arXiv",
    primaryClass = "hep-th",
    doi = "10.1016/j.nuclphysb.2011.08.014",
    journal = "Nucl. Phys. B",
    volume = "854",
    pages = "1--27",
    year = "2012"
}

@article{Kuzenko:2008ep,
    author = "Kuzenko, S. M. and Lindstrom, U. and Rocek, M. and Tartaglino-Mazzucchelli, G.",
    title = "{4D N = 2 Supergravity and Projective Superspace}",
    eprint = "0805.4683",
    archivePrefix = "arXiv",
    primaryClass = "hep-th",
    reportNumber = "UUITP-07-08, YITP-SB-08-27",
    doi = "10.1088/1126-6708/2008/09/051",
    journal = "JHEP",
    volume = "09",
    pages = "051",
    year = "2008"
}

@article{Kuzenko:2008qw,
    author = "Kuzenko, Sergei M. and Tartaglino-Mazzucchelli, Gabriele",
    title = "{Field theory in 4D N=2 conformally flat superspace}",
    eprint = "0807.3368",
    archivePrefix = "arXiv",
    primaryClass = "hep-th",
    doi = "10.1088/1126-6708/2008/10/001",
    journal = "JHEP",
    volume = "10",
    pages = "001",
    year = "2008"
}

@article{Butter:2011ym,
    author = "Butter, Daniel and Kuzenko, Sergei M.",
    title = "{N=2 AdS supergravity and supercurrents}",
    eprint = "1104.2153",
    archivePrefix = "arXiv",
    primaryClass = "hep-th",
    doi = "10.1007/JHEP07(2011)081",
    journal = "JHEP",
    volume = "07",
    pages = "081",
    year = "2011"
}

@article{Kuzenko:2023qkg,
    author = "Kuzenko, Sergei M. and Raptakis, Emmanouil S. N.",
    title = "{$ \mathcal{N} $ = 3 conformal superspace in four dimensions}",
    eprint = "2312.07242",
    archivePrefix = "arXiv",
    primaryClass = "hep-th",
    doi = "10.1007/JHEP03(2024)026",
    journal = "JHEP",
    volume = "03",
    pages = "026",
    year = "2024"
}

@article{Howe:1980sy,
    author = "Howe, Paul S.",
    title = "{A Superspace Approach to Extended Conformal Supergravity}",
    reportNumber = "CERN-TH-2996",
    doi = "10.1016/0370-2693(81)90143-X",
    journal = "Phys. Lett. B",
    volume = "100",
    pages = "389--392",
    year = "1981"
}

@article{Kuzenko:2009zu,
    author = "Kuzenko, S. M. and Lindstrom, U. and Rocek, M. and Tartaglino-Mazzucchelli, G.",
    title = "{On conformal supergravity and projective superspace}",
    eprint = "0905.0063",
    archivePrefix = "arXiv",
    primaryClass = "hep-th",
    reportNumber = "UUITP-13-09, YITP-SB-09-09, UMD-PP-09-035",
    doi = "10.1088/1126-6708/2009/08/023",
    journal = "JHEP",
    volume = "08",
    pages = "023",
    year = "2009"
}

@article{Grimm:1977kp,
    author = "Grimm, R. and Wess, J. and Zumino, B.",
    title = "{Consistency Checks on the Superspace Formulation of Supergravity}",
    reportNumber = "Print-78-0011 (KARLSRUHE)",
    doi = "10.1016/0370-2693(78)90753-0",
    journal = "Phys. Lett. B",
    volume = "73",
    pages = "415--417",
    year = "1978"
}

@article{Siegel:1978fc,
    author = "Siegel, Warren",
    title = "{Superconformal Invariance of Superspace With Nonminimal Auxiliary Fields}",
    reportNumber = "HUTP-78/A023",
    doi = "10.1016/0370-2693(79)90203-X",
    journal = "Phys. Lett. B",
    volume = "80",
    pages = "224--227",
    year = "1979"
}

@article{Volkov:1972jx,
    author = "Volkov, D. V. and Akulov, V. P.",
    editor = "Wess, J. and Akulov, V. P.",
    title = "{Possible universal neutrino interaction}",
    journal = "JETP Lett.",
    volume = "16",
    pages = "438--440",
    year = "1972"
}

@article{Volkov:1973ix,
    author = "Volkov, D. V. and Akulov, V. P.",
    title = "{Is the Neutrino a Goldstone Particle?}",
    doi = "10.1016/0370-2693(73)90490-5",
    journal = "Phys. Lett. B",
    volume = "46",
    pages = "109--110",
    year = "1973"
}

@article{Akulov:1974xz,
    author = "Akulov, V. P. and Volkov, D. V.",
    title = "{Goldstone fields with spin 1/2}",
    doi = "10.1007/BF01036922",
    journal = "Theor. Math. Phys.",
    volume = "18",
    pages = "28",
    year = "1974"
}

@article{KLT-M12,
    author = "Kuzenko, Sergei M. and Lindstrom, Ulf and Tartaglino-Mazzucchelli, Gabriele",
    title = "{Three-dimensional (p,q) AdS superspaces and matter couplings}",
    eprint = "1205.4622",
    archivePrefix = "arXiv",
    primaryClass = "hep-th",
    reportNumber = "UUITP-11-12",
    doi = "10.1007/JHEP08(2012)024",
    journal = "JHEP",
    volume = "08",
    pages = "024",
    year = "2012"
}

@article{Butter:2012jj,
    author = "Butter, Daniel and Kuzenko, Sergei M. and Lindstrom, Ulf and Tartaglino-Mazzucchelli, Gabriele",
    title = "{Extended supersymmetric sigma models in AdS$_4$ from projective superspace}",
    eprint = "1203.5001",
    archivePrefix = "arXiv",
    primaryClass = "hep-th",
    reportNumber = "UUITP-07-12",
    doi = "10.1007/JHEP05(2012)138",
    journal = "JHEP",
    volume = "05",
    pages = "138",
    year = "2012"
}

@article{kappa1,
    author = "Siegel, W.",
    title = "{Hidden Local Supersymmetry in the Supersymmetric Particle Action}",
    reportNumber = "UCB-PTH-83/7",
    doi = "10.1016/0370-2693(83)90924-3",
    journal = "Phys. Lett. B",
    volume = "128",
    pages = "397--399",
    year = "1983"
}

@article{kappa2,
    author = "Green, Michael B. and Schwarz, John H.",
    title = "{Covariant Description of Superstrings}",
    reportNumber = "QMC-83-7",
    doi = "10.1016/0370-2693(84)92021-5",
    journal = "Phys. Lett. B",
    volume = "136",
    pages = "367--370",
    year = "1984"
}

@inproceedings{Szg,
    author = "Sezgin, E.",
    title = "{Aspects of kappa symmetry}",
    booktitle = "{Conference on Highlights of Particle and Condensed Matter Physics (SALAMFEST)}",
    eprint = "hep-th/9310126",
    archivePrefix = "arXiv",
    reportNumber = "CTP-TAMU-28-93",
    month = "10",
    year = "1993"
}

@article{Haag:1974qh,
    author = "Haag, Rudolf and Lopuszanski, Jan T. and Sohnius, Martin",
    title = "{All Possible Generators of Supersymmetries of the s Matrix}",
    reportNumber = "Print-74-1630 (KARLSRUHE)",
    doi = "10.1016/0550-3213(75)90279-5",
    journal = "Nucl. Phys. B",
    volume = "88",
    pages = "257",
    year = "1975"
}

@article{Coleman:1967ad,
    author = "Coleman, Sidney R. and Mandula, J.",
    editor = "Zichichi, A.",
    title = "{All Possible Symmetries of the S Matrix}",
    doi = "10.1103/PhysRev.159.1251",
    journal = "Phys. Rev.",
    volume = "159",
    pages = "1251--1256",
    year = "1967"
}

@article{Fradkin:1985am,
    author = "Fradkin, E. S. and Tseytlin, Arkady A.",
    title = "{Conformal Supergravity}",
    doi = "10.1016/0370-1573(85)90138-3",
    journal = "Phys. Rept.",
    volume = "119",
    pages = "233--362",
    year = "1985"
}

@article{Park:1999pd,
    author = "Park, Jeong-Hyuck",
    title = "{Superconformal symmetry and correlation functions}",
    eprint = "hep-th/9903230",
    archivePrefix = "arXiv",
    reportNumber = "KIAS-99019",
    doi = "10.1016/S0550-3213(99)00432-0",
    journal = "Nucl. Phys. B",
    volume = "559",
    pages = "455--501",
    year = "1999"
}

@article{Kuzenko:1999pi,
    author = "Kuzenko, Sergei M. and Theisen, Stefan",
    title = "{Correlation functions of conserved currents in N=2 superconformal theory}",
    eprint = "hep-th/9907107",
    archivePrefix = "arXiv",
    reportNumber = "LMU-TPW-99-14, LMU-TPW-99-14",
    doi = "10.1088/0264-9381/17/3/307",
    journal = "Class. Quant. Grav.",
    volume = "17",
    pages = "665--696",
    year = "2000"
}

@article{Ferrara:1976fu,
    author = "Ferrara, S. and van Nieuwenhuizen, P.",
    title = "{Consistent Supergravity with Complex Spin 3/2 Gauge Fields}",
    reportNumber = "ITP-SB-76-48",
    doi = "10.1103/PhysRevLett.37.1669",
    journal = "Phys. Rev. Lett.",
    volume = "37",
    pages = "1669",
    year = "1976"
}

@article{Freedman:1976aw,
    author = "Freedman, Daniel Z. and Das, Ashok K.",
    title = "{Gauge Internal Symmetry in Extended Supergravity}",
    reportNumber = "ITP-SB-76-64",
    doi = "10.1016/0550-3213(77)90041-4",
    journal = "Nucl. Phys. B",
    volume = "120",
    pages = "221--230",
    year = "1977"
}

@article{K-compactified12,
    author = "Kuzenko, Sergei M.",
    title = "{Conformally compactified Minkowski superspaces revisited}",
    eprint = "1206.3940",
    archivePrefix = "arXiv",
    primaryClass = "hep-th",
    doi = "10.1007/JHEP10(2012)135",
    journal = "JHEP",
    volume = "10",
    pages = "135",
    year = "2012"
}

@article{Kuzenko:2014yia,
    author = "Kuzenko, Sergei M. and Sorokin, D.",
    title = "{Superconformal structures on the three-sphere}",
    eprint = "1406.7090",
    archivePrefix = "arXiv",
    primaryClass = "hep-th",
    doi = "10.1007/JHEP10(2014)080",
    journal = "JHEP",
    volume = "10",
    pages = "080",
    year = "2014"
}

@article{Binder:2020raz,
    author = "Binder, Damon J. and Freedman, Daniel Z. and Pufu, Silviu S.",
    title = "{A bispinor formalism for spinning Witten diagrams}",
    eprint = "2003.07448",
    archivePrefix = "arXiv",
    primaryClass = "hep-th",
    reportNumber = "PUPT-2617, MIT-CTP/5190",
    doi = "10.1007/JHEP02(2022)040",
    journal = "JHEP",
    volume = "02",
    pages = "040",
    year = "2022"
}

@article{Howe:1981gz,
    author = "Howe, Paul S.",
    title = "{Supergravity in Superspace}",
    reportNumber = "CERN-TH-3117",
    doi = "10.1016/0550-3213(82)90349-2",
    journal = "Nucl. Phys. B",
    volume = "199",
    pages = "309--364",
    year = "1982"
}

@article{Grimm:1978ch,
    author = "Grimm, R. and Wess, J. and Zumino, B.",
    title = "{A Complete Solution of the Bianchi Identities in Superspace}",
    reportNumber = "PRINT-78-1005 (KARLSRUHE)",
    doi = "10.1016/0550-3213(79)90102-0",
    journal = "Nucl. Phys. B",
    volume = "152",
    pages = "255--265",
    year = "1979"
}

@article{HT,
    author = "Howe, Paul S. and Tucker, R. W.",
    title = "{Scale Invariance in Superspace}",
    reportNumber = "CERN-TH-2524",
    doi = "10.1016/0370-2693(78)90327-1",
    journal = "Phys. Lett. B",
    volume = "80",
    pages = "138--140",
    year = "1978"
}

@inbook{Grimm1980,
    Address = {Boston, MA},
    Author = {Grimm, Richard},
    Booktitle = {Unification of the Fundamental Particle Interactions},
    Doi = {10.1007/978-1-4613-3171-1_27},
    Editor = {Ferrara, Sergio and Ellis, John and van Nieuwenhuizen, Peter},
    Isbn = {978-1-4613-3171-1},
    Pages = {509--523},
    Publisher = {Springer US},
    Title = {Solution of the Bianchi Identities in SU(2) Extended Superspace with Constraints},
    Url = {https://doi.org/10.1007/978-1-4613-3171-1_27},
    Year = {1980},
}

@book{DeWitt,
    author = "DeWitt, Bryce S.",
    title = "{Supermanifolds}",
    isbn = "978-1-139-24051-2, 978-0-521-42377-9",
    publisher = "Cambridge Univ. Press",
    address = "Cambridge, UK",
    series = "Cambridge Monographs on Mathematical Physics",
    month = "5",
    year = "2012"
}

@book{Castellani:1991et,
    author = "Castellani, L. and D'Auria, R. and Fre, P.",
    title = "{Supergravity and superstrings: A Geometric perspective. Vol. 1: Mathematical foundations}",
    year = "1991",
    publisher= "World Scientific Publishing Company"
}

@article{Claus:2000ka,
    author = "Claus, Piet and Rahmfeld, J. and Robins, Harlan and Tannenhauser, Jonathan and Zunger, Yonatan",
    title = "{Isometries in anti-de Sitter and conformal superspaces}",
    eprint = "hep-th/0007099",
    archivePrefix = "arXiv",
    reportNumber = "CALT-68-2284, CITUSC-00-037, SU-ITP-00-18",
    doi = "10.1088/1126-6708/2000/07/047",
    journal = "JHEP",
    volume = "07",
    pages = "047",
    year = "2000"
}

@article{Kallosh:1998qs,
    author = "Kallosh, Renata and Rajaraman, Arvind",
    title = "{Vacua of M theory and string theory}",
    eprint = "hep-th/9805041",
    archivePrefix = "arXiv",
    reportNumber = "SLAC-PUB-7820, SU-ITP-98-33",
    doi = "10.1103/PhysRevD.58.125003",
    journal = "Phys. Rev. D",
    volume = "58",
    pages = "125003",
    year = "1998"
}

@article{Kleppe,
    author = "Kleppe, A. F. and Wainwright, Chris",
    title = "{Super coset space geometry}",
    eprint = "hep-th/0610039",
    archivePrefix = "arXiv",
    reportNumber = "DAMTP-2006-83",
    doi = "10.1063/1.2735814",
    journal = "J. Math. Phys.",
    volume = "48",
    pages = "053511",
    year = "2007"
}

@article{K-lec,
    author = "Kuzenko, Sergei M.",
    editor = "Cadek, Martin",
    title = "{Lectures on nonlinear sigma-models in projective superspace}",
    eprint = "1004.0880",
    archivePrefix = "arXiv",
    primaryClass = "hep-th",
    doi = "10.1088/1751-8113/43/44/443001",
    journal = "J. Phys. A",
    volume = "43",
    pages = "443001",
    year = "2010"
}

@book{s-diff,
  title={Differential Geometry: Cartan's Generalization of Klein's Erlangen Program},
  author={Sharpe, R.W. and Chern, S.S.},
  isbn={9780387947327},
  lccn={lc96013757},
  series={Graduate Texts in Mathematics},
  url={https://books.google.com.au/books?id=d7w6AKaD_DkC},
  year={2000},
  publisher={Springer New York}
}

@book{f-diff,
  title={The Geometry of Physics: An Introduction},
  author={Frankel, T.},
  isbn={9780521539272},
  lccn={2003044030},
  url={https://books.google.com.au/books?id=DUnjs6nEn8wC},
  year={2004},
  publisher={Cambridge University Press}
}

@book{n-diff,
  title={Geometry, Topology and Physics, Second Edition},
  author={Nakahara, M.},
  isbn={9780750306065},
  lccn={2003282202},
  series={Graduate student series in physics},
  url={https://books.google.com.au/books?id=cH-XQB0Ex5wC},
  year={2003},
  publisher={Taylor \& Francis}
}

@article{A-twist,
    author = "Adamo, Tim",
    title = "{Lectures on twistor theory}",
    eprint = "1712.02196",
    archivePrefix = "arXiv",
    primaryClass = "hep-th",
    doi = "10.22323/1.323.0003",
    journal = "PoS",
    volume = "Modave2017",
    pages = "003",
    year = "2018"
}

@book{PR,
    author = "Penrose, Roger and Rindler, Wolfgang",
    title = "{Spinors and Space-Time}",
    isbn = "978-0-521-33707-6, 978-0-511-86766-8, 978-0-521-33707-6",
    publisher = "Cambridge Univ. Press",
    address = "Cambridge, UK",
    series = "Cambridge Monographs on Mathematical Physics",
    month = "4",
    year = "2011"
}

@book{Huggett:1986fs,
    author = "Huggett, S. A. and Tod, K. P.",
    title = "{An Introduction to Twistor Theory}",
    isbn = "978-0-521-45689-0",
    series = "London Mathematical Society Student Texts",
    publisher = "Cambridge Univ. Press",
    year = "1994"
}

@book{Ward:1990vs,
    author = "Ward, R. S. and Wells, R. O.",
    title = "{Twistor geometry and field theory}",
    doi = "10.1017/CBO9780511524493",
    isbn = "978-0-521-42268-0, 978-0-521-42268-0, 978-0-511-86977-8",
    publisher = "Cambridge University Press",
    series = "Cambridge Monographs on Mathematical Physics",
    month = "8",
    year = "1991"
}

@article{BF1,
    author = "Buchbinder, I. L. and Fedoruk, S. A.",
    title = "{Continuous spin superparticle model}",
    eprint = "2506.19709",
    archivePrefix = "arXiv",
    primaryClass = "hep-th",
    doi = "10.1016/j.physletb.2025.139810",
    journal = "Phys. Lett. B",
    volume = "868",
    pages = "139810",
    year = "2025"
}

@article{BF2,
    author = "Buchbinder, I. L. and Fedoruk, S. A.",
    title = "{Continuous spin superparticle in $4D$, ${\cal N}=1$ curved superspace}",
    eprint = "2507.18524",
    archivePrefix = "arXiv",
    primaryClass = "hep-th",
    month = "7",
    year = "2025"
}

@article{DZ,
    author = "Deser, Stanley and Zumino, B.",
    editor = "Salam, A. and Sezgin, E.",
    title = "{Consistent Supergravity}",
    reportNumber = "CERN-TH-2164",
    doi = "10.1016/0370-2693(76)90089-7",
    journal = "Phys. Lett. B",
    volume = "62",
    pages = "335",
    year = "1976"
}

@article{FvNF,
    author = "Freedman, Daniel Z. and van Nieuwenhuizen, P. and Ferrara, S.",
    title = "{Progress Toward a Theory of Supergravity}",
    reportNumber = "ITP-SB-76-23",
    doi = "10.1103/PhysRevD.13.3214",
    journal = "Phys. Rev. D",
    volume = "13",
    pages = "3214--3218",
    year = "1976"
}

@article{D-history,
    author = "Deser, S.",
    title = "{A brief history (and geography) of Supergravity: the first 3 weeks... and after}",
    eprint = "1704.05886",
    archivePrefix = "arXiv",
    primaryClass = "physics.hist-ph",
    reportNumber = "BRX TH 6813, CALT-TH 2017-18, BRX-TH-6813, CALT-TH-2017-18",
    doi = "10.1140/epjh/e2018-90005-3",
    journal = "Eur. Phys. J. H",
    volume = "43",
    number = "3",
    pages = "281--291",
    year = "2018"
}

@article{Ivanov:1979ft,
    author = "Ivanov, E. A. and Sorin, Alexander Savelievich",
    title = "{{Wess-Zumino} Model as Linear Sigma Model of Spontaneously Broken Conformal and Osp(1,4) Supersymmetries}",
    reportNumber = "JINR-E2-12331",
    journal = "Sov. J. Nucl. Phys.",
    volume = "30",
    pages = "440",
    year = "1979"
}

@book{W-grav,
    author = "Weinberg, Steven",
    title = "{Gravitation and Cosmology}: {Principles and Applications of the General Theory of Relativity}",
    isbn = "978-0-471-92567-5, 978-0-471-92567-5",
    publisher = "John Wiley and Sons",
    address = "New York",
    year = "1972"
}

@book{HE-grav,
    author = "Hawking, Stephen W. and Ellis, George F. R.",
    title = "{The Large Scale Structure of Space-Time}",
    isbn = "978-1-009-25316-1, 978-1-009-25315-4, 978-0-521-20016-5, 978-0-521-09906-6, 978-0-511-82630-6, 978-0-521-09906-6",
    publisher = "Cambridge University Press",
    series = "Cambridge Monographs on Mathematical Physics",
    month = "2",
    year = "2023"
}

@article{Wigner:1939cj,
    author = "Wigner, Eugene P.",
    editor = "Kim, Y. S. and Zachary, W. W.",
    title = "{On Unitary Representations of the Inhomogeneous Lorentz Group}",
    doi = "10.2307/1968551",
    journal = "Annals Math.",
    volume = "40",
    pages = "149--204",
    year = "1939"
}

@article{Cunningham:1910pxu,
    author = "Cunningham, E.",
    title = "{The Principle of Relativity in Electrodynamics and an Extension Thereof}",
    doi = "10.1112/plms/s2-8.1.77",
    journal = "Proc. Lond. Math. Soc. s",
    volume = "2-8",
    number = "1",
    pages = "77--98",
    year = "1910"
}

@article{Bateman:1910mvi,
    author = "Bateman, H.",
    title = "{The Transformation of the Electrodynamical Equations}",
    doi = "10.1112/plms/s2-8.1.223",
    journal = "Proc. Lond. Math. Soc. s",
    volume = "2-8",
    number = "1",
    pages = "223--264",
    year = "1910"
}

@article{Bateman:1909pyp,
    author = "Bateman, H.",
    title = "{The Conformal Transformations of a Space of Four Dimensions and Their Applications to Geometrical Optics}",
    doi = "10.1112/plms/s2-7.1.70",
    journal = "Proc. Lond. Math. Soc. s",
    volume = "2-7",
    number = "1",
    pages = "70--89",
    year = "1909"
}

@article{Maldacena:1997re,
    author = "Maldacena, Juan Martin",
    title = "{The Large $N$ limit of superconformal field theories and supergravity}",
    eprint = "hep-th/9711200",
    archivePrefix = "arXiv",
    reportNumber = "HUTP-97-A097, HUTP-98-A097",
    doi = "10.4310/ATMP.1998.v2.n2.a1",
    journal = "Adv. Theor. Math. Phys.",
    volume = "2",
    pages = "231--252",
    year = "1998"
}

@article{Gubser:1998bc,
    author = "Gubser, S. S. and Klebanov, Igor R. and Polyakov, Alexander M.",
    title = "{Gauge theory correlators from noncritical string theory}",
    eprint = "hep-th/9802109",
    archivePrefix = "arXiv",
    reportNumber = "PUPT-1767",
    doi = "10.1016/S0370-2693(98)00377-3",
    journal = "Phys. Lett. B",
    volume = "428",
    pages = "105--114",
    year = "1998"
}

@article{Witten:1998qj,
    author = "Witten, Edward",
    title = "{Anti de Sitter space and holography}",
    eprint = "hep-th/9802150",
    archivePrefix = "arXiv",
    reportNumber = "IASSNS-HEP-98-15",
    doi = "10.4310/ATMP.1998.v2.n2.a2",
    journal = "Adv. Theor. Math. Phys.",
    volume = "2",
    pages = "253--291",
    year = "1998"
}

@article{Aharony:1999ti,
    author = "Aharony, Ofer and Gubser, Steven S. and Maldacena, Juan Martin and Ooguri, Hirosi and Oz, Yaron",
    title = "{Large N field theories, string theory and gravity}",
    eprint = "hep-th/9905111",
    archivePrefix = "arXiv",
    reportNumber = "CERN-TH-99-122, HUTP-99-A027, LBNL-43113, RU-99-18, UCB-PTH-99-16, LBL-43113",
    doi = "10.1016/S0370-1573(99)00083-6",
    journal = "Phys. Rept.",
    volume = "323",
    pages = "183--386",
    year = "2000"
}

@book{Blumenhagen:2013fgp,
    author = {Blumenhagen, Ralph and L{\"u}st, Dieter and Theisen, Stefan},
    title = "{Basic concepts of string theory}",
    isbn = "978-3-642-29496-9",
    publisher = "Springer",
    address = "Heidelberg, Germany",
    series = "Theoretical and Mathematical Physics",
    year = "2013"
}

@book{Green:2012oqa,
    author = "Green, Michael B. and Schwarz, John H. and Witten, Edward",
    title = "{Superstring Theory Vol. 1}: {25th Anniversary Edition}",
    isbn = "978-1-139-53477-2, 978-1-107-02911-8",
    publisher = "Cambridge University Press",
    series = "Cambridge Monographs on Mathematical Physics",
    month = "11",
    year = "2012"
}

@book{Green:2012pqa,
    author = "Green, Michael B. and Schwarz, John H. and Witten, Edward",
    title = "{Superstring Theory Vol. 2}: {25th Anniversary Edition}",
    isbn = "978-1-139-53478-9, 978-1-107-02913-2",
    publisher = "Cambridge University Press",
    series = "Cambridge Monographs on Mathematical Physics",
    month = "11",
    year = "2012"
}

@book{Blagojevic:2013xpa,
    editor = "Blagojevi{\'c}, Milutin and Hehl, Friedrich W.",
    title = "{Gauge Theories of Gravitation}: {A Reader with Commentaries}",
    isbn = "978-1-84816-726-1",
    publisher = "World Scientific",
    address = "Singapore",
    year = "2013"
}

@article{Fcp1,
    author = "Fronsdal, C.",
    title = "{Elementary Particles in a Curved Space}",
    doi = "10.1103/RevModPhys.37.221",
    journal = "Rev. Mod. Phys.",
    volume = "37",
    pages = "221--224",
    year = "1965"
}

@article{Fcp2,
    author = "Fronsdal, C.",
    title = "{Elementary Particles in a Curved Space. ii}",
    doi = "10.1103/PhysRevD.10.589",
    journal = "Phys. Rev. D",
    volume = "10",
    pages = "589--598",
    year = "1974"
}

@article{Fcp3,
    author = "Fronsdal, C. and Haugen, R. B.",
    title = "{Elementary Particles in a Curved Space. 3}",
    doi = "10.1103/PhysRevD.12.3810",
    journal = "Phys. Rev. D",
    volume = "12",
    pages = "3810--3818",
    year = "1975"
}

@article{Fcp4,
    author = "Fronsdal, Christian",
    title = "{Elementary Particles in a Curved Space. 4. Massless Particles}",
    reportNumber = "UCLA/75/TEP/4",
    doi = "10.1103/PhysRevD.12.3819",
    journal = "Phys. Rev. D",
    volume = "12",
    pages = "3819",
    year = "1975"
}

@article{AdSqft1,
    author = "Avis, S. J. and Isham, C. J. and Storey, D.",
    title = "{Quantum Field Theory in anti-De Sitter Space-Time}",
    reportNumber = "ICTP-77-78-4",
    doi = "10.1103/PhysRevD.18.3565",
    journal = "Phys. Rev. D",
    volume = "18",
    pages = "3565",
    year = "1978"
}

@article{AdSqft2,
    author = "Dusedau, Dieter W. and Freedman, Daniel Z.",
    title = "{Lehmann Spectral Representation for Anti-de Sitter Quantum Field Theory}",
    reportNumber = "MIT-CTP-1290",
    doi = "10.1103/PhysRevD.33.389",
    journal = "Phys. Rev. D",
    volume = "33",
    pages = "389",
    year = "1986"
}

@article{deWit:1999ui,
    author = "de Wit, Bernard and Herger, Ivan",
    editor = "Kowalski-Glikman, J.",
    title = "{Anti-de Sitter supersymmetry}",
    eprint = "hep-th/9908005",
    archivePrefix = "arXiv",
    reportNumber = "THU-99-21",
    journal = "Lect. Notes Phys.",
    volume = "541",
    pages = "79--100",
    year = "2000"
}

@book{Birrell:1982ix,
    author = "Birrell, N. D. and Davies, P. C. W.",
    title = "{Quantum Fields in Curved Space}",
    isbn = "978-0-511-62263-2, 978-0-521-27858-4",
    publisher = "Cambridge University Press",
    address = "Cambridge, UK",
    series = "Cambridge Monographs on Mathematical Physics",
    year = "1982"
}

@article{Golfand:1971iw,
    author = "Golfand, Yu. A. and Likhtman, E. P.",
    editor = "Salam, A. and Sezgin, E.",
    title = "{Extension of the Algebra of Poincare Group Generators and Violation of p Invariance}",
    doi = "10.1142/9789814542340_0001",
    journal = "JETP Lett.",
    volume = "13",
    pages = "323--326",
    year = "1971"
}

@article{Wess:1974tw,
    author = "Wess, J. and Zumino, B.",
    editor = "Salam, A. and Sezgin, E.",
    title = "{Supergauge Transformations in Four-Dimensions}",
    doi = "10.1016/0550-3213(74)90355-1",
    journal = "Nucl. Phys. B",
    volume = "70",
    pages = "39--50",
    year = "1974"
}

@article{Wess:1973kz,
    author = "Wess, J. and Zumino, B.",
    title = "{A Lagrangian Model Invariant Under Supergauge Transformations}",
    reportNumber = "CERN-TH-1794",
    doi = "10.1016/0370-2693(74)90578-4",
    journal = "Phys. Lett. B",
    volume = "49",
    pages = "52",
    year = "1974"
}

@book{W-3,
    author = "Weinberg, Steven",
    title = "{The quantum theory of fields. Vol. 3: Supersymmetry}",
    isbn = "978-0-521-67055-5, 978-1-139-63263-8, 978-0-521-67055-5",
    publisher = "Cambridge University Press",
    month = "6",
    year = "2013"
}

@book{W-1,
    author = "Weinberg, Steven",
    title = "{The Quantum theory of fields. Vol. 1: Foundations}",
    isbn = "978-0-521-67053-1, 978-0-511-25204-4",
    publisher = "Cambridge University Press",
    month = "6",
    year = "2005"
}

@book{W-2,
    author = "Weinberg, Steven",
    title = "{The quantum theory of fields. Vol. 2: Modern applications}",
    isbn = "978-1-139-63247-8, 978-0-521-67054-8, 978-0-521-55002-4",
    publisher = "Cambridge University Press",
    month = "8",
    year = "2013"
}

@book{Fsg,
    author = "Freedman, Daniel Z. and Van Proeyen, Antoine",
    title = "{Supergravity}",
    isbn = "978-1-139-36806-3, 978-0-521-19401-3",
    publisher = "Cambridge Univ. Press",
    address = "Cambridge, UK",
    month = "5",
    year = "2012"
}

@book{Polchinski:1998rq,
    author = "Polchinski, J.",
    title = "{String theory. Vol. 1: An introduction to the bosonic string}",
    isbn = "978-0-511-25227-3, 978-0-521-67227-6, 978-0-521-63303-1",
    publisher = "Cambridge University Press",
    series = "Cambridge Monographs on Mathematical Physics",
    month = "12",
    year = "2007"
}

@book{Polchinski:1998rr,
    author = "Polchinski, J.",
    title = "{String theory. Vol. 2: Superstring theory and beyond}",
    isbn = "978-0-511-25228-0, 978-0-521-63304-8, 978-0-521-67228-3",
    publisher = "Cambridge University Press",
    series = "Cambridge Monographs on Mathematical Physics",
    month = "12",
    year = "2007"
}

@book{Castellani:1991eu,
    author = "Castellani, L. and D'Auria, R. and Fre, P.",
    title = "{Supergravity and superstrings: A Geometric perspective. Vol. 2: Supergravity}",
    year = "1991",
    publisher= "World Scientific Publishing Company"
}

@book{Castellani:1991ev,
    author = "Castellani, L. and D'Auria, R. and Fre, P.",
    title = "{Supergravity and superstrings: A Geometric perspective. Vol. 3: Superstrings}",
    year = "1991",
    publisher= "World Scientific Publishing Company"
}

@article{Salam:1974yz,
    author = "Salam, Abdus and Strathdee, J. A.",
    title = "{Supergauge Transformations}",
    reportNumber = "IC/74/11",
    doi = "10.1016/0550-3213(74)90537-9",
    journal = "Nucl. Phys. B",
    volume = "76",
    pages = "477--482",
    year = "1974"
}

@article{Galperin:1984av,
    author = "Galperin, A. and Ivanov, E. and Kalitzin, S. and Ogievetsky, V. and Sokatchev, E.",
    title = "{Unconstrained N=2 Matter, Yang-Mills and Supergravity Theories in Harmonic Superspace}",
    reportNumber = "IC-84-43",
    doi = "10.1088/0264-9381/1/5/004",
    journal = "Class. Quant. Grav.",
    volume = "1",
    pages = "469--498",
    year = "1984",
    note = "[Erratum: Class.Quant.Grav. 2, 127 (1985)]"
}

@book{Galperin:2001seg,
    author = "Galperin, A. S. and Ivanov, E. A. and Ogievetsky, V. I. and Sokatchev, E. S.",
    title = "{Harmonic superspace}",
    isbn = "978-0-511-53510-9, 978-0-521-02042-8, 978-0-521-80164-5, 978-0-511-03236-3",
    publisher = "Cambridge University Press",
    series = "Cambridge Monographs on Mathematical Physics",
    year = "2007"
}

@article{Karlhede:1984vr,
    author = "Karlhede, Anders and Lindstrom, Ulf and Rocek, Martin",
    title = "{Selfinteracting Tensor Multiplets in $N=2$ Superspace}",
    reportNumber = "ITP-SB-84-54",
    doi = "10.1016/0370-2693(84)90120-5",
    journal = "Phys. Lett. B",
    volume = "147",
    pages = "297--300",
    year = "1984"
}

@article{Gates:1984nk,
    author = "Gates, Jr., S. J. and Hull, C. M. and Rocek, M.",
    title = "{Twisted Multiplets and New Supersymmetric Nonlinear Sigma Models}",
    reportNumber = "ITP-SB-84-53",
    doi = "10.1016/0550-3213(84)90592-3",
    journal = "Nucl. Phys. B",
    volume = "248",
    pages = "157--186",
    year = "1984"
}

@article{LR1,
author = {Lindstrom, Ulf and Roček, Martin},
year = {1988},
month = {01},
pages = {21-29},
title = {New hyperkähler metrics and new supermultiplets},
volume = {115},
journal = {Communications in Mathematical Physics},
doi = {10.1007/BF01238851}
}

@article{Lindstrom:1989ne,
    author = "Lindstrom, U. and Rocek, M.",
    title = "{$N=2$ Superyang-mills Theory in Projective Superspace}",
    reportNumber = "ITP-SB-89-42",
    doi = "10.1007/BF02097052",
    journal = "Commun. Math. Phys.",
    volume = "128",
    pages = "191",
    year = "1990"
}

@article{Siegel:1977hn,
    author = "Siegel, Warren",
    title = "{A Polynomial Action for a Massive, Selfinteracting Chiral Superfield Coupled to Supergravity}",
    journal = "Harvard preprint HUTP-77/A077",
    month = "12",
    year = "1977"
}

@article{Freedman:1976uk,
    author = "Freedman, Daniel Z.",
    title = "{Supergravity with Axial Gauge Invariance}",
    reportNumber = "ITP-SB-76-50",
    doi = "10.1103/PhysRevD.15.1173",
    journal = "Phys. Rev. D",
    volume = "15",
    pages = "1173",
    year = "1977"
}

@article{deWit:1981yv,
    author = "de Wit, B. and Nicolai, H.",
    title = "{Extended Supergravity With Local SO(5) Invariance}",
    reportNumber = "CERN-TH-3044",
    doi = "10.1016/0550-3213(81)90107-3",
    journal = "Nucl. Phys. B",
    volume = "188",
    pages = "98--108",
    year = "1981"
}

@article{Gates:1982ct,
    author = "Gates, Jr., S. James and Zwiebach, Barton",
    title = "{Gauged $N=4$ Supergravity Theory With a New Scalar Potential}",
    reportNumber = "CALT-68-969",
    doi = "10.1016/0370-2693(83)90422-7",
    journal = "Phys. Lett. B",
    volume = "123",
    pages = "200--204",
    year = "1983"
}

@article{Pernici:1984xx,
    author = "Pernici, M. and Pilch, K. and van Nieuwenhuizen, P.",
    title = "{Gauged Maximally Extended Supergravity in Seven-dimensions}",
    reportNumber = "ITP-SB-84-31",
    doi = "10.1016/0370-2693(84)90813-X",
    journal = "Phys. Lett. B",
    volume = "143",
    pages = "103--107",
    year = "1984"
}

@article{Romans:1985tw,
    author = "Romans, L. J.",
    editor = "Salam, A. and Sezgin, E.",
    title = "{The F(4) Gauged Supergravity in Six-dimensions}",
    reportNumber = "NSF-ITP-85-137",
    doi = "10.1016/0550-3213(86)90517-1",
    journal = "Nucl. Phys. B",
    volume = "269",
    pages = "691",
    year = "1986"
}

@article{Allen:1985wd,
    author = "Allen, Bruce and Jacobson, Theodore",
    title = "{Vector Two Point Functions in Maximally Symmetric Spaces}",
    reportNumber = "UCSB-TH-4-1985",
    doi = "10.1007/BF01211169",
    journal = "Commun. Math. Phys.",
    volume = "103",
    pages = "669",
    year = "1986"
}

@article{Allen:1986qj,
    author = "Allen, Bruce and Lutken, C. A.",
    title = "{Spinor Two Point Functions in Maximally Symmetric Spaces}",
    reportNumber = "TUTP 86-2",
    doi = "10.1007/BF01454972",
    journal = "Commun. Math. Phys.",
    volume = "106",
    pages = "201",
    year = "1986"
}

@article{Atiyah:2017erd,
    author = "Atiyah, Michael and Dunajski, Maciej and Mason, Lionel",
    title = "{Twistor theory at fifty: from contour integrals to twistor strings}",
    eprint = "1704.07464",
    archivePrefix = "arXiv",
    primaryClass = "hep-th",
    reportNumber = "DAMTP-2017-17",
    doi = "10.1098/rspa.2017.0530",
    journal = "Proc. Roy. Soc. Lond. A",
    volume = "473",
    number = "2206",
    pages = "20170530",
    year = "2017"
}

@article{BENGTSSON198881,
title = {Particles, twistors and the division algebras},
journal = {Nuclear Physics B},
volume = {302},
number = {1},
pages = {81-103},
year = {1988},
issn = {0550-3213},
doi = {https://doi.org/10.1016/0550-3213(88)90667-0},
url = {https://www.sciencedirect.com/science/article/pii/0550321388906670},
author = {Ingemar Bengtsson and Martin Cederwall},
}

@article{ManinNi,
    Author = {Manin, Yu.  I. },
    Da = {1985/07/01},
    Doi = {10.1007/BF02105859},
    Id = {Manin1985},
    Isbn = {1573-8795},
    Journal = {Journal of Soviet Mathematics},
    Number = {2},
    Pages = {1927--1975},
    Title = {Holomorphic supergeometry and Yang-Mills superfields},
    Ty = {JOUR},
    Url = {https://doi.org/10.1007/BF02105859},
    Volume = {30},
    Year = {1985}}

@inbook{Kuzenko:2022skv,
    author = "Kuzenko, Sergei M. and Raptakis, Emmanouil S. N. and Tartaglino-Mazzucchelli, Gabriele",
    title = "{Superspace Approaches to $\mathscr {N} = \text{1}$ Supergravity}",
    eprint = "2210.17088",
    archivePrefix = "arXiv",
    primaryClass = "hep-th",
    doi = "10.1007/978-981-19-3079-9_40-1",
    year = "2023"
}

@inbook{Kuzenko:2022ajd,
    author = "Kuzenko, S. M. and Raptakis, E. S. N. and Tartaglino-Mazzucchelli, G.",
    title = "{Covariant Superspace Approaches to $\mathscr {N}=\text{2}$ Supergravity}",
    eprint = "2211.11162",
    archivePrefix = "arXiv",
    primaryClass = "hep-th",
    doi = "10.1007/978-981-19-3079-9_44-1",
    year = "2023"
}

@inproceedings{Rosly:1983ya,
    author = "Rosly, A. A.",
    title = "{Super Yang-Mills constraints as integrability conditions}",
    year = "1983",
    booktitle = "Proc. Int. Seminar on Group Theoretical Methods in Physics",
    address = "Zvenigrod, USSR, 1982",
    volume = "1",
    editor = "M A Markov",
    pages = "263"
}

@book{Schwartz:2014sze,
    author = "Schwartz, Matthew D.",
    title = "{Quantum Field Theory and the Standard Model}",
    isbn = "978-1-107-03473-0, 978-1-107-03473-0",
    publisher = "Cambridge University Press",
    month = "3",
    year = "2014"
}

@inbook{Sohnius:1976pa,
    address = "M{\"u}nchen",
    author = "Sohnius, Martin F.",
    booktitle = "{Quantum Theory and the Structures of Time and Space}",
    editor = "L Castell and M Drieschner and C F von Weizs{\"a}cker",
    pages = "241-252",
    publisher = "Carl Hanser Verlag",
    title = "{The Conformal Group in Superspace}",
    year = "1977"
}

@article{H2D,
doi = {10.1088/0264-9381/4/1/005},
url = {https://doi.org/10.1088/0264-9381/4/1/005},
year = {1987},
month = {jan},
publisher = {},
volume = {4},
number = {1},
pages = {11},
author = {P S Howe and G Papadopoulos},
title = {N=2, d=2 supergeometry},
journal = {Classical and Quantum Gravity},
}

@article{Grisaru:1994dm,
    author = "Grisaru, Marcus T. and Wehlau, Marcia E.",
    title = "{Prepotentials for (2,2) supergravity}",
    eprint = "hep-th/9409043",
    archivePrefix = "arXiv",
    reportNumber = "BRX-TH-360",
    doi = "10.1142/S0217751X95000358",
    journal = "Int. J. Mod. Phys. A",
    volume = "10",
    pages = "753--766",
    year = "1995"
}

@article{Grisaru:1995dr,
    author = "Grisaru, Marcus T. and Wehlau, M. E.",
    title = "{Superspace measures, invariant actions, and component projection formulae for (2,2) supergravity}",
    eprint = "hep-th/9508139",
    archivePrefix = "arXiv",
    reportNumber = "BRX-TH-378",
    doi = "10.1016/0550-3213(95)00529-3",
    journal = "Nucl. Phys. B",
    volume = "457",
    pages = "219--239",
    year = "1995"
}

@article{Gates:1995du,
    author = "Gates, Jr., S. J. and Grisaru, Marcus T. and Wehlau, M. E.",
    title = "{A Study of general 2-D, N=2 matter coupled to supergravity in superspace}",
    eprint = "hep-th/9509021",
    archivePrefix = "arXiv",
    reportNumber = "BRX-TH-379, UMDEPP-96-21",
    doi = "10.1016/0550-3213(95)00648-6",
    journal = "Nucl. Phys. B",
    volume = "460",
    pages = "579--614",
    year = "1996"
}

@article{TM2D,
    author = "Tartaglino-Mazzucchelli, Gabriele",
    title = "{2D N = (4,4) superspace supergravity and bi-projective superfields}",
    eprint = "0911.2546",
    archivePrefix = "arXiv",
    primaryClass = "hep-th",
    reportNumber = "UMD-PP-09-046",
    doi = "10.1007/JHEP04(2010)034",
    journal = "JHEP",
    volume = "04",
    pages = "034",
    year = "2010"
}

@article{Bellucci:2000yx,
    author = "Bellucci, S. and Ivanov, E.",
    title = "{N=(4,4), 2-D supergravity in SU(2) x SU(2) harmonic superspace}",
    eprint = "hep-th/0003154",
    archivePrefix = "arXiv",
    reportNumber = "LNF-00-009-P, JINR-E2-99-341",
    doi = "10.1016/S0550-3213(00)00347-3",
    journal = "Nucl. Phys. B",
    volume = "587",
    pages = "445--480",
    year = "2000"
}

@article{Evans:1986ada,
    author = "Evans, Mark and Ovrut, Burt A.",
    title = "{The World Sheet Supergravity of the Heterotic String}",
    reportNumber = "RU/86/149",
    doi = "10.1016/0370-2693(86)91527-3",
    journal = "Phys. Lett. B",
    volume = "171",
    pages = "177--181",
    year = "1986"
}

@article{Brooks:1986uh,
    author = "Brooks, R. and Muhammad, F. and Gates, S. J.",
    title = "{Unidexterous D=2 Supersymmetry in Superspace}",
    doi = "10.1016/0550-3213(86)90261-0",
    journal = "Nucl. Phys. B",
    volume = "268",
    pages = "599--620",
    year = "1986"
}

@article{GOVINDARAJAN1992251,
title = {The anomaly structure of (2,0) heterotic world-sheet supergravity with gauged R-invariance},
journal = {Nuclear Physics B},
volume = {385},
number = {1},
pages = {251-275},
year = {1992},
issn = {0550-3213},
doi = {https://doi.org/10.1016/0550-3213(92)90101-G},
url = {https://www.sciencedirect.com/science/article/pii/055032139290101G},
author = {Suresh Govindarajan and Burt A. Ovrut},
}

@article{Howe:1978ia,
    author = "Howe, Paul S.",
    title = "{Super Weyl Transformations in Two-Dimensions}",
    reportNumber = "Print-78-0538 (LANCASTER)",
    doi = "10.1088/0305-4470/12/3/015",
    journal = "J. Phys. A",
    volume = "12",
    pages = "393--402",
    year = "1979"
}

@article{Binder:2021euo,
    author = "Binder, Damon J. and Freedman, Daniel Z. and Pufu, Silviu S. and Zan, Bernardo",
    title = "{The holographic contributions to the sphere free energy}",
    eprint = "2107.12382",
    archivePrefix = "arXiv",
    primaryClass = "hep-th",
    reportNumber = "PUPT-2626, MIT-CTP/5317",
    doi = "10.1007/JHEP01(2022)171",
    journal = "JHEP",
    volume = "01",
    pages = "171",
    year = "2022"
}

@article{Poland:2018epd,
    author = "Poland, David and Rychkov, Slava and Vichi, Alessandro",
    title = "{The Conformal Bootstrap: Theory, Numerical Techniques, and Applications}",
    eprint = "1805.04405",
    archivePrefix = "arXiv",
    primaryClass = "hep-th",
    doi = "10.1103/RevModPhys.91.015002",
    journal = "Rev. Mod. Phys.",
    volume = "91",
    pages = "015002",
    year = "2019"
}

@phdthesis{Stone:2023ctp,
    author = "Stone, Benjamin J.",
    title = "{Correlation functions of conserved currents in (super)conformal field theory}",
    eprint = "2407.17384",
    archivePrefix = "arXiv",
    primaryClass = "hep-th",
    school = "Western Australia U.",
    year = "2023"
}

@article{Gazeau:2006gq,
    author = "Gazeau, Jean Pierre and Lachieze Rey, Marc",
    editor = "Bytsenko, Andrey A. and Dias, Sebastiao Alves and Helayel-Neto, Jose A. and Guimaraes, Maria E. Xavier",
    title = "{Quantum field theory in de Sitter space: A Survey of recent approaches}",
    eprint = "hep-th/0610296",
    archivePrefix = "arXiv",
    doi = "10.22323/1.031.0007",
    journal = "PoS",
    volume = "IC2006",
    pages = "007",
    year = "2006"
}

@article{Akhmedov:2013vka,
    author = "Akhmedov, E. T.",
    title = "{Lecture notes on interacting quantum fields in de Sitter space}",
    eprint = "1309.2557",
    archivePrefix = "arXiv",
    primaryClass = "hep-th",
    reportNumber = "ITEP-TH-32-13",
    doi = "10.1142/S0218271814300018",
    journal = "Int. J. Mod. Phys. D",
    volume = "23",
    pages = "1430001",
    year = "2014"
}

@article{Aizawa:2023lhx,
    author = "Aizawa, Naruhiko and Ito, Ren and Kuznetsova, Zhanna and Toppan, Francesco",
    title = "{New aspects of the Z2{\texttimes}Z2-graded 1D superspace: Induced strings and 2D relativistic models}",
    eprint = "2301.06089",
    archivePrefix = "arXiv",
    primaryClass = "hep-th",
    reportNumber = "CBPF-NF-004/22",
    doi = "10.1016/j.nuclphysb.2023.116202",
    journal = "Nucl. Phys. B",
    volume = "991",
    pages = "116202",
    year = "2023"
}

@article{Kuznetsova:2021byk,
    author = "Kuznetsova, Zhanna and Toppan, Francesco",
    title = "{Beyond the 10-fold Way: 13 Associative $ {\mathbb Z}_2\times {\mathbb Z}_2$-Graded Superdivision Algebras}",
    eprint = "2112.00840",
    archivePrefix = "arXiv",
    primaryClass = "math-ph",
    reportNumber = "Preprint CBPF-NF-004/21",
    doi = "10.1007/s00006-023-01263-1",
    journal = "Adv. Appl. Clifford Algebras",
    volume = "33",
    number = "2",
    pages = "24",
    year = "2023"
}

@article{Aizawa:2020ovt,
    author = "Aizawa, N. and Kuznetsova, Z. and Toppan, F.",
    title = "{${\mathbb Z}_2\times {\mathbb Z}_2$-graded mechanics: the classical theory}",
    eprint = "2003.06470",
    archivePrefix = "arXiv",
    primaryClass = "hep-th",
    reportNumber = "CBPF-NF-002/20",
    doi = "10.1140/epjc/s10052-020-8242-x",
    journal = "Eur. Phys. J. C",
    volume = "80",
    number = "7",
    pages = "668",
    year = "2020"
}

@article{Aizawa:2020xni,
    author = "Aizawa, N. and Kuznetsova, Z. and Toppan, F.",
    title = "{${\mathbb Z}_2\times {\mathbb Z}_2$-graded mechanics: The quantization}",
    eprint = "2005.10759",
    archivePrefix = "arXiv",
    primaryClass = "hep-th",
    reportNumber = "Preprint CBPF-NF-003/20",
    doi = "10.1016/j.nuclphysb.2021.115426",
    journal = "Nucl. Phys. B",
    volume = "967",
    pages = "115426",
    year = "2021"
}

@article{Toppan:2020rjz,
    author = "Toppan, Francesco",
    title = "{${\mathbb Z}_2\times {\mathbb Z}_2$-graded parastatistics in multiparticle quantum Hamiltonians}",
    eprint = "2008.11554",
    archivePrefix = "arXiv",
    primaryClass = "hep-th",
    reportNumber = "CBPF-NF-007/20",
    doi = "10.1088/1751-8121/abe2f2",
    journal = "J. Phys. A",
    volume = "54",
    number = "11",
    pages = "115203",
    year = "2021"
}

@article{Vasiliev:1985vtv,
    author = "Vasiliev, Mikhail A.",
    title = "{De Sitter Supergravity With Positive Cosmological Constant and Generalized Lie Superalgebras}",
    doi = "10.1088/0264-9381/2/5/007",
    journal = "Class. Quant. Grav.",
    volume = "2",
    pages = "645--659",
    year = "1985"
}

@article{Pilch:1984aw,
    author = "Pilch, K. and van Nieuwenhuizen, P. and Sohnius, M. F.",
    title = "{De Sitter Superalgebras and Supergravity}",
    reportNumber = "ITP-SB-84-46",
    doi = "10.1007/BF01211046",
    journal = "Commun. Math. Phys.",
    volume = "98",
    pages = "105",
    year = "1985"
}

@article{Bergshoeff:2015tra,
    author = "Bergshoeff, Eric A. and Freedman, Daniel Z. and Kallosh, Renata and Van Proeyen, Antoine",
    title = "{Pure de Sitter Supergravity}",
    eprint = "1507.08264",
    archivePrefix = "arXiv",
    primaryClass = "hep-th",
    doi = "10.1103/PhysRevD.93.069901",
    journal = "Phys. Rev. D",
    volume = "92",
    number = "8",
    pages = "085040",
    year = "2015",
    note = "[Erratum: Phys.Rev.D 93, 069901 (2016)]"
}

@inproceedings{Howe:1981ev,
    author = "Howe, Paul S.",
    title = "{Off-shell N=2 and N=4 Supergravity in Five-Dimensions}",
    booktitle = "{Nuffield Workshop on Quantum Structure of Space and Time}",
    reportNumber = "CERN-TH-3181",
    month = "10",
    year = "1981"
}

@inproceedings{deWit:2002vz,
    author = "de Wit, Bernard",
    title = "{Supergravity}",
    booktitle = "{Les Houches Summer School: Session 76: Euro Summer School on Unity of Fundamental Physics: Gravity, Gauge Theory and Strings}",
    eprint = "hep-th/0212245",
    archivePrefix = "arXiv",
    reportNumber = "ITP-UU-02-56, SPIN-02-35",
    pages = "1--135",
    month = "12",
    year = "2002"
}

@inbook{VanProeyen:2025ylh,
    author = "Van Proeyen, Antoine",
    title = "{Matter Couplings in Supergravity. The first 10 years}",
    eprint = "2503.12510",
    archivePrefix = "arXiv",
    primaryClass = "hep-th",
    month = "3",
    year = "2025"
}

@article{Coleman:1969sm,
    author = "Coleman, Sidney R. and Wess, J. and Zumino, Bruno",
    title = "{Structure of phenomenological Lagrangians. 1.}",
    doi = "10.1103/PhysRev.177.2239",
    journal = "Phys. Rev.",
    volume = "177",
    pages = "2239--2247",
    year = "1969"
}

@article{Callan:1969sn,
    author = "Callan, Jr., Curtis G. and Coleman, Sidney R. and Wess, J. and Zumino, Bruno",
    title = "{Structure of phenomenological Lagrangians. 2.}",
    doi = "10.1103/PhysRev.177.2247",
    journal = "Phys. Rev.",
    volume = "177",
    pages = "2247--2250",
    year = "1969"
}

@article{Volkov:1973vd,
    author = "Volkov, Dmitri V.",
    title = "{Phenomenological Lagrangians}",
    journal = "Fiz. Elem. Chast. Atom. Yadra",
    volume = "4",
    pages = "3--41",
    year = "1973"
}

@inproceedings{ogiev-non,
    author = "Ogievetsky, V. I.",
    title = "{Nonlinear realisations of internal and space-time symmetries}",
    booktitle = "{Proceeding of the 10th Karpacz Winter School of Theoretical Physics}",
    volume = "1",
    address = "Wroslaw",
    pages = "117-132",
    year = "1974"
}

@article{Isham:1969ci,
    author = "Isham, C. J.",
    title = "{A group-theoretic approach to chiral transformations}",
    doi = "10.1007/BF02755023",
    journal = "Nuovo Cim. A",
    volume = "59",
    pages = "356--376",
    year = "1969"
}

@article{Salam:1969rq,
    author = "Salam, Abdus and Strathdee, J. A.",
    title = "{Nonlinear realizations. 1: The Role of Goldstone bosons}",
    reportNumber = "IC-68-105",
    doi = "10.1103/PhysRev.184.1750",
    journal = "Phys. Rev.",
    volume = "184",
    pages = "1750--1759",
    year = "1969"
}

@article{Salam:1969bwb,
    author = "Salam, Abdus and Strathdee, J. A.",
    title = "{Nonlinear realizations. 2. Conformal symmetry}",
    reportNumber = "IC-68-107",
    doi = "10.1103/PhysRev.184.1760",
    journal = "Phys. Rev.",
    volume = "184",
    pages = "1760--1768",
    year = "1969"
}

@article{Isham:1970gz,
    author = "Isham, C. J. and Salam, Abdus and Strathdee, J. A.",
    title = "{Spontaneous breakdown of conformal symmetry}",
    reportNumber = "IC-70-3",
    doi = "10.1016/0370-2693(70)90177-2",
    journal = "Phys. Lett. B",
    volume = "31",
    pages = "300--302",
    year = "1970"
}

@article{Isham:1971dv,
    author = "Isham, C. J. and Salam, Abdus and Strathdee, J. A.",
    title = "{Nonlinear realizations of space-time symmetries. Scalar and tensor gravity}",
    reportNumber = "IC-70-5",
    doi = "10.1016/0003-4916(71)90269-7",
    journal = "Annals Phys.",
    volume = "62",
    pages = "98--119",
    year = "1971"
}

@article{Gargett:2025xcg,
    author = "Gargett, Timothy and Samsonov, Igor",
    title = "{Analytic action principle in $\mathcal{N}=2$ AdS$_4$ harmonic superspace}",
    eprint = "2510.08905",
    archivePrefix = "arXiv",
    primaryClass = "hep-th",
    month = "10",
    year = "2025"
}

@article{Ivanov:2025jdp,
    author = "Ivanov, Evgeny and Zaigraev, Nikita",
    title = "{$\mathcal{N}=2$ AdS hypermultiplets in harmonic superspace}",
    eprint = "2509.01406",
    archivePrefix = "arXiv",
    primaryClass = "hep-th",
    doi = "10.1016/j.physletb.2025.139964",
    journal = "Phys. Lett. B",
    volume = "871",
    pages = "139964",
    year = "2025"
}

@article{Craig:2022eqo,
    author = "Craig, Nathaniel",
    title = "{Naturalness: past, present, and future}",
    eprint = "2205.05708",
    archivePrefix = "arXiv",
    primaryClass = "hep-ph",
    doi = "10.1140/epjc/s10052-023-11928-7",
    journal = "Eur. Phys. J. C",
    volume = "83",
    number = "9",
    pages = "825",
    year = "2023"
}

@article{Uvarov:2007vs,
    author = "Uvarov, D. V.",
    title = "{Supertwistor formulation for higher dimensional superstrings}",
    eprint = "hep-th/0703051",
    archivePrefix = "arXiv",
    doi = "10.1088/0264-9381/24/22/004",
    journal = "Class. Quant. Grav.",
    volume = "24",
    pages = "5383--5400",
    year = "2007"
}

@article{Boffo:2024lwd,
    author = "Boffo, Eugenia and Grassi, Pietro Antonio and Hulik, Ondrej and Sachs, Ivo",
    title = "{On superparticles and their partition functions}",
    eprint = "2402.09868",
    archivePrefix = "arXiv",
    primaryClass = "hep-th",
    doi = "10.1007/JHEP05(2025)140",
    journal = "JHEP",
    volume = "05",
    pages = "140",
    year = "2025"
}

@article{Bertan:2018afl,
    author = "Bertan, Igor and Sachs, Ivo and Skvortsov, Evgeny D.",
    title = "{Quantum $\phi^4$ Theory in AdS${}_4$ and its CFT Dual}",
    eprint = "1810.00907",
    archivePrefix = "arXiv",
    primaryClass = "hep-th",
    reportNumber = "LMU-ASC 63/18",
    doi = "10.1007/JHEP02(2019)099",
    journal = "JHEP",
    volume = "02",
    pages = "099",
    year = "2019"
}

@article{Park:1997bq,
    author = "Park, Jeong-Hyuck",
    title = "{N=1 superconformal symmetry in four-dimensions}",
    eprint = "hep-th/9703191",
    archivePrefix = "arXiv",
    reportNumber = "DAMTP-97-27",
    doi = "10.1142/S0217751X98000755",
    journal = "Int. J. Mod. Phys. A",
    volume = "13",
    pages = "1743--1772",
    year = "1998"
}

@article{Park:1998nra,
    author = "Park, Jeong-Hyuck",
    title = "{Superconformal symmetry in six-dimensions and its reduction to four-dimensions}",
    eprint = "hep-th/9807186",
    archivePrefix = "arXiv",
    reportNumber = "DAMTP-98-88",
    doi = "10.1016/S0550-3213(98)00720-2",
    journal = "Nucl. Phys. B",
    volume = "539",
    pages = "599--642",
    year = "1999"
}

@article{Park:1999cw,
    author = "Park, Jeong-Hyuck",
    title = "{Superconformal symmetry in three-dimensions}",
    eprint = "hep-th/9910199",
    archivePrefix = "arXiv",
    reportNumber = "KIAS-99-101, KIAS-99101",
    doi = "10.1063/1.1290056",
    journal = "J. Math. Phys.",
    volume = "41",
    pages = "7129--7161",
    year = "2000"
}

@article{Lee:2004jx,
    author = "Lee, Sangmin and Park, Jeong-Hyuck",
    title = "{Noncentral extension of the AdS(5) x S**5 superalgebra: Supermultiplet of brane charges}",
    eprint = "hep-th/0404051",
    archivePrefix = "arXiv",
    reportNumber = "CERN-PH-TH-2004-064",
    doi = "10.1088/1126-6708/2004/06/038",
    journal = "JHEP",
    volume = "06",
    pages = "038",
    year = "2004"
}

@article{Lindstrom:2009afn,
    author = "Lindstrom, Ulf and Rocek, Martin",
    title = "{Properties of hyperkahler manifolds and their twistor spaces}",
    eprint = "0807.1366",
    archivePrefix = "arXiv",
    primaryClass = "hep-th",
    reportNumber = "UUITP-03-06, YITP-SB-06-07",
    doi = "10.1007/s00220-009-0923-0",
    journal = "Commun. Math. Phys.",
    volume = "293",
    pages = "257--278",
    year = "2010"
}

@article{CSG2,
      author        = "Kaku, M and Townsend, P K and van Nieuwenhuizen, P",
      title         = "{Properties of conformal supergravity}",
      reportNumber  = "ITP-SB-78-2",
      journal       = "Phys. Rev. D",
      volume        = "17",
      pages         = "3179-3187",
      year          = "1978",            
      url           = "https://cds.cern.ch/record/133560",
      doi           = "10.1103/PhysRevD.17.3179",
}

@article{Gunaydin:1984qu,
    author = "Gunaydin, M. and Romans, L. J. and Warner, N. P.",
    title = "{Gauged N=8 Supergravity in Five-Dimensions}",
    reportNumber = "CALT-68-1217",
    doi = "10.1016/0370-2693(85)90361-2",
    journal = "Phys. Lett. B",
    volume = "154",
    pages = "268--274",
    year = "1985"
}

@article{SGLBST1,
    author = "Siegel, W.",
    title = "{Green-Schwarz formulation of selfdual superstring}",
    eprint = "hep-th/9210008",
    archivePrefix = "arXiv",
    reportNumber = "ITP-SB-92-53",
    doi = "10.1103/PhysRevD.47.2512",
    journal = "Phys. Rev. D",
    volume = "47",
    pages = "2512--2516",
    year = "1993"
}

@article{SGLBST2,
    author = "Siegel, W.",
    title = "{Supermulti - instantons in conformal chiral superspace}",
    eprint = "hep-th/9412011",
    archivePrefix = "arXiv",
    reportNumber = "ITP-SB-94-66",
    doi = "10.1103/PhysRevD.52.1042",
    journal = "Phys. Rev. D",
    volume = "52",
    pages = "1042--1050",
    year = "1995"
}

@article{Mezincescu:2013nta,
    author = "Mezincescu, Luca and Routh, Alasdair J. and Townsend, Paul K.",
    title = "{Supertwistors and massive particles}",
    eprint = "1312.2768",
    archivePrefix = "arXiv",
    primaryClass = "hep-th",
    reportNumber = "DAMTP-2013-64",
    doi = "10.1016/j.aop.2014.04.007",
    journal = "Annals Phys.",
    volume = "346",
    pages = "66--90",
    year = "2014"
}

@article{ramondstring,
    author = "Ramond, Pierre",
    title = "{Dual Theory for Free Fermions}",
    reportNumber = "FERMILAB-PUB-70-008-T, FERMILAB-PUB-70-008-THY, NAL-THY-8",
    doi = "10.1103/PhysRevD.3.2415",
    journal = "Phys. Rev. D",
    volume = "3",
    pages = "2415--2418",
    year = "1971"
}

@article{lindcomponents,
    author = "Rocek, M. and Lindstrom, U.",
    title = "{Components of Superspace}",
    doi = "10.1016/0370-2693(78)90226-5",
    journal = "Phys. Lett. B",
    volume = "79",
    pages = "217--218",
    year = "1978"
}

@article{Siegelsugra,
    author = "Siegel, Warren and Gates, Jr., S. James",
    title = "{Superfield Supergravity}",
    reportNumber = "HUTP-78/A019",
    doi = "10.1016/0550-3213(79)90416-4",
    journal = "Nucl. Phys. B",
    volume = "147",
    pages = "77--104",
    year = "1979"
}

@article{Howe:1996yn,
    author = "Howe, Paul S. and Sezgin, E.",
    title = "{D = 11, p = 5}",
    eprint = "hep-th/9611008",
    archivePrefix = "arXiv",
    reportNumber = "KCL-TH-96-16, CTP-TAMU-55-96",
    doi = "10.1016/S0370-2693(96)01672-3",
    journal = "Phys. Lett. B",
    volume = "394",
    pages = "62--66",
    year = "1997"
}

@article{Achucarro:1986uwr,
    author = "Achucarro, A. and Townsend, P. K.",
    editor = "Salam, A. and Sezgin, E.",
    title = "{A Chern-Simons Action for Three-Dimensional anti-De Sitter Supergravity Theories}",
    reportNumber = "Print-87-0078 (CAMBRIDGE)",
    doi = "10.1016/0370-2693(86)90140-1",
    journal = "Phys. Lett. B",
    volume = "180",
    pages = "89",
    year = "1986"
}

@article{Achucarro:1989gm,
    author = "Achucarro, A. and Townsend, P. K.",
    title = "{Extended Supergravities in $d$ = (2+1) as {Chern-Simons} Theories}",
    reportNumber = "PRINT-89-0627 (CAMBRIDGE)",
    doi = "10.1016/0370-2693(89)90423-1",
    journal = "Phys. Lett. B",
    volume = "229",
    pages = "383--387",
    year = "1989"
}

@article{Capozziello:2025qmh,
    author = "Capozziello, Salvatore and Chaudhary, Himanshu and Harko, Tiberiu and Mustafa, Ghulam",
    title = "{Is dark energy dynamical in the DESI era? A critical review}",
    eprint = "2512.10585",
    archivePrefix = "arXiv",
    primaryClass = "astro-ph.CO",
    doi = "10.1016/j.dark.2025.102196",
    journal = "Phys. Dark Univ.",
    volume = "51",
    pages = "102196",
    year = "2026"
}

@article{DESI:2025wyn,
    author = "Gu, Gan and others",
    collaboration = "DESI",
    title = "{Dynamical dark energy in light of the DESI DR2 baryonic acoustic oscillations measurements}",
    eprint = "2504.06118",
    archivePrefix = "arXiv",
    primaryClass = "astro-ph.CO",
    reportNumber = "FERMILAB-PUB-25-0235-PPD",
    doi = "10.1038/s41550-025-02669-6",
    journal = "Nature Astron.",
    volume = "9",
    number = "12",
    pages = "1879--1889",
    year = "2025",
    note = "[Erratum: Nature Astron. 9, 1898 (2025)]"
}

@article{Breitenlohner:1982jf,
    author = "Breitenlohner, Peter and Freedman, Daniel Z.",
    title = "{Stability in Gauged Extended Supergravity}",
    reportNumber = "Print-82-0500 (MIT)",
    doi = "10.1016/0003-4916(82)90116-6",
    journal = "Annals Phys.",
    volume = "144",
    pages = "249",
    year = "1982"
}

@article{Dirac:1936fq,
    author = "Dirac, Paul A. M.",
    title = "{Wave equations in conformal space}",
    doi = "10.2307/1968455",
    journal = "Annals Math.",
    volume = "37",
    pages = "429--442",
    year = "1936"
}
\bibliographystyle{chapters/bibliography/JHEP}
\end{footnotesize}

\end{document}